\documentclass[longauth]{aa}

\usepackage{txfonts,graphicx,xfrac}
\usepackage{hyperref}
\hypersetup{colorlinks,linkcolor=red,citecolor=blue,urlcolor=blue}
\usepackage{array}

\usepackage{natbib}
\bibpunct{(}{)}{;}{a}{}{,}

\newcommand{\hi}{H\,\textsc{i}}
\newcommand{\lii}{Li\,\textsc{i}}
\newcommand{\beii}{Be\,\textsc{ii}}
\newcommand{\ci}{C\,\textsc{i}}
\newcommand{\oi}{O\,\textsc{i}}
\newcommand{\nai}{Na\,\textsc{i}}
\newcommand{\mgi}{Mg\,\textsc{i}}
\newcommand{\mgii}{Mg\,\textsc{ii}}
\newcommand{\ali}{Al\,\textsc{i}}
\newcommand{\sii}{Si\,\textsc{i}}
\newcommand{\siii}{Si\,\textsc{ii}}
\newcommand{\phosphori}{P\,\textsc{i}}
\newcommand{\si}{S\,\textsc{i}}
\newcommand{\ki}{K\,\textsc{i}}
\newcommand{\cai}{Ca\,\textsc{i}}
\newcommand{\caii}{Ca\,\textsc{ii}}
\newcommand{\sci}{Sc\,\textsc{i}}
\newcommand{\scii}{Sc\,\textsc{ii}}
\newcommand{\tii}{Ti\,\textsc{i}}
\newcommand{\tiii}{Ti\,\textsc{ii}}
\newcommand{\vi}{V\,\textsc{i}}
\newcommand{\vii}{V\,\textsc{ii}}
\newcommand{\cri}{Cr\,\textsc{i}}
\newcommand{\crii}{Cr\,\textsc{ii}}
\newcommand{\mni}{Mn\,\textsc{i}}
\newcommand{\mnii}{Mn\,\textsc{ii}}
\newcommand{\fei}{Fe\,\textsc{i}}
\newcommand{\feii}{Fe\,\textsc{ii}}
\newcommand{\coi}{Co\,\textsc{i}}
\newcommand{\coii}{Co\,\textsc{ii}}
\newcommand{\nii}{Ni\,\textsc{i}}
\newcommand{\niii}{Ni\,\textsc{ii}}
\newcommand{\cui}{Cu\,\textsc{i}}
\newcommand{\zni}{Zn\,\textsc{i}}
\newcommand{\sri}{Sr\,\textsc{i}}
\newcommand{\srii}{Sr\,\textsc{ii}}
\newcommand{\asi}{As\,\textsc{i}}
\newcommand{\rbi}{Rb\,\textsc{i}}
\newcommand{\yi}{Y\,\textsc{i}}
\newcommand{\yii}{Y\,\textsc{ii}}
\newcommand{\zri}{Zr\,\textsc{i}}
\newcommand{\zrii}{Zr\,\textsc{ii}}
\newcommand{\nbi}{Nb\,\textsc{i}}
\newcommand{\moi}{Mo\,\textsc{i}}
\newcommand{\rui}{Ru\,\textsc{i}}
\newcommand{\rhi}{Rh\,\textsc{i}}
\newcommand{\pdi}{Pd\,\textsc{i}}
\newcommand{\bai}{Ba\,\textsc{i}}
\newcommand{\baii}{Ba\,\textsc{ii}}
\newcommand{\lai}{La\,\textsc{i}}
\newcommand{\laii}{La\,\textsc{ii}}
\newcommand{\cei}{Ce\,\textsc{i}}
\newcommand{\ceii}{Ce\,\textsc{ii}}
\newcommand{\pri}{Pr\,\textsc{i}}
\newcommand{\prii}{Pr\,\textsc{ii}}
\newcommand{\priii}{Pr\,\textsc{iii}}
\newcommand{\ndi}{Nd\,\textsc{i}}
\newcommand{\ndii}{Nd\,\textsc{ii}}
\newcommand{\ndiii}{Nd\,\textsc{iii}}
\newcommand{\smi}{Sm\,\textsc{i}}
\newcommand{\smii}{Sm\,\textsc{ii}}
\newcommand{\eui}{Eu\,\textsc{i}}
\newcommand{\euii}{Eu\,\textsc{ii}}
\newcommand{\gdi}{Gd\,\textsc{i}}
\newcommand{\gdii}{Gd\,\textsc{ii}}
\newcommand{\tbii}{Tb\,\textsc{ii}}
\newcommand{\dyi}{Dy\,\textsc{i}}
\newcommand{\dyii}{Dy\,\textsc{ii}}
\newcommand{\hoi}{Ho\,\textsc{i}}
\newcommand{\eri}{Er\,\textsc{i}}
\newcommand{\erii}{Er\,\textsc{ii}}
\newcommand{\tmi}{Tm\,\textsc{i}}
\newcommand{\tmii}{Tm\,\textsc{ii}}
\newcommand{\ybi}{Yb\,\textsc{i}}
\newcommand{\ybii}{Yb\,\textsc{ii}}
\newcommand{\lui}{Lu\,\textsc{i}}
\newcommand{\luii}{Lu\,\textsc{ii}}
\newcommand{\hfi}{Hf\,\textsc{i}}
\newcommand{\hfii}{Hf\,\textsc{ii}}
\newcommand{\tai}{Ta\,\textsc{i}}
\newcommand{\rei}{Re\,\textsc{i}}
\newcommand{\osi}{Os\,\textsc{i}}
\newcommand{\iri}{Ir\,\textsc{i}}
\newcommand{\pti}{Pt\,\textsc{i}}
\newcommand{\thi}{Th\,\textsc{i}}
\newcommand{\thii}{Th\,\textsc{ii}}

\newcommand{\gfflag}{\emph{gf\_flag}}
\newcommand{\synflag}{\emph{synflag}}

\newcommand{\Yes}{\emph{Y}}
\newcommand{\Un}{\emph{U}}
\newcommand{\No}{\emph{N}}

\newcommand{\Teff}{$T_{\rm eff}$}
\newcommand{\logg}{$\log g$}
\newcommand{\ulogg}{cm~s$^{-2}$}
\newcommand{\FeH}{[Fe/H]}

\newcommand{\betGem}{$\beta$~Gem}

\newcommand{\betVir}{$\beta$~Vir}
\newcommand{\delEri}{$\delta$~Eri}
\newcommand{\epsEri}{$\epsilon$~Eri}

\newcommand{\epsVir}{$\epsilon$~Vir}

\newcommand{\muLeo}{$\mu$~Leo}

\newcommand{\textIntro}{Appendices~\ref{sect:aux} to \ref{sect:broadrec}}

\newcommand{\refSectGf}{Appendix~\ref{sect:gf}}
\newcommand{\refSectGfT}{Appendix~\ref{sect:gf}}
\newcommand{\refSectGfIntroT}{Appendices~\ref{sect:gf} to \ref{sect:background}}

\newcommand{\refSectHcoll}{Appendices~\ref{sect:theory} to \ref{sect:broadrec}}
\newcommand{\refSectHcollT}{Appendices~\ref{sect:theory} to \ref{sect:broadrec}}

\newcommand{\refSectFe}{Appendix~\ref{sect:Fe}}
\newcommand{\refSectFeT}{Appendix~\ref{sect:Fe}}
\newcommand{\refSectHydrogen}{Appendix~\ref{sect:hydrogen}}
\newcommand{\refSectLi}{Appendix~\ref{sect:Li}}
\newcommand{\refSectCa}{Appendix~\ref{sect:Ca}}
\newcommand{\refSectBa}{Appendix~\ref{sect:Ba}}

\newcommand{\refSectPresel}{Sect.~\ref{sect:preselection}}

\newcommand{\refSectGfflag}{\ref{sect:gfflag}}

\newcommand{\refSectSpectra}{\ref{sect:spectra}}
\newcommand{\refSectSpectraT}{\ref{sect:spectra}}

\newcommand{\refSectImpact}{\ref{sect:impact}}
\newcommand{\refSectDataneeds}{\ref{sect:dataneeds}}

\newcommand{\refTabFlagstats}{\ref{tab:flagstats}}
\newcommand{\refTabFlagstatsT}{\ref{tab:flagstats}}
\newcommand{\refTabIsotopes}{\ref{tab:isotopes}}

\begin{document}
\raggedbottom

\title{Atomic data for the Gaia-ESO Survey\thanks{The atomic and molecular data are only available in electronic form at the CDS via anonymous ftp to cdsarc.u-strasbg.fr (130.79.128.5) or via \url{http://cdsweb.u-strasbg.fr/cgi-bin/qcat?J/A+A/...}}}

\titlerunning{Atomic data for the Gaia-ESO Survey}

\author{
Ulrike Heiter\inst{\ref{inst:Uppsala}}
\and
Karin Lind\inst{\ref{inst:Stockholm}}
\and
Maria Bergemann\inst{\ref{inst:Heidelberg}}
\and
Martin Asplund\inst{\ref{inst:Australia}}
\and
\v{S}arunas Mikolaitis\inst{\ref{inst:Vilnius}}
\and
Paul~S. Barklem\inst{\ref{inst:Uppsala}}
\and
Thomas Masseron\inst{\ref{inst:Tenerife1},\ref{inst:Tenerife2}}
\and
Patrick de Laverny\inst{\ref{inst:Nice}}
\and
Laura Magrini\inst{\ref{inst:Arcetri}}
\and
Bengt Edvardsson\inst{\ref{inst:Uppsala}}
\and
Henrik J\"onsson\inst{\ref{inst:Malmoe},\ref{inst:Lund}}
\and
Juliet~C. Pickering\inst{\ref{inst:London}}
\and
Nils Ryde\inst{\ref{inst:Lund}}
\and
Amelia Bayo Ar\'an\inst{\ref{inst:Valparaiso_1},\ref{inst:Valparaiso_2}}
\and
Thomas Bensby\inst{\ref{inst:Lund}}
\and
Andrew~R. Casey\inst{\ref{inst:Cambridge},\ref{inst:Monash},\ref{inst:Astro3D}}
\and
Sofia Feltzing\inst{\ref{inst:Lund}}
\and
Paula Jofr\'e\inst{\ref{inst:DiegoPortales}}
\and
Andreas~J. Korn\inst{\ref{inst:Uppsala}}
\and
Elena Pancino\inst{\ref{inst:Arcetri},\ref{inst:Roma}}
\and
Francesco Damiani\inst{\ref{inst:Palermo}}
\and
Alessandro Lanzafame\inst{\ref{inst:Catania}}
\and
Carmela Lardo\inst{\ref{inst:Lausanne}}
\and
Lorenzo Monaco\inst{\ref{inst:AndresBello}}
\and
Lorenzo Morbidelli\inst{\ref{inst:Arcetri}}
\and
Rodolfo Smiljanic\inst{\ref{inst:Warsaw}}
\and
Clare Worley\inst{\ref{inst:Cambridge}}
\and
Simone Zaggia\inst{\ref{inst:Padova}}
\and
Sofia Randich\inst{\ref{inst:Arcetri}}
\and
Gerard~F. Gilmore\inst{\ref{inst:Cambridge}}
}

\institute{
Observational Astrophysics, Department of Physics and Astronomy, Uppsala University, Box 516, 751 20 Uppsala, Sweden,
\email{ulrike.heiter@physics.uu.se}
\label{inst:Uppsala}
\and
Department of Astronomy, Stockholm University, AlbaNova, Roslagstullbacken 21, 106 91 Stockholm, Sweden
\label{inst:Stockholm}
\and
Max-Planck Institut f\"{u}r Astronomie (MPIA), K\"{o}nigstuhl 17, 69117 Heidelberg, Germany
\label{inst:Heidelberg}
\and
Research School of Astronomy \& Astrophysics, Australian National University, Cotter Road, Weston Creek, ACT 2611, Australia
\label{inst:Australia}
\and
Institute of Theoretical Physics and Astronomy, Vilnius University, Saul\.{e}tekio av. 3, 10257 Vilnius, Lithuania
\label{inst:Vilnius}
\and
Instituto de Astrof\'{\i}sica de Canarias, 38205 La Laguna, Tenerife, Spain
\label{inst:Tenerife1}
\and
Universidad de La Laguna, Dept. Astrof\'{\i}sica, 38206 La Laguna, Tenerife, Spain
\label{inst:Tenerife2}
\and
Universit\'e C\^ote d'Azur, Observatoire de la C\^ote d'Azur, CNRS, Laboratoire Lagrange, Blvd de l'Observatoire, 06304 Nice, France
\label{inst:Nice}
\and
INAF - Osservatorio Astrofisico di Arcetri, Largo E. Fermi 5, 50125, Florence, Italy
\label{inst:Arcetri}
\and
Materials Science and Applied Mathematics, Malm\"o University, 205 06 Malm\"o, Sweden
\label{inst:Malmoe}
\and
Lund Observatory, Department of Astronomy and Theoretical Physics, Lund University, Box 43, 221 00 Lund, Sweden
\label{inst:Lund}
\and
Blackett Laboratory, Imperial College London, London SW7 2BW, United Kingdom
\label{inst:London}
\and
Instituto de F\'isica y Astronom\'ia, Facultad de Ciencias, Universidad de Valpara\'iso, Av. Gran Breta\~na 1111, 5030 Casilla, Valpara\'iso, Chile
\label{inst:Valparaiso_1}
\and
N\'ucleo Milenio de Formaci\'on Planetaria - NPF, Universidad de Valpara\'iso, Av. Gran Breta\~na 1111, Valpara\'iso, Chile
\label{inst:Valparaiso_2}
\and
Institute of Astronomy, University of Cambridge, Madingley Road, Cambridge CB3 0HA, United Kingdom
\label{inst:Cambridge}
\and
School of Physics \& Astronomy, Monash University, Wellington Rd, Clayton 3800, Victoria, Australia
\label{inst:Monash}
\and
Center of Excellence for Astrophysics in Three Dimensions (ASTRO-3D), Australia
\label{inst:Astro3D}
\and
N\'ucleo de Astronom\'{i}a, Facultad de Ingenier\'{i}a, Universidad Diego Portales, Av. Ej\'ercito 441, Santiago, Chile
\label{inst:DiegoPortales}
\and
Space Science Data Center - Agenzia Spaziale Italiana, via del Politecnico, s.n.c., 00133, Roma, Italy
\label{inst:Roma}
\and
INAF - Osservatorio Astronomico di Palermo, Piazza del Parlamento 1, 90134, Palermo, Italy
\label{inst:Palermo}
\and
Dipartimento di Fisica e Astronomia, Sezione Astrofisica, Universit\'{a} di Catania, via S. Sofia 78, 95123, Catania, Italy
\label{inst:Catania}
\and
Laboratoire d'astrophysique, Ecole Polytechnique F\'ed\'erale de Lausanne (EPFL), Observatoire de Sauverny, 1290 Versoix, Switzerland
\label{inst:Lausanne}
\and
Departamento de Ciencias Fisicas, Universidad Andres Bello, Fernandez Concha 700, Las Condes, Santiago, Chile
\label{inst:AndresBello}
\and
Nicolaus Copernicus Astronomical Center, Polish Academy of Sciences, ul. Bartycka 18, 00-716, Warsaw, Poland
\label{inst:Warsaw}
\and
INAF - Padova Observatory, Vicolo dell'Osservatorio 5, 35122 Padova, Italy
\label{inst:Padova}
}

\authorrunning{Heiter et al.}

\date{Received 11 July 2019 / Accepted 14 October 2020}

\abstract
% Context
{
We describe the atomic and molecular data that were used for the abundance analyses of FGK-type stars carried out within the Gaia-ESO Public Spectroscopic Survey in the years 2012 to 2019. The Gaia-ESO survey is one among several current and future stellar spectroscopic surveys producing abundances for Milky-Way stars on an industrial scale.
}
% Aims
{
We present an unprecedented effort to create a homogeneous common line list, which was used by several abundance analysis groups using different radiative transfer codes to calculate synthetic spectra and equivalent widths.
The atomic data are accompanied by quality indicators and detailed references to the sources. The atomic and molecular data are made publicly available in electronic form.
}
% Methods
{
In general experimental transition probabilities were preferred but theoretical values were also used. Astrophysical $gf$-values were avoided due to the model-dependence of such a procedure.
For elements whose lines are significantly affected by hyperfine structure or isotopic splitting
a concerted effort has been made to collate the necessary data for the individual line components.
Synthetic stellar spectra calculated for the Sun and Arcturus were used to assess the blending properties of the lines.
We also performed a detailed investigation of available data for line broadening due to collisions with neutral hydrogen atoms.
}
% Results
{
Among a subset of over 1300 lines of 35 elements in the wavelength ranges from 475~nm to 685~nm and from 850~nm to 895~nm we identified about 200 lines of 24 species which have accurate $gf$-values and are free of blends in the spectra of the Sun and Arcturus.
For the broadening due to collisions with neutral hydrogen, we recommend data based on Anstee-Barklem-O'Mara theory, where possible. We recommend to avoid lines of neutral species for which these are not available. Theoretical broadening data by R.L.~Kurucz should be used for \scii, \tiii, and \yii\ lines, and for ionised rare-earth species the Uns\"old approximation with an enhancement factor of 1.5 for the line width can be used.
}
% Conclusions
{
The line list has proven to be a useful tool for abundance determinations based on the spectra obtained within the Gaia-ESO Survey, as well as other spectroscopic projects.
Accuracies below 0.2~dex are regularly achieved, where part of the uncertainties are due to differences in the employed analysis methods.
Desirable improvements in atomic data were identified for a number of species, most importantly \ali, \si, and \crii, but also \nai, \sii, \caii, and \nii.
}


\maketitle

%----------------------------------------------------------------------------
% Introduction
%----------------------------------------------------------------------------

\section{Introduction}
\label{sect:intro}

The Gaia-ESO Public Spectroscopic Survey (GES, \citealt{2012Msngr.147...25G,2013Msngr.154...47R}) % Gilmore et al., Randich et al.
started in 2011 and was completed in 2018. High quality spectra were obtained for of order $10^5$ stars in the Milky Way, predominantly of F-, G-, and K-type. Spectra were obtained with the FLAMES multi fiber facility using the UVES and GIRAFFE spectrographs at the Very Large Telescope at the Paranal Observatory, Chile. Several wavelength regions were covered, mostly 480--680~nm and 850--900~nm at different resolutions ($R=\lambda/\Delta\lambda=$47\,000 and $\sim$20\,000). As part of the GES, these spectra are analysed to determine radial velocities and stellar parameters: effective temperatures, surface gravities, and elemental abundances. To perform this analysis, corresponding atomic and molecular data are needed for the observed spectral regions, as well as to determine which lines are suitable for analysis across the range of spectral types. Within the GES the spectra are analysed independently by different groups, which has the advantage of providing checks on the analysis. However, in order to limit the potential sources of differences between the analyses, it was decided to use standard input data as far as possible, in particular a common list of lines with corresponding atomic and molecular data \citep[see][]{Pancino2017}. A common line list is also desirable from the point of view that accurate atomic and molecular data are needed to ensure the best possible results from the survey. The critical compilation of such data is a significant and time-consuming task often requiring specialised knowledge and tracking of recent advances in the field \citep[e.g.,][]{2016A&ARv..24....9B}. % Barklem

The purpose of this paper is to describe the common line list and the process of building it for the GES. The line list is also expected to be useful for other surveys with overlapping spectral regions, and for the analysis of F-, G-, and K-type stars in general.
The paper is organised as follows.
Section~\ref{sect:data} describes the general procedure for critical selection and assessment of the lines for analysis, and their corresponding data.
In Sect.~\ref{sect:molecules}, the molecular data are described.
In Sect.~\ref{sect:discussion}, the line list is discussed in general terms, future data needs are identified, and access to the data is described (Sect.~\ref{sect:access}).

Specific aspects of the line data are described in more detail
in \textIntro. 
In \refSectGfIntroT\ 
%In \citet{Heiter_etal_2020_S1}
%In Sect.~\ref{sect:gf}
the sources for the fundamental properties of the atomic lines (energies, wavelengths, oscillator strengths, etc.), and the quality assessment for the lines in stellar spectra, are discussed element by element.
In \refSectHcollT\ 
%In \citet{Heiter_etal_2020_S2}
%In Sect.~\ref{sect:H-broadening},
the methods for calculation of data for collisional broadening by neutral hydrogen, of particular relevance for these spectral types, are described and discussed.

%----------------------------------------------------------------------------
% Overview / Background
%----------------------------------------------------------------------------

\section{Data selection and assessment}
\label{sect:data}

%%%%%%%%%%%%%%%%%%%%%%%%%%%%%%%%%%%%%%%%%%%%%%%%%%%%%%%%%%%%%%%%%%%%%%%%%%%%%

\subsection{Preselected line list}
\label{sect:preselection}

The task of defining a standard line list for stellar parameter determination and abundance analysis was started in May 2012. We asked all of the groups participating in the Gaia-ESO analysis of FGK-type stars to provide us with their ``favorite'' line lists.
In this way we collected lists of lines contained within the standard UVES-580 and GIRAFFE HR21 settings that the groups found particularly appropriate for spectral analysis of the FGK-type stars among Gaia-ESO targets.
Note that the UVES-580 setting covers the wavelength ranges of the GIRAFFE HR10 and HR15N settings that were also employed in the GES for cool stars.
This resulted in a unique set of 1341 lines for 35 elements (44 species comprising neutral and singly-ionised atoms), which we refer to as the \emph{preselected line list}.
The total number of lines included for each species can be seen in Table~\ref{tab:flagstats}.
These numbers do not include hyperfine-structure or isotopic components (cf. Sect.~\ref{sect:hfs}). When those are included the total number of transitions is 2631.
The lines cover the wavelength ranges from 475~nm to 685~nm and from 850~nm to 895~nm. The \caii\ NIR triplet line at 849.8~nm is included as well.
The first range is somewhat larger than the nominal wavelength region of the UVES-580 setting (480~nm to 680~nm) in order to account for objects with large radial velocities.
The second range corresponds to the GIRAFFE HR21 setting.

\begin{table*}
\renewcommand{\tabcolsep}{0.9mm}
\caption{List of species included in the preselected line list.}
\label{tab:flagstats}
\centering
\begin{tabular}{rlrrrrrrrrrrrrrrr}
\hline\hline\noalign{\smallskip}
 $Z$ & Species & $N_{\rm tot}$ & ABO$_{\rm tot}$ & Y/Y & ABO & Y/U & ABO & U/Y & ABO & U/U & ABO &  N/Y & N/U &  Y/N & U/N & N/N \\
\noalign{\smallskip}\hline\noalign{\smallskip}
  3 & \lii &   2 &   2 &   0 &   0 &   2 &   2 &   0 &   0 &   0 &   0 &   0 &   0 &   0 &   0 &   0 \\
  6 & \ci &   4 &   2 &   1 &   1 &   3 &   1 &   0 &   0 &   0 &   0 &   0 &   0 &   0 &   0 &   0 \\
  8 & \oi &   4 &   1 &   0 &   0 &   3 &   1 &   0 &   0 &   0 &   0 &   0 &   0 &   1 &   0 &   0 \\
 11 & \nai &  11 &   4 &   2 &   2 &   0 &   0 &   3 &   1 &   1 &   0 &   0 &   0 &   0 &   5 &   0 \\
 12 & \mgi &  12 &   5 &   3 &   3 &   1 &   1 &   6 &   1 &   2 &   0 &   0 &   0 &   0 &   0 &   0 \\
 13 & \ali &   5 &   2 &   0 &   0 &   0 &   0 &   1 &   0 &   4 &   2 &   0 &   0 &   0 &   0 &   0 \\
 14 & \sii &  45 &   3 &   3 &   2 &   7 &   1 &   2 &   0 &   2 &   0 &  10 &  11 &   2 &   2 &   6 \\
 14 & \siii &   2 &   0 &   0 &   0 &   2 &   0 &   0 &   0 &   0 &   0 &   0 &   0 &   0 &   0 &   0 \\
 16 & \si &   8 &   1 &   0 &   0 &   0 &   0 &   0 &   0 &   8 &   1 &   0 &   0 &   0 &   0 &   0 \\
 20 & \cai &  31 &  26 &  12 &  11 &   8 &   5 &   1 &   1 &   0 &   0 &   0 &   0 &   8 &   2 &   0 \\
 20 & \caii &   8 &   3 &   2 &   2 &   1 &   1 &   0 &   0 &   4 &   0 &   0 &   0 &   0 &   1 &   0 \\
 21 & \sci &   7 &   3 &   2 &   2 &   3 &   1 &   0 &   0 &   0 &   0 &   0 &   0 &   2 &   0 &   0 \\
 21 & \scii &  17 &   0 &   2 &   0 &   4 &   0 &   0 &   0 &   0 &   0 &   4 &   1 &   6 &   0 &   0 \\
 22 & \tii & 105 &  84 &  23 &  22 &  47 &  40 &   0 &   0 &   0 &   0 &   3 &   3 &  24 &   0 &   5 \\
 22 & \tiii &  23 &   0 &   1 &   0 &  10 &   0 &   0 &   0 &   0 &   0 &   1 &   3 &   7 &   0 &   1 \\
 23 & \vi &  49 &  46 &  15 &  15 &  16 &  16 &   0 &   0 &   0 &   0 &   3 &   6 &   5 &   0 &   4 \\
 23 & \vii &   3 &   0 &   0 &   0 &   0 &   0 &   0 &   0 &   0 &   0 &   0 &   1 &   0 &   0 &   2 \\
 24 & \cri &  64 &  52 &   9 &   9 &  15 &  15 &   0 &   0 &   0 &   0 &   9 &  11 &  11 &   0 &   9 \\
 24 & \crii &  24 &  24 &   0 &   0 &   1 &   1 &   0 &   0 &   0 &   0 &   1 &   7 &   5 &   0 &  10 \\
 25 & \mni &  27 &  12 &   3 &   2 &  15 &   9 &   0 &   0 &   0 &   0 &   3 &   4 &   1 &   0 &   1 \\
 26 & \fei & 545 & 442 &  83 &  79 & 120 & 104 &  44 &  43 &  84 &  79 &  33 &  75 &  45 &  34 &  27 \\
 26 & \feii &  42 &  42 &   6 &   6 &   7 &   7 &   6 &   6 &   7 &   7 &   0 &   0 &   6 &  10 &   0 \\
 27 & \coi &  34 &  22 &   8 &   8 &  14 &  12 &   0 &   0 &   0 &   0 &   1 &   8 &   2 &   0 &   1 \\
 28 & \nii &  99 &  86 &  16 &  14 &  14 &  11 &   1 &   1 &   2 &   2 &  19 &  31 &   3 &   1 &  12 \\
 29 & \cui &   6 &   0 &   1 &   0 &   4 &   0 &   0 &   0 &   0 &   0 &   0 &   0 &   1 &   0 &   0 \\
 30 & \zni &   2 &   1 &   0 &   0 &   2 &   1 &   0 &   0 &   0 &   0 &   0 &   0 &   0 &   0 &   0 \\
 38 & \sri &  10 &   0 &   0 &   0 &   1 &   0 &   0 &   0 &   1 &   0 &   0 &   0 &   4 &   3 &   1 \\
 39 & \yi &   6 &   0 &   0 &   0 &   0 &   0 &   0 &   0 &   0 &   0 &   0 &   2 &   0 &   0 &   4 \\
 39 & \yii &  19 &   0 &   4 &   0 &   7 &   0 &   0 &   0 &   0 &   0 &   0 &   0 &   8 &   0 &   0 \\
 40 & \zri &  13 &   3 &   5 &   3 &   8 &   0 &   0 &   0 &   0 &   0 &   0 &   0 &   0 &   0 &   0 \\
 40 & \zrii &   3 &   0 &   0 &   0 &   1 &   0 &   0 &   0 &   0 &   0 &   0 &   0 &   1 &   1 &   0 \\
 41 & \nbi &   8 &   0 &   0 &   0 &   0 &   0 &   0 &   0 &   0 &   0 &   0 &   0 &   8 &   0 &   0 \\
 42 & \moi &   6 &   1 &   2 &   1 &   3 &   0 &   0 &   0 &   0 &   0 &   0 &   0 &   1 &   0 &   0 \\
 44 & \rui &   1 &   0 &   0 &   0 &   1 &   0 &   0 &   0 &   0 &   0 &   0 &   0 &   0 &   0 &   0 \\
 56 & \baii &   4 &   4 &   1 &   1 &   2 &   2 &   0 &   0 &   0 &   0 &   0 &   0 &   1 &   0 &   0 \\
 57 & \laii &   6 &   0 &   0 &   0 &   4 &   0 &   1 &   0 &   0 &   0 &   0 &   0 &   1 &   0 &   0 \\
 58 & \ceii &  12 &   0 &   2 &   0 &   4 &   0 &   0 &   0 &   3 &   0 &   0 &   0 &   2 &   1 &   0 \\
 59 & \prii &   7 &   0 &   1 &   0 &   1 &   0 &   0 &   0 &   0 &   0 &   0 &   0 &   5 &   0 &   0 \\
 60 & \ndii &  53 &   0 &   0 &   0 &  18 &   0 &   2 &   0 &   8 &   0 &   0 &   0 &  17 &   8 &   0 \\
 62 & \smii &   5 &   0 &   0 &   0 &   4 &   0 &   0 &   0 &   0 &   0 &   0 &   0 &   1 &   0 &   0 \\
 63 & \euii &   5 &   0 &   0 &   0 &   4 &   0 &   0 &   0 &   0 &   0 &   0 &   0 &   1 &   0 &   0 \\
 64 & \gdii &   1 &   0 &   0 &   0 &   1 &   0 &   0 &   0 &   0 &   0 &   0 &   0 &   0 &   0 &   0 \\
 66 & \dyii &   1 &   0 &   0 &   0 &   1 &   0 &   0 &   0 &   0 &   0 &   0 &   0 &   0 &   0 &   0 \\

\noalign{\smallskip}\hline\hline
\end{tabular}
\tablefoot{
The two preselected hydrogen lines, H$\alpha$ and H$\beta$, were omitted from the table. $N_{\rm tot}$ is the total number of transitions (hyperfine-structure or isotopic components are not included in the count). ABO$_{\rm tot}$ is the total number of transitions with line broadening data from the Anstee-Barklem-O'Mara theory (ABO data,
see \refSectHcoll).
%see \citealt{Heiter_etal_2020_S2}).
%see Sect.~\ref{sect:H-broadening}).
The remaining columns give the number of lines with the respective combinations of quality flags (\gfflag/\synflag, see Sects.~\ref{sect:gfflag} and \ref{sect:synflag}) and the corresponding subset of lines with ABO data.
}
\end{table*}

Note that there are a few lines with fine structure components (levels with different values of the total angular momentum quantum number $J$) that have wavelengths within $\sim$0.01~\AA\ from each other.
The individual transitions are thus not resolved in stellar spectra.
To simplify for equivalent-width based analysis methods, we merged these transitions and added their $gf$-values. The lines in question are listed in Table~\ref{tab:fine} in the appendix.

Our objective was to select the best available atomic data for the preselected lines, and to provide critical assessments of the atomic data quality and the blending properties of these lines in selected benchmark stars.
The intended primary use of this information was the line selection for a homogeneous abundance analysis within the GES, with the best possible accuracy, but it should be useful for other spectroscopic projects as well.
To summarise and communicate the information about the quality of the transition probabilities and the blending properties, we developed an easy-to-use flagging system.
For each of these two aspects each line was assigned one of three possible flags for recommended use: \Yes, for ``yes, we recommend to use this line'', \No, for ``not recommended'', or \Un, for ``undecided''.
The general approach for data compilation and assessment is described in Sects.~\ref{sect:gfflag} and \ref{sect:synflag}, while an in-depth description on an element-by-element basis can be found
in \refSectGfT.
%in \citet{Heiter_etal_2020_S1}.
%in Sect.~\ref{sect:gf}.

In summary, we highly recommend lines flagged with \Yes/\Yes\ for $gf$-value quality/blending property, while we strongly advise against using lines flagged with \No/\No. Lines with other combinations need to be examined and decided upon on a case-by-case basis.
Note that we have been more restrictive with \Yes-flags for elements with more spectral lines, such as Fe, compared to, e.g., O, with very few lines.

%%%%%%%%%%%%%%%%%%%%%%%%%%%%%%%%%%%%%%%%%%%%%%%%%%%%%%%%%%%%%%%%%%%%%%%%%%%%%

\subsection{Data compilation and quality assessment for transition probabilities}
\label{sect:gfflag}

We set out to assemble the best possible transition probabilities for all lines, using all literature at hand. In brief, we generally gave highest priority to laboratory measurements, followed by advanced quantum mechanical calculations such as those provided by the Opacity Project\footnote{\url{http://cdsweb.u-strasbg.fr/topbase/topbase.html}} and the MCHF project\footnote{\url{http://nlte.nist.gov/MCHF/}}. When data were not found in such sources, we used the semi-empirical calculations by R.L.~Kurucz\footnote{\url{http://kurucz.harvard.edu}}.
Measurements of astrophysical nature have not been derived or included at this point.
Transition probabilities derived by fitting synthetic to observed stellar spectra are inherently associated with the specific reference object(s) and models used, and would not be applicable to all targets and analysis groups.
An exception was made for a few atomic lines located in the vicinity of the \lii\ 670.8~nm line, for which astrophysical $gf$-values were derived
(see \refSectLi).
%(see Sect.~\sectLi in \citealt{Heiter_etal_2020_S1}).
%(see Sect.~\ref{sect:Li}).

Transition probabilities are given in the form of $\log gf$, and the flags for their quality (hereafter \gfflag) were assigned according to the following general scheme:
\begin{description}
\item[\Yes:] data which are considered highly accurate, or which were the most accurate ones available for the element under consideration at the time of compilation,
\item[\Un:] data for which the quality is not decided,
\item[\No:] data which are considered to have low accuracy.
\end{description}

The assignment of the flags for different elements is described in the respective sub-sections of \refSectGfT.
The starting point for the list of references was the literature sources used in the series of articles on the elemental composition of the Sun by \citet{2015A&A...573A..25S,2015A&A...573A..26S} and \citet{2015A&A...573A..27G}. % Scott et al. 2015a, 2015b and Grevesse et al. 2015
This was complemented by further sources as needed.
For the quality assessment we were guided by the uncertainties given by atomic data producers for the life-times, branching fractions, etc. measured in the laboratory.
Our intention was to make a homogeneous selection of sources for each element. To this effect, lines with data from one and the same source were assigned the same \gfflag, with the exception of a few Fe lines (see discussion in \refSectFe).
The number of lines to which the different \gfflag s were assigned for each species is indicated in Table~\ref{tab:flagstats}.
We emphasise that the only purpose of the flags was to provide a qualitative guideline for usage within the Gaia-ESO survey. They were used to help decide on the usage of a line in case of doubt.
However, for future applications the flags should be carefully re-evaluated and replaced by the user's personal assessment of data quality.

%%%%%%%%%%%%%%%%%%%%%%%%%%%%%%%%%%%%%%%%%%%%%%%%%%%%%%%%%%%%%%%%%%%%%%%%%%%%%

\subsection{Reference spectra for illustration of spectral lines}
\label{sect:spectra}

In this section and
in \refSectGfT\ 
%in \citet{Heiter_etal_2020_S1}
we use both calculated and observed spectra of benchmark stars to illustrate the behaviour of spectral lines associated with atomic properties. Here we give some information on these spectra, including stellar parameters, other assumptions, and references to spectroscopic data.

With the optimised set of transition probabilities we computed synthetic spectra of all preselected lines for the parameters of the Sun and Arcturus at a spectral resolution of $R=\lambda/\Delta\lambda=$47\,000, which is roughly the resolution of the UVES spectra obtained in the GES.
For the Sun we used (\Teff\ [K], \logg\ [\ulogg])=(5777, 4.44), similar to the recommended values of (5771, 4.4380) given in \citet{2015A&A...582A..49H}, see also \citet{2016AJ....152...41P}, % Heiter et al. 2015, Prsa et al. 2016
microturbulence=1~kms$^{-1}$, rotational broadening with $\varv \sin i$=2~kms$^{-1}$, macroturbulence=2~kms$^{-1}$, and abundances from \citet{2007SSRv..130..105G}\footnote{These are the reference abundances that are used throughout the GES analyses. They are also the abundances which are used in the MARCS model atmosphere grid used by GES. Therefore they are not the most recent set of abundances published for the Sun.}. % Grevesse et al.
For Arcturus we used (\Teff\ [K], \logg\ [\ulogg], \FeH, [$\alpha$/Fe])=(4286, 1.6, $-0.52$~dex, +0.24~dex), where $\alpha$-elements are those with even atomic numbers from 8 to 22, from \citet{2015A&A...582A..49H} and \citet{2014A&A...564A.133J,2015A&A...582A..81J}, % Heiter et al., Jofre et al.
microturbulence=1.71~kms$^{-1}$, $\varv \sin i$=1~kms$^{-1}$, and macroturbulence=4.5~kms$^{-1}$.

The synthesis was performed with the radiative transfer code SME \citep{Valenti1996,Piskunov2017} based on interpolated MARCS atmospheric models \citep{Gust:08}, which are the same as those employed in the Gaia-ESO analysis. % Gustafsson et al.
The observational data for the Sun and Arcturus are the Kitt Peak Fourier Transform Spectrometer (FTS) solar and Arcturus flux atlases \citep{1984sfat.book.....K, 2000vnia.book.....H}, % Kurucz et al. and Hinkle et al.
degraded to a spectral resolution of 47\,000.

\begin{figure*}
   \begin{center}
      \resizebox{0.63\hsize}{!}{\includegraphics{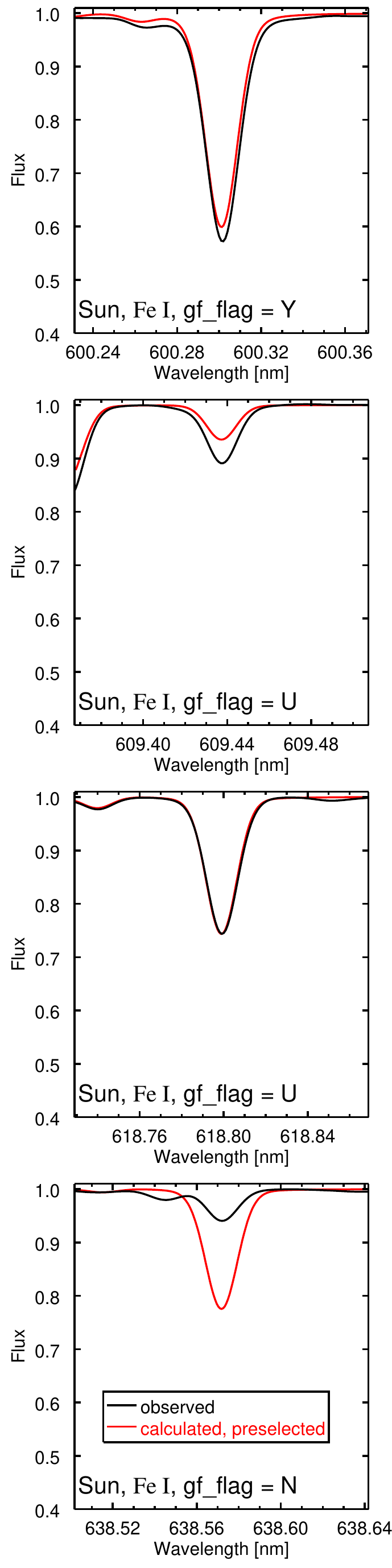}\includegraphics{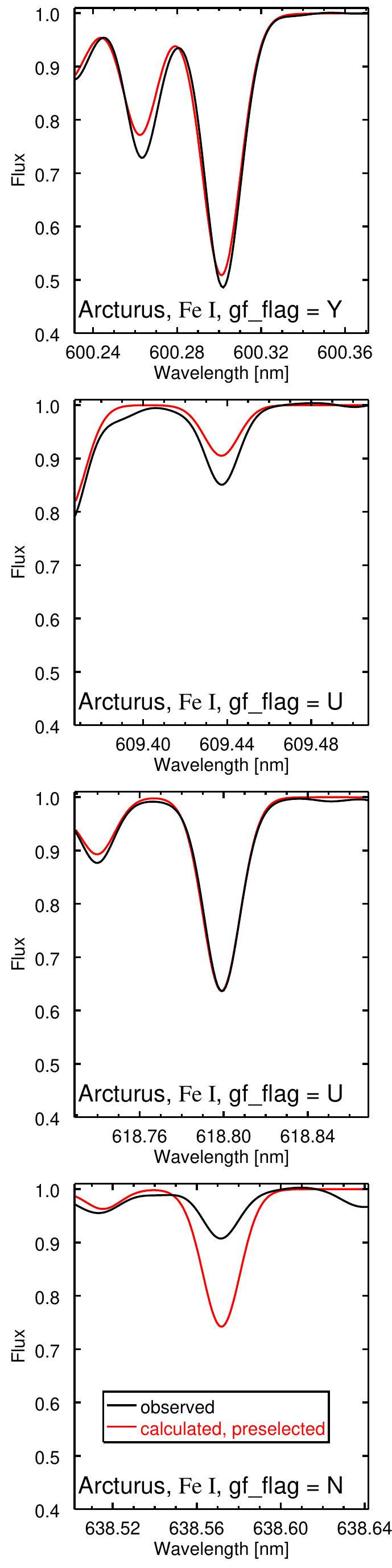}}
   \end{center}
   \caption{Comparison of observed and calculated line profiles around four of the preselected \fei\ lines with different \gfflag\ assignments for the Sun (left) and Arcturus (right). Black lines: observations, red lines: calculations including preselected spectral lines only. All of these lines are flagged with \Yes\ with respect to their blending properties. Note that the \gfflag\ assignments were based purely on the type of the $gf$-value sources. The performance of the lines in syntheses of the Sun and Arcturus were not taken into account. These are shown here for illustrative purposes only.
   }
   \label{fig:Fe1_profiles_gfflag}
\end{figure*}

Figure~\ref{fig:Fe1_profiles_gfflag} shows typical example line profiles for four largely unblended \fei\ lines with different \gfflag\ assignments.
See \refSectFeT\ 
%See Sect.~\sectFe in \citet{Heiter_etal_2020_S1}
%See Sect.~\ref{sect:Fe}
for details on the \fei\ data.
The figure illustrates the indicative and statistical nature of the flags.
The observed and synthetic spectra agree for both stars in the case of the line with the \Yes\ flag (top row in Fig.~\ref{fig:Fe1_profiles_gfflag}).
For \gfflag=\Un\ or \No\ the synthetic profiles often deviate from the observed ones to different degrees.
For the unblended \fei\ lines this is the case for about 40\% of the \gfflag=\Un\ (e.g., second row in Fig.~\ref{fig:Fe1_profiles_gfflag}) and 60\% of the \gfflag=\No\ lines (e.g., bottom row in Fig.~\ref{fig:Fe1_profiles_gfflag}), while the remaining lines provide a good fit between observed and synthetic spectra (e.g., third row in Fig.~\ref{fig:Fe1_profiles_gfflag}).
Examples for some of the other elements are given
in \refSectGfT.
%in \citet{Heiter_etal_2020_S1}.
%
We point out that the \gfflag\ assignments were based purely on the type of the $gf$-value sources. The performance of the lines in syntheses of the Sun and Arcturus were not taken into account. These are described here for illustrative purposes only.

% Profile plots for selected Fe 1 lines (ABO)

In addition we use observations of a subset of the Gaia FGK benchmark stars \citep{2015A&A...582A..49H,2014A&A...564A.133J}, including the solar twin 18~Sco and the Arcturus-like star HD~107328. Their spectra were taken from the library of \citet{2014A&A...566A..98B}\footnote{\url{http://www.blancocuaresma.com/s/benchmarkstars/}, version 2016-05-29}. They were normalised to the continuum and convolved to $R=$47\,000.
The stellar parameters are given in Table~\ref{tab:GBS}.

\begin{table}
\caption{Stellar parameters for selected Gaia FGK benchmark stars from \citet{2015A&A...582A..49H} and \citet{2014A&A...564A.133J}.}
\label{tab:GBS}
\centering
\begin{tabular}{rrrl}
\hline\hline\noalign{\smallskip}
\Teff\ [K] & \logg\ [\ulogg] & \FeH\ [dex] & Name \\
\noalign{\smallskip}\hline\noalign{\smallskip}
4374 & 4.63 & $-$0.33 & 61~Cyg~A    \\
4474 & 2.51 & $ $0.25 & \muLeo     \\
4496 & 2.09 & $-$0.33 & HD~107328  \\
4587 & 1.61 & $-$2.64 & HD~122563  \\
4858 & 2.90 & $ $0.13 & \betGem    \\
4954 & 3.76 & $ $0.06 & \delEri    \\
4983 & 2.77 & $ $0.15 & \epsVir    \\
5076 & 4.61 & $-$0.09 & \epsEri    \\
5810 & 4.44 & $ $0.03 & 18~Sco     \\
5868 & 4.27 & $-$0.86 & HD~22879   \\
6083 & 4.10 & $ $0.24 & \betVir    \\
6356 & 4.06 & $-$2.03 & HD~84937   \\
6554 & 4.00 & $ $0.01 & Procyon   \\
6635 & 4.20 & $-$0.41 & HD~49933   \\
\noalign{\smallskip}\hline\hline
\end{tabular}
\end{table}

To illustrate the effect on abundance determination when using the quality flags for line selection we computed line abundances for four elements, which have a sufficient number of lines for a statistical analysis, for the Sun and three other benchmark stars (Arcturus, the metal-poor dwarf star HD~22879, and 61~Cyg~A).
Equivalent widths were measured with DAOSPEC \citep{2008PASP..120.1332S,2010ascl.soft11002S} from the spectra used for calibration within the GES, at a spectral resolution of $R=$47\,000. Abundances were determined from these using the MOOG code \citep{1973PhDT.......180S}.
The results for the species \sii, \cri, \fei, and \nii\ are presented and discussed in the respective subsections
in \refSectGfT.
% in \citet{Heiter_etal_2020_S1}.
% (Sects.~\ref{sect:Si}, \ref{sect:Cr}, \ref{sect:Fe}, and \ref{sect:Ni}).
The observed spreads in line abundances generally support the quality assessment for $gf$-values, although the statistical significance is low for most of these elements.

%%%%%%%%%%%%%%%%%%%%%%%%%%%%%%%%%%%%%%%%%%%%%%%%%%%%%%%%%%%%%%%%%%%%%%%%%%%%%

\subsection{Background line list}
\label{sect:master}

Even though the work on the Gaia-ESO line list is focused on the preselected lines, these data are not sufficient for a thorough analysis. We need complete information, as far as possible, on all transitions visible in the observed wavelength ranges in the stars of interest. These data allow us to identify blends for the preselected lines, to include those blends in synthetic spectrum calculations, and to evaluate the quality of spectrum processing (e.g., continuum normalization).
Therefore, the preselected lines were complemented with data for additional atomic lines extracted from the VALD database\footnote{\url{http://vald.astro.uu.se}} \citep{Pisk:95,2015PhyS...90e4005R}, % Piskunov et al., Ryabchikova et al.
as well as data for 27 molecular species (see Sect.~\ref{sect:molecules}).

The VALD extraction was done on 2014-09-02 using version~820 of the VALD3 database and software, and the default configuration, slightly modified to exclude molecular data and to use line lists without isotopic splitting.
The number of lines was limited to those relevant for the GES by using the ``Extract Stellar'' mode for stellar parameters encompassing those of the target stars.
We used a metallicity of +0.5~dex, a microturbulence of 2~kms$^{-1}$, and two combinations of \Teff\ and \logg: 6500~K and 4.0, and 4000~K and 1.0.
Filtering by a minimum estimated central line depth %$R_c$
of 0.001 (without applying any instrumental or rotational broadening) and removing duplicates between the two \Teff/\logg\ extractions resulted in a total number of about 71\,000 and 8\,000 atomic lines contained in the UVES-580 and HR21 wavelength ranges, respectively.
The atomic part of the background line list corresponding to these wavelength ranges\footnote{from 4750~\AA\ to 6850~\AA\ and from 8488~\AA\ to 8950~\AA} is provided together with the preselected line list in electronic form (see Sect.~\ref{sect:access}).

\begin{figure*}
   \begin{center}
      \resizebox{0.85\hsize}{!}{\includegraphics{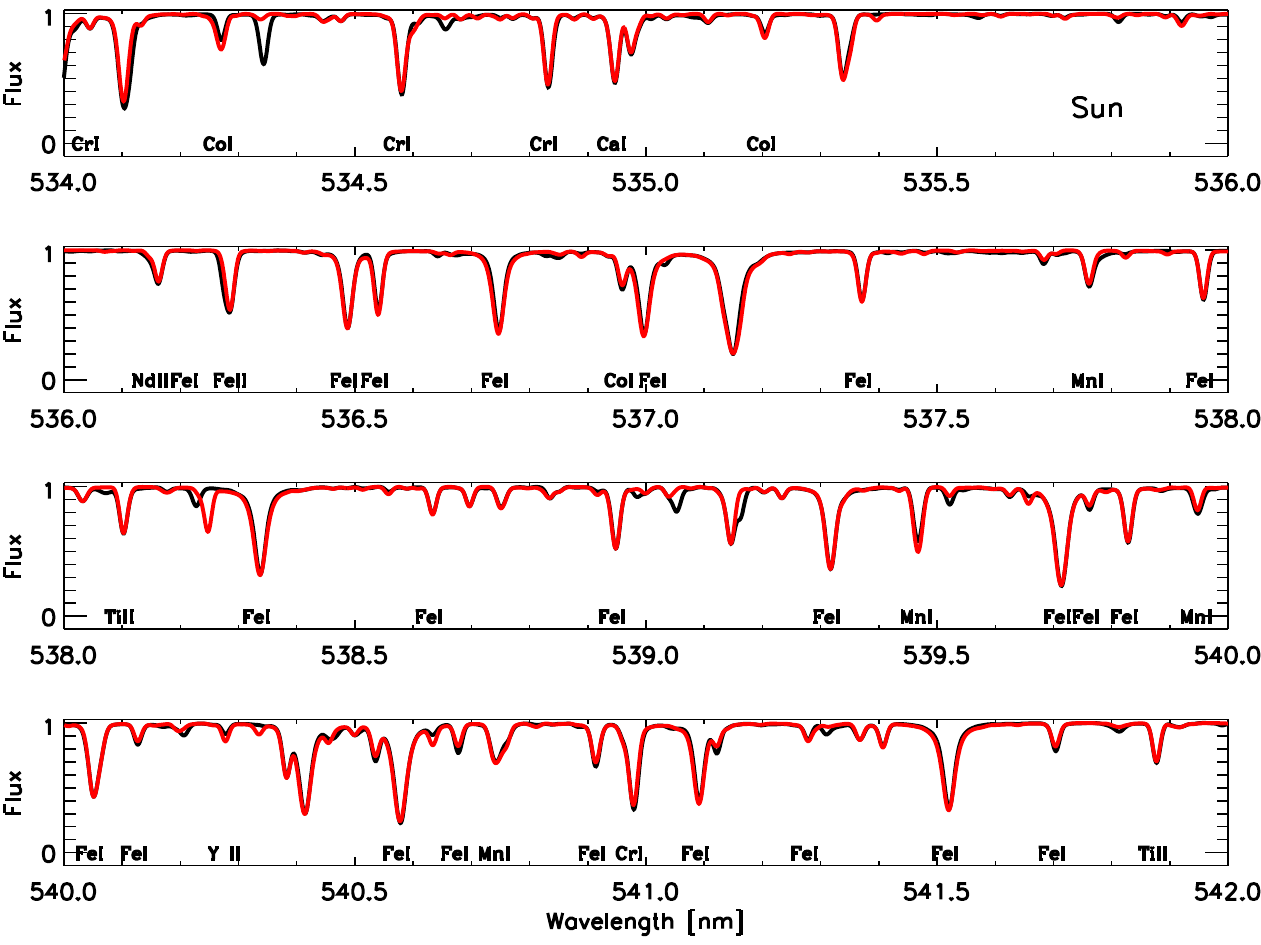}}
      \caption{Observed (black) and calculated (red) spectra for the Sun for an 8~nm-wide interval in the optical region. The Gaia-ESO bulk line list was used as input for the calculations, which includes preselected and background lines. Some of the strongest preselected lines are labelled by their species.
      }
      \label{fig:master_Sun}
   \end{center}
\end{figure*}

\begin{figure*}
   \begin{center}
      \resizebox{0.85\hsize}{!}{\includegraphics{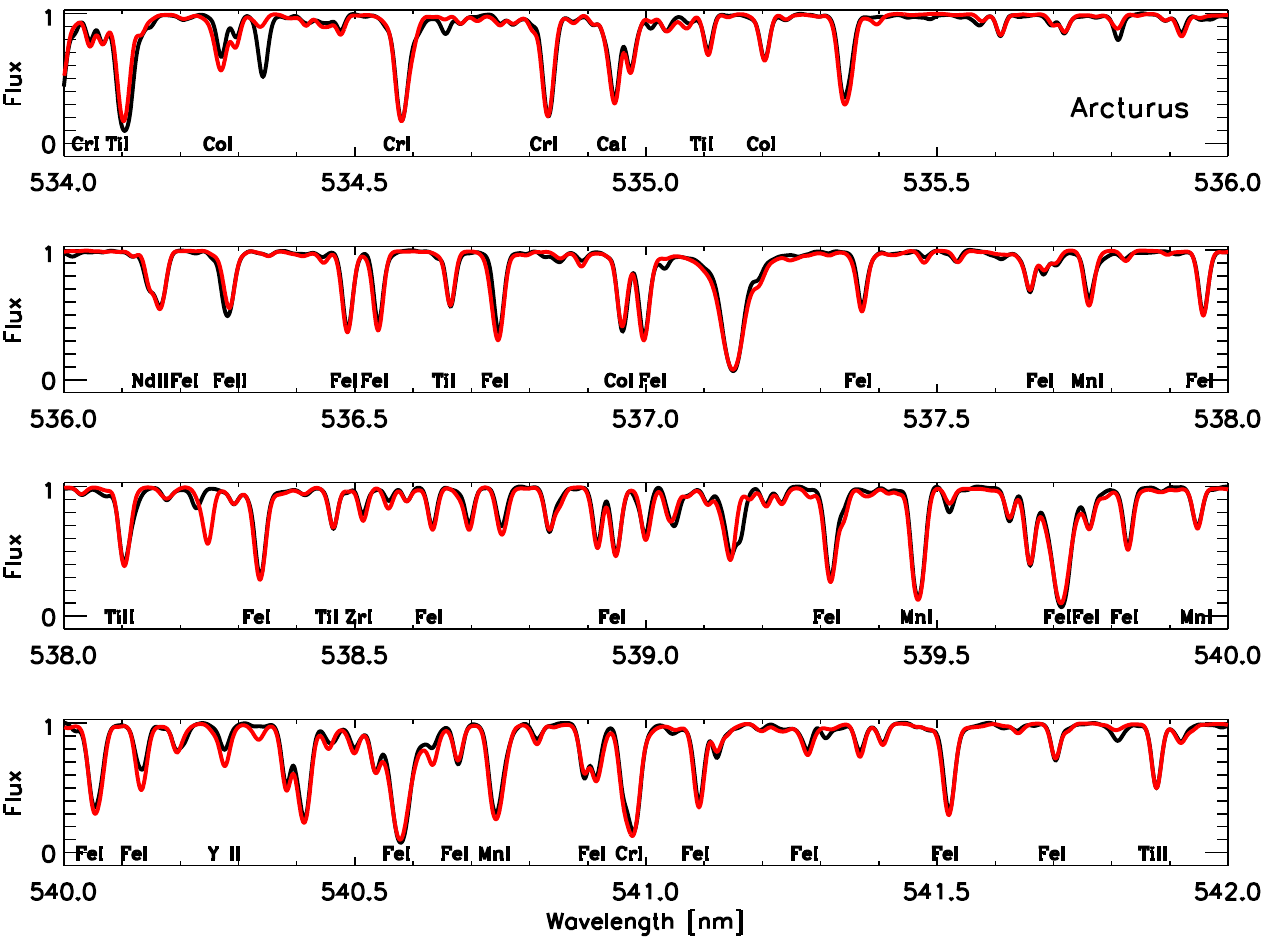}}
      \caption{Same as Fig.~\ref{fig:master_Sun} for Arcturus.
      }
      \label{fig:master_Arc}
   \end{center}
\end{figure*}

Figures~\ref{fig:master_Sun} and \ref{fig:master_Arc} show the observed and calculated spectra for the Sun and Arcturus for an interval of 8~nm in the optical region, using the bulk line list (i.e., including preselected and background lines) and the parameters and method described in Sect.~\ref{sect:spectra}. Most of the observed features are reproduced by the calculations, but we caution that deviations occur in several places.
Calculated lines may be too weak or completely missing (e.g., at 534.35~nm), or may be too strong compared to the observed lines (e.g., at 538.25~nm). This indicates incorrect or lacking atomic data.
Note that we do not provide quality flags for the lines in the background line list (except for some of the \fei\ lines,
see \refSectFe).
%see Sect.~\sectFe in \citealt{Heiter_etal_2020_S1}).
%see Sect.~\ref{sect:Fe}).
However, the VALD extraction followed the quality ranking of the sources in the database recommended by the VALD team.

%%%%%%%%%%%%%%%%%%%%%%%%%%%%%%%%%%%%%%%%%%%%%%%%%%%%%%%%%%%%%%%%%%%%%%%%%%%%%

\subsection{Hyperfine structure components and isotopic splitting}
\label{sect:hfs}

In the case of species with non-zero nuclear spin $I$\footnote{That is, all isotopes with odd baryon number, or with even baryon number and odd proton number, $^6$Li being the only one representing the latter case among the stable isotopes with non-negligible Solar System abundances.} the interaction between the nucleus and the electrons may cause a splitting of the fine structure levels into several hyperfine levels. The corresponding hyperfine transitions can be seen in high-resolution spectra as several hyperfine structure (hereafter HFS) components for individual atomic lines. Even when the HFS components are not resolved, as is the case for the Gaia-ESO spectra, they must be taken into account in the abundance analysis. In this case, HFS can be regarded as an additional broadening mechanism, altering both the shape of the line profile and the total line intensity.
Table~\ref{tab:isotopes} lists isotope information for the elements included in the preselected line list.

\begin{table*}
\caption{Isotope information for the elements included in the preselected line list.}
\label{tab:isotopes}
\centering
\begin{tabular}{rlccllll}
\hline\hline\noalign{\smallskip}
 $Z$ & El. & I   & H   & Baryon numbers & Rel. abund. [\%] & Reference & Nucl. spin \\
 (1) & (2) & (3) & (4) & (5)            & (6)              & (7)       & (8) \\
\noalign{\smallskip}\hline\noalign{\smallskip}
3 & Li & y & n & 6:7 & 7.6:92.4 & \citet{1997IJMSI.171..263Q} & 1, 1.5 \\
6 & C & n & n & 12:13 & 99:1 & \citet{Chan:1990} & 0.5 \\
8 & O &  &  & 16 & 99.8 & \citet{1976EaPSL..31..341B,Li:1988} &  \\
11 & Na &  & n & 23 & 100 & \citet{1956PhRv..101.1786W} & 1.5 \\
12 & Mg & n & n & 24:25:26 & 79:10:11 & \citet{C0JA00190B} & 2.5 \\
13 & Al &  & n & 27 & 100 & \citet{1956PhRv..101.1786W} & 2.5 \\
14 & Si & n & n & 28:29:30 & 92:5:3 & \citet{571916} & 0.5 \\
16 & S & n & n & 32:33:34 & 95:1:4 & \citet{2001GeCoA..65.2433D} & 1.5 \\
20 & Ca & n &  & 40:42:44 & 97:1:2 & \citet{doi:10.1021/ac60322a014} &  \\
21 & Sc &  & y & 45 & 100 & \citet{1950PhRv...77..634L} & 3.5 \\
22 & Ti & n & n & 46:47:48:49:50 & 8:7:74:5:5 & \citet{1993IJMSI.123...29S} & 2.5, 3.5 \\
23 & V &  & y & 51 & 99.8 & \citet{Fles:1966} & 3.5 \\
24 & Cr & n & n & 50:52:53:54 & 4:84:10:2 & \citet{Shie:1966} & 1.5 \\
25 & Mn &  & y & 55 & 100 & \citet{1963ApSpe..17..158L} & 2.5 \\
26 & Fe & n & n & 54:56:57 & 6:92:2 & \citet{1992IJMSI.121..111T} & 0.5 \\
27 & Co &  & y & 59 & 100 & \citet{1963ApSpe..17..158L} & 3.5 \\
28 & Ni & n & n & 58:60:61:62:64 & 68:26:1:4:1 & \citet{Gram:1989} & 1.5 \\
29 & Cu & y & y & 63:65 & 69:31 & \citet{JGR:JGR4707} & 1.5, 1.5 \\
30 & Zn & n & n & 64:66:67:68:70 & 49:28:4:18:1 & \citet{PONZEVERA20061413} & 2.5 \\
38 & Sr & n & n & 86:87:88 & 10:7:83 & \citet{Moor:1982} & 4.5 \\
39 & Y &  & n & 89 & 100 & \citet{1957PhRv..105..196C} & 0.5 \\
40 & Zr & n & n & 90:91:92:94:96 & 51:11:17:17:3 & \citet{1983IJMSI..50..219N} & 2.5 \\
41 & Nb &  & n & 93 & 100 & \citet{1956PhRv..101.1786W} & 4.5 \\
42 & Mo & n & n & 92:94:95:96:97:98:100 & 15:9:16:17:10:24:10 & \citet{C3JA50164G} & 2.5, 2.5 \\
44 & Ru & n & n & 96:98:99:100:101:102:104 & 5:2:13:13:17:31:19 & \citet{doi:10.1021/ac960648n} & 2.5, 2.5 \\
56 & Ba & y & y & 134:135:136:137:138 & 2:7:8:11:72 & \citet{1969JGR....74.3897E} & 1.5, 1.5 \\
57 & La &  & y & 139 & 100 & \citet{2005IJMSp.244...91D} & 3.5 \\
58 & Ce & n &  & 140:142 & 88:11 & \citet{1995IJMSI.142..125C} &  \\
59 & Pr &  & y & 141 & 100 & \citet{1957PhRv..105..196C} & 2.5 \\
60 & Nd & y & y & 142:143:144:145:146:148:150 & 27:12:24:8:17:6:6 & \citet{2005IJMSp.245...36Z} & 3.5, 3.5 \\
62 & Sm & n & y & 144:147:148:149:150:152:154 & 3:15:11:14:7:27:23 & \citet{2002IJMSp.218..167C} & 3.5, 3.5 \\
63 & Eu & n & y & 151:153 & 48:52 & \citet{1994IJMSI.139...95C} & 2.5, 2.5 \\
64 & Gd & n & n & 154:155:156:157:158:160 & 2:15:20:16:25:22 & \citet{1970JGR....75.2753E} & 1.5, 1.5 \\
66 & Dy & n & n & 160:161:162:163:164 & 2:19:25:25:28 & \citet{2001IJMSp.207...13C} & 2.5, 2.5 \\

\noalign{\smallskip}\hline\hline
\end{tabular}
\tablefoot{
Columns~3 and 4 indicate whether isotopic and/or HFS components are included in the line list (y) or not (n). These fields are empty when the effect does not apply. Column~5 gives the baryon numbers of stable isotopes contributing at least 1\% to the abundance in the Solar System. Column~6 gives the corresponding relative isotopic abundances in per cent, with reference in Column~7. Column~8 gives the nuclear spins of the isotopes with odd baryon numbers (except for Li).
}
\end{table*}

The HFS part of the Gaia-ESO line list was constructed in the following way.
The difference in energy of the hyperfine levels from the fine structure level with a given total electronic angular momentum quantum number $J$ was calculated with the Casimir equation (\citealt{Casi:1936}, cf. \citealt{kopfermann1958} and Eq.~1 in \citealt{1996ApJS..107..811P}). % Pickering 1996
The energy difference depends only on the quantum numbers $J$, $I$, and $F$, where the latter is associated with the total angular momentum of the atom. The equation consists of two terms corresponding to the magnetic dipole and the electric quadrupole interactions between electron and nucleus. The respective contributions of these interactions are parametrized by the HFS constants $A$ and $B$, which can be empirically determined for any given fine structure level.
The number of components for a particular species and transition are governed by selection rules for the $F$ values of the levels involved.
The relative intensities of the HFS components were calculated from the line strength formulae derived in the 1920s for fine-structure multiplets in the Russel-Saunders (LS) coupling scheme (e.g., Eq.~2 in Chapter~IX.2 in \citealt{Cond:1935}). % Condon \& Shortley 1935
To use these formulae for HFS the electron spin quantum number $S$ is replaced by $I$, the orbital angular momentum quantum number $L$ by $J$, and $J$ by $F$ (the resulting equations are given for example in Appendix~B in \citealt{2000AJ....120.2513P}\footnote{with an obvious typo in their Eq.~(B9), where $L$ should be replaced by $F$}). % Prochaska et al.

For the current work HFS splittings were taken into account for the lower and upper levels of the transitions included in the preselected line list for those elements for which an impact on abundance analysis is expected, whenever laboratory data for the $A$ and $B$ constants were available.
In the cases where the available HFS data were incomplete ($A$ and $B$ constants available only for one of the two levels), the missing $A$ and $B$ constants were set to zero in the computation of the HFS components (twelve \vi\ lines, three \cui\ lines, 30 \ndii\ lines, one \smii\ line).
For each transition the HFS components were co-added within bins of 0.01~\AA.
Detailed comments on the selection of $A$ and $B$ values for \sci, \vi, \mni, \coi, \cui, \baii, \laii, \prii, \ndii, \smii, and \euii,
as well as data tables can be found
in \refSectGfT.
%in \citet{Heiter_etal_2020_S1}.
%in the respective subsections of Sect.~\ref{sect:gf}.  
%
\paragraph{Isotopic splitting}
For species with several stable isotopes of non-negligible abundances we provide transition data for each isotope separately, where available. For a given electronic state the different atomic masses of the isotopes result in different energy levels. Thus, a given transition can be regarded as split into several lines with different wavelengths for different isotopes.
Note that the transition probabilities given for each isotopic component are the same. Accordingly, line list users need to scale the $gf$-values for isotopes by their relative abundances in the Solar System \citep{IUPAC13} % Meija et al. (2016), Isotopic compositions of the elements 2013 (IUPAC Technical Report)
for normal stars, or as applicable for other isotopic compositions.
The data used to calculate the isotopic splitting for the transitions under investigation are described
in \refSectGfT.
%in \citet{Heiter_etal_2020_S1}.
%in the respective subsections for each atomic species.

%%%%%%%%%%%%%%%%%%%%%%%%%%%%%%%%%%%%%%%%%%%%%%%%%%%%%%%%%%%%%%%%%%%%%%%%%%%%%

\subsection{Other atomic data}
\label{sect:other}

So far, we have discussed the atomic data needed to model the strengths of radiative transitions of the neutral and singly ionised atoms dominating the photospheres of FGK stars. However, to solve the radiative transfer problem and produce a synthetic spectrum, a wealth of additional data are needed.
These include data to describe the intrinsic widths and shapes of spectral lines (damping profile parameters),
ionisation energies and partition functions to determine level populations under the assumption of LTE, and continuous opacities.
Most of these are provided as part of the Gaia-ESO line list together with the transition probabilities, and for others we refer to recent publications.

% Radiative damping
Natural or radiative broadening is caused by the limited life-times of the atomic states involved in the transitions. The width of the resulting damping profile is given by the sum of all transition rates for spontaneous deexcitations of both the upper level and the lower level.
R.L.~Kurucz provides radiative damping widths as part of his atomic structure calculations\footnote{\url{http://kurucz.harvard.edu/atoms.html}}, and these were included in the Gaia-ESO line list (via the VALD database).
These data are available for many of the preselected lines. % 88 percent
The exceptions are all of the \ali, \zni, and \sri\ lines, some of the \nai, \mgi, and \yii\ lines, and all lines for elements with $Z>40$. % 163 lines
For the lines without calculated radiative damping widths one can resort to using the classical description of a spectral line as a damped harmonic oscillator in a two-level atom\footnote{Radiative damping width $\gamma_{\rm rad} = 8 \pi^2 e^2 10^{-7} c/ (3 m_e \lambda^2)$, at wavelength $\lambda$, where all quantities are in SI units. It is not recommended to use this approximation for lines where calculated $\gamma_{\rm rad}$ values are available. The approximate value can be up to 1~dex smaller or larger than the calculated value, e.g., for the \caii\ IR triplet lines, $\log\gamma_{\rm rad}$(calculated)=8.2, while $\log\gamma_{\rm rad}$(approximate)=7.5.}.

% Collisional broadening
Further broadening of spectral lines is caused by elastic collisions between particles in the stellar atmosphere.
Collisional broadening of hydrogen lines is addressed
in \refSectHydrogen.
%in Sect.~\sectHydrogen in \citet{Heiter_etal_2020_S1}.
%in Sect.~\ref{sect:hydrogen}.
%
Broadening of metal lines via the quadratic Stark effect due to impacting electrons and ions results in a damping profile, which is parametrised by the Stark damping parameter. This effect is in general not very important in the atmospheres of FGK-type stars, containing few charged particles (for exceptions see discussion in \citealt{2016A&ARv..24....9B}, Sect.~4.1.2). For completeness, Stark broadening data were extracted from the VALD database and included in the Gaia-ESO line list, where available. In the same way as for radiative broadening, they come as a by-product of the calculations by R.L.~Kurucz who computed them from sums over all possible transitions to a given level as described in \citet[p.~75]{1981SAOSR.390.....K}. % Kurucz, SAO Special Report 390,1981

Collisional broadening by neutral hydrogen atoms is important for many metal lines, and is discussed in detail separately
in \refSectHcollT.
%in \citet{Heiter_etal_2020_S2}.
%in Sect.~\ref{sect:H-broadening}.
In summary, collisional line widths for neutral and ionised Fe lines computed with the Anstee-Barklem-O'Mara (ABO) theory (\citealt{1991MNRAS.253..549A} % Anstee & O’Mara
         and successive expansions by P.S.\ Barklem and collaborators) were compared to those computed by R.L.\ Kurucz and with the Unsöld recipe, which are based on Lindholm-Foley theory.
The ratios between line widths from different theoretical approaches show a large spread, in particular for high values of the excitation energy.
As the ABO theory is considered the most reliable theory \citep{2016A&ARv..24....9B}, % Barklem
new broadening data were calculated according to the ABO theory for 41 lines of \fei\ and eight other neutral species.
These were included in the Gaia-ESO line list, together with previously available data for all other lines based on the ABO theory or provided by Kurucz, which had been extracted from the VALD database.
Based on the analysis
in \refSectHcollT\ 
%in \citet{Heiter_etal_2020_S2}
we recommend to avoid lines of neutral species for which ABO data are not available (cf. Table~\ref{tab:flagstats}). For lines of ionised species without ABO data that have low excitation energies, data by Kurucz should be used where available (\scii, \tiii, and \yii\ lines), otherwise   the Unsöld approximation with an enhancement factor of 1.5 for the line width can be used (lines of rare-earth species).

% Ionisation energies
For ionisation energies for atoms we refer to the NIST Atomic Spectra Database\footnote{NIST ASD: \url{http://physics.nist.gov/PhysRefData/ASD/ionEnergy.html}} \citep{NIST_ASD}, % first to third for all elements: ”H-U I-III”
or Table~4 in \citet{2016A&A...588A..96B}. % Barklem & Collet
%
% Partition functions
\citeauthor{2016A&A...588A..96B} calculated partition functions for all elements from H to U and the first three ionisation stages, for temperatures up to 10\,000~K (their Table~8), based on excitation energies from the NIST ASD.
Their data agree very well with those of \citet{1987A&A...182..348I} % Irwin
for temperatures in common (i.e., above 1000~K) for most species.
However, for some of the rare-earth elements, differences of up to 50\% are seen (their Figs.~5 and 6). For \laii, the new partition functions are lower than \citet{1987A&A...182..348I} at low temperatures, for \smii\ and \euii\ they are higher at low temperatures, and for \prii\ and \dyii\ they are higher at high temperatures.

% Continuous opacities
Finally, calculations of continuous fluxes are needed to be able to compare synthetic spectra to observations normalised to the continuum. This requires a large amount of input data on continuous opacities (bound-free and free-free transitions, as well as scattering processes for numerous species which are abundant in cool stellar atmospheres).
We did not define a standard set of data to be used for this aspect within GES.
Instead, we refer to the data commonly used by the codes employed for Gaia-ESO data analysis.
For example, the SME package and the MOOG code compute continuous opacities using adapted versions of the subroutines embedded in the ATLAS9 code by R.L.~Kurucz\footnote{\url{http://kurucz.harvard.edu/programs/atlas9/atlas9.for}} \citep[p. 73]{1970SAOSR.309.....K}. % Kurucz 1970
Obviously, the same routines are used in the SYNTHE code by Kurucz.
For radiative transfer codes associated with the MARCS model atmosphere package (e.g., Turbospectrum) references for continuous opacity data are given in Table~1 of \citet{Gust:08}, and are discussed in their Sect.~4.
For other codes see the references given in \citet{2014A&A...570A.122S}, % Smiljanic et al.
who describe most of the Gaia-ESO analysis methods.

%%%%%%%%%%%%%%%%%%%%%%%%%%%%%%%%%%%%%%%%%%%%%%%%%%%%%%%%%%%%%%%%%%%%%%%%%%%%%

\subsection{Blending properties for the Sun and Arcturus}
\label{sect:synflag}

In order to assess the blending properties of the preselected lines, two spectra each were calculated for parameters of the Sun and Arcturus (as described in Sect.~\ref{sect:spectra}) -- one using only the preselected lines as input, and another one including all the blending atomic and molecular lines from the background line list (Sects.~\ref{sect:master} and \ref{sect:molecules}).

The flags for blending properties (hereafter \synflag) were assigned after visual inspection of the three overplotted line profiles (the two synthetic ones and the observed one) according to the following general scheme:
\begin{description}
\item[\Yes:] Line is unblended or only blended with line of same species in both stars.
\item[\Un:] Line may be inappropriate in at least one of the stars.
\item[\No:] Line is strongly blended with line(s) of different species in both stars.
\end{description}

\begin{figure*}
   \begin{center}
      \resizebox{0.63\hsize}{!}{\includegraphics{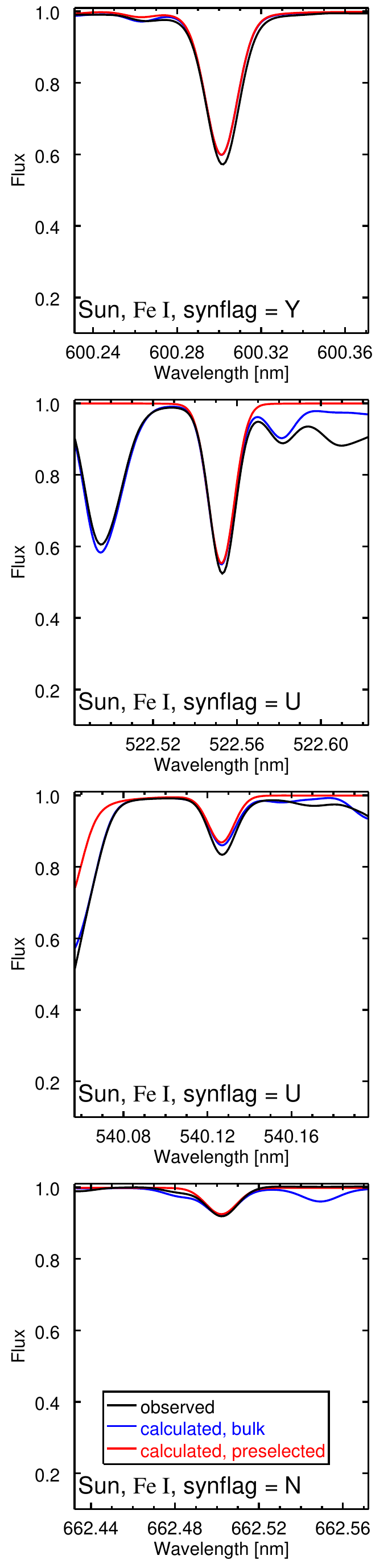}\includegraphics{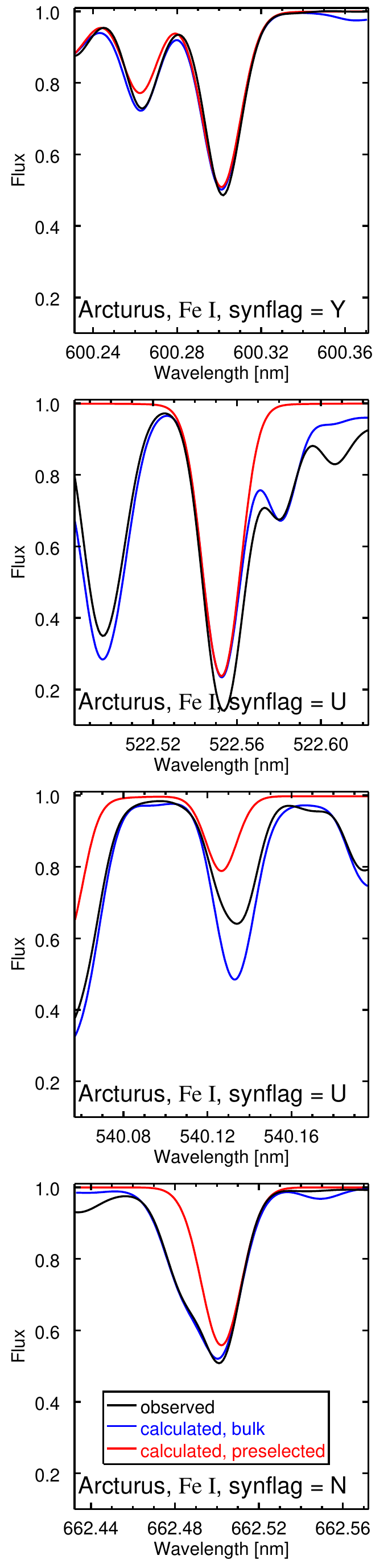}}
   \end{center}
   \caption{Comparison of observed and calculated line profiles around four of the preselected \fei\ lines with different \synflag\ assignments for the Sun (left) and Arcturus (right). Black lines: observations, red lines: calculations including preselected spectral lines only, blue lines: calculations using the Gaia-ESO bulk line list, including preselected and background lines. All of these lines have \gfflag=\Yes.
   }
   \label{fig:Fe1_profiles_synflag}
\end{figure*}

The number of lines to which the different \synflag s were assigned for each species is indicated in Table~\ref{tab:flagstats}.
%
% Profile plots for Sun and Arcturus
Figure~\ref{fig:Fe1_profiles_synflag} shows typical example line profiles for four \fei\ lines with accurate $gf$-values (\gfflag=\Yes) and with different \synflag\ assignments.
The top row shows the same line at 600.3~nm as in the top row of Figure~\ref{fig:Fe1_profiles_gfflag}. It is assessed to be unblended (\synflag=\Yes), because all three spectra lie on top of each other.
The second and third rows of Fig.~\ref{fig:Fe1_profiles_synflag} show examples for lines with undecided blending properties (\synflag=\Un).
The line at 522.5~nm has a weak blend at the red side in both stars, and a possible additional unidentified blend as seen from the comparison with the observed Arcturus spectrum.
The line at 540.1~nm is almost blend-free in the Sun, but strongly blended in Arcturus. In addition, the data from the background line list around this line are incorrect, as is obvious from the comparison with the observed Arcturus spectrum.
The bottom row of Fig.~\ref{fig:Fe1_profiles_synflag} shows an example for a line which is clearly blended in both stars (\synflag=\No).

We did not consider whether the strength of a line is appropriate for analysis in a specific star (i.e., not too weak or too strong), since this will vary much between survey targets. The blending assessment of a line should thus be valid only in comparison to other lines of the same species and line strength.
This means for example that \synflag=\Yes\ has been assigned to unblended lines in Arcturus even when they were not detectable in the Sun. 

\begin{figure*}
   \begin{center}
      \resizebox{\hsize}{!}{\includegraphics{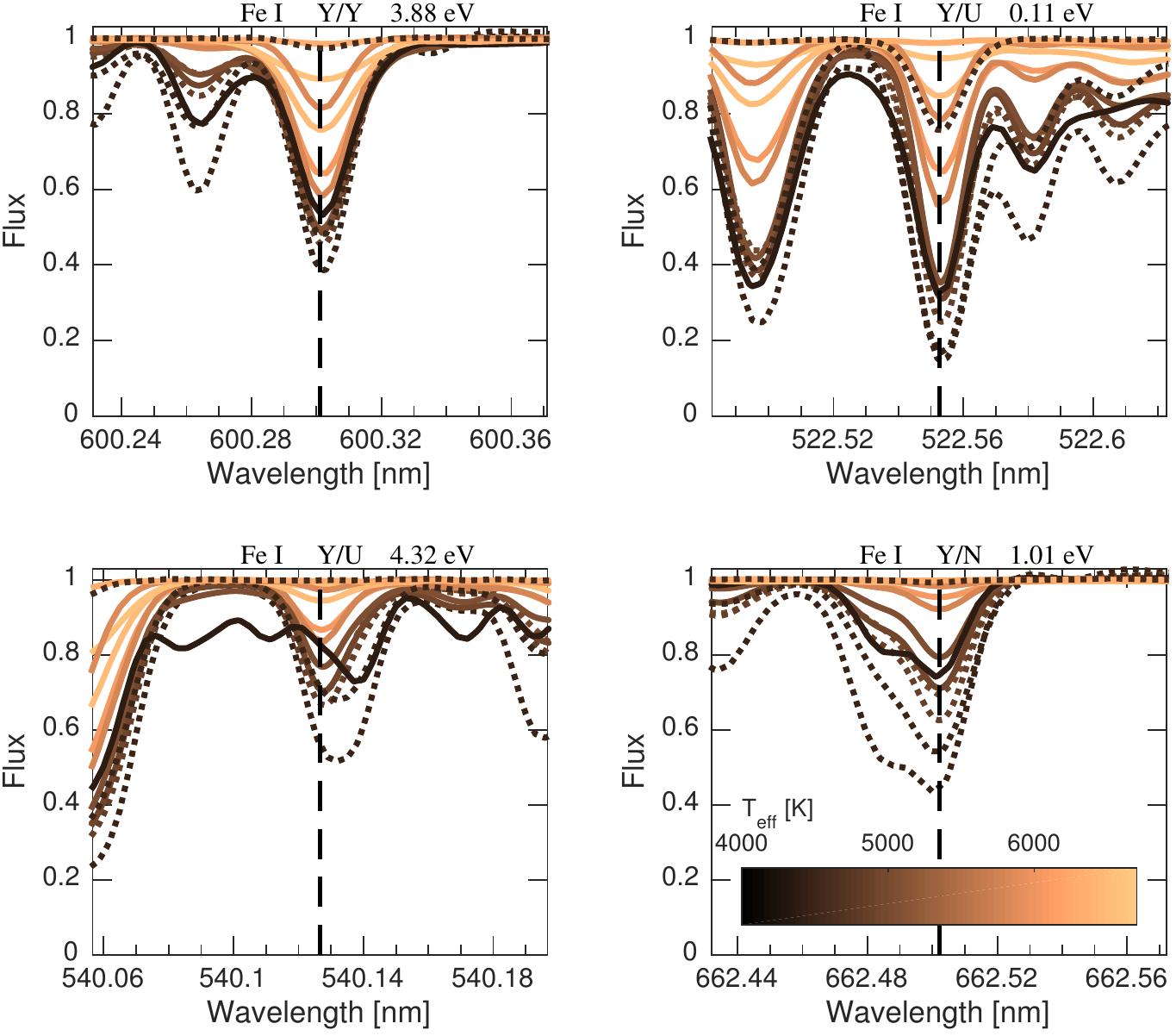}}
   \end{center}
   \caption{
   Line profiles for four preselected \fei\ lines generated from observed spectra of selected Gaia FGK benchmark stars (see Sect.~\ref{sect:spectra}). Quality flags and lower level energy are given at the top of each panel. The vertical dashed line indicates central wavelength. \Teff\ is coded by colour, solid lines are dwarfs, and dotted lines are giants. 
   }
   \label{fig:Fe1_profiles_GBS}
\end{figure*}

% Profile plots for benchmark stars
Observed line profiles for other benchmark stars with a wider range of stellar parameters are shown for the same four \fei\ lines in Fig.~\ref{fig:Fe1_profiles_GBS} (see Sect.~\ref{sect:spectra} for information on the spectra).
Note that for two stars of similar temperature and gravity (e.g., dark brown dotted lines representing cool giants) the variation in line strength is due to the difference in metallicity.
The \synflag=\Yes\ line seems blend-free in all stars (except the coolest dwarf star 61~Cyg~A). The \synflag=\Un\ and \No\ lines are unblended in the warmer dwarfs and the metal-poor giant HD~122563, but they are blended in giant stars and cooler dwarfs.
Further examples of \fei\ lines with \synflag=\No\ are those at 516.6~nm and 517.2~nm, which lie on the wings of the \mgi~b 516.7~nm and 517.3~nm lines.
The line at 559.5~nm is a blend with a preselected \cai\ line.
Several further \fei\ lines have weak blends in the Sun, but strong blends in Arcturus (e.g., 547.3~nm, 623.1~nm), and the reverse case is also encountered (e.g., 625.4~nm, 635.9~nm).
Examples for other species are given
in \refSectGfT.
%in \citet{Heiter_etal_2020_S1}.
%in some of the subsections for individual elements.

% Profile plots for selected Fe 1 lines (ABO)

%%%%%%%%%%%%%%%%%%%%%%%%%%%%%%%%%%%%%%%%%%%%%%%%%%%%%%%%%%%%%%%%%%%%%%%%%%%%%

% place all figures and tables from this section here
% (prevents LaTeX Error ”Too many unprocessed floats”)
\clearpage

%----------------------------------------------------------------------------
% Molecules
%----------------------------------------------------------------------------

\section{Molecular data}
\label{sect:molecules}

In addition to atomic data, we also include molecular data. In fact, the molecular transitions are much more numerous than the atomic transitions. Thus, it is crucial to include extensive molecular line lists for determining stellar parameters and identifying and fitting atomic line blends, although they also may be used for abundance determination and to derive isotopic ratios.
Priority was given to molecules which contribute significantly to the absorption in the spectra of G or K-type stars. This includes CH, NH, OH, C$_2$, CN, MgH, SiH, CaH, FeH, TiO, VO, and ZrO (and their isotopologues). Although the best line lists available in the literature were used (Table~\ref{tab:molref}), the quality of the molecular data varies from one molecule to the other.

For CH, NH, OH, and MgH, improved line lists were computed using high-quality laboratory line positions (better than $10^{-3}$~\AA) combined with an accurate computation of transition moments (used to derive $\log gf$ for molecules).
The laboratory line positions were used as input in the program for simulating molecular spectra PGopher\footnote{\url{http://pgopher.chm.bris.ac.uk}} \citep{WESTERN2017221} % Western 2017
in parallel with the programs RKR1 and LEVEL\footnote{\url{http://leroy.uwaterloo.ca/programs.html}} \citep{LEROY2017158,LEROY2017167} % Le Roy 2017ab
for derivation of the line intensities.
Radiative broadening parameters were also computed whenever the information of the full electronic structure of the molecule was available.
An illustration of the procedure for the case of CH can be found in \citet{2014AA...571A..47M}. % Masseron et al. 
The line lists for the remaining molecules from the literature in general make use of the Born-Oppenheimer approximation to determine the positions and assume identical oscillator strengths for all isotopologues. In particular, the line lists for SiH, CaH, FeH, TiO, VO, and ZrO suffer from larger uncertainties, which for positions can sometimes reach several \AA. 

In addition to line positions and oscillator strengths, dissociation energies (Table~\ref{tab:molref}) and partition functions are required to compute the molecular equilibrium.
The partition functions used in this work are given in Table~\ref{tab:molpfs} in the appendix.

Recently, \citet{2016A&A...588A..96B} % Barklem & Collet (2016) 
reviewed partition functions and dissociation energies for a large number of diatomic molecules. Regarding dissociation energies, our selected values agree within 1\% with the selection of \citeauthor{2016A&A...588A..96B}, except for CaH. However, because this molecule only appears in the coolest stars, this discrepancy should only have a marginal impact on the Gaia-ESO analysis. Concerning partition functions, an agreement better than 10\% for temperatures lower than 5000~K is found between our adopted values and the compilation of \citeauthor{2016A&A...588A..96B}, except for FeH. We note that this latter molecule contributes only in the HR21 setting of the GIRAFFE instrument and only for very cool stars.

The molecular data are available in electronic form at the CDS for the same wavelength ranges as the atomic data, in a format described in Sect.~\ref{sect:access}.

\begin{table*}[h]
\renewcommand{\tabcolsep}{1.0mm}
\caption{List of molecular species, recommended dissociation energies $D_{00}$, and references for molecular transitions.}
\label{tab:molref}
\centering
\begin{tabular}{lllll}
\hline\hline\noalign{\smallskip}
 & Isotopologues & $D_{00}$ & $D_{00}$ references & Line list references \\
 & & [eV] & & \\
\noalign{\smallskip}\hline\noalign{\smallskip}
CH    & $^{12}$CH, $^{13}$CH    & 3.466 & \citet{Kuma:1998} % Kumar et al. (1997)
                                        & \citet{2014AA...571A..47M} \\ % Masseron et al. (2014)
NH    &                         & 3.420 & \citet{Tarr:1997} % Tarroni et al. (1997) 
                                        & T. Masseron (priv. comm.) \\
OH    &                         & 4.392 & \citet{Huber1979} % Huber & Herzberg
                                        &  T. Masseron (priv. comm.) \\
C$_2$\tablefootmark{$\dagger$} & $^{12}$C$^{12}$C & 6.371 & \citet{luo2007} % Luo et al. (2007)
                                                  &    \citet[Swan]{2013JQSRT.124...11B} \\ % Brooke et al.
      & $^{12}$C$^{13}$C        &                 &  & \citet[Swan]{2014ApJS..211....5R} \\ % Ram et al.
      & $^{13}$C$^{13}$C        &                 &  & F. Quercy (priv. comm., Swan) \\
CN    & $^{12}$C$^{14}$N & 7.738  & \citet{Huan:1992} % Huang et al. (1992)
                                                      & \citet{2014ApJS..210...23B} \\ % Brooke et al. (2014)
      & $^{12}$C$^{15}$N, $^{13}$C$^{14}$N &        & & \citet{2014ApJS..214...26S} \\ % Sneden et al. (2014)
MgH   & $^{24}$MgH, $^{25}$MgH, $^{26}$MgH & 1.285  & \citet{Shay:2007} % Shayesteh et al. (2007)
                                                    & \citet[A-X]{2013ApJS..207...26H}, \\ % Hinkle et al. (2013)
                                              & & & & T. Masseron (priv. comm., B-X) \\
SiH   &                         & 3.060 & \citet{Huber1979} % Huber & Herzberg
                                        & \citet{KSiH} \\ % Kurucz 2010
CaH   &                         & 1.700 & \citet{Huber1979} % Huber & Herzberg
                                        & B. Plez (priv. comm.) \\
FeH   &                         & 1.598 & \citet{2003ApJ...594..651D} % Dulick et al.
                                        & \citet{2003ApJ...594..651D} \\
TiO   & $^{46}$TiO, $^{47}$TiO, $^{48}$TiO, $^{49}$TiO, $^{50}$TiO & 6.87 & \citet{1997CPL...266..335N} % Naulin et al.
                                                                          & B. Plez (priv. comm.) \\
VO    &                         & 6.437 & B. Plez (priv. comm.)            & B. Plez (priv. comm.) \\
ZrO   & $^{90}$ZrO, $^{91}$ZrO, $^{92}$ZrO, $^{94}$ZrO, $^{96}$ZrO  & 7.890 & B. Plez (priv. comm.) & B. Plez (priv. comm.) \\
\noalign{\smallskip}\hline\hline
\end{tabular}
\tablefoot{
\tablefoottext{$\dagger$}{Excitation potentials were normalized to 0.0256~eV.}
}
\end{table*}

% place all figures and tables from this section here
\clearpage

%----------------------------------------------------------------------------
% Discussion
%----------------------------------------------------------------------------

\section{Discussion and outlook}
\label{sect:discussion}

% ----- Discussion -----

\subsection{Selected examples for the application of the line list}
\label{sect:impact}

% 1 Introduction
The Gaia-ESO line list has been used within the consortium for the determination of atmospheric parameters and abundances of calibration stars, stars across Galactic populations, and stars in several clusters.
The results have been presented in over 50 refereed articles.
Other spectroscopic surveys have also started to use the Gaia-ESO line list as a basis for their analyses. Examples for these are the EMBLA \citep{2016MNRAS.460..884H}, % Howes et al., see also 2014MNRAS.445.4241H
GALAH \citep{2015MNRAS.449.2604D}, % De Silva et al.
and OCCASO \citep{2016MNRAS.458.3150C} surveys. % Casamiquela et al.
Here, we briefly mention some of these works.

% 2 GBS
The Gaia FGK benchmark stars consist of about 30 well-known stars and were the main calibrators in the GES \citep{Pancino2017}.
Reference values for metallicity (\citealt{2014A&A...564A.133J,2016A&A...592A..70H}) and abundances of ten $\alpha$- and iron-peak elements \citep{2015A&A...582A..81J} were determined from an analysis of high-resolution spectra.
These works used a subset of the GES line list, where line selection was based on the quality flags\footnote{All lines with \gfflag=\No\ and Fe lines with \synflag=\No\ were discarded.}, but also on low method-to-method dispersion.
The standard deviations of the abundances of the \fei\ lines derived by six different methods for each star were between 0.01 and 0.03~dex.
The abundances of the remaining elements were derived by eight different methods differentially (line-by-line) to different reference stars after grouping the stars by atmospheric parameters. The decrease in dispersion for the differential line abundances compared to the absolute abundances confirmed, among other things, the importance of hyperfine structure effects.
For each element, lines commonly used for stars within groups of similar spectral types were identified, and were referred to as ``golden lines''. In the case of Fe lines these were lines used for all stars within a group, and for the other elements a line was defined to be a golden line when it was analysed in at least 50\% of the stars in the group.
These include on the order of 100 \fei\ lines, and the \sii\ line at 568.448~nm, which was used for all stars except two, to give some examples.
For V and Co, no golden lines were identified in metal-poor stars.
A detailed discussion of golden lines can be found in
\citet[Sect.~6.3, Tables~4 and 5]{2014A&A...564A.133J}
and
\citet[Sect.~4.5 and on-line table\footnote{\url{http://cdsarc.u-strasbg.fr/vizier/ftp/cats/J/A+A/582/A81/infoline.dat}}]{2015A&A...582A..81J}.

% 3 GES synthetic spectra grid
For the analysis of the UVES and GIRAFFE spectra observed within GES the different groups either used their own radiative transfer codes or a pre-computed grid of synthetic spectra. In both cases the Gaia-ESO line list was adopted. The grid contains 13\,784 high-resolution synthetic spectra for FGKM-type stars over the spectral ranges 420--690~nm and 845--895~nm. A wide range of metallicities (from [M/H]=$-5.0$ to +1.0~dex) and [$\alpha$/Fe] enrichments (five values for each metallicity) is covered. For details see \citet{deLaverny12}. % de Laverny et al.

% 4 General analysis papers: UVES - general + young, GIRAFFE
An overview of the analysis procedure of UVES spectra observed within the GES is given in \citet{2014A&A...570A.122S} % Smiljanic et al.
for the case of FGK stars in the field and in older open clusters,
and by \citet{2015A&A...576A..80L} % Lanzafame et al.
for the case of F- to M-type stars in the fields of young open clusters (age $<$100~Myr).
Atmospheric parameters (mainly based on Fe lines) and abundances of up to 24 or 26 elements were derived by up to 13 and four different groups, respectively. Their results were subsequently combined into a homogeneous set of recommended values. The groups made their individual choices of line sub-sets to be used with their methods. For an illustration of the variation in line selection for \cai\ see Fig.~4 in \citet{2019ARA&A..57..571J}.
In the first case, spectral lines measured by at least three groups were included in the combined abundances, and lines affected by blends as indicated by \synflag\ were removed for species with a large number of lines (20 or more).
In the second case, one of the groups based their selection on \synflag\ for elements with $Z>28$, % CAUP, Sect. 6.2 of Lanzafame et al.
while the homogenisation was done without any further line selection.
The precision of the abundances as indicated by the method-to-method dispersions was found to be similar in both cases, ranging from below 0.15~dex to 0.35~dex depending on the element.
Stellar parameters and abundances of up to eleven chemical species were derived from GIRAFFE spectra observed within the GES by five different groups (A.~Recio-Blanco, priv. comm., see also \citealt{Worley_etal_2020}).
In this case the variation in line selection between groups is expected to be small because the number of lines in the relevant spectral range is low.
Typical method-to-method dispersions in the combined and homogenised abundances were 0.04~dex.

% 5 Galactic populations
Examples for studies of the structure and evolution of the Galactic bulge and disc based on the Gaia-ESO recommended metallicities and abundances are given in
\citet{2016ApJ...824L..29W}, % Williams et al.
\citet{2014A&A...565A..89B}, % Bergemann et al.
\citet{RecioBlanco14}, % Recio-Blanco et al. 2014
and \citet{2014A&A...572A..33M}. % Mikolaitis et al.
Typical mean uncertainties in abundances reported in these works are around 0.1~dex, which reflect the adoption of the Gaia-ESO line list, among other things. The abundance data allowed the authors to clearly distinguish between different sub-components in terms of metallicity and $\alpha$-element abundances, and to derive trends of abundances with other stellar properties such as age or galactocentric radius.
\citet{2015A&A...575L..12L} % Lind et al.
identified one star among a few hundred halo stars that has most likely been ejected from a globular cluster, based on a difference in [Mg/Fe] abundance  of 0.8~dex at a 4$\sigma$ significance compared to typical halo stars.

% 6 Open clusters
An example for advances in the area of open clusters made possible by GES spectra and the Gaia-ESO line list is given by three inner-disk clusters (with ages of 0.3 to 1.5~Gyr). C, N, and O abundances with a typical precision of 0.05~dex were determined by \citet{2015A&A...573A..55T}, % Tautvaišienė et al.
and abundances of eleven elements with average uncertainties of about 0.1~dex were determined by \citet{2015A&A...580A..85M}. % Magrini et al.
In these works predictions made by models for stellar evolution and for Galactic chemical evolution were confirmed, and one cluster was found to be locally enriched by the Type II supernova explosion of a single star.

% 7 Other survey (EMBLA)
The EMBLA survyey -- an investigation of metal-poor stars in the Galactic bulge -- is based on spectra obtained with a different instrument\footnote{the MIKE high-resolution spectrograph on Magellan's 6.5m Clay telescope}, with similar resolution as the GES but covering a larger wavelength region \citep{2016MNRAS.460..884H}. % Howes et al.
For the abundance analysis, the Gaia-ESO preselected line list was used as a starting point and was complemented by lines and data from other sources.
There was a large overlap in line data between the two surveys for \feii\ (mostly lines with \gfflag=\Yes), resulting in an average standard error of Fe line abundances of 0.06~dex. Other species with lines in common or with the same source for the $gf$-values were \cai, \scii, and \zni, with mean uncertainties in abundances of $\sim$0.1~dex, and \cri\ and \baii\ with $\sim$0.2~dex.
Carbon abundances were determined from CH band heads using the same molecular data as the GES, with uncertainties of $\sim$0.2~dex.
These data allowed the authors to conclude that the fraction of carbon-enhanced stars might be lower in the bulge compared to the halo, and that some of the other elements behaved differently in the bulge stars than in halo stars.

% ----- Outlook -----

\subsection{Data needs and recent developments}
\label{sect:dataneeds}

This article describes version~6 of the Gaia-ESO line list, which was the last version produced within the GES consortium (mainly in 2014, with minor changes in 2016), and this is the version used for the final release of GES data products. The list is complete in the sense that it contains all atomic and molecular data available at the time of compilation for those transitions widely used for abundance analysis of FGK-type stars in the wavelength region of interest at the resolution of the FLAMES-UVES spectrograph. This includes hyperfine structure and isotope splitting data, as well as references to the original sources for the data. 
The list contains about 200 lines (not counting HFS or IS components) of 24 species which have accurate $gf$-values and are free of blends in the spectra of the Sun and Arcturus, i.e., both \gfflag\ and \synflag=\Yes\ (see column ``Y/Y'' in Table~\ref{tab:flagstats}).

Prospective users of the line list are advised to update the $gf$-values with new data that may have become available since the compilation was done. Also, HFS and IS data are available for more species than considered relevant in the context of the GES, and users should add these according to the needs of their specific application. Note that for several species more recent calculations by R.L. Kurucz are available, in particular for \ci, \siii, \cai, \tii, \tiii, \vi, \vii, \cri, \crii, \fei\ (see discussion
in \refSectFe\ 
%in Sect.~\sectFe in \citealt{Heiter_etal_2020_S1}
%in Sect.~\ref{sect:Fe}
for the latter). The Kurucz website provides HFS and IS components for many species.

Based on the presentation of available data
in \refSectGfT\ 
%in \citet{Heiter_etal_2020_S1}
%in Sect.~\ref{sect:gf}
we comment on those atomic species and lines for which future improvements of transition probability data should have the highest priority.
We focus on the preselected lines with \synflag=\Yes\ and \Un.
A number of species have less than five such lines (see Table~\ref{tab:flagstats}). 
Most of these have high-quality $gf$-values (\gfflag=\Yes), including the light species \lii, \ci, \oi, and \siii, and the heavy species \zni, \zrii, \rui, \baii, \prii, \smii, \euii, \gdii, and \dyii.
The exceptions are \vii, \sri, and \yi, which have one or two lines with theoretical $gf$-values. However, these lines are extremely weak and probably blended in most stellar spectra and thus of low priority for abundance analysis.

Among the species with at least five lines with \synflag=\Yes\ or \Un\ several have high-quality $gf$-values for all of those lines (\sci, \cui, \yii, \zri, \moi), or for the vast majority (>90\%, \cai\ and \tii).
Another group of species has low-quality $gf$-values for more than 10\%, but less than 50\% of the \synflag=\Yes\ or \Un\ lines (\scii, \tiii, \vi, \cri, \mni, \coi, \laii, \ceii, \ndii). These should be considered for laboratory measurements in the long term, but they are of lower immediate priority.
Note that for the five \scii\ lines with \gfflag=\No,
% 5552.224, 6245.6366, 6279.753, 6300.6977, 6320.8513 Å
new theoretical $gf$-values based on branching fractions calculated using the relativistic Hartree–Fock method and life-times measured by \citet{1988JOSAB...5..606M} % Marsden et al.
are available in \citet{2017MNRAS.472.3337P}. % Pehlivan Rhodin et al.
Finally, there are a number of species with low-quality $gf$-values for 50\% or more of the \synflag=\Yes\ or \Un\ lines, which should be given highest priority in current and future laboratory experiments.

The percentage is about 50 for both \fei\ and \feii.
% Fe I
Specifically, 236 of 439 \fei\ lines would need improved $gf$-values. These are roughly evenly distributed over the whole wavelength range considered here and comprise a wide variety of transitions. They originate from lower levels which are preferentially odd (70\%) and belong to 36 different terms, of which the most frequent are $y^5F^o$ and $y^3F^o$ with energies of $\sim$4.25 and $\sim$4.60~eV, respectively. The upper levels are more diverse with 75 different terms, of which the most frequent are $g^5F$ and $f^5G$ with energies around 6.6~eV. One third of these lines have \synflag=\Yes. Almost half of them have theoretical $gf$-values by \citet[which should be replaced by more recent calculations available at the Kurucz website]{K07}, and most of the remaining ones have $gf$-values measured by \citet{MRW}.
% Fe II
In the case of \feii\ 13 of 26 lines have purely theoretical $gf$-values. These are listed in Table~\ref{tab:gfneed1} in the appendix and half of them have \synflag=\Yes.

About 60 to 70\% of the \synflag=\Yes\ or \Un\ lines need improved $gf$-values for the species \nai, \mgi, \sii, \caii, \nii. These lines are also listed in Table~\ref{tab:gfneed1} except for \mgi\ and \nii.
For \mgi, new experimental oscillator strengths were recently published by \citet{2017A&A...598A.102P}, % Pehlivan Rhodin et al.
combining branching fractions measured from an FTS spectrum of a hollow cathode discharge lamp with radiative life-times from the literature and from their own calculations. They also published new theoretical oscillator strengths based on the multiconfiguration Hartree-Fock method. All but one of the eight \mgi\ lines in the Gaia-ESO line list needing improvement are included in this work, as well as the four lines which already had high-quality $gf$-values. Table~\ref{tab:Mg1} in the appendix lists both the data in the GES line list and the new data for all of these lines.

Numerous \nii\ lines found over the whole wavelength range considered here are candidates for new experimental transition probabilities (53 of 83 with \synflag=\Yes\ or \Un).
These lines are listed in Table~\ref{tab:Ni1} in the appendix. Almost all of them currently have theoretical $gf$-values by \citet{K08}.
Nearly all of them originate from odd lower levels, which belong to 16 different terms, the most frequent one being $y^3F^o$ with energies of $\sim$4.2~eV. The upper levels belong to 14 different terms, the most frequent one being $3d^9(^2D_{5/2})4d \enspace ^2[\sfrac{7}{2}]$ with energies around 6.1~eV. About 40\% of these lines have \synflag=\Yes.

\begin{table}[ht]
   \caption{Lines of \ali, \si, and \crii\ with \synflag=\Yes\ or \Un, and with \gfflag=\Un\ or \No.}
\label{tab:gfneed2}
\centering
\begin{tabular}{lcll}
\hline\hline\noalign{\smallskip}
% header
Species & Transition & Wavelength & Flags \\
        &            &     [\AA]  & $gf$/syn \\
\noalign{\smallskip}\hline\noalign{\smallskip}
% data
\ali & $ a^2S_{0.5} - w^2P^o_{1.5} $ & 5557.063 & \emph{U/U} \\
\ali & $ a^2S_{0.5} - x^2P^o_{1.5} $ & 6696.023 & \emph{U/U} \\
\ali & $ a^2S_{0.5} - x^2P^o_{0.5} $ & 6698.673 & \emph{U/Y} \\
\ali & $ a^2D_{1.5} - y^2F^o_{2.5} $ & 8772.865 & \emph{U/U} \\
\ali & $ a^2D_{2.5} - y^2F^o_{6.5} $ & 8773.896 & \emph{U/U} \\
\si & $ a^5P_{1} - x^5D^o_{0} $ & 6743.483 & \emph{U/U} \\
\si & $ a^5P_{1} - x^5D^o_{2} $ & 6743.540 & \emph{U/U} \\
\si & $ a^5P_{1} - x^5D^o_{1} $ & 6743.580 & \emph{U/U} \\
\si & $ a^5P_{2} - x^5D^o_{1} $ & 6748.570 & \emph{U/U} \\
\si & $ a^5P_{2} - x^5D^o_{2} $ & 6748.580 & \emph{U/U} \\
\si & $ a^5P_{2} - x^5D^o_{3} $ & 6748.790 & \emph{U/U} \\
\si & $ a^5P_{3} - x^5D^o_{4} $ & 6757.150 & \emph{U/U} \\
\si & $ a^5P_{3} - y^5D^o_{4} $ & 8694.710 & \emph{U/U} \\
\crii & $ a^4F_{3.5} - z^4F^o_{3.5} $ & 4848.235 & \emph{N/U} \\
\crii & $ a^4F_{2.5} - z^4F^o_{1.5} $ & 4884.607 & \emph{N/U} \\
\crii & $ b^4F_{4.5} - z^4F^o_{4.5} $ & 5237.328 & \emph{N/U} \\
\crii & $ b^4F_{4.5} - z^4F^o_{3.5} $ & 5279.876 & \emph{N/U} \\
\crii & $ b^4P_{2.5} - z^4P^o_{2.5} $ & 5305.853 & \emph{N/Y} \\
\crii & $ b^4F_{1.5} - z^4F^o_{2.5} $ & 5310.686 & \emph{N/U} \\
\crii & $ b^4F_{2.5} - z^4F^o_{2.5} $ & 5313.563 & \emph{N/U} \\
\crii & $ b^4G_{4.5} - z^4F^o_{3.5} $ & 5502.067 & \emph{N/U} \\

\noalign{\smallskip}\hline\hline
\end{tabular}
\end{table}

For \ali, \si, and \crii\ none of the preselected lines with \synflag=\Yes\ or \Un\ have high-quality $gf$-values (except for one \crii\ line at 524.68~nm). Experimental work on these species is highly needed and the 21 lines concerned are listed in Table~\ref{tab:gfneed2}.

\subsection{Access to data}
\label{sect:access}

The data comprising the Gaia-ESO line list in the wavelength ranges from 4750~\AA\ to 6850~\AA\ and from 8488~\AA\ to 8950~\AA\ are made available in electronic form at the CDS.
The atomic data are stored in a single table with one record for each transition. Hyperfine structure components and different isotopes are included as separate transitions, where applicable (see Sect.~\ref{sect:hfs}).
HFS components belonging to the same fine structure transition can be identified by having the exact same label and $J$ value for both the lower and the upper levels.
Both the preselected lines and the background line list are included. Preselected lines can be identified by having both non-empty \gfflag\ and \synflag\ entries.
Here we describe the contents of the data fields included for each transition.

\begin{description}
   \item{\texttt{Element}} Element symbol (e.g., Fe).
   \item{\texttt{Ion}} Ionisation stage (1=neutral, 2=singly ionised, 3=doubly ionised).
   \item{\texttt{Isotope}} Isotope information for \texttt{Element}: 0 if only one isotope is present in the list, otherwise the baryon number is given.
   \item{\texttt{lambda}} Wavelength of the transition in air, in units of \AA.
   \item{\texttt{r\_lambda}} Reference code for \texttt{lambda}.
   \item{\texttt{loggf}} Logarithm (base 10) of the product of the oscillator strength of the transition and the statistical weight of the lower level.
   \item{\texttt{e\_loggf}} Uncertainty in \texttt{loggf} for experimental $gf$-values if available.
   \item{\texttt{r\_loggf}} Reference code for \texttt{loggf}. This field may contain several labels, combined with + or |. When the labels are combined with + then \texttt{loggf} is the average from more than one source, while | means that relative $gf$-values from the first source were re-normalised to an absolute scale using accurate lifetime measurements from the second source
       (see the respective subsection on Si, Ti, Fe, Cu, and Zn
       in \refSectGf).
      %in \citealt{Heiter_etal_2020_S1}).
      %(see Sects.~\ref{sect:Si}, \ref{sect:Ti}, \ref{sect:Fe}, \ref{sect:Cu}, \ref{sect:Zn}).
   \item{\gfflag} Flag indicating the relative quality for \texttt{loggf} (usage recommendation, values \emph{Y/U/N}, see Sect.~\ref{sect:gfflag}), for preselected lines only.
   \item{\synflag} Flag indicating the blending quality of the spectral line for synthesis (usage recommendation based on spectra of the Sun and Arcturus, values \emph{Y/U/N}, see Sect.~\ref{sect:synflag}), for preselected lines only.
   \item{\texttt{Label\_low}} A string of characters specifying the electronic configuration and the term designation for the lower energy level. Taken from the VALD database, which follows the notation adopted by the NIST Atomic Spectra Database\footnote{see \url{http://www.astro.uu.se/valdwiki/AtomicLevel}} .
   \item{\texttt{J\_low}} Total angular momentum quantum number $J$ for the lower level.
   \item{\texttt{E\_low}} Lower level energy in units of eV.
   \item{\texttt{r\_E\_low}} Reference code for \texttt{E\_low}.
   \item{\texttt{Label\_up}} Configuration and term label for the upper energy level.
   \item{\texttt{J\_up}} Total angular momentum quantum number $J$ for the upper level.
   \item{\texttt{E\_up}} Upper level energy in units of eV.
   \item{\texttt{r\_E\_up}} Reference code for \texttt{E\_up}.
   \item{\texttt{Rad\_damp}} Logarithm of the radiative damping width in units of rad/s (see Sect.~\ref{sect:other}).
   \item{\texttt{r\_Rad\_damp}} Reference code for \texttt{Rad\_damp}.
   \item{\texttt{Sta\_damp}} Logarithm of the Stark broadening width per unit perturber number density at 10\,000~K, in units of rad s$^{-1}$ cm$^{3}$ (see Sect.~\ref{sect:other}).
   \item{\texttt{r\_Sta\_damp}} Reference code for \texttt{Sta\_damp}.
   \item{\texttt{Vdw\_damp}} Van der Waals broadening parameter
      (see Sect.~\ref{sect:other} and
      \refSectHcoll).
      %\citealt{Heiter_etal_2020_S2}).
      %(see Sect.~\ref{sect:H-broadening}).
      Values greater than zero were obtained from ABO theory and are expressed in a packed notation where the integer component is the broadening cross-section, $\sigma$, in atomic units, and the decimal component is the dimensionless velocity parameter, $\alpha$. Values less than zero are the logarithm of the broadening width per unit perturber number density at 10\,000~K in units of rad s$^{-1}$ cm$^{3}$.
   \item{\texttt{r\_Vdw\_damp}} Reference code for \texttt{Vdw\_damp}.
\end{description}

The atomic data are available in their entirety in a machine-readable form at the CDS. Tables~\ref{tab:atomdat1} and \ref{tab:atomdat2} list excerpted data fields
from the single electronic data table containing all the atomic line parameters, for guidance regarding the content of the electronic table.
When the values for the fields \texttt{e\_loggf}, \texttt{Rad\_damp}, \texttt{Sta\_damp}, and \texttt{Vdw\_damp} are equal to zero, this means that they are not available for the respective transition.
Field names starting with \texttt{r\_} contain reference codes, i.e., labels to be used with the provided BibTeX file (see below).
Tables~\ref{tab:atomdat1} and \ref{tab:atomdat2} do not contain the fields \texttt{Label\_low} and \texttt{Label\_up}, which are however present for each line in the electronic table. 
As a single example,
the strings given in fields \texttt{Label\_low} and \texttt{Label\_up} for the \ci\ line
listed in Tables~\ref{tab:atomdat1} and \ref{tab:atomdat2}
 are \texttt{'LS 2s2.2p.3p 1P'} and \texttt{'LS 2s2.2p.4d 1P*'}, respectively (where multiple white spaces were collapsed into one).

The molecular data are stored in similar form, with two data fields for both element symbol and isotope information (\texttt{Element\_1, Element\_2, Isotope\_1, Isotope\_2}), a subset of the remaining data fields
(\texttt{lambda, loggf, E\_low, E\_up, Rad\_damp}), and the following additional data fields:
\begin{description}
   \item{\texttt{State\_low}} Lower level electronic state symbol.
   \item{\texttt{State\_up}} Upper level electronic state symbol.
   \item{\texttt{v\_low}} Lower level vibrational quantum number.
   \item{\texttt{v\_up}} Upper level vibrational quantum number.
   \item{\texttt{Branch}} Label indicating branch -- a string of characters in most cases consisting of the branch designation (e.g. P, Q, R), the spin components (1, 2, ...), the rotationless parity (e or f) whenever lambda doubling has been computed, and in parentheses the total angular momentum quantum number ($J$) for the lower level. For further explanations we refer to the references for the sources of molecular data (see Table~\ref{tab:molref}).
   \item{\texttt{r\_mol}} Reference code.
\end{description}
Note that the quantum numbers were not used for the calculation of synthetic spectra within the GES.
The data are available in their entirety in a machine-readable form at the CDS. An excerpt is shown in Table~\ref{tab:moldat} for guidance regarding the content of the electronic table.

\begin{table*}
\renewcommand{\tabcolsep}{0.9mm}
\caption{Examples for atomic data, part 1.}
\label{tab:atomdat1}
\centering
\begin{tabular}{lcrrlrrlcc}
\hline\hline\noalign{\smallskip}
E & \texttt{Ion} & I & \texttt{lambda   } & \texttt{r\_lambda   } & \texttt{loggf    } & \texttt{e\_loggf    } & \texttt{r\_loggf    } & gf & syn \\
\noalign{\smallskip}\hline\noalign{\smallskip}
\texttt{Li} & \texttt{1} & \texttt{7} & \texttt{6707.7635} & \texttt{LN} & \texttt{-0.002} & \texttt{0.000} & \texttt{1998PhRvA..57.1652Y} & \texttt{Y} & \texttt{U} \\
\texttt{C} & \texttt{1} & \texttt{0} & \texttt{6587.6100} & \texttt{NIST10} & \texttt{-1.021} & \texttt{0.000} & \texttt{1993AAS...99..179H} & \texttt{Y} & \texttt{Y} \\
\texttt{O} & \texttt{1} & \texttt{0} & \texttt{6158.1858} & \texttt{NIST10} & \texttt{-0.296} & \texttt{0.000} & \texttt{1991JPhB...24.3943H} & \texttt{Y} & \texttt{U} \\
\texttt{Na} & \texttt{1} & \texttt{0} & \texttt{5889.9509} & \texttt{NIST10} & \texttt{0.108} & \texttt{0.001} & \texttt{1996PhRvL..76.2862V} & \texttt{Y} & \texttt{Y} \\
\texttt{Mg} & \texttt{1} & \texttt{0} & \texttt{5172.6843} & \texttt{NIST10} & \texttt{-0.450} & \texttt{0.040} & \texttt{ATJL} & \texttt{Y} & \texttt{Y} \\
\texttt{Al} & \texttt{1} & \texttt{0} & \texttt{6698.6730} & \texttt{WSM} & \texttt{-1.870} & \texttt{0.000} & \texttt{1995JPhB...28.3485M} & \texttt{U} & \texttt{Y} \\
\texttt{Si} & \texttt{1} & \texttt{0} & \texttt{5690.4250} & \texttt{GARZ} & \texttt{-1.773} & \texttt{0.000} & \texttt{GARZ|BL} & \texttt{Y} & \texttt{Y} \\
\texttt{Si} & \texttt{2} & \texttt{0} & \texttt{6347.1087} & \texttt{K12} & \texttt{0.169} & \texttt{0.000} & \texttt{S-G+BBC+MER} & \texttt{Y} & \texttt{U} \\
\texttt{S} & \texttt{1} & \texttt{0} & \texttt{6743.4832} & \texttt{K04} & \texttt{-1.310} & \texttt{0.000} & \texttt{2006JPhB...39.2861Z+GESMCHF} & \texttt{U} & \texttt{U} \\
\texttt{Ca} & \texttt{1} & \texttt{0} & \texttt{5260.3870} & \texttt{SR+Sm} & \texttt{-1.719} & \texttt{0.011} & \texttt{SR} & \texttt{Y} & \texttt{Y} \\
\texttt{Ca} & \texttt{2} & \texttt{0} & \texttt{8542.0910} & \texttt{T} & \texttt{-0.463} & \texttt{0.000} & \texttt{T} & \texttt{Y} & \texttt{Y} \\
\texttt{Sc} & \texttt{1} & \texttt{0} & \texttt{5356.0868} & \texttt{LD} & \texttt{-0.189} & \texttt{0.000} & \texttt{LD} & \texttt{Y} & \texttt{Y} \\
\texttt{Sc} & \texttt{2} & \texttt{0} & \texttt{5657.8960} & \texttt{LD} & \texttt{-0.603} & \texttt{0.000} & \texttt{LD} & \texttt{Y} & \texttt{Y} \\
\texttt{Ti} & \texttt{1} & \texttt{0} & \texttt{4758.1178} & \texttt{LGWSC} & \texttt{0.510} & \texttt{0.000} & \texttt{2013ApJS..205...11L} & \texttt{Y} & \texttt{Y} \\
\texttt{Ti} & \texttt{2} & \texttt{0} & \texttt{5418.7678} & \texttt{WLSC} & \texttt{-2.130} & \texttt{0.000} & \texttt{2013ApJS..208...27W} & \texttt{Y} & \texttt{Y} \\
\texttt{V} & \texttt{1} & \texttt{0} & \texttt{5604.9012} & \texttt{K09} & \texttt{-1.644} & \texttt{0.000} & \texttt{1985AA...153..109W} & \texttt{Y} & \texttt{Y} \\
\texttt{V} & \texttt{2} & \texttt{0} & \texttt{6028.2680} & \texttt{K10} & \texttt{-2.122} & \texttt{0.000} & \texttt{K10} & \texttt{N} & \texttt{U} \\
\texttt{Cr} & \texttt{1} & \texttt{0} & \texttt{4936.3350} & \texttt{WLHK} & \texttt{-0.250} & \texttt{0.000} & \texttt{SLS} & \texttt{Y} & \texttt{Y} \\
\texttt{Cr} & \texttt{2} & \texttt{0} & \texttt{5246.7680} & \texttt{PGBH} & \texttt{-2.466} & \texttt{0.000} & \texttt{PGBH} & \texttt{Y} & \texttt{U} \\
\texttt{Mn} & \texttt{1} & \texttt{0} & \texttt{5394.6191} & \texttt{K07} & \texttt{-4.070} & \texttt{0.000} & \texttt{1984MNRAS.208..147B} & \texttt{Y} & \texttt{Y} \\
\texttt{Fe} & \texttt{1} & \texttt{0} & \texttt{4802.8797} & \texttt{BWL} & \texttt{-1.514} & \texttt{0.051} & \texttt{BWL} & \texttt{Y} & \texttt{Y} \\
\texttt{Fe} & \texttt{2} & \texttt{0} & \texttt{4923.9212} & \texttt{K13} & \texttt{-1.260} & \texttt{0.000} & \texttt{2009AA...497..611M} & \texttt{Y} & \texttt{Y} \\
\texttt{Co} & \texttt{1} & \texttt{0} & \texttt{5331.4121} & \texttt{K08} & \texttt{-2.461} & \texttt{0.000} & \texttt{1999ApJS..122..557N} & \texttt{Y} & \texttt{Y} \\
\texttt{Ni} & \texttt{1} & \texttt{0} & \texttt{5424.6450} & \texttt{K08} & \texttt{-2.770} & \texttt{0.000} & \texttt{1985JQSRT..33..307D} & \texttt{Y} & \texttt{Y} \\
\texttt{Cu} & \texttt{1} & \texttt{65} & \texttt{5782.0385} & \texttt{K12} & \texttt{-2.817} & \texttt{0.000} & \texttt{KR|1989ZPhyD..11..287C} & \texttt{Y} & \texttt{Y} \\
\texttt{Zn} & \texttt{1} & \texttt{0} & \texttt{4810.5280} & \texttt{Wa} & \texttt{-0.160} & \texttt{0.000} & \texttt{1980AA....84..361B|1980ZPhyA.298..249K} & \texttt{Y} & \texttt{U} \\
\texttt{Sr} & \texttt{1} & \texttt{0} & \texttt{6791.0160} & \texttt{GC} & \texttt{-0.730} & \texttt{0.000} & \texttt{GC} & \texttt{Y} & \texttt{U} \\
\texttt{Y} & \texttt{1} & \texttt{0} & \texttt{6222.5775} & \texttt{K06} & \texttt{-1.452} & \texttt{0.000} & \texttt{K06} & \texttt{N} & \texttt{U} \\
\texttt{Y} & \texttt{2} & \texttt{0} & \texttt{4883.6821} & \texttt{K11} & \texttt{0.190} & \texttt{0.000} & \texttt{BBEHL} & \texttt{Y} & \texttt{Y} \\
\texttt{Zr} & \texttt{1} & \texttt{0} & \texttt{6127.4400} & \texttt{BGHL} & \texttt{-1.060} & \texttt{0.000} & \texttt{BGHL} & \texttt{Y} & \texttt{Y} \\
\texttt{Zr} & \texttt{2} & \texttt{0} & \texttt{5112.2700} & \texttt{LNAJ} & \texttt{-0.850} & \texttt{0.000} & \texttt{LNAJ} & \texttt{Y} & \texttt{U} \\
\texttt{Nb} & \texttt{1} & \texttt{0} & \texttt{5095.2930} & \texttt{DLa} & \texttt{-1.048} & \texttt{0.000} & \texttt{1986JQSRT..35..281D} & \texttt{Y} & \texttt{N} \\
\texttt{Mo} & \texttt{1} & \texttt{0} & \texttt{5751.4080} & \texttt{WBb} & \texttt{-1.014} & \texttt{0.000} & \texttt{WBb} & \texttt{Y} & \texttt{Y} \\
\texttt{Ru} & \texttt{1} & \texttt{0} & \texttt{4869.1530} & \texttt{WSL} & \texttt{-0.830} & \texttt{0.000} & \texttt{WSL} & \texttt{Y} & \texttt{U} \\
\texttt{Ba} & \texttt{2} & \texttt{135} & \texttt{5853.6663} & \texttt{MW} & \texttt{-0.907} & \texttt{0.000} & \texttt{1992AA...255..457D} & \texttt{Y} & \texttt{Y} \\
\texttt{La} & \texttt{2} & \texttt{139} & \texttt{4804.0020} & \texttt{LBS} & \texttt{-2.092} & \texttt{0.000} & \texttt{LBS} & \texttt{Y} & \texttt{U} \\
\texttt{Ce} & \texttt{2} & \texttt{0} & \texttt{5274.2290} & \texttt{LSCI} & \texttt{0.130} & \texttt{0.000} & \texttt{LSCI} & \texttt{Y} & \texttt{Y} \\
\texttt{Pr} & \texttt{2} & \texttt{141} & \texttt{5322.6729} & \texttt{ILW} & \texttt{-2.870} & \texttt{0.000} & \texttt{2007PhyS...76..577L} & \texttt{Y} & \texttt{Y} \\
\texttt{Nd} & \texttt{2} & \texttt{143} & \texttt{4914.3624} & \texttt{HLSC} & \texttt{-1.226} & \texttt{0.000} & \texttt{HLSC} & \texttt{Y} & \texttt{U} \\
\texttt{Sm} & \texttt{2} & \texttt{149} & \texttt{4836.6422} & \texttt{LD-HS} & \texttt{-2.758} & \texttt{0.000} & \texttt{LD-HS} & \texttt{Y} & \texttt{U} \\
\texttt{Eu} & \texttt{2} & \texttt{153} & \texttt{5818.7119} & \texttt{LWHS} & \texttt{-2.572} & \texttt{0.000} & \texttt{LWHS} & \texttt{Y} & \texttt{U} \\
\texttt{Gd} & \texttt{2} & \texttt{0} & \texttt{4865.0410} & \texttt{DLSC} & \texttt{-0.870} & \texttt{0.000} & \texttt{DLSC} & \texttt{Y} & \texttt{U} \\
\texttt{Dy} & \texttt{2} & \texttt{0} & \texttt{5169.6900} & \texttt{WLN} & \texttt{-1.950} & \texttt{0.000} & \texttt{WLN} & \texttt{Y} & \texttt{U} \\

\noalign{\smallskip}\hline\hline
\end{tabular}
\tablefoot{
Full table available at CDS. Column headers give field names, where E, I, gf, and syn are \texttt{Element, Isotope, gf\_flag}, and \texttt{synflag}, respectively. See Sect.~\ref{sect:access} and ReadMe file at CDS for description of fields. 
}
\end{table*}

\begin{table*}
\renewcommand{\tabcolsep}{0.9mm}
\caption{Examples for atomic data, part 2.}
\label{tab:atomdat2}
\centering
\begin{tabular}{lcrrlrrlrlrlrl}
\hline\hline\noalign{\smallskip}
E & \texttt{Ion} & \texttt{J\_low} & \texttt{E\_low} & \texttt{r\_E\_low} & \texttt{J\_up} & \texttt{E\_up} & \texttt{r\_E\_up} & R & \texttt{r\_Rad\_damp} & S & \texttt{r\_Sta\_damp} & V & \texttt{r\_Vdw\_damp} \\
\noalign{\smallskip}\hline\noalign{\smallskip}
\texttt{Li} & \texttt{1} & \texttt{ 0.5} & \texttt{ 0.000} & \texttt{LN      } & \texttt{ 1.5} & \texttt{ 1.848} & \texttt{LN      } & \texttt{ 7.56} & \texttt{CDROM18 } & \texttt{-5.78} & \texttt{CDROM18 } & \texttt{ 346.236} & \texttt{BA-J;BPM}    \\
\texttt{C} & \texttt{1} & \texttt{ 1.0} & \texttt{ 8.537} & \texttt{NIST10  } & \texttt{ 1.0} & \texttt{10.419} & \texttt{NIST10  } & \texttt{ 8.10} & \texttt{K10     } & \texttt{-3.44} & \texttt{K10     } & \texttt{1953.319} & \texttt{BA-J;BPM}    \\
\texttt{O} & \texttt{1} & \texttt{ 3.0} & \texttt{10.741} & \texttt{NIST10  } & \texttt{10.0} & \texttt{12.754} & \texttt{NIST10  } & \texttt{ 7.62} & \texttt{CDROM18 } & \texttt{-3.96} & \texttt{CDROM18 } & \texttt{1915.322} & \texttt{BA-J;BPM}    \\
\texttt{Na} & \texttt{1} & \texttt{ 0.5} & \texttt{ 0.000} & \texttt{NIST10  } & \texttt{ 1.5} & \texttt{ 2.104} & \texttt{NIST10  } & \texttt{ 7.80} & \texttt{CDROM18 } & \texttt{-5.64} & \texttt{CDROM18 } & \texttt{ 407.273} & \texttt{BA-J;BPM}    \\
\texttt{Mg} & \texttt{1} & \texttt{ 1.0} & \texttt{ 2.712} & \texttt{NIST10  } & \texttt{ 1.0} & \texttt{ 5.108} & \texttt{NIST10  } & \texttt{ 7.99} & \texttt{CDROM18 } & \texttt{-5.47} & \texttt{CDROM18 } & \texttt{ 729.238} & \texttt{BA-J;BPM}    \\
\texttt{Al} & \texttt{1} & \texttt{ 0.5} & \texttt{ 3.143} & \texttt{WSM     } & \texttt{ 0.5} & \texttt{ 4.993} & \texttt{WSM     } & \texttt{ 0.00} & \texttt{        } & \texttt{ 0.00} & \texttt{        } & \texttt{   0.000} & \texttt{        }    \\
\texttt{Si} & \texttt{1} & \texttt{ 1.0} & \texttt{ 4.930} & \texttt{GARZ    } & \texttt{ 1.0} & \texttt{ 7.108} & \texttt{GARZ    } & \texttt{ 8.54} & \texttt{K07     } & \texttt{-4.57} & \texttt{K07     } & \texttt{1770.220} & \texttt{BA-J;BPM}    \\
\texttt{Si} & \texttt{2} & \texttt{ 0.5} & \texttt{ 8.121} & \texttt{K12     } & \texttt{ 1.5} & \texttt{10.074} & \texttt{K12     } & \texttt{ 9.08} & \texttt{K12     } & \texttt{-5.68} & \texttt{K12     } & \texttt{  -7.690} & \texttt{K12     }    \\
\texttt{S} & \texttt{1} & \texttt{ 1.0} & \texttt{ 7.866} & \texttt{K04     } & \texttt{ 0.0} & \texttt{ 9.704} & \texttt{K04     } & \texttt{ 7.60} & \texttt{K04     } & \texttt{-4.54} & \texttt{K04     } & \texttt{   0.000} & \texttt{        }    \\
\texttt{Ca} & \texttt{1} & \texttt{ 1.0} & \texttt{ 2.521} & \texttt{SR+Sm   } & \texttt{ 2.0} & \texttt{ 4.878} & \texttt{SR+Sm   } & \texttt{ 7.90} & \texttt{K07     } & \texttt{-5.76} & \texttt{K07     } & \texttt{ 421.260} & \texttt{BA-J;BPM}    \\
\texttt{Ca} & \texttt{2} & \texttt{ 2.5} & \texttt{ 1.700} & \texttt{T       } & \texttt{ 1.5} & \texttt{ 3.151} & \texttt{T       } & \texttt{ 8.21} & \texttt{K10     } & \texttt{-5.70} & \texttt{K10     } & \texttt{ 291.275} & \texttt{BA-J;BPM}    \\
\texttt{Sc} & \texttt{1} & \texttt{ 3.5} & \texttt{ 1.865} & \texttt{LD      } & \texttt{ 2.5} & \texttt{ 4.179} & \texttt{LD      } & \texttt{ 8.19} & \texttt{K09     } & \texttt{-5.82} & \texttt{K09     } & \texttt{ 412.271} & \texttt{BA-J;BPM}    \\
\texttt{Sc} & \texttt{2} & \texttt{ 2.0} & \texttt{ 1.507} & \texttt{LD      } & \texttt{ 2.0} & \texttt{ 3.698} & \texttt{LD      } & \texttt{ 8.18} & \texttt{K09     } & \texttt{-6.55} & \texttt{K09     } & \texttt{  -7.860} & \texttt{K09     }    \\
\texttt{Ti} & \texttt{1} & \texttt{ 5.0} & \texttt{ 2.249} & \texttt{LGWSC   } & \texttt{ 5.0} & \texttt{ 4.854} & \texttt{LGWSC   } & \texttt{ 8.08} & \texttt{K10     } & \texttt{-6.04} & \texttt{K10     } & \texttt{ 326.246} & \texttt{BA-J;BPM}    \\
\texttt{Ti} & \texttt{2} & \texttt{ 2.5} & \texttt{ 1.582} & \texttt{WLSC    } & \texttt{ 2.5} & \texttt{ 3.869} & \texttt{WLSC    } & \texttt{ 8.15} & \texttt{K10     } & \texttt{-6.59} & \texttt{K10     } & \texttt{  -7.850} & \texttt{K10     }    \\
\texttt{V} & \texttt{1} & \texttt{ 0.5} & \texttt{ 1.043} & \texttt{K09     } & \texttt{ 1.5} & \texttt{ 3.255} & \texttt{K09     } & \texttt{ 8.02} & \texttt{K09     } & \texttt{-5.96} & \texttt{K09     } & \texttt{ 343.237} & \texttt{BA-J;BPM}    \\
\texttt{V} & \texttt{2} & \texttt{ 1.0} & \texttt{ 2.491} & \texttt{K10     } & \texttt{ 2.0} & \texttt{ 4.547} & \texttt{K10     } & \texttt{ 8.33} & \texttt{K10     } & \texttt{-6.45} & \texttt{K10     } & \texttt{  -7.870} & \texttt{K10     }    \\
\texttt{Cr} & \texttt{1} & \texttt{ 3.0} & \texttt{ 3.113} & \texttt{WLHK    } & \texttt{ 4.0} & \texttt{ 5.624} & \texttt{WLHK    } & \texttt{ 7.91} & \texttt{K10     } & \texttt{-5.97} & \texttt{K10     } & \texttt{ 342.245} & \texttt{BA-J;BPM}    \\
\texttt{Cr} & \texttt{2} & \texttt{ 0.5} & \texttt{ 3.714} & \texttt{PGBH    } & \texttt{ 1.5} & \texttt{ 6.076} & \texttt{PGBH    } & \texttt{ 8.32} & \texttt{K10     } & \texttt{-6.49} & \texttt{K10     } & \texttt{ 186.227} & \texttt{BA-J;BPM}    \\
\texttt{Mn} & \texttt{1} & \texttt{ 2.5} & \texttt{ 0.000} & \texttt{K07     } & \texttt{ 3.5} & \texttt{ 2.298} & \texttt{K07     } & \texttt{ 4.29} & \texttt{K07     } & \texttt{-6.27} & \texttt{K07     } & \texttt{ 219.252} & \texttt{BA-J;BPM}    \\
\texttt{Fe} & \texttt{1} & \texttt{ 3.0} & \texttt{ 3.642} & \texttt{K07     } & \texttt{ 2.0} & \texttt{ 6.222} & \texttt{K07     } & \texttt{ 7.96} & \texttt{K07     } & \texttt{-6.03} & \texttt{K07     } & \texttt{ 356.244} & \texttt{BA-J;BPM}    \\
\texttt{Fe} & \texttt{2} & \texttt{ 2.5} & \texttt{ 2.891} & \texttt{K13     } & \texttt{ 1.5} & \texttt{ 5.408} & \texttt{K13     } & \texttt{ 8.49} & \texttt{K13     } & \texttt{-6.53} & \texttt{K13     } & \texttt{ 175.202} & \texttt{BA-J;BPM}    \\
\texttt{Co} & \texttt{1} & \texttt{ 0.5} & \texttt{ 1.785} & \texttt{K08     } & \texttt{ 1.5} & \texttt{ 4.110} & \texttt{K08     } & \texttt{ 8.12} & \texttt{K08     } & \texttt{-6.15} & \texttt{K08     } & \texttt{ 303.260} & \texttt{BA-J;BPM}    \\
\texttt{Ni} & \texttt{1} & \texttt{ 1.0} & \texttt{ 1.951} & \texttt{K08     } & \texttt{ 2.0} & \texttt{ 4.236} & \texttt{K08     } & \texttt{ 8.07} & \texttt{K08     } & \texttt{-6.16} & \texttt{K08     } & \texttt{ 232.270} & \texttt{BA-J;BPM}    \\
\texttt{Cu} & \texttt{1} & \texttt{ 1.5} & \texttt{ 1.642} & \texttt{K12     } & \texttt{ 0.5} & \texttt{ 3.786} & \texttt{K12     } & \texttt{ 8.18} & \texttt{K12     } & \texttt{-6.07} & \texttt{K12     } & \texttt{  -7.790} & \texttt{K12     }    \\
\texttt{Zn} & \texttt{1} & \texttt{ 2.0} & \texttt{ 4.078} & \texttt{Wa      } & \texttt{ 1.0} & \texttt{ 6.655} & \texttt{Wa      } & \texttt{ 0.00} & \texttt{        } & \texttt{ 0.00} & \texttt{        } & \texttt{ 676.238} & \texttt{BA-J;BPM}    \\
\texttt{Sr} & \texttt{1} & \texttt{ 0.0} & \texttt{ 1.775} & \texttt{GC      } & \texttt{ 1.0} & \texttt{ 3.600} & \texttt{GC      } & \texttt{ 0.00} & \texttt{        } & \texttt{ 0.00} & \texttt{        } & \texttt{   0.000} & \texttt{        }    \\
\texttt{Y} & \texttt{1} & \texttt{ 1.5} & \texttt{ 0.000} & \texttt{K06     } & \texttt{ 2.5} & \texttt{ 1.992} & \texttt{K06     } & \texttt{ 6.96} & \texttt{K06     } & \texttt{-5.97} & \texttt{K06     } & \texttt{  -7.740} & \texttt{K06     }    \\
\texttt{Y} & \texttt{2} & \texttt{ 4.0} & \texttt{ 1.084} & \texttt{K11     } & \texttt{ 3.0} & \texttt{ 3.622} & \texttt{K11     } & \texttt{ 8.47} & \texttt{K11     } & \texttt{-6.40} & \texttt{K11     } & \texttt{  -7.760} & \texttt{K11     }    \\
\texttt{Zr} & \texttt{1} & \texttt{ 4.0} & \texttt{ 0.154} & \texttt{BGHL    } & \texttt{ 4.0} & \texttt{ 2.177} & \texttt{BGHL    } & \texttt{ 0.00} & \texttt{        } & \texttt{ 0.00} & \texttt{        } & \texttt{ 260.244} & \texttt{BA-J;BPM}    \\
\texttt{Zr} & \texttt{2} & \texttt{ 1.5} & \texttt{ 1.665} & \texttt{LNAJ    } & \texttt{ 1.5} & \texttt{ 4.090} & \texttt{LNAJ    } & \texttt{ 0.00} & \texttt{        } & \texttt{ 0.00} & \texttt{        } & \texttt{   0.000} & \texttt{        }    \\
\texttt{Nb} & \texttt{1} & \texttt{ 3.5} & \texttt{ 0.086} & \texttt{DLa     } & \texttt{ 3.5} & \texttt{ 2.519} & \texttt{DLa     } & \texttt{ 0.00} & \texttt{        } & \texttt{ 0.00} & \texttt{        } & \texttt{   0.000} & \texttt{        }    \\
\texttt{Mo} & \texttt{1} & \texttt{ 2.0} & \texttt{ 1.420} & \texttt{WBb     } & \texttt{ 2.0} & \texttt{ 3.575} & \texttt{WBb     } & \texttt{ 0.00} & \texttt{        } & \texttt{ 0.00} & \texttt{        } & \texttt{   0.000} & \texttt{        }    \\
\texttt{Ru} & \texttt{1} & \texttt{ 4.0} & \texttt{ 0.928} & \texttt{WSL     } & \texttt{ 4.0} & \texttt{ 3.473} & \texttt{WSL     } & \texttt{ 0.00} & \texttt{        } & \texttt{ 0.00} & \texttt{        } & \texttt{   0.000} & \texttt{        }    \\
\texttt{Ba} & \texttt{2} & \texttt{ 1.5} & \texttt{ 0.604} & \texttt{MW      } & \texttt{ 1.5} & \texttt{ 2.722} & \texttt{MW      } & \texttt{ 0.00} & \texttt{        } & \texttt{ 0.00} & \texttt{        } & \texttt{ 365.264} & \texttt{BA-J;BPM}    \\
\texttt{La} & \texttt{2} & \texttt{ 1.0} & \texttt{ 0.235} & \texttt{LBS     } & \texttt{ 1.0} & \texttt{ 2.815} & \texttt{LBS     } & \texttt{ 0.00} & \texttt{        } & \texttt{ 0.00} & \texttt{        } & \texttt{   0.000} & \texttt{        }    \\
\texttt{Ce} & \texttt{2} & \texttt{ 6.5} & \texttt{ 1.044} & \texttt{LSCI    } & \texttt{ 5.5} & \texttt{ 3.395} & \texttt{LSCI    } & \texttt{ 0.00} & \texttt{        } & \texttt{ 0.00} & \texttt{        } & \texttt{   0.000} & \texttt{        }    \\
\texttt{Pr} & \texttt{2} & \texttt{ 6.0} & \texttt{ 0.483} & \texttt{ILW     } & \texttt{ 5.0} & \texttt{ 2.811} & \texttt{ILW     } & \texttt{ 0.00} & \texttt{        } & \texttt{ 0.00} & \texttt{        } & \texttt{   0.000} & \texttt{        }    \\
\texttt{Nd} & \texttt{2} & \texttt{ 5.5} & \texttt{ 0.380} & \texttt{HLSC    } & \texttt{ 4.5} & \texttt{ 2.902} & \texttt{HLSC    } & \texttt{ 0.00} & \texttt{        } & \texttt{ 0.00} & \texttt{        } & \texttt{   0.000} & \texttt{        }    \\
\texttt{Sm} & \texttt{2} & \texttt{ 2.5} & \texttt{ 0.104} & \texttt{LD-HS   } & \texttt{ 1.5} & \texttt{ 2.667} & \texttt{LD-HS   } & \texttt{ 0.00} & \texttt{        } & \texttt{ 0.00} & \texttt{        } & \texttt{   0.000} & \texttt{        }    \\
\texttt{Eu} & \texttt{2} & \texttt{ 2.0} & \texttt{ 1.230} & \texttt{LWHS    } & \texttt{ 3.0} & \texttt{ 3.361} & \texttt{LWHS    } & \texttt{ 0.00} & \texttt{        } & \texttt{ 0.00} & \texttt{        } & \texttt{   0.000} & \texttt{        }    \\
\texttt{Gd} & \texttt{2} & \texttt{ 2.5} & \texttt{ 1.157} & \texttt{DLSC    } & \texttt{ 1.5} & \texttt{ 3.704} & \texttt{DLSC    } & \texttt{ 0.00} & \texttt{        } & \texttt{ 0.00} & \texttt{        } & \texttt{   0.000} & \texttt{        }    \\
\texttt{Dy} & \texttt{2} & \texttt{ 7.5} & \texttt{ 0.103} & \texttt{WLN     } & \texttt{ 7.5} & \texttt{ 2.500} & \texttt{WLN     } & \texttt{ 0.00} & \texttt{        } & \texttt{ 0.00} & \texttt{        } & \texttt{   0.000} & \texttt{        }    \\

\noalign{\smallskip}\hline\hline
\end{tabular}
\tablefoot{
Full table available at CDS. Column headers give field names, where E, R, S, and V are \texttt{Element}, \texttt{Rad\_damp}, \texttt{Sta\_damp}, and \texttt{Vdw\_damp}, respectively. See Sect.~\ref{sect:access} and ReadMe file at CDS for description of fields. 
}
\end{table*}

\begin{table*}
\renewcommand{\tabcolsep}{0.9mm}
\caption{Examples for molecular data.}
\label{tab:moldat}
\centering
\begin{tabular}{lcrrrrrrrrrrrrl}
\hline\hline\noalign{\smallskip}
E1 & E2 & I1 & I2 & \texttt{lambda   } & \texttt{loggf    } & \texttt{E\_low    } & \texttt{E\_up     } & R & Sl & Su & \texttt{v\_low} & \texttt{v\_up} & \texttt{Branch} & \texttt{r\_mol    } \\
\noalign{\smallskip}\hline\noalign{\smallskip}
\texttt{C} & \texttt{H} & \texttt{12} & \texttt{1} & \texttt{4750.010} & \texttt{-3.113} & \texttt{1.208} & \texttt{3.818} & \texttt{6.522} & \texttt{X} & \texttt{A} & \texttt{3} & \texttt{2} & \texttt{R2f(11.5)} & \texttt{2014AA...571A..47M} \\
\texttt{C} & \texttt{H} & \texttt{12} & \texttt{1} & \texttt{4750.088} & \texttt{-3.117} & \texttt{1.208} & \texttt{3.818} & \texttt{6.522} & \texttt{X} & \texttt{A} & \texttt{3} & \texttt{2} & \texttt{R1e(12.5)} & \texttt{2014AA...571A..47M} \\
\texttt{C} & \texttt{H} & \texttt{12} & \texttt{1} & \texttt{4750.050} & \texttt{-5.854} & \texttt{1.198} & \texttt{3.808} & \texttt{14.283} & \texttt{X} & \texttt{B} & \texttt{2} & \texttt{1} & \texttt{R12(17.5)} & \texttt{2014AA...571A..47M} \\
\texttt{C} & \texttt{H} & \texttt{13} & \texttt{1} & \texttt{4750.085} & \texttt{-3.482} & \texttt{2.095} & \texttt{4.705} & \texttt{6.290} & \texttt{X} & \texttt{A} & \texttt{4} & \texttt{3} & \texttt{R1e(24.5)} & \texttt{2014AA...571A..47M} \\
\texttt{N} & \texttt{H} & \texttt{14} & \texttt{1} & \texttt{4750.087} & \texttt{-6.597} & \texttt{1.136} & \texttt{3.746} & \texttt{6.391} & \texttt{X} & \texttt{A} & \texttt{3} & \texttt{0} & \texttt{R1e(5.0)} & \texttt{T.Masseron.priv.comm} \\
\texttt{N} & \texttt{H} & \texttt{14} & \texttt{1} & \texttt{4750.587} & \texttt{-5.503} & \texttt{1.815} & \texttt{4.424} & \texttt{6.336} & \texttt{X} & \texttt{A} & \texttt{3} & \texttt{0} & \texttt{P3e(19.0)} & \texttt{T.Masseron.priv.comm} \\
\texttt{N} & \texttt{H} & \texttt{14} & \texttt{1} & \texttt{4750.777} & \texttt{-7.782} & \texttt{1.815} & \texttt{4.424} & \texttt{6.330} & \texttt{X} & \texttt{A} & \texttt{3} & \texttt{0} & \texttt{Q23e(19.0)} & \texttt{T.Masseron.priv.comm} \\
\texttt{N} & \texttt{H} & \texttt{14} & \texttt{1} & \texttt{4750.806} & \texttt{-5.460} & \texttt{1.815} & \texttt{4.424} & \texttt{6.330} & \texttt{X} & \texttt{A} & \texttt{3} & \texttt{0} & \texttt{P2f(20.0)} & \texttt{T.Masseron.priv.comm} \\
\texttt{O} & \texttt{H} & \texttt{16} & \texttt{1} & \texttt{4751.112} & \texttt{-6.767} & \texttt{2.617} & \texttt{5.226} & \texttt{8.960} & \texttt{X} & \texttt{A} & \texttt{5} & \texttt{2} & \texttt{P12e(17.5)} & \texttt{T.Masseron.priv.comm} \\
\texttt{O} & \texttt{H} & \texttt{16} & \texttt{1} & \texttt{4751.586} & \texttt{-6.010} & \texttt{3.566} & \texttt{6.175} & \texttt{9.294} & \texttt{X} & \texttt{A} & \texttt{4} & \texttt{0} & \texttt{R1e(33.5)} & \texttt{T.Masseron.priv.comm} \\
\texttt{O} & \texttt{H} & \texttt{16} & \texttt{1} & \texttt{4751.861} & \texttt{-7.328} & \texttt{2.498} & \texttt{5.107} & \texttt{7.083} & \texttt{X} & \texttt{A} & \texttt{4} & \texttt{1} & \texttt{P12e(20.5)} & \texttt{T.Masseron.priv.comm} \\
\texttt{O} & \texttt{H} & \texttt{16} & \texttt{1} & \texttt{4752.954} & \texttt{-8.063} & \texttt{3.566} & \texttt{6.174} & \texttt{9.149} & \texttt{X} & \texttt{A} & \texttt{4} & \texttt{0} & \texttt{Q21e(33.5)} & \texttt{T.Masseron.priv.comm} \\
\texttt{C} & \texttt{C} & \texttt{12} & \texttt{12} & \texttt{4750.079} & \texttt{-3.539} & \texttt{1.858} & \texttt{4.468} & \texttt{0.000} & \texttt{a} & \texttt{d} & \texttt{9} & \texttt{10} & \texttt{sR32(29)} & \texttt{2013JQSRT.124...11B} \\
\texttt{C} & \texttt{C} & \texttt{12} & \texttt{12} & \texttt{4750.130} & \texttt{-6.061} & \texttt{1.608} & \texttt{4.218} & \texttt{0.000} & \texttt{a} & \texttt{d} & \texttt{8} & \texttt{9} & \texttt{pR13(20)} & \texttt{2013JQSRT.124...11B} \\
\texttt{C} & \texttt{C} & \texttt{12} & \texttt{13} & \texttt{4750.040} & \texttt{-6.833} & \texttt{0.012} & \texttt{2.622} & \texttt{0.000} & \texttt{a} & \texttt{d} & \texttt{0} & \texttt{1} & \texttt{oQ13(06)} & \texttt{2014ApJS..211....5R} \\
\texttt{C} & \texttt{C} & \texttt{13} & \texttt{13} & \texttt{4750.000} & \texttt{-1.086} & \texttt{0.193} & \texttt{2.803} & \texttt{0.000} & \texttt{a} & \texttt{d} & \texttt{0} & \texttt{1} & \texttt{P2(6)} & \texttt{Quercy.priv.comm} \\
\texttt{C} & \texttt{N} & \texttt{12} & \texttt{14} & \texttt{4750.026} & \texttt{-4.015} & \texttt{0.886} & \texttt{3.496} & \texttt{0.000} & \texttt{X} & \texttt{A} & \texttt{11} & \texttt{3} & \texttt{Qc(235)} & \texttt{2014ApJS..210...23B} \\
\texttt{C} & \texttt{N} & \texttt{12} & \texttt{14} & \texttt{4750.043} & \texttt{-5.373} & \texttt{0.007} & \texttt{2.617} & \texttt{0.000} & \texttt{X} & \texttt{A} & \texttt{7} & \texttt{0} & \texttt{Qc(045)} & \texttt{2014ApJS..210...23B} \\
\texttt{C} & \texttt{N} & \texttt{12} & \texttt{15} & \texttt{4879.879} & \texttt{-3.393} & \texttt{1.537} & \texttt{4.077} & \texttt{0.000} & \texttt{X} & \texttt{B} & \texttt{0} & \texttt{3} & \texttt{Rb(595)} & \texttt{2014ApJS..214...26S} \\
\texttt{C} & \texttt{N} & \texttt{13} & \texttt{14} & \texttt{4750.006} & \texttt{-3.602} & \texttt{2.119} & \texttt{4.729} & \texttt{0.000} & \texttt{X} & \texttt{A} & \texttt{19} & \texttt{9} & \texttt{Rc(015)} & \texttt{2014ApJS..214...26S} \\
\texttt{Mg} & \texttt{H} & \texttt{24} & \texttt{1} & \texttt{4750.116} & \texttt{-0.939} & \texttt{0.467} & \texttt{3.077} & \texttt{0.000} & \texttt{X} & \texttt{A} & \texttt{3} & \texttt{2} & \texttt{R2f(13.5)} & \texttt{2013ApJS..207...26H} \\
\texttt{Mg} & \texttt{H} & \texttt{24} & \texttt{1} & \texttt{4750.144} & \texttt{-1.713} & \texttt{1.086} & \texttt{3.696} & \texttt{0.000} & \texttt{X} & \texttt{B} & \texttt{5} & \texttt{9} & \texttt{Re(22.5)} & \texttt{T.Masseron.priv.comm} \\
\texttt{Mg} & \texttt{H} & \texttt{25} & \texttt{1} & \texttt{5007.964} & \texttt{0.412} & \texttt{0.662} & \texttt{3.137} & \texttt{0.000} & \texttt{X} & \texttt{A} & \texttt{0} & \texttt{0} & \texttt{R2e(31.5)} & \texttt{2013ApJS..207...26H} \\
\texttt{Mg} & \texttt{H} & \texttt{26} & \texttt{1} & \texttt{5008.187} & \texttt{0.412} & \texttt{0.662} & \texttt{3.137} & \texttt{0.000} & \texttt{X} & \texttt{A} & \texttt{0} & \texttt{0} & \texttt{R2e(31.5)} & \texttt{2013ApJS..207...26H} \\
\texttt{Si} & \texttt{H} & \texttt{28} & \texttt{1} & \texttt{4750.010} & \texttt{-2.852} & \texttt{1.375} & \texttt{3.985} & \texttt{0.000} & \texttt{X} & \texttt{A} & \texttt{4} & \texttt{4} & \texttt{Q2e(22.5)} & \texttt{Kurucz.database} \\
\texttt{Si} & \texttt{H} & \texttt{28} & \texttt{1} & \texttt{4750.026} & \texttt{-3.517} & \texttt{1.362} & \texttt{3.971} & \texttt{0.000} & \texttt{X} & \texttt{A} & \texttt{6} & \texttt{8} & \texttt{R2f(3.5)} & \texttt{Kurucz.database} \\
\texttt{Si} & \texttt{H} & \texttt{28} & \texttt{1} & \texttt{4750.041} & \texttt{-3.517} & \texttt{1.362} & \texttt{3.971} & \texttt{0.000} & \texttt{X} & \texttt{A} & \texttt{6} & \texttt{8} & \texttt{R2e(3.5)} & \texttt{Kurucz.database} \\
\texttt{Si} & \texttt{H} & \texttt{28} & \texttt{1} & \texttt{4750.119} & \texttt{-1.745} & \texttt{1.405} & \texttt{4.015} & \texttt{0.000} & \texttt{X} & \texttt{A} & \texttt{5} & \texttt{6} & \texttt{Q1f(18.5)} & \texttt{Kurucz.database} \\
\texttt{Ca} & \texttt{H} & \texttt{40} & \texttt{1} & \texttt{4840.955} & \texttt{-8.362} & \texttt{0.141} & \texttt{2.701} & \texttt{0.000} & \texttt{X} & \texttt{A} & \texttt{0} & \texttt{5} & \texttt{SR21e(016)} & \texttt{B.Plez.priv.comm} \\
\texttt{Ca} & \texttt{H} & \texttt{40} & \texttt{1} & \texttt{4841.032} & \texttt{-8.416} & \texttt{0.158} & \texttt{2.718} & \texttt{0.000} & \texttt{X} & \texttt{A} & \texttt{0} & \texttt{5} & \texttt{SR21e(017)} & \texttt{B.Plez.priv.comm} \\
\texttt{Ca} & \texttt{H} & \texttt{40} & \texttt{1} & \texttt{4748.793} & \texttt{-6.341} & \texttt{0.411} & \texttt{3.021} & \texttt{0.000} & \texttt{X} & \texttt{B} & \texttt{0} & \texttt{5} & \texttt{P2e(027)} & \texttt{B.Plez.priv.comm} \\
\texttt{Ca} & \texttt{H} & \texttt{40} & \texttt{1} & \texttt{4753.848} & \texttt{-9.479} & \texttt{0.411} & \texttt{3.018} & \texttt{0.000} & \texttt{X} & \texttt{B} & \texttt{0} & \texttt{5} & \texttt{PQ12e(027)} & \texttt{B.Plez.priv.comm} \\
\texttt{Fe} & \texttt{H} & \texttt{56} & \texttt{1} & \texttt{4748.793} & \texttt{-6.341} & \texttt{0.411} & \texttt{3.021} & \texttt{0.000} & \texttt{X} & \texttt{F} & \texttt{0} & \texttt{5} & \texttt{P2e(027)} & \texttt{B.Plez.priv.comm} \\
\texttt{Fe} & \texttt{H} & \texttt{56} & \texttt{1} & \texttt{4753.848} & \texttt{-9.479} & \texttt{0.411} & \texttt{3.018} & \texttt{0.000} & \texttt{X} & \texttt{F} & \texttt{0} & \texttt{5} & \texttt{PQ12e(027)} & \texttt{B.Plez.priv.comm} \\
\texttt{Fe} & \texttt{H} & \texttt{56} & \texttt{1} & \texttt{4754.121} & \texttt{-6.333} & \texttt{0.411} & \texttt{3.018} & \texttt{0.000} & \texttt{X} & \texttt{F} & \texttt{0} & \texttt{5} & \texttt{P1e(028)} & \texttt{B.Plez.priv.comm} \\
\texttt{Fe} & \texttt{H} & \texttt{56} & \texttt{1} & \texttt{4754.306} & \texttt{-6.543} & \texttt{0.877} & \texttt{3.484} & \texttt{0.000} & \texttt{X} & \texttt{F} & \texttt{0} & \texttt{5} & \texttt{R1e(042)} & \texttt{B.Plez.priv.comm} \\
\texttt{Ti} & \texttt{O} & \texttt{46} & \texttt{16} & \texttt{4750.674} & \texttt{0.046} & \texttt{4.105} & \texttt{4.694} & \texttt{6.521} & \texttt{d} & \texttt{b} & \texttt{11} & \texttt{8} & \texttt{Q(184.0)} & \texttt{B.Plez.priv.comm} \\
\texttt{Ti} & \texttt{O} & \texttt{47} & \texttt{16} & \texttt{4750.672} & \texttt{-8.717} & \texttt{3.193} & \texttt{3.782} & \texttt{6.831} & \texttt{X} & \texttt{A} & \texttt{13} & \texttt{3} & \texttt{SR21(168.0)} & \texttt{B.Plez.priv.comm} \\
\texttt{Ti} & \texttt{O} & \texttt{48} & \texttt{16} & \texttt{4750.672} & \texttt{-8.992} & \texttt{3.104} & \texttt{3.693} & \texttt{7.084} & \texttt{E} & \texttt{B} & \texttt{3} & \texttt{4} & \texttt{Q1(146.0)} & \texttt{B.Plez.priv.comm} \\
\texttt{Ti} & \texttt{O} & \texttt{49} & \texttt{16} & \texttt{4750.675} & \texttt{-4.167} & \texttt{4.584} & \texttt{5.173} & \texttt{7.643} & \texttt{E} & \texttt{B} & \texttt{9} & \texttt{11} & \texttt{P3(192.0)} & \texttt{B.Plez.priv.comm} \\
\texttt{Ti} & \texttt{O} & \texttt{50} & \texttt{16} & \texttt{4750.673} & \texttt{-5.596} & \texttt{2.472} & \texttt{3.061} & \texttt{7.210} & \texttt{E} & \texttt{B} & \texttt{6} & \texttt{7} & \texttt{QR12(74.0)} & \texttt{B.Plez.priv.comm} \\
\texttt{V} & \texttt{O} & \texttt{51} & \texttt{16} & \texttt{5025.989} & \texttt{-6.632} & \texttt{0.291} & \texttt{2.757} & \texttt{0.000} & \texttt{X} & \texttt{A} & \texttt{2} & \texttt{14} & \texttt{Q4e(023)} & \texttt{B.Plez.priv.comm} \\
\texttt{V} & \texttt{O} & \texttt{51} & \texttt{16} & \texttt{5027.160} & \texttt{-6.601} & \texttt{0.294} & \texttt{2.760} & \texttt{0.000} & \texttt{X} & \texttt{A} & \texttt{2} & \texttt{14} & \texttt{Q4e(024)} & \texttt{B.Plez.priv.comm} \\
\texttt{V} & \texttt{O} & \texttt{51} & \texttt{16} & \texttt{5054.819} & \texttt{-5.435} & \texttt{0.478} & \texttt{2.930} & \texttt{0.000} & \texttt{X} & \texttt{B} & \texttt{3} & \texttt{12} & \texttt{Q4e(038)} & \texttt{B.Plez.priv.comm} \\
\texttt{V} & \texttt{O} & \texttt{51} & \texttt{16} & \texttt{4750.684} & \texttt{-3.953} & \texttt{0.526} & \texttt{3.135} & \texttt{0.000} & \texttt{X} & \texttt{C} & \texttt{1} & \texttt{6} & \texttt{RQ32e(077)} & \texttt{B.Plez.priv.comm} \\
\texttt{Zr} & \texttt{O} & \texttt{90} & \texttt{16} & \texttt{4750.673} & \texttt{0.260} & \texttt{1.518} & \texttt{4.127} & \texttt{0.000} & \texttt{X} & \texttt{H} & \texttt{10} & \texttt{11} & \texttt{R2e(068)} & \texttt{B.Plez.priv.comm} \\
\texttt{Zr} & \texttt{O} & \texttt{91} & \texttt{16} & \texttt{4750.672} & \texttt{-4.525} & \texttt{0.174} & \texttt{2.783} & \texttt{0.000} & \texttt{X} & \texttt{G} & \texttt{0} & \texttt{4} & \texttt{P2e(009)} & \texttt{B.Plez.priv.comm} \\
\texttt{Zr} & \texttt{O} & \texttt{92} & \texttt{16} & \texttt{4750.675} & \texttt{0.012} & \texttt{1.834} & \texttt{4.443} & \texttt{0.000} & \texttt{X} & \texttt{H} & \texttt{7} & \texttt{8} & \texttt{R1e(136)} & \texttt{B.Plez.priv.comm} \\
\texttt{Zr} & \texttt{O} & \texttt{94} & \texttt{16} & \texttt{4750.673} & \texttt{-2.682} & \texttt{0.390} & \texttt{2.999} & \texttt{0.000} & \texttt{X} & \texttt{H} & \texttt{1} & \texttt{1} & \texttt{OP12e(045)} & \texttt{B.Plez.priv.comm} \\
\texttt{Zr} & \texttt{O} & \texttt{96} & \texttt{16} & \texttt{4750.672} & \texttt{-2.568} & \texttt{1.284} & \texttt{3.893} & \texttt{0.000} & \texttt{X} & \texttt{G} & \texttt{4} & \texttt{9} & \texttt{R2e(115)} & \texttt{B.Plez.priv.comm} \\

\noalign{\smallskip}\hline\hline
\end{tabular}
\tablefoot{
Full table available at CDS. Column headers give field names, where E1, E2, I1, I2, R, Sl, and Su are \texttt{Element\_1, Element\_2, Isotope\_1, Isotope\_2, Rad\_damp, State\_low}, and \texttt{State\_up}, respectively. See Sect.~\ref{sect:access} and ReadMe file at CDS for description of fields. 
}
\end{table*}

We strongly encourage users of the Gaia-ESO line list to cite, in addition to this overview article, the individual sources for the atomic and molecular data used in a particular work.
It is important that providers of atomic data receive credit for their work by citing the original publications. This is also a prerequisite for the continued funding of this type of research.
To facilitate citations of original sources we provide, together with the data tables, a BibTeX file with the relevant entries.

% place all figures and tables from this section here
\clearpage

%----------------------------------------------------------------------------

\begin{acknowledgements}
%
% contributions by others
%
We are thankful for the contributions of Enrico Maiorca, Matthew~P. Ruffoni, and Jennifer Sobeck to the line list work.
We thank Robert~L. Kurucz for information on his calculations.
%
% individual authors
%
U.H. and A.J.K. acknowledge support from the Swedish National Space Agency (SNSA/Rymdstyrelsen).
K.L. acknowledges funds from the European Research Council (ERC) under the European Union's Horizon 2020 research and innovation programme (Grant agreement No. 852977).
\v{S}.M. acknowledges support from the Research Council of Lithuania (LMT) through grant LAT-08/2016.
T.M. acknowledges support from the State Research Agency (AEI) of the Spanish Ministry of Science, Innovation and Universities (MCIU) and the European Regional Development Fund (FEDER) under grant AYA2017-88254-P.
J.C.P. acknowledges support from the STFC of the UK.
T.B. and U.H. were supported by the project grant ``The New Milky Way'' from the Knut and Alice Wallenberg Foundation.
A.R.C. is supported in part by the Australian Research Council through a Discovery Early Career Researcher Award (DE190100656).
Parts of this research were supported by the Australian Research Council Centre of Excellence for All Sky Astrophysics in 3 Dimensions (ASTRO 3D), through project number CE170100013.
P.J. acknowledges funds from FONDECYT Iniciaci\'on Grant number 11170174.
R.S. acknowledges support from NCN through grant 2014/15/B/ST9/03981 and from the Polish Ministry of Science and Higher Education.
%
% databases etc.
%
This work has made use of the VALD database, operated at Uppsala University, the Institute of Astronomy RAS in Moscow, and the University of Vienna.
Based on data products from observations made with ESO Telescopes at the La Silla Paranal Observatory under programme ID 188.B-3002. These data products have been processed by the Cambridge Astronomy Survey Unit (CASU) at the Institute of Astronomy, University of Cambridge, and by the FLAMES/UVES reduction team at INAF/Osservatorio Astrofisico di Arcetri. These data have been obtained from the Gaia-ESO Survey Data Archive, prepared and hosted by the Wide Field Astronomy Unit, Institute for Astronomy, University of Edinburgh, which is funded by the UK Science and Technology Facilities Council.
This work was partly supported by the European Union FP7 programme through ERC grant number 320360 and by the Leverhulme Trust through grant RPG-2012-541. We acknowledge the support from INAF and Ministero dell' Istruzione, dell' Universit\`a' e della Ricerca (MIUR) in the form of the grant "Premiale VLT 2012". The results presented here benefit from discussions held during the Gaia-ESO workshops and conferences supported by the ESF (European Science Foundation) through the GREAT Research Network Programme.

\end{acknowledgements}

%----------------------------------------------------------------------------

\bibliographystyle{aa}
\bibliography{GES-Linelists,GESreferencesv5all,GES-publications,broadening,GaiaESO,isotopes}

%----------------------------------------------------------------------------

\begin{appendix}

%----------------------------------------------------------------------------
% Appendix: Auxiliary tables
%----------------------------------------------------------------------------

\section{Auxiliary tables}
\label{sect:aux}

This appendix provides five tables with auxiliary data referred to in Sects.~\ref{sect:preselection}, ~\ref{sect:molecules}, and ~\ref{sect:dataneeds}.

\begin{table}[ht]
\renewcommand{\tabcolsep}{1.5mm}
\caption{List of transitions for which multiple fine-structure components were merged into one line in the preselected line list.}
\label{tab:fine}
\centering
\begin{tabular}{lllclrr}
\hline\hline\noalign{\smallskip}
Species & Wavelength & $N$ & \multicolumn{2}{c}{Actual} & \multicolumn{2}{c}{GES line list} \\
        & [\AA]      &   & $J_{\rm low}$ & $J_{\rm upp}$ & $J_{\rm low}$ & $J_{\rm upp}$ \\
\noalign{\smallskip}\hline\noalign{\smallskip}
\oi     & 6158.1858  & 3 & 3.0 & 2.0, 3.0, 4.0 & 3.0 & 10.0 \\ 
\nai    & 4982.814   & 2 & 1.5 & 1.5, 2.5      & 1.5 &  4.5 \\
\nai    & 5688.205   & 2 & 1.5 & 1.5, 2.5      & 1.5 &  4.5 \\
\mgi    & 8717.825   & 3 & 2.0 & 1.0, 2.0, 3.0 & 2.0 &  7.0 \\
\mgi    & 8736.019   & 6 & 1.0 & 2.0           & 7.0 & 10.0 \\
        &            &   & 2.0 & 2.0, 3.0      &     &      \\
        &            &   & 3.0 & 2.0, 3.0, 4.0 &     &      \\
\ali    & 8773.896   & 2 & 2.5 & 2.5, 3.5      & 2.5 & 6.5  \\
\caii   & 5339.188   & 3 & 2.5 & 3.5           & 6.5 & 8.5  \\
        &            &   & 3.5 & 3.5, 4.5      &     &      \\
\caii   & 6456.875   & 3 & 2.5 & 3.5           & 6.5 & 8.5  \\
        &            &   & 3.5 & 3.5, 4.5      &     &      \\
\caii   & 8927.356   & 2 & 2.5 & 2.5, 3.5      & 2.5 & 6.5  \\
\noalign{\smallskip}\hline\hline
\end{tabular}
\tablefoot{
$N$ \ldots\ Number of components merged. The following columns give the $J$ values of the lower and upper levels of the original transitions, and the $J$ values quoted in the line list. The latter were obtained by summing up the $J$ values occurring in the respective level for the merged transitions (ignoring repetitions), and adding 1 in the case of integer $J$ values and 0.5 in the case of half-integer $J$ values.
}
\end{table}

\begin{table*}[ht]
\renewcommand{\tabcolsep}{1.0mm}
\caption{Polynomial coefficients for partition functions ($Q$) for molecules.}
\label{tab:molpfs}
\centering
\begin{tabular}{lrrrrrr}
\hline\hline\noalign{\smallskip}
Mol. & $a_0$ & $a_1$ & $a_2$ & $a_3$ & $a_4$ & $a_5$ \\
\noalign{\smallskip}\hline\noalign{\smallskip}
CH    &  $-$4.91887806$\times 10^{+2}$ &    3.09155097$\times 10^{+2}$  & $-$7.70038741$\times 10^{+1}$  &    9.57241011                 & $-$5.93380866$\times 10^{-1}$ &    1.47420199$\times 10^{-2}$ \\
NH    &  $-$2.70001339$\times 10^{+2}$ &    1.77303226$\times 10^{+2}$  & $-$4.62157841$\times 10^{+1}$  &    6.04541228                 & $-$3.95886800$\times 10^{-1}$ &    1.04339634$\times 10^{-2}$ \\
OH    &  $-$4.56875469$\times 10^{+2}$ &    2.87960316$\times 10^{+2}$  & $-$7.20525364$\times 10^{+1}$  &    9.01193168                 & $-$5.62538083$\times 10^{-1}$ &    1.40661068$\times 10^{-2}$ \\
C$_2$ &  $-$1.37611619$\times 10^{+2}$ &    7.52473987$\times 10^{+1}$  & $-$1.60701760$\times 10^{+1}$  &    1.72410801                 & $-$9.08948951$\times 10^{-2}$ &    1.90678635$\times 10^{-3}$ \\
CN    &     6.81165938$\times 10^{+2}$ & $-$4.46981105$\times 10^{+2}$  &    1.17584738$\times 10^{+2}$  & $-$1.53931948$\times 10^{+1}$ &    1.00366669                 & $-$2.60084236$\times 10^{-2}$ \\
MgH   &     6.53454485$\times 10^{+2}$ & $-$4.13828321$\times 10^{+2}$  &    1.04888640$\times 10^{+2}$  & $-$1.32317730$\times 10^{+1}$ &    8.31982710$\times 10^{-1}$ & $-$2.07967780$\times 10^{-2}$ \\
SiH   &     9.01120422$\times 10^{+1}$ & $-$4.42299680$\times 10^{+1}$  &    8.35706604                  & $-$6.74290104$\times 10^{-1}$ &    1.83429417$\times 10^{-2}$ & $ $1.99216485$\times 10^{-4}$ \\
CaH   &     1.57677958$\times 10^{+3}$ & $-$9.82846806$\times 10^{+2}$  &    2.44267425$\times 10^{+2}$  & $-$3.01777064$\times 10^{+1}$ &    1.85392515                 & $-$4.52292569$\times 10^{-2}$ \\
FeH   &     0.1552109                  &    0.3983233                   &    0.6073527                   & $-$0.198406                   &    2.47056   $\times 10^{-2}$ & $-$9.90570   $\times 10^{-4}$ \\
TiO   &     5.92027276$\times 10^{+2}$ & $-$3.65351492$\times 10^{+2}$  &    9.03939514$\times 10^{+1}$  & $-$1.10869716$\times 10^{+1}$ &    6.75722876$\times 10^{-1}$ & $-$1.63144071$\times 10^{-2}$ \\
VO    &     6.62090157$\times 10^{+2}$ & $-$4.03350494$\times 10^{+2}$  &    9.82836218$\times 10^{+1}$  & $-$1.18526504$\times 10^{+1}$ &    7.08429905$\times 10^{-1}$ &                               \\
ZrO   &     4.27195765$\times 10^{+2}$ & $-$2.51905561$\times 10^{+2}$  &    5.85682500$\times 10^{+1}$  & $-$6.63032743                 &    3.67462428$\times 10^{-1}$ & $-$7.92597014$\times 10^{-3}$ \\
\noalign{\smallskip}\hline\hline
\end{tabular}
\tablefoot{
Partition functions are obtained as a function of temperature $T$ from the polynomial equation
$ \displaystyle \ln (Q) = \sum_{i=0}^{5}a_i \ln(T)^i$.
}
\end{table*}

\begin{table*}[ht]
   \caption{Lines of \feii, \nai, \sii, and \caii\ with \synflag=\Yes\ or \Un, and with \gfflag=\Un\ or \No.}
\label{tab:gfneed1}
\centering
\begin{tabular}{lcll}
\hline\hline\noalign{\smallskip}
% header
        & Transition & Wavelength & Flags \\
        &            &     [\AA]  & $gf$/syn \\
\noalign{\smallskip}\hline\noalign{\smallskip}
% data
\feii & $ b^4F_{4.5} - z^6P^o_{3.5} $ & 4993.350 & \emph{U/Y} \\
\feii & $ b^4F_{4.5} - z^6F^o_{4.5} $ & 5132.661 & \emph{U/U} \\
\feii & $ a^6S_{2.5} - z^6F^o_{2.5} $ & 5256.932 & \emph{U/U} \\
\feii & $ a^6S_{2.5} - z^6F^o_{3.5} $ & 5284.103 & \emph{U/U} \\
\feii & $ b^4G_{5.5} - w^4F^o_{4.5} $ & 5427.816 & \emph{U/Y} \\
\feii & $ b^2H_{5.5} - z^4F^o_{4.5} $ & 5534.838 & \emph{U/U} \\
\feii & $ a^4G_{5.5} - z^6F^o_{4.5} $ & 5991.371 & \emph{U/Y} \\
\feii & $ a^4G_{4.5} - z^6F^o_{3.5} $ & 6084.102 & \emph{U/Y} \\
\feii & $ a^4G_{3.5} - z^6F^o_{2.5} $ & 6113.319 & \emph{U/U} \\
\feii & $ b^4D_{0.5} - z^4P^o_{0.5} $ & 6149.246 & \emph{U/Y} \\
\feii & $ b^4D_{2.5} - z^4P^o_{1.5} $ & 6247.557 & \emph{U/U} \\
\feii & $ z^4F^o_{3.5} - c^4D_{3.5} $ & 6442.958 & \emph{U/U} \\
\feii & $ b^4D_{3.5} - z^4P^o_{2.5} $ & 6456.380 & \emph{U/Y} \\
\nai & $ z^2P^o_{1.5} - e^2S_{0.5} $ & 4751.822 & \emph{U/U} \\
\nai & $ z^2P^o_{1.5} - b^2D_{4.5} $ & 5688.205 & \emph{U/Y} \\
\nai & $ z^2P^o_{0.5} - c^2S_{0.5} $ & 6154.225 & \emph{U/Y} \\
\nai & $ z^2P^o_{1.5} - c^2S_{0.5} $ & 6160.747 & \emph{U/Y} \\
\sii & $ z^1P^o_{1} - 3s^2 3p(^2P^o_{3/2})4f\enspace ^2[\sfrac{5}{2}]_{2} $ & 5517.533 & \emph{N/U} \\
\sii & $ z^3D^o_{1} - 3s^2 3p(^2P^o_{1/2})6f\enspace ^2[\sfrac{5}{2}]_{2} $ & 5747.667 & \emph{N/U} \\
\sii & $ z^3D^o_{2} - 3s^2 3p(^2P^o_{1/2})6f\enspace ^2[\sfrac{5}{2}]_{3} $ & 5753.623 & \emph{N/U} \\
\sii & $ z^3D^o_{1} - 3s^2 3p(^2P^o_{3/2})5f\enspace ^2[\sfrac{5}{2}]_{2} $ & 6125.021 & \emph{N/U} \\
\sii & $ z^3D^o_{2} - 3s^2 3p(^2P^o_{3/2})5f\enspace ^2[\sfrac{5}{2}]_{2} $ & 6131.573 & \emph{N/Y} \\
\sii & $ z^3D^o_{2} - 3s^2 3p(^2P^o_{3/2})5f\enspace ^2[\sfrac{5}{2}]_{3} $ & 6131.852 & \emph{N/Y} \\
\sii & $ z^3D^o_{3} - 3s^2 3p(^2P^o_{3/2})5f\enspace ^2[\sfrac{5}{2}]_{3} $ & 6142.483 & \emph{N/Y} \\
\sii & $ z^3D^o_{2} - 3s^2 3p(^2P^o_{3/2})5f\enspace ^2[\sfrac{7}{2}]_{3} $ & 6145.016 & \emph{N/U} \\
\sii & $ z^3D^o_{3} - 3s^2 3p(^2P^o_{3/2})5f\enspace ^2[\sfrac{7}{2}]_{4} $ & 6155.134 & \emph{N/U} \\
\sii & $ z^3D^o_{3} - 3s^2 3p(^2P^o_{3/2})5f\enspace ^2[\sfrac{7}{2}]_{3} $ & 6155.693 & \emph{N/Y} \\
\sii & $ z^1D^o_{2} - 3s^2 3p(^2P^o_{1/2})7f\enspace ^2[\sfrac{7}{2}]_{3} $ & 6195.433 & \emph{N/U} \\
\sii & $ a^1D_{3} - 3s^2 3p(^2P^o_{3/2})10s_{1/2} \enspace (\sfrac{3}{2},\sfrac{1}{2})^o_{2} $ & 6208.541 & \emph{N/U} \\
\sii & $ z^3D^o_{1} - 3s^2 3p(^2P^o_{1/2})5f\enspace ^2[\sfrac{5}{2}]_{2} $ & 6237.319 & \emph{N/Y} \\
\sii & $ z^3D^o_{2} - 3s^2 3p(^2P^o_{1/2})5f\enspace ^2[\sfrac{7}{2}]_{3} $ & 6243.815 & \emph{N/U} \\
\sii & $ z^3D^o_{2} - 3s^2 3p(^2P^o_{1/2})5f\enspace ^2[\sfrac{5}{2}]_{3} $ & 6244.466 & \emph{N/Y} \\
\sii & $ z^1D^o_{2} - 3s^2 3p(^2P^o_{3/2})6f\enspace ^2[\sfrac{7}{2}]_{3} $ & 6414.980 & \emph{N/U} \\
\sii & $ a^1P_{1} - w^1D^o_{2} $ & 6721.848 & \emph{U/U} \\
\sii & $ z^1D^o_{2} - 3s^2 3p(^2P^o_{3/2})4f\enspace ^2[\sfrac{7}{2}]_{3} $ & 8556.776 & \emph{N/Y} \\
\sii & $ z^3F^o_{3} - 3s^2 3p(^2P^o_{3/2})5f\enspace ^2[\sfrac{9}{2}]_{4} $ & 8556.805 & \emph{N/Y} \\
\sii & $ z^3F^o_{4} - 3s^2 3p(^2P^o_{3/2})5f\enspace ^2[\sfrac{9}{2}]_{5} $ & 8648.465 & \emph{N/U} \\
\sii & $ z^3F^o_{2} - 3s^2 3p(^2P^o_{1/2})5f\enspace ^2[\sfrac{7}{2}]_{3} $ & 8728.010 & \emph{N/Y} \\
\sii & $ z^1D^o_{2} - 3s^2 3p(^2P^o_{1/2})4f\enspace ^2[\sfrac{7}{2}]_{3} $ & 8742.446 & \emph{N/Y} \\
\sii & $ a^1D_{3} - 3s^2 3p(^2P^o_{3/2})6s_{1/2} \enspace (\sfrac{3}{2},\sfrac{1}{2})^o_{2} $ & 8892.720 & \emph{U/Y} \\
\sii & $ b^1D_{2} - x^1F^o_{3} $ & 8899.231 & \emph{U/U} \\
\sii & $ a^1D_{2} - 3s^2 3p(^2P^o_{1/2})6s_{1/2} \enspace (\sfrac{1}{2},\sfrac{1}{2})^o_{1} $ & 8949.091 & \emph{U/Y} \\
\caii & $ y^2P^o_{0.5} - d^2D_{1.5} $ & 5001.479 & \emph{U/U} \\
\caii & $ z^2F^o_{6.5} - b^2G_{8.5} $ & 6456.875 & \emph{U/U} \\
\caii & $ b^2D_{1.5} - z^2F^o_{2.5} $ & 8912.068 & \emph{U/U} \\
\caii & $ b^2D_{2.5} - z^2F^o_{6.5} $ & 8927.356 & \emph{U/U} \\

\noalign{\smallskip}\hline\hline
\end{tabular}
\end{table*}

\begin{table*}[ht]
   \caption{Atomic data for \mgi\ lines with \synflag=\Yes\ or \Un.}
\label{tab:Mg1}
\centering
\begin{tabular}{cllllll}
\hline\hline\noalign{\smallskip}
% header
Transition & Wavelength & log($gf$) & Ref. & Flags & log($gf$)\tablefootmark{$a$} & log($gf$)\tablefootmark{$a$} \\
           &     [\AA]  & GES v6    &      & $gf$/syn & Exp. & Calc. \\
\noalign{\smallskip}\hline\noalign{\smallskip}
% data
$z^3P^o_0     - a^3S_1        $ & 5167.322 & $-0.931$ & (1) & \emph{Y/U} & $-0.854 \pm 0.05$ & $-0.865$ \\
$z^3P^o_1     - a^3S_1        $ & 5172.684 & $-0.450$ & (1) & \emph{Y/Y} & $-0.363 \pm 0.04$ & $-0.387$ \\
$z^3P^o_2     - a^3S_1        $ & 5183.604 & $-0.239$ & (1) & \emph{Y/Y} & $-0.168 \pm 0.04$ & $-0.166$ \\
$z^1P^o_1     - b^1D_2        $ & 5528.405 & $-0.498$ & (2) & \emph{U/Y} & $-0.547 \pm 0.02$ & $-0.513$ \\
$z^1P^o_1     - c^1S_0        $ & 5711.088 & $-1.724$ & (2) & \emph{U/Y} & $-1.842 \pm 0.05$ & $-1.742$ \\
$a^3S_1       - w^3P^o_2      $ & 6318.717 & $-2.103$ & (3) & \emph{U/U} &                   & $-2.020$ \\
$a^3S_1       - w^3P^o_1      $ & 6319.237 & $-2.324$ & (3) & \emph{U/Y} &                   & $-2.242$ \\
$a^3S_1       - w^3P^o_0      $ & 6319.493 & $-2.803$ & (3) & \emph{U/Y} &                   & $-2.719$ \\
$y^3P^o_2     - e^3D_{1,2,3}  $ & 8717.825 & $-0.866$ & (3) & \emph{U/U} &                   & $-0.930$ \\
$a^3D_{1,2,3} - w^1F^o_{2,3,4}$ & 8736.019 & $-0.356$ & (3) & \emph{U/Y} &                   &   \\
$z^1P^o_1     - a^1D_2        $ & 8806.757 & $-0.134$ & (4) & \emph{Y/Y} & $-0.144 \pm 0.03$ & $-0.128$ \\
$b^1S_0       - x^1P^o_1      $ & 8923.569 & $-1.678$ & (2) & \emph{U/Y} &                   & $-1.679$ \\
\noalign{\smallskip}\hline\hline
\end{tabular}
\tablefoot{
References: (1) \citet{ATJL}, uncertainties: 0.04~dex, (2) \citet{1990JQSRT..43..207C}, (3) \citet{1993JPhB...26.4409B}, (4) \citet{GESMCHF}.
\tablefoottext{$a$}{\citet{2017A&A...598A.102P}}
}
\end{table*}

\begin{table}[ht]
   \caption{\nii\ lines with \synflag=\Yes\ or \Un, and with \gfflag=\Un\ or \No.}
\label{tab:Ni1}
\centering
\begin{tabular}{cll}
\hline\hline\noalign{\smallskip}
% header
Transition & Wavelength & Flags \\
           &     [\AA]  & $gf$/syn \\
\noalign{\smallskip}\hline\noalign{\smallskip}
% data
$ z^3D^o_{3} - f^3F_{4} $ & 4806.987 & \emph{N/U} \\
$ z^3P^o_{2} - 3d^9(^2D_{5/2})4d\enspace ^2[\sfrac{5}{2}]_{3} $ & 4829.023 & \emph{N/Y} \\
$ z^3P^o_{2} - 3d^9(^2D_{5/2})4d\enspace ^2[\sfrac{1}{2}]_{1} $ & 4904.412 & \emph{N/U} \\
$ z^5F^o_{1} - e^5F_{1} $ & 4912.018 & \emph{N/U} \\
$ z^3P^o_{0} - 3d^9(^2D_{3/2})4d\enspace ^2[\sfrac{1}{2}]_{1} $ & 4913.973 & \emph{N/Y} \\
$ z^3G^o_{4} - f^3F_{3} $ & 4918.364 & \emph{N/Y} \\
$ z^3G^o_{3} - f^3F_{2} $ & 4935.831 & \emph{N/Y} \\
$ z^3F^o_{2} - 3d^9(^2D_{3/2})4d\enspace ^2[\sfrac{5}{2}]_{3} $ & 4946.032 & \emph{N/Y} \\
$ y^3F^o_{2} - d^1F_{3} $ & 4976.697 & \emph{U/U} \\
$ z^3F^o_{3} - 3d^9(^2D_{5/2})4d\enspace ^2[\sfrac{7}{2}]_{4} $ & 4995.650 & \emph{N/U} \\
$ z^3F^o_{3} - 3d^9(^2D_{5/2})4d\enspace ^2[\sfrac{5}{2}]_{3} $ & 5010.938 & \emph{N/Y} \\
$ z^1D^o_{2} - f^3F_{3} $ & 5032.727 & \emph{N/U} \\
$ z^1F^o_{3} - 3d^9(^2D_{3/2})4d\enspace ^2[\sfrac{7}{2}]_{4} $ & 5081.110 & \emph{N/U} \\
$ z^3D^o_{3} - 3d^9(^2D_{5/2})4d\enspace ^2[\sfrac{7}{2}]_{4} $ & 5084.096 & \emph{N/Y} \\
$ z^3D^o_{1} - 3d^9(^2D_{3/2})4d\enspace ^2[\sfrac{1}{2}]_{1} $ & 5094.411 & \emph{N/U} \\
$ z^1D^o_{2} - 3d^9(^2D_{3/2})4d\enspace ^2[\sfrac{5}{2}]_{2} $ & 5155.126 & \emph{N/U} \\
$ z^1D^o_{2} - 3d^9(^2D_{3/2})4d\enspace ^2[\sfrac{5}{2}]_{3} $ & 5155.764 & \emph{N/U} \\
$ z^1D^o_{2} - 3d^9(^2D_{3/2})4d\enspace ^2[\sfrac{3}{2}]_{2} $ & 5176.560 & \emph{N/U} \\
$ z^3F^o_{2} - 3d^9(^2D_{5/2})4d\enspace ^2[\sfrac{7}{2}]_{3} $ & 5347.708 & \emph{N/U} \\
$ y^3D^o_{3} - f^3F_{2} $ & 5392.331 & \emph{N/U} \\
$ z^3D^o_{2} - f^3F_{2} $ & 5424.536 & \emph{N/Y} \\
$ z^1F^o_{3} - 3d^9(^2D_{5/2})4d\enspace ^2[\sfrac{7}{2}]_{4} $ & 5462.493 & \emph{N/U} \\
$ z^1F^o_{3} - 3d^9(^2D_{5/2})4d\enspace ^2[\sfrac{7}{2}]_{3} $ & 5468.104 & \emph{N/U} \\
$ z^3D^o_{1} - 3d^9(^2D_{5/2})4d\enspace ^2[\sfrac{3}{2}]_{1} $ & 5475.429 & \emph{N/U} \\
$ z^1D^o_{2} - 3d^9(^2D_{5/2})4d\enspace ^2[\sfrac{5}{2}]_{2} $ & 5589.358 & \emph{N/Y} \\
$ z^1D^o_{2} - 3d^9(^2D_{5/2})4d\enspace ^2[\sfrac{7}{2}]_{3} $ & 5593.735 & \emph{N/Y} \\
$ z^1P^o_{1} - 3d^9(^2D_{3/2})4d\enspace ^2[\sfrac{3}{2}]_{2} $ & 5625.317 & \emph{N/Y} \\
$ z^1P^o_{1} - e^5F_{1} $ & 5628.342 & \emph{N/U} \\
$ z^1D^o_{2} - 3d^9(^2D_{5/2})4d\enspace ^2[\sfrac{3}{2}]_{1} $ & 5638.747 & \emph{N/U} \\
$ y^3F^o_{3} - 3d^9(^2D_{3/2})4d\enspace ^2[\sfrac{5}{2}]_{3} $ & 5641.881 & \emph{N/U} \\
$ z^1G^o_{4} - f^3F_{3} $ & 5643.078 & \emph{N/U} \\
$ y^3F^o_{3} - 3d^9(^2D_{3/2})4d\enspace ^2[\sfrac{7}{2}]_{4} $ & 5682.199 & \emph{N/Y} \\
$ z^1P^o_{1} - 3d^9(^2D_{3/2})4d\enspace ^2[\sfrac{1}{2}]_{1} $ & 5694.983 & \emph{N/U} \\
$ y^3F^o_{3} - f^3F_{4} $ & 5760.830 & \emph{N/U} \\
$ z^3D^o_{2} - 3d^9(^2D_{3/2})4d\enspace ^2[\sfrac{5}{2}]_{3} $ & 5805.217 & \emph{N/Y} \\
$ y^3F^o_{2} - 3d^9(^2D_{3/2})4d\enspace ^2[\sfrac{5}{2}]_{2} $ & 5996.730 & \emph{N/U} \\
$ y^3F^o_{2} - 3d^9(^2D_{3/2})4d\enspace ^2[\sfrac{3}{2}]_{2} $ & 6025.754 & \emph{N/U} \\
$ y^3D^o_{1} - 3d^9(^2D_{3/2})4d\enspace ^2[\sfrac{5}{2}]_{2} $ & 6086.282 & \emph{N/Y} \\
$ y^3F^o_{4} - 3d^9(^2D_{5/2})4d\enspace ^2[\sfrac{7}{2}]_{4} $ & 6111.070 & \emph{N/Y} \\
$ y^3D^o_{1} - 3d^9(^2D_{3/2})4d\enspace ^2[\sfrac{3}{2}]_{1} $ & 6130.135 & \emph{N/U} \\
$ y^3F^o_{4} - 3d^9(^2D_{5/2})4d\enspace ^2[\sfrac{5}{2}]_{3} $ & 6133.963 & \emph{N/U} \\
$ z^1P^o_{1} - 3d^9(^2D_{5/2})4d\enspace ^2[\sfrac{3}{2}]_{1} $ & 6175.367 & \emph{N/Y} \\
$ y^3F^o_{3} - 3d^9(^2D_{5/2})4d\enspace ^2[\sfrac{5}{2}]_{3} $ & 6186.711 & \emph{N/Y} \\
$ y^3F^o_{3} - 3d^9(^2D_{5/2})4d\enspace ^2[\sfrac{3}{2}]_{2} $ & 6230.089 & \emph{U/U} \\
$ y^3D^o_{3} - 3d^9(^2D_{5/2})4d\enspace ^2[\sfrac{7}{2}]_{3} $ & 6322.166 & \emph{N/U} \\
$ z^3P^o_{2} - 3d^9(^2D_{3/2})5s\enspace ^2[\sfrac{3}{2}]_{2} $ & 6370.346 & \emph{N/U} \\
$ z^3D^o_{2} - 3d^9(^2D_{5/2})4d\enspace ^2[\sfrac{3}{2}]_{1} $ & 6424.851 & \emph{N/U} \\
$ a^3P_{2} - z^1F^o_{3} $ & 6482.798 & \emph{U/Y} \\
$ y^3F^o_{2} - 3d^9(^2D_{5/2})4d\enspace ^2[\sfrac{7}{2}]_{3} $ & 6598.598 & \emph{N/Y} \\
$ y^1F^o_{3} - 3d^9(^2D_{3/2})4d\enspace ^2[\sfrac{7}{2}]_{4} $ & 6635.122 & \emph{N/U} \\
$ z^3P^o_{1} - 3d^9(^2D_{3/2})5s\enspace ^2[\sfrac{3}{2}]_{2} $ & 6772.315 & \emph{N/Y} \\
$ z^3P^o_{1} - 3d^9(^2D_{3/2})5s\enspace ^2[\sfrac{3}{2}]_{1} $ & 6842.037 & \emph{N/U} \\
$ a^1G_{4} - y^3D^o_{3} $ & 8770.678 & \emph{N/U} \\

\noalign{\smallskip}\hline\hline
\end{tabular}
\end{table}

% place all figures and tables from this section here
\clearpage
\clearpage

%\begin{comment}
% include this for single manuscript
% comment out for 3-part manuscript
%----------------------------------------------------------------------------
% Appendix: Atomic data - element by element
%----------------------------------------------------------------------------

\section{Detailed description of atomic data}
\label{sect:gf}

We discuss the data sources and quality aspects for the lines of each atomic species element by element.
For each element included in the preselected line list, the sources for transition probabilities (oscillator strengths), and the assignment of the quality flag for $gf$-values (\gfflag) are discussed.
This is supplemented by a discussion of the blending properties and the assignment of the corresponding quality flag (\synflag).
Finally, if applicable, the sources for hyperfine structure (HFS) and isotopic shift (IS) data are presented.
Sources for the $gf$-values of elements appearing only in the background line list are summarised in Appendix~\ref{sect:background}.

The following species are included in the preselected line list and linked to their respective subsections
(see Table~\refTabFlagstats\ for an overview):
%(see Table~\tabFlagstats in \citealt{Heiter_etal_2020} for an overview):
\hyperref[sect:hydrogen]{\hi}, 
\hyperref[sect:Li]{\lii}, 
\hyperref[sect:C]{\ci}, 
\hyperref[sect:O]{\oi}, 
\hyperref[sect:Na]{\nai}, 
\hyperref[sect:Mg]{\mgi}, 
\hyperref[sect:Al]{\ali}, 
\hyperref[sect:Si]{\sii}, 
\hyperref[sect:Si]{\siii}, 
\hyperref[sect:S]{\si}, 
\hyperref[sect:Ca]{\cai}, 
\hyperref[sect:Ca]{\caii}, 
\hyperref[sect:Sc]{\sci}, 
\hyperref[sect:Sc]{\scii}, 
\hyperref[sect:Ti]{\tii}, 
\hyperref[sect:Ti]{\tiii}, 
\hyperref[sect:V]{\vi}, 
\hyperref[sect:V]{\vii}, 
\hyperref[sect:Cr]{\cri}, 
\hyperref[sect:Cr]{\crii}, 
\hyperref[sect:Mn]{\mni}, 
\hyperref[sect:Fe]{\fei}, 
\hyperref[sect:Fe]{\feii}, 
\hyperref[sect:Co]{\coi}, 
\hyperref[sect:Ni]{\nii}, 
\hyperref[sect:Cu]{\cui}, 
\hyperref[sect:Zn]{\zni}, 
\hyperref[sect:Sr]{\sri}, 
\hyperref[sect:Y]{\yi}, 
\hyperref[sect:Y]{\yii}, 
\hyperref[sect:Zr]{\zri}, 
\hyperref[sect:Zr]{\zrii}, 
\hyperref[sect:Nb]{\nbi}, 
\hyperref[sect:Mo]{\moi}, 
\hyperref[sect:Ru]{\rui}, 
\hyperref[sect:Ba]{\baii}, 
\hyperref[sect:La]{\laii}, 
\hyperref[sect:Ce]{\ceii}, 
\hyperref[sect:Pr]{\prii}, 
\hyperref[sect:Nd]{\ndii}, 
\hyperref[sect:Sm]{\smii}, 
\hyperref[sect:Eu]{\euii}, 
\hyperref[sect:Gd]{\gdii}, 
\hyperref[sect:Dy]{\dyii}.

%
% Martin 1: H, Li, C, O, Na, Mg, Al, Si, S
%

\enlargethispage{-2\baselineskip}

\subsection{Hydrogen (Z=1)}
\label{sect:hydrogen}

Hydrogen being the simplest of atoms, the transition probabilities can be calculated from first principles \citep[e.g.,][p.~236]{2008oasp.book.....G}, with the source for the hydrogen Balmer and Paschen lines used here being \citet[\gfflag=\Yes]{2009JPCRD..38..565W}. %Wiese \& Fuhr 2009
% Ref. in Wiese & Fuhr: J. Baker, NIST Technical Note No. 1612, 2008.
An extended list for Paschen lines, including highly excited transitions from principal quantum numbers $>$20 can be found in \citet{CDROM18}\footnote{Available on-line at \url{http://kurucz.harvard.edu/linelists/gfall/gf0100.all}.}.

Broadening of hydrogen lines due to the presence of electrons and ions via the linear Stark effect needs to be taken into account, using for example the refined calculations by 
\citet{1999A&AS..140...93S}, % Stehle & Hutcheon (1999)
which however in practise are quite similar to the seminal work by 
\citet{1973ApJS...25...37V}. % Vidal et al. (1973)
An accurate description of resonance broadening of H lines due to collisions with other hydrogen atoms (self-broadening) is given in \citet{2000A&A...363.1091B}, %Barklem et al. 2000
which has further been improved by 
\citet{2008A&A...480..581A} % Allard et al. 2008
for the case of the H$\alpha$ 656.2~nm line. 
A recent review on hydrogen Balmer lines can be found in \citet[Sect.~4.1.1]{2016A&ARv..24....9B}, % Barklem 2016
and a ready-to-use implementation is made available by 
P. Barklem \& N. Piskunov\footnote{\url{http://ascl.net/1507.008}}. 
%\citet{2015ascl.soft07008B}. % Barklem & Piskunov, HLINOP: Hydrogen LINe OPacity in stellar atmospheres

\subsection{Lithium (Z=3)}
\label{sect:Li}

The $gf$-values for the two \lii\ 670.8~nm resonance fine structure transitions were taken from the theoretical calculation of \citet{1998PhRvA..57.1652Y}. %Yan et al. (1998)
These are essentially identical to the theoretical values quoted in \citet{YD} and \citet{2009JPCRD..38..565W}.
The predicted life-time of the $2p~^2P^o$ upper level of these transitions from these calculations agrees extremely well with the experimental measurements of \citet{1996PhST...65...48V} and \citet{1996PhRvA..54....5M}. %Volz \& Schmoranzer (1996); McAlexander et al. 1996
The transition probabilities are considered accurate (\gfflag=\Yes). 

We adopted the highly accurate isotopic splitting data of \citet{sansonetti95} from frequency-modulation spectroscopy, %Sansonetti et al. 1995
which are in excellent agreement with the Fourier-transform spectrometry measurements of \citet{REB}. %Radziemski et al. 1995
The latter authors also report HFS constants from the literature. HFS components were however not computed for the Gaia-ESO line list.
The transition probabilities for the \lii\ 610.3 nm subordinate lines in the background line list come from \citet{LN}.

\begin{figure}
   \begin{center}
      \resizebox{\hsize}{!}{\includegraphics{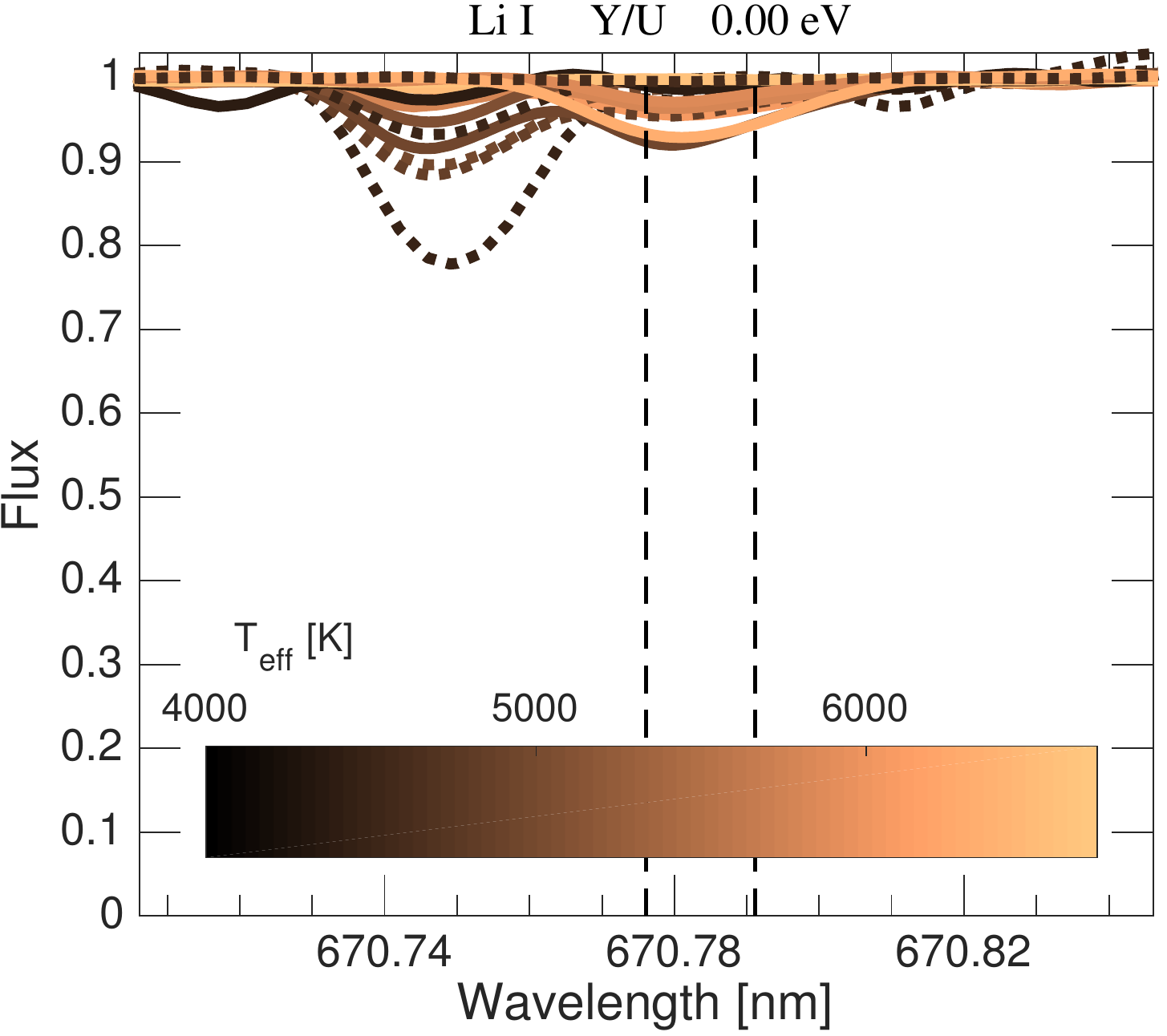}}
   \end{center}
   \caption{Observed spectra of selected Gaia FGK benchmark stars around the \lii\ 670.8~nm feature
   (see Sect.~\refSectSpectra).}
%   (see Sect.~\sectSpectra in \citealt{Heiter_etal_2020}).}
%   (see Sect.~\ref{sect:spectra})
   \label{fig:Li1GBS}
\end{figure}

The \lii\ 670.8~nm feature is often weak and blended to varying degrees in different stars (\synflag=\Un, see Fig.~\ref{fig:Li1GBS}).
Lithium abundances are therefore best derived using spectral synthesis. Unfortunately, the available atomic data in the region around the \lii\ feature are quite poor. They cannot be used directly for abundance determination and require an astrophysical calibration.
This concerns in particular two \fei\ and two \vi\ lines located within 0.4~\AA\ from the \lii\ feature.
The two lines bluewards of \lii\ are by far too weak in a synthetic spectrum of the Sun and Arcturus, while the two lines redwards of \lii\ are by far too strong, when using the original $gf$-values, which are from \citet{K07} and \citet{K08} for \fei\ and \vi, respectively.

\begin{figure}
   \begin{center}
      \resizebox{0.8\hsize}{!}{\includegraphics{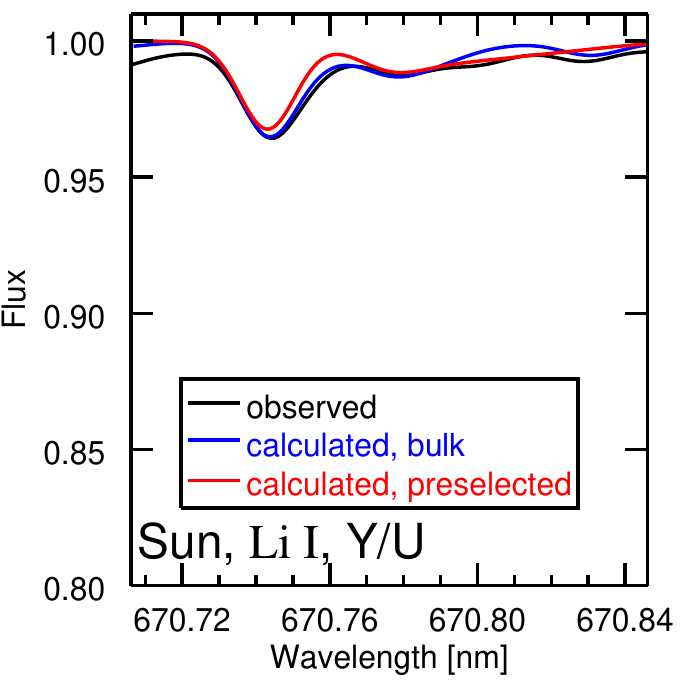}}
      \resizebox{0.8\hsize}{!}{\includegraphics{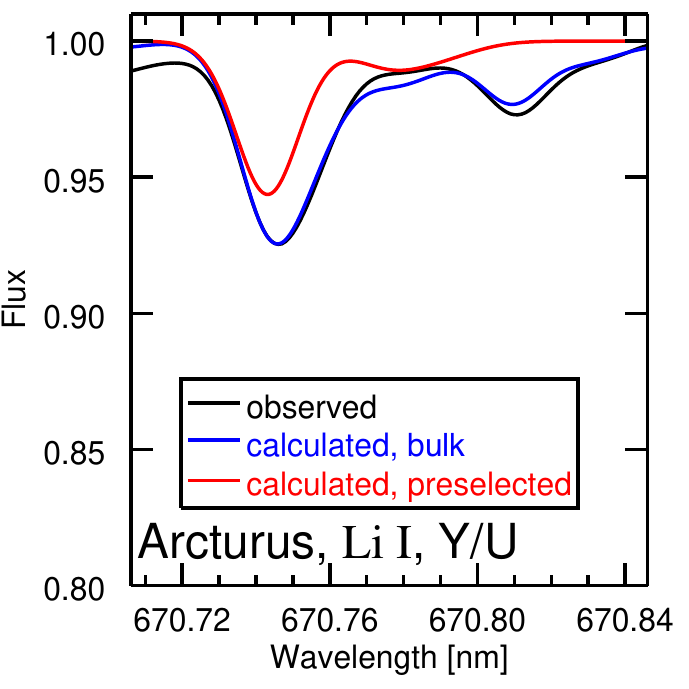}}
   \end{center}
   \caption{Line profiles around the \lii\ feature for the Sun and Arcturus. Black lines: observations, red lines: calculations including preselected spectral lines only, blue lines: calculations including blends from background line list. For Arcturus the Li abundance was set to $\log(\varepsilon_{\rm Li})+12 = -0.8$~dex, where $\varepsilon_{\rm Li}=N_{\rm Li}/N_{\rm H}$ \citep[e.g.,][]{1989ApJS...71..293B,2016A&A...595A..18G}.}
   \label{fig:Li1SunArc}
\end{figure}

In order to provide a good basis for the Gaia-ESO analysis the following changes were made ($\log gf$-old $\rightarrow$ $\log gf$-new): \fei\ 670.743~nm, $-3.917 \rightarrow -2.2$; \vi\ 670.752~nm, $-2.938 \rightarrow -0.8$; \vi\ 670.809~nm, $-2.443 \rightarrow -2.75$; \fei\ 670.828~nm, $-1.280 \rightarrow -2.85$.
The \fei\ 670.743~nm is one of the preselected lines (quality \No/\Un), while the remaining lines are part of the background line list.
These modifications result in a much improved synthesis for the Sun and Arcturus in this region (see Fig.~\ref{fig:Li1SunArc}).

\subsection{Carbon (Z=6)}
\label{sect:C}

For three permitted and one forbidden \ci\ lines in the preselected line list, we adopted the theoretical transition probabilities of \citet[]{1993AAS...99..179H} %Hibbert et al. (1993),
-- using the Length values rather than the Velocity values as recommended by the authors,
which are in good agreement with those calculated as part of the Opacity Project \citep{1989JPhB...22.3377L} %(Luo \& Pradhan 1993)
when assuming $LS$ coupling.
Unfortunately no recent accurate experimental measurements exist for these transitions but
the theoretical values are considered reliable (\gfflag=\Yes).
For the [\ci] 872.7~nm line, the here adopted $\log gf$ value by \citet{1993AAS...99..179H} is 0.03~dex larger than the recommended value by \citet{2007JPCRD..36.1287W}, %Wiese \& Fuhr 2007
which stems from the calculations of \citet{2001CaJPh..79..955T}. %Tachiev \& Froese Fischer 2001
The transition probabilities for the other \ci\ lines in the background line list come
from \citet{K10}, % Kurucz
and from NIST \citep{NIST10}, %NIST: Ralchenko et al.
which are almost exclusively based on \citet{1993AAS...99..179H} and \citet{1989JPhB...22.3377L}. %Hibbert et al. (1993), Luo \& Pradhan 1993)

The four \ci\ lines primarily used for abundance purposes are typically weak and partly blended (\synflag=\Un) except for the 658.7~nm line which is considered largely clean  (\synflag=\Yes).
Considering the low natural abundance of the $^{13}$C isotope
(see Table~\refTabIsotopes)
%(see Table~\tabIsotopes in \citealt{Heiter_etal_2020})
%(see Table~\ref{tab:isotopes})
neither isotopic nor HFS components were included.

\subsection{Oxygen (Z=8)}
\label{sect:O}

The adopted $gf$-values for the [\oi ]630.0 and 636.4~nm forbidden lines are the mean of the theoretical predictions of \citet{2000MNRAS.312..813S} and \citet{GESMCHF}: % Storey \& Zeippen (2000) and Froese Fischer \& Tachiev (MCHF)
$\log gf(630.3)=-9.715$ and $\log gf(636.3)=-10.190$.
For the [\oi ] 557.7~nm forbidden line, the predictions from \citet{1988JPhB...21.1455B}, \citet{1997A&AS..123..159G} and \citet{GESMCHF} agree well and can furthermore be put on an accurate absolute scale using the life-time measurement of the $2p~^1S$ upper level by \citet{1972JPhB....5..686C} to yield $\log gf(557.7)=-8.241$.
For the \oi\ 615.8~nm line we adopted the transition probability from \citet{1991JPhB...24.3943H} assuming $LS$ coupling.
All four theoretical transition probabilities are considered reliable (\gfflag=\Yes).
For the other \oi\ lines in the background line list the transition probabilities were adopted from the NIST database \citep{NIST10}, which are largely based on \citet{1991JPhB...24.3943H} and the Opacity Project \citep{Butl:1991} % Butler \& Zeippen 1991
assuming $LS$ coupling.

As discussed for example in \citet{2009ARA&A..47..481A}, % Asplund et al. (2009)
all of these four \oi\ lines are partly blended (\synflag=\Un\ with \synflag=\No\ for the case of 557.7~nm).
In particular, the \nii\ 630.0342~nm line in the background line list is very close to the [\oi ] line at 630.0304~nm, and its $gf$-value of $-2.11$ was explicitly taken from \citet{2003ApJ...584L.107J} % Johansson et al. 2003
to replace the value of $-2.674$ from \citet{K08} contained in the VALD database.

\subsection{Sodium (Z=11)}
\label{sect:Na}

Accurate (\gfflag=\Yes) experimental transition probabilities exist for the \nai\ 589~nm doublet from \citet{1996PhRvL..76.2862V}.
In the absence of reliable experimental data for the other preselected \nai\ lines we adopted the theoretical $gf$-values from \citet{GESMCHF}, % Froese Fischer \& Tachiev (MCHF) 
which have been classified as \gfflag=\Un.
The data for other \nai\ lines in the background line list come from the NIST database \citep{NIST10}.
Sodium is exclusively in the form of $^{23}$Na with a nuclear spin of 3/2. Sodium is thus prone to
hyperfine splitting which however has not been accounted for in the Gaia-ESO line list.
Suitable HFS data are available in 
\citet{2008JPhB...41c5001D}, %Das & Natarajan 2008
including the \nai\ D lines at 589~nm.

Besides the \nai\ D lines, three \nai\ lines are considered largely clean (\synflag=\Yes): 568.8, 615.4 and 616.0~nm.

\subsection{Magnesium (Z=12)}
\label{sect:Mg}

The $gf$-values for optical \mgi\ lines are notoriously uncertain with few experimental data to rely on until recently (see discussion of new data by \citealt{2017A&A...598A.102P} in 
Sect.~\refSectDataneeds).
%Sect.~\sectDataneeds in \citealt{Heiter_etal_2020}).
%Sect.~\ref{sect:dataneeds}).
An exception is found for the \mgi\ b triplet lines, 
which have accurate transition probabilities provided by \citet{ATJL}  % Aldenius et al. (2007)
from measurements of life-times and branching fractions (BFs).
For the other \mgi\ lines in the preselected line list we adopted theoretical values from \citet[line at 880.7~nm, \gfflag=\Yes]{GESMCHF}, % Froese Fischer & Tachiev (2012)
from the Opacity Project \citep{1993JPhB...26.4409B} %Butler et al. (1993)
under the assumption of $LS$ coupling, or from \citet{1990JQSRT..43..207C}, % Chang \& Tang (1990)
the latter two with \gfflag=\Un.
Several of these lines are considered largely clean (\synflag=\Yes).
For \mgi\ lines only appearing in the background line list we rely on values given by  \citet{NIST10} if available and otherwise by \citet{KP}.
Isotopic splitting or HFS components are not measurable in \mgi\ spectra of natural isotopic composition \citep{2017A&A...598A.102P}.

\subsection{Aluminium (Z=13)}
\label{sect:Al}

No reliable experimental data exist for the \ali\ lines in the Gaia-ESO line list. We therefore resorted to using the theoretical calculations by the Opacity Project \citep{1995JPhB...28.3485M} %Mendoza et al. (1995)
under the assumption of $LS$ coupling with a \gfflag=\Un\ rating for the five \ali\ lines in the preselected line list. 
The same $gf$-values were adopted by \citet{2015A&A...573A..25S} %Scott et al. (2015)
in their recent analysis of the solar chemical composition. 
For other \ali\ lines in the background line list we make use of \citet{K75} and \citet{WSM}.
Aluminium consists entirely of $^{27}$Al, which has nuclear spin 5/2. 
While not accounted for explicitly in the Gaia-ESO line list, good HFS data are available in for example
\citet{2007JaJAP..46..815N} %Nakai et al. 2007
and 
\citet{2005JPhB...38.4185S}, % Sur et al. 2005
see \citet[][their Table~A.2]{2017A&A...607A..75N}. % Nordlander & Lind

Of the available \ali\ lines, only 669.867~nm is considered largely unblended (\synflag=\Yes).

\subsection{Silicon (Z=14)}
\label{sect:Si}

When available, we adopted the experimental transition probabilities of \citet{GARZ} %Garz 1973
for \sii .
Those are however only reliable in a relative sense and therefore we re-normalised them to an improved absolute scale with the highly accurate, laser-induced fluorescence (LIF) life-times of the $4s~^3P_{0,1,2}$ levels measured by \citet{BL}, %O'Brian \& Lawler 1991
in the same manner as \citet{2015A&A...573A..25S} and resulting in a \gfflag=\Yes\ rating.
For a few \sii\ lines not available in \citet{GARZ}, %Garz 1973
we rely on the Opacity Project calculations of \citet{1993PhyS...48..297N}, 
which were obtained under the close-coupling approximation with the R-matrix method (\gfflag=\Un).
These data were complemented by the extensive calculations of \citet[\gfflag=\No]{K07} for a considerable number of lines.

For the two \siii\ lines at 634.711 and 637.137~nm in the preselected line list we adopted the same $gf$-values as in \citet{2015A&A...573A..25S}, % Scott et al. 2015
which were obtained from taking the mean of \citet{S-G}, \citet{BBC} and \citet{MER} % Schulz-Gulde (1969), Blanco et al. (1995), Matheron et al. (2001)
and given a \gfflag=\Yes\ evaluation.
For \sii\ lines only appearing in the background line list we made use of \citet{K07}.

\begin{figure*}
   \begin{center}
      \resizebox{0.9\hsize}{!}{\includegraphics{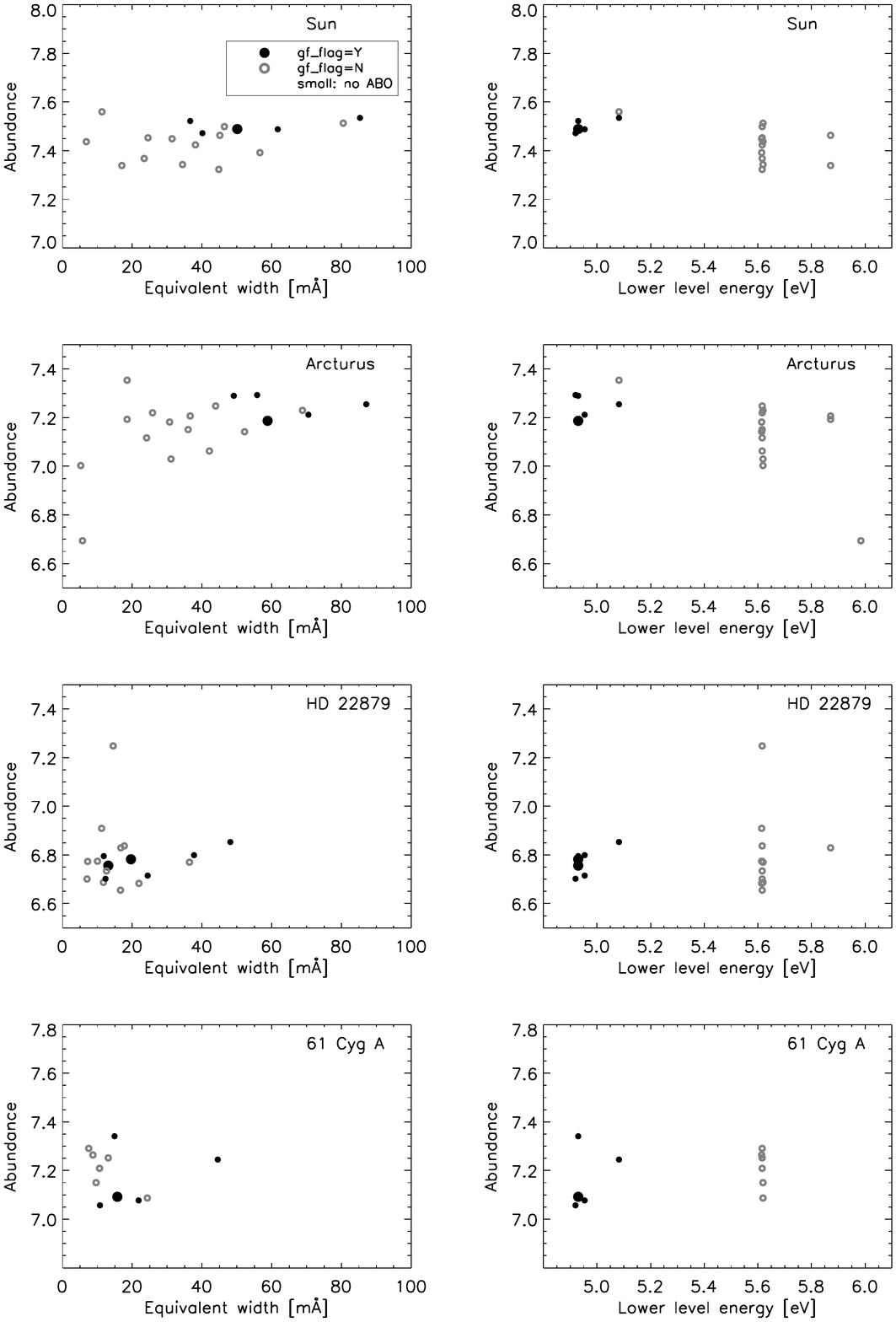}}
   \end{center}
\caption{Line-by-line \sii\ abundances determined for four benchmark stars for two groups of lines with different quality assessment of their transition probabilities, as a function of equivalent width and lower level energy. Abundances are given as $\log(\varepsilon_{\rm Si})+12$, where $\varepsilon_{\rm Si}=N_{\rm Si}/N_{\rm H}$. Only lines with \synflag=\Yes\ or \Un\ and with equivalent widths $>5$~m\AA\ are included.
}
\label{fig:abuSi} 
\end{figure*}

There are three \sii\ lines with a simultaneous \gfflag=\Yes\ and \synflag=\Yes\ rating:
569.043, 570.110 and 594.854~nm, while the remainder of the \citet{GARZ} lines in the preselected line list are partly blended.
A comparison of line abundances derived for four benchmark stars
(see Sect.~\refSectSpectra)
%(see Sect.~\sectSpectra in \citealt{Heiter_etal_2020})
%(see Sect.~\ref{sect:spectra})
for two different sets of lines (with \gfflag=\Yes\ and \gfflag=\No) can be seen in Fig.~\ref{fig:abuSi}, while mean abundances are given in Table~\ref{tab:abustat}.
The \gfflag=\No\ lines generally result in a larger scatter of line abundances than the \gfflag=\Yes\ lines, although the statistical significance is low owing to the small number of \gfflag=\Yes\ lines.

\begin{table*}
   \caption{Abundance statistics (number of lines, mean and standard deviation of $\log\varepsilon_{\rm X}+12$, where $\varepsilon_{\rm X}=N_{\rm X}/N_{\rm H}$) for four species and four Gaia FGK benchmark stars. Only lines with \synflag=\Yes\ or \Un\ and with equivalent widths $>5$~m\AA\ and $\le100$~m\AA\ were included.
   See Sect.~\refSectSpectraT, and Sects.~\ref{sect:Si}, \ref{sect:Cr}, \ref{sect:Fe}, and \ref{sect:Ni}.
   %See Sect.~\sectSpectra in \citet{Heiter_etal_2020}, and Sects.~\ref{sect:Si}, \ref{sect:Cr}, \ref{sect:Fe}, and \ref{sect:Ni}.
   % See Sects.~\ref{sect:spectra}, \ref{sect:Si}, \ref{sect:Cr}, \ref{sect:Fe}, and \ref{sect:Ni}.
   } 
\label{tab:abustat}
\centering
\begin{tabular}{lrrrrrrrrrrrr}
\hline\hline\noalign{\smallskip}
              & \multicolumn{3}{c}{Sun} & \multicolumn{3}{c}{Arcturus} & \multicolumn{3}{c}{HD 22879} & \multicolumn{3}{c}{61 Cyg A} \\
              & N & mean & stdd & N & mean & stdd & N & mean & stdd & N & mean & stdd \\
\noalign{\smallskip}\hline\noalign{\smallskip}
\noalign{\smallskip}
\sii \\
\gfflag=\Yes  &   5 & 7.50 & 0.03  &   5 & 7.25 & 0.05  &   7 & 6.77 & 0.05  &   5 & 7.16 & 0.12 \\
\gfflag=\No   &  13 & 7.43 & 0.07  &  14 & 7.13 & 0.16  &  12 & 6.80 & 0.16  &   6 & 7.21 & 0.08 \\
\noalign{\smallskip}
\cri \\
\gfflag=\Yes  &  13 & 5.61 & 0.11  &  11 & 4.96 & 0.08  &   9 & 4.53 & 0.07  &   8 & 5.26 & 0.26 \\
\gfflag=\No   &   9 & 5.75 & 0.20  &  10 & 5.01 & 0.19  &   3 & 4.58 & 0.06  &  11 & 5.32 & 0.21 \\
\noalign{\smallskip}
\fei \\
\gfflag=\Yes  &  81 & 7.52 & 0.14  &  39 & 6.88 & 0.15  & 149 & 6.42 & 0.13  &  59 & 7.07 & 0.27 \\
\gfflag=\Un   &  84 & 7.53 & 0.14  &  78 & 6.95 & 0.20  &  69 & 6.54 & 0.13  &  72 & 7.27 & 0.28 \\
\gfflag=\No   &  46 & 7.69 & 0.66  &  55 & 7.04 & 0.61  &  16 & 6.52 & 0.31  &  34 & 7.43 & 0.74 \\
\noalign{\smallskip}
\nii \\
\gfflag=\Yes  &  20 & 6.32 & 0.20  &  10 & 5.65 & 0.29  &  23 & 5.23 & 0.08  &  17 & 5.93 & 0.16 \\
\gfflag=\No   &  36 & 6.16 & 0.15  &  39 & 5.57 & 0.21  &  28 & 5.17 & 0.15  &  29 & 6.01 & 0.45 \\

\noalign{\smallskip}\hline\hline
\end{tabular}
\end{table*}

\subsection{Sulphur (Z=16)}
\label{sect:S}

For the \si\ triplets at 674.3, 674.8 and 675.7~nm and the line at 869.4~nm only theoretical $gf$-values are available. We adopted the mean transition probabilities of \citet{GESMCHF} %Froese Fischer \& Tachiev (2012)
and 
\citet{2006JPhB...39.2861Z}, %Zatsarinny \& Bartschat (2006)
which are given a \gfflag=\Un\ rating.
All of these are also partly blended (\synflag=\Un ).
The $gf$-values for other \si\ lines in the background line list mainly come from the theoretical calculations by \citet{BQZ} and \citet{K04}.

% place all figures and tables here
%\clearpage

%
% Karin: Ca
%

\subsection{Calcium (Z=20)}
\label{sect:Ca}

\begin{figure*}
   \begin{center}
      \resizebox{0.89\hsize}{!}{\includegraphics{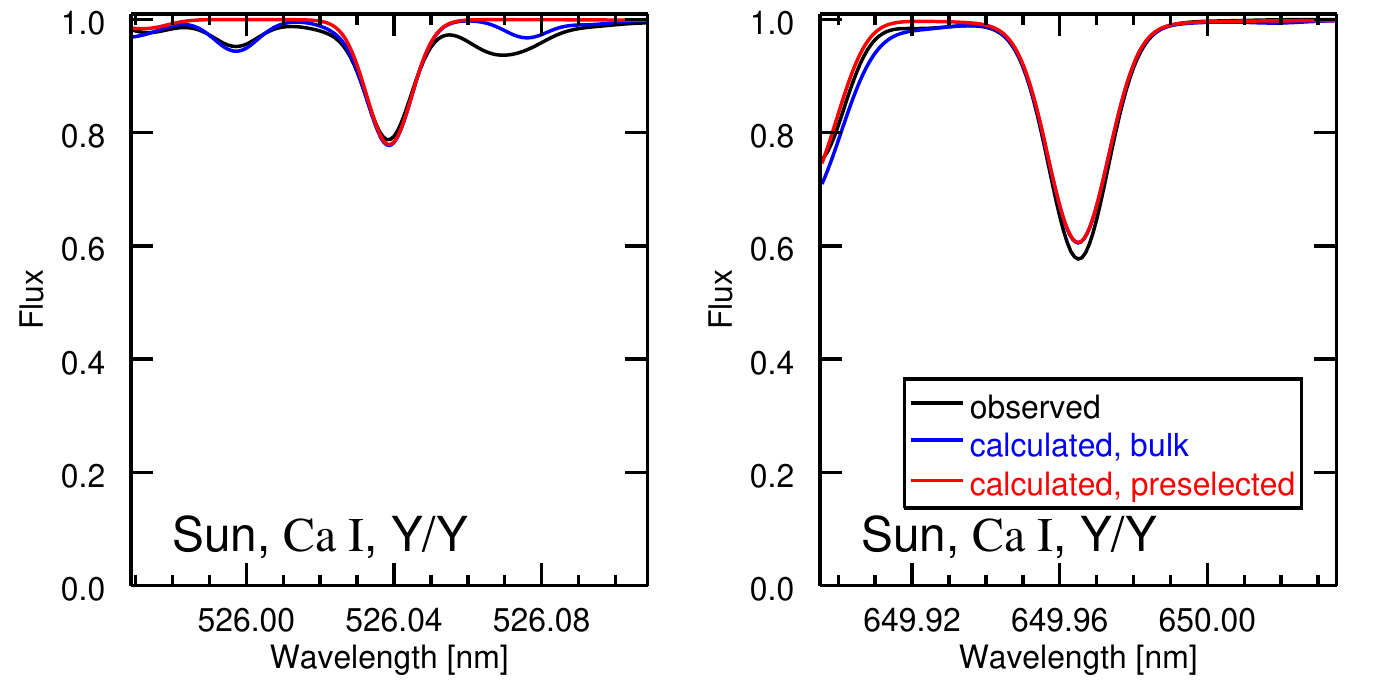}}
      \resizebox{0.89\hsize}{!}{\includegraphics{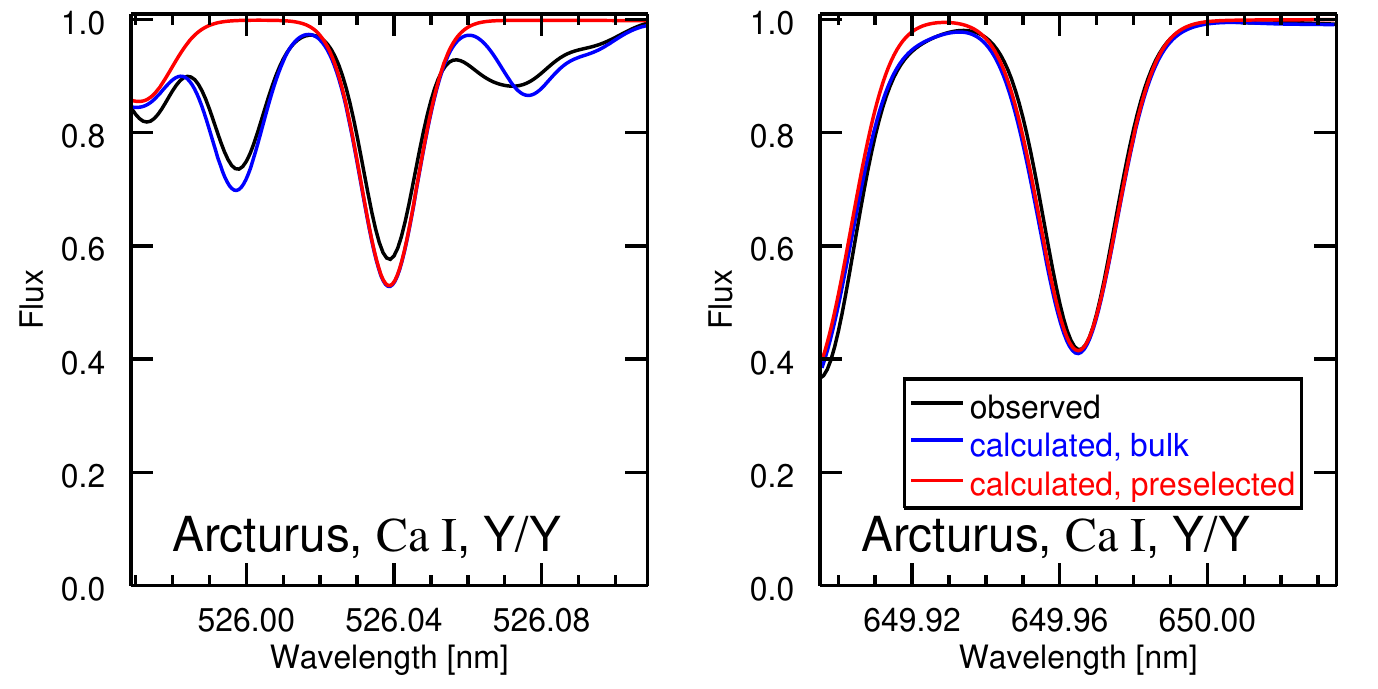}}
      \resizebox{0.89\hsize}{!}{\includegraphics{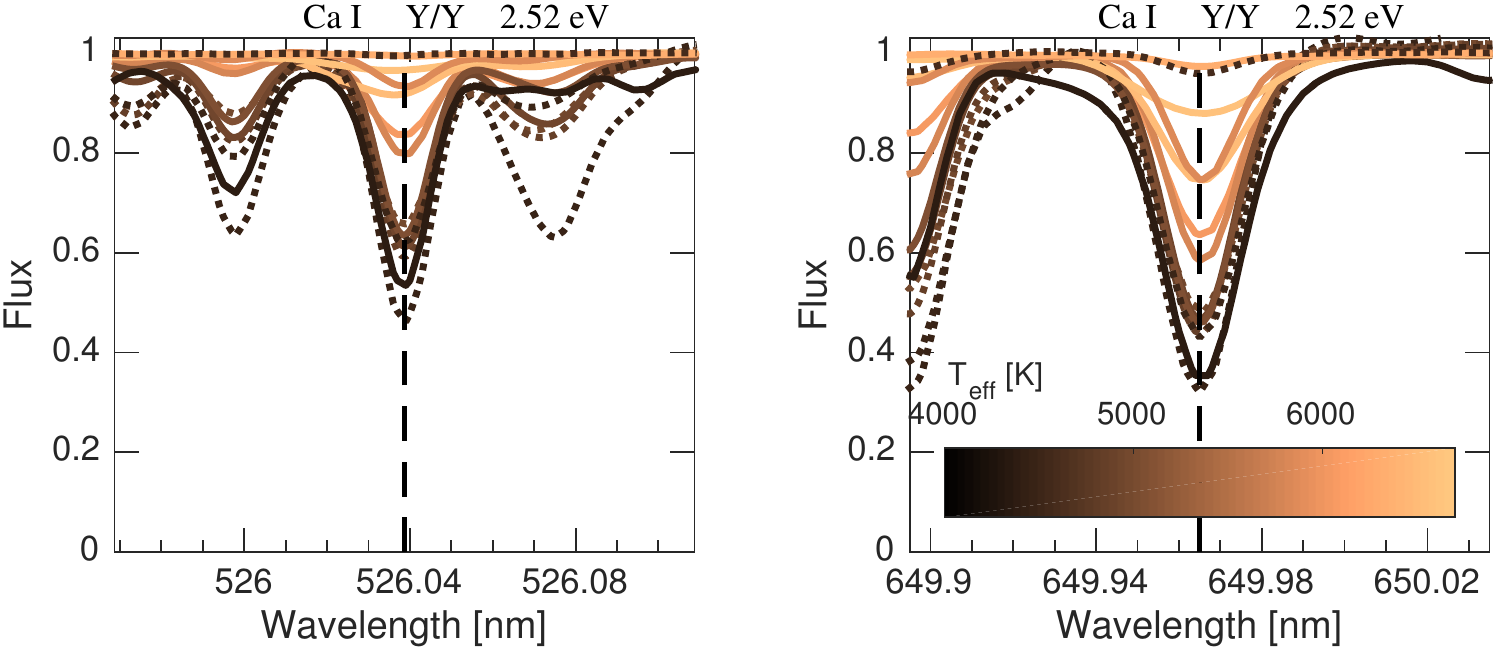}}
   \end{center}
   \caption{
   Observed and calculated line profiles around two preselected \cai\ lines for the Sun (top row) and Arcturus (middle row). Black lines: observations, red lines: calculations including preselected spectral lines only, blue lines: calculations including blends from background line list.
   Bottom row: Line profiles generated from observed spectra of selected Gaia FGK benchmark stars.
   See Sect.~\refSectSpectraT.
   %See Sect.~\sectSpectra in \citet{Heiter_etal_2020}.
   %See Sect.~\ref{sect:spectra} for description.
   }
   \label{fig:Ca1}
\end{figure*}

\begin{figure*}
\centering
\resizebox{0.9\hsize}{!}{\includegraphics{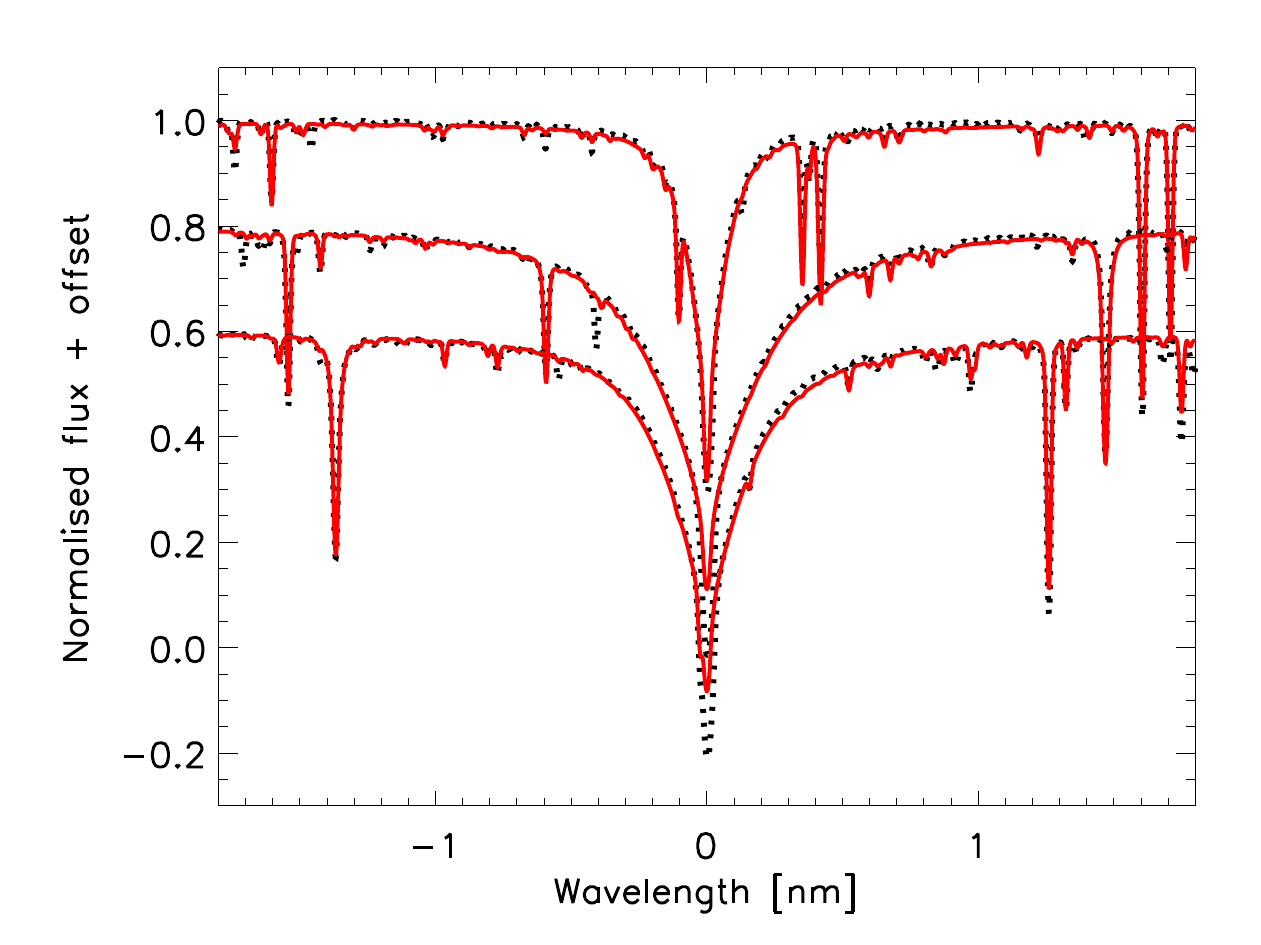}}
\caption{
Observed (black dotted lines) and synthesised (red solid lines) solar spectra of the \caii\ NIR triplet lines at 849.8, 854.2, and 866.2~nm, from top to bottom. The lines are shifted with respect to their central wavelength and offsets are added to the normalised fluxes for clarity.
}
\label{fig:caii}
\end{figure*}

Highly accurate experimental $gf$-values are available for most of the preselected \cai\ lines (\gfflag=\Yes). The majority of these have been determined by 
\citet{SR}. % Smith \& Raggett (1981) 
Others have been published by  
\citet{S}, \citet{DIKH}, and \citet{2009AA...502..989A}. % Smith (1988), Drozdowski et al. (1997), Aldenius et al. (2009)
We also included a few lines without experimental $gf$-values, for which we used the calculations by \citet{GESMCHF} with \gfflag=\Un.
About one third of these lines are largely blend-free in the Sun and Arcturus (\synflag=\Yes,
see Table~\refTabFlagstats).
%see Table~\tabFlagstats in \citealt{Heiter_etal_2020}).
%see Table~\ref{tab:flagstats}).
Two examples are illustrated in Fig.~\ref{fig:Ca1}. The line at 526.039~nm was used for most of the FGK dwarfs and giants in the abundance determination for benchmark stars by \citet{2015A&A...582A..81J}, and the line at 649.965~nm for most of the FG dwarfs and the metal-poor stars in the same study
(see also Sect.~\refSectImpact).
%(see also Sect.~\sectImpact in \citealt{Heiter_etal_2020}).
%(see also Sect.~\ref{sect:impact}).

For \caii\ only few experimental $gf$-values exist. We instead rely on theoretical data from the Opacity Project \citep[\gfflag=\Un]{GESOP} under the assumption of $LS$ coupling, as discussed in detail in \citet{2007A&A...461..261M}, % Mashonkina et al. (2007)
and on calculations by \citet{T} for the NIR triplet lines (\gfflag=\Yes). % Theodosiou (1989)
For two of the NIR triplet lines the calculations by \citet{T} show excellent agreement with the experimental data by \citet{1967PhRv..157...24G}. % Gallagher (1967)
Note that the values of \citet{T} are approximately 0.05\,dex higher than those from  \citet{GESOP}, which were included in version~4 of the Gaia-ESO line list.
Figure~\ref{fig:caii} shows the solar observed and synthetic spectrum
(see Sect.~\refSectSpectra)
%(see Sect.~\sectSpectra in \citealt{Heiter_etal_2020})
%(see Sect.~\ref{sect:spectra})
for the NIR triplet lines, which are the most important \caii\ lines in the GIRAFFE setting used by the GES. The pressure sensitivity of the lines provides an excellent gravity constraint for dwarf stars, in particular the lines at 854.2 and 866.2~nm (\synflag=\Yes). All other preselected \caii\ lines are blended to some degree (\synflag=\Un\ or \No).

Data for \cai\ lines in the background line list were also taken from \citet{Sm} and \citet{K07}, in addition to the references above.
For \caii\ lines we used additional data from \citet{TB} and \citet{K99}.
The three \cai\ autoionising lines at 631.811, 634.331, and 636.175~nm lie within the Gaia-ESO wavelength range and are included in the background line list. To enable a realistic modelling of their Fano profiles the radiative damping parameters were assigned values derived from the  Shore parameters provided by R.L. Kurucz\footnote{http://kurucz.harvard.edu/atoms/2000/gf2000.all}. The first Shore parameter is the radiative width \citep{1967JOSA...57..881S}, which is given in frequency units by Kurucz, i.e., log$_{10}(\Gamma_f)$, where $\Gamma_f$ is the FWHM in Hz. The radiative damping parameter in the Gaia-ESO line list (and in the Kurucz lists for lines other than autoionising lines) is log$_{10}(\Gamma)$, where $\Gamma$ is the FWHM in angular frequency units (rad s$^{-1}$). Thus, to convert to the usual radiative damping parameter, one uses $\Gamma = 2\pi\Gamma_f$, and thus log$_{10}(\Gamma)$=log$_{10}(\Gamma_f)$+0.80, and this value is given in the line list.

The remaining two parameters given by Kurucz are the Shore parameters describing the profile shape, $a$ and $b$.
The three lines have an asymmetry parameter $a$ of practically zero (log$_{10}(a) = -30$~cm$^2$g$^{-1}$) which implies a Lorentz absorption profile with no asymmetry (i.e., the equivalent Fano parameter $q = \infty$; see \citealt{1967JOSA...57..881S}).
Thus an appropriate radiative damping parameter achieves the same result as the profile according to the Shore parameterisation of the Fano profile. Tests indicate that these data reproduce profiles in standard stars reasonably well.
The presence of the autoionising lines affects the derivation of abundances from other lines that fall in this region, an example being Zn (see Sect.~\ref{sect:Zn}).

% place all figures and tables here
%\clearpage

%
% Maria 1: Sc, Ti, V, Cr, Mn
%

\subsection{Scandium (Z=21)}
\label{sect:Sc}

For both \sci\ and \scii\ lines the $gf$-values of \citet{LD} %Lawler \& Dakin (1989)  
are to be preferred (\gfflag=\Yes, all of the preselected \sci\ lines).
These authors determined BFs using emission FTS measurements, while the absolute scale was obtained using the time-resolved laser-induced fluorescence (TRLIF) life-times of 
\citet{1988JOSAB...5..606M}.  %Marsden et al. (1988)
For a few preselected \scii\ lines without experimental $gf$-values we used the calculations by \citet[\gfflag=\No]{K09}.
The background line list also contains $gf$-values from \citet{MFW}.

\begin{figure*}
   \begin{center}
      \resizebox{0.89\hsize}{!}{\includegraphics{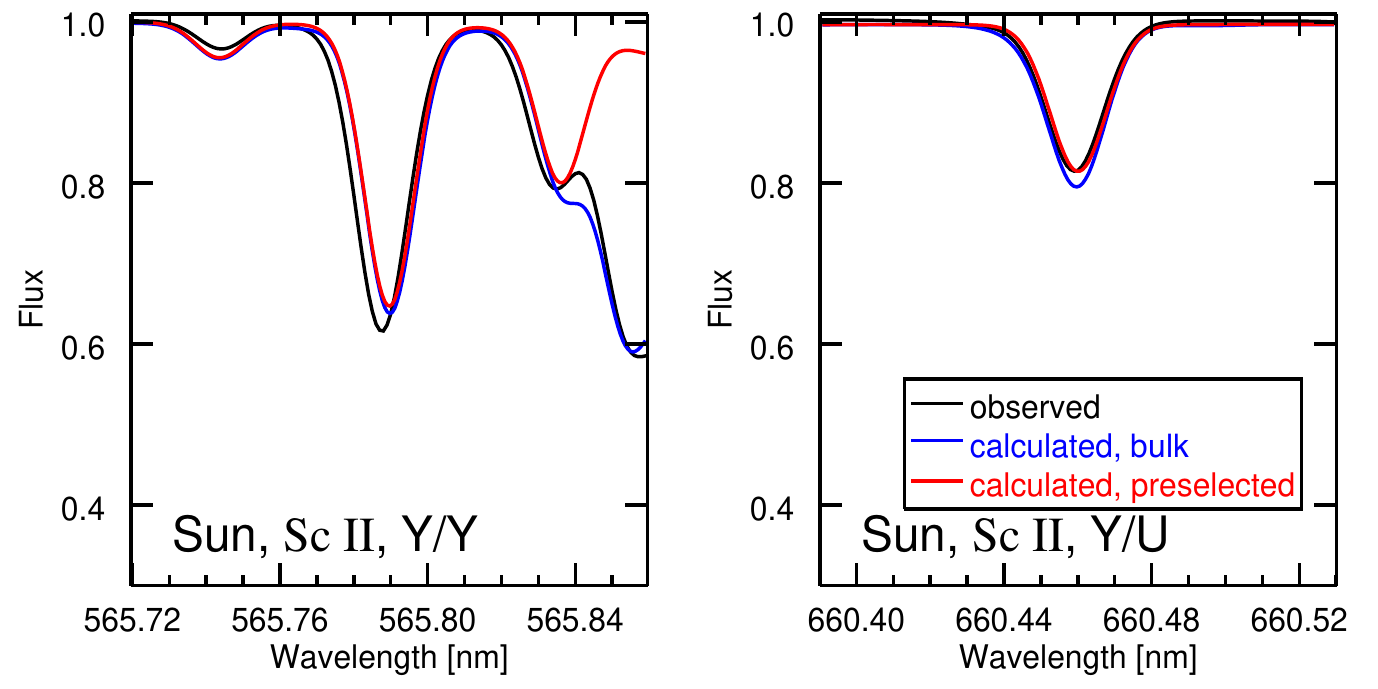}}
      \resizebox{0.89\hsize}{!}{\includegraphics{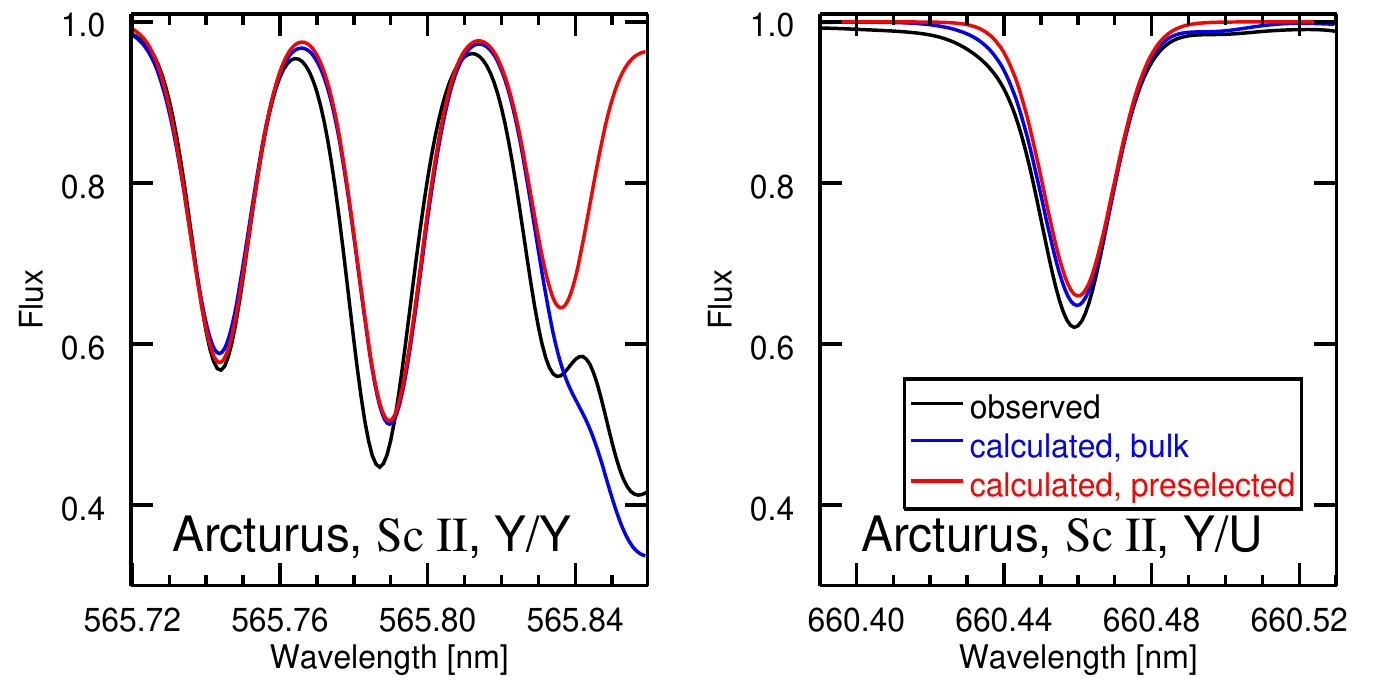}}
      \resizebox{0.89\hsize}{!}{\includegraphics{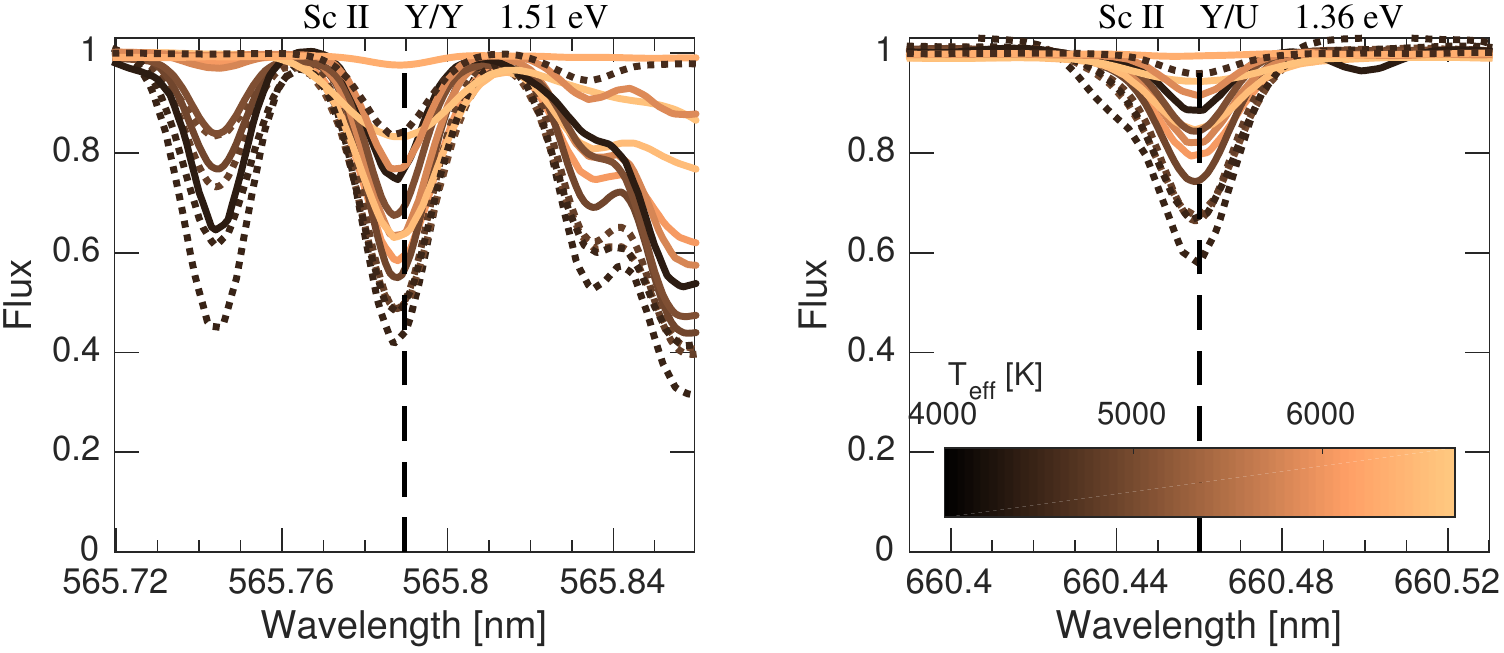}}
   \end{center}
   \caption{
   Observed and calculated line profiles around two preselected \scii\ lines for the Sun (top row) and Arcturus (middle row). Black lines: observations, red lines: calculations including preselected spectral lines only, blue lines: calculations including blends from background line list.
   Bottom row: Line profiles generated from observed spectra of selected Gaia FGK benchmark stars.
   See Sect.~\refSectSpectraT.
   % See Sect.~\sectSpectra in \citet{Heiter_etal_2020}.
   % See Sect.~\ref{sect:spectra} for description.
   }
   \label{fig:Sc2}
\end{figure*}

Only the two \sci\ lines at 535.6 and 621.1~nm and the two \scii\ lines with high-quality $gf$-values at 565.8 and 566.7~nm are blend-free (\synflag=\Yes).
Examples for \scii\ line profiles are shown in Fig.~\ref{fig:Sc2}. The line at 565.790~nm was used for most of the FGK dwarfs and the metal-poor stars in the abundance determination for benchmark stars by \citet{2015A&A...582A..81J}, and the line at 660.460~nm for most of the FGK dwarfs and giants
(see also Sect.~\refSectImpact).
%(see also Sect.~\sectImpact in \citealt{Heiter_etal_2020}).
%(see also Sect.~\ref{sect:impact}).

\paragraph{HFS data}

For \sci\ a wealth of HFS data from both experimental and theoretical studies are available.
\citet{1971PhRvA...4.1767C} employed the atomic-beam magnetic-resonance technique (ABMR) resulting in highly accurate data for the $3d4s^2 \enspace ^2D_{3/2,5/2}$ levels.
ABMR data were also presented by \citet{1976ZPhyA.276....9E} and \citet{1976PhLA...55..405Z} for the $(^3F)4s \enspace ^4F_{3/2,5/2,7/2,9/2}$ levels, which we employ.
For several others, including the $(^3F)4s \enspace ^2F_{5/2,7/2}$ levels, the only source of data are the calculations published by \citet{2004PhyS...69..189B}.
The high-excitation levels are represented by the theoretical predictions by \citet{1402-4896-75-5-006}. % Öztürk et al.
The complete set of HFS data is given in Table~\ref{HFSsc}.
For \scii\ no HFS components were computed, but HFS data for most of the preselected \scii\ lines can be found in \citet[their Sect.~6.1.2 and Table~2]{2015A&A...573A..26S}. % Scott et al. 2015a

\subsection{Titanium (Z=22)}
\label{sect:Ti}

For \tii\ we adopted transition probabilities primarily from
\citet{NWL} and \citet{2013ApJS..205...11L}, % Nitz et al. (1998) and Lawler et al. (2013)
and for \tiii\ we preferred \citet{2013ApJS..208...27W}. % Wood et al. (2013)
\citet{NWL} %Nitz et al. (1998)
measured BFs for \tii\ from FTS spectra and combined those with accurate TRLIF life-times from \citet{1990A&A...239..407S}, % Salih & Lawler
and \citet{2013ApJS..205...11L} expanded on that work.
In addition, we used accurate $gf$-values from \citet{1989AA...208..157G}, % Grevesse et al.
produced by re-normalising the relative oscillator strengths from the Oxford group \citep{1982MNRAS.199...21B,1983MNRAS.204..883B,GESB86} % Blackwell et al.
with the absolute scale fixed by using the TRLIF life-times of \citet{1982JPhB...15L.599R}. % Rudolph & Helbig
We also used data by \citet{SK} for one line.
All lines with data from these sources were assigned a \gfflag\ of \Yes.
For the remaining few \tii\ lines we used the semi-empirical calculations by \citet[\gfflag=\No]{K10}.

Most of the \tiii\ lines in the preselected line list are covered by the high-quality FTS and echelle work by \citet{2013ApJS..208...27W}. % Wood et al. (2013)
Another FTS study by \citet{PTP} %Pickering et al. (2001)
produced $gf$-values for many lines but relied on calculated oscillator strengths of weak lines to complete the fractions for some branches.
To put their BFs on an absolute scale \citeauthor{PTP} %Pickering et al. (2001)
used life-times from \citet{BHN} %Bizzarri et al. (1993)
or life-times derived from the theoretical transition probabilities of \citet{K10}.
These data were used for two of the lines. All lines from these two sources were assigned a \gfflag\ of \Yes, while the data for the remaining few \tiii\ lines were taken from \citet{K10} with \gfflag=\No.

\begin{figure*}
   \begin{center}
      \resizebox{0.89\hsize}{!}{\includegraphics{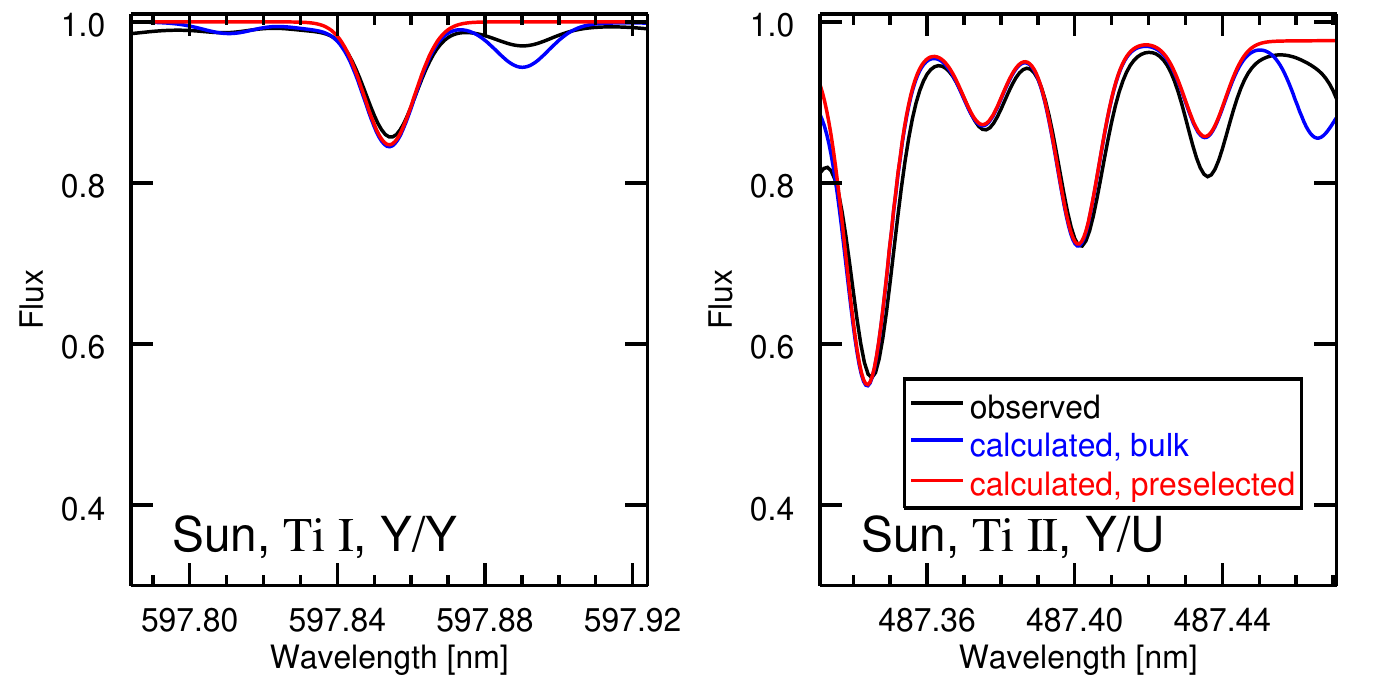}}
      \resizebox{0.89\hsize}{!}{\includegraphics{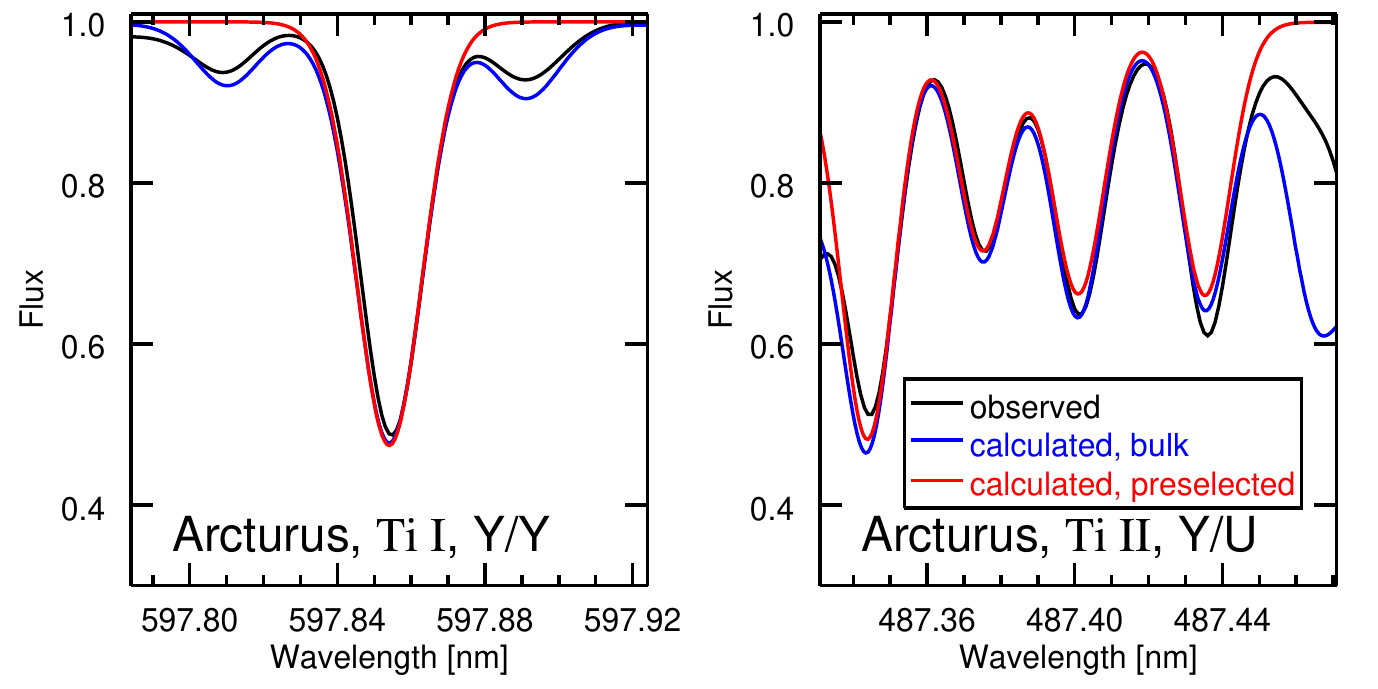}}
      \resizebox{0.89\hsize}{!}{\includegraphics{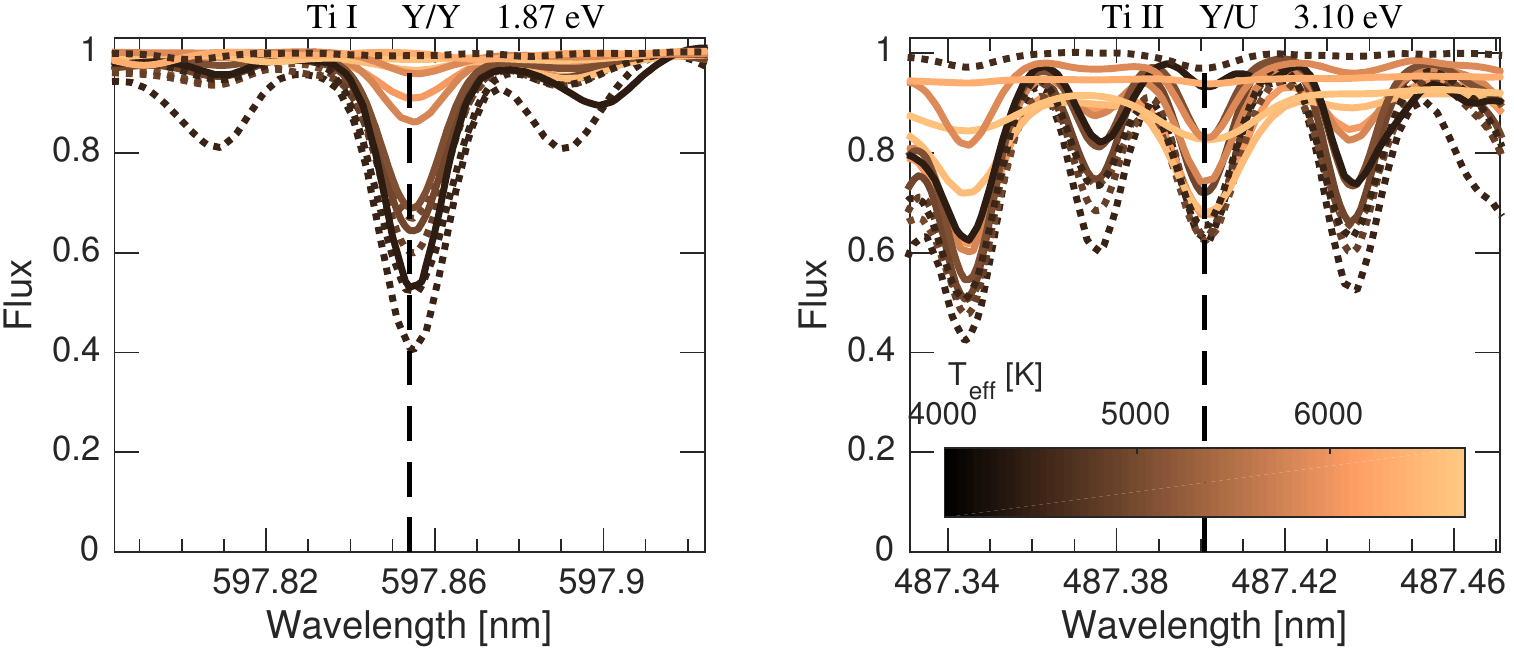}}
   \end{center}
   \caption{
   Observed and calculated line profiles around two preselected \tii\ and \tiii\ lines for the Sun (top row) and Arcturus (middle row). Black lines: observations, red lines: calculations including preselected spectral lines only, blue lines: calculations including blends from background line list.
   Bottom row: Line profiles generated from observed spectra of selected Gaia FGK benchmark stars.
   See Sect.~\refSectSpectraT.
   % See Sect.~\sectSpectra in \citet{Heiter_etal_2020}.
   % See Sect.~\ref{sect:spectra} for description.
   }
   \label{fig:Ti_profiles}
\end{figure*}

The \tii\ lines are the second most numerous (after \fei) among the preselected lines 
(see Table~\refTabFlagstats).
%(see Table~\tabFlagstats in \citealt{Heiter_etal_2020}).
%(see Table~\ref{tab:flagstats}).
90\% of them have high-quality $gf$-values and 24\% of those are also unblended in the Sun and Arcturus (\synflag=\Yes).
In contrast, there is only one \tiii\ line among the 18 with \gfflag=\Yes, which also is blend-free (541.877~nm).
Examples for line profiles are shown in Fig.~\ref{fig:Ti_profiles}. The \tii\ line at 597.854~nm was used for most of the FGK dwarfs and giants in the abundance determination for benchmark stars by \citet{2015A&A...582A..81J}, and the \tiii\ line at 487.401~nm for most of the FG dwarfs, the FGK giants, and the metal-poor stars in the same study 
(see also Sect.~\refSectImpact).
%(see also Sect.~\sectImpact in \citealt{Heiter_etal_2020}).
%(see also Sect.~\ref{sect:impact}).

The background line list relies mainly on calculations by \citet{K10}, and for \tii, on data from \citet{MFW}, \citet{NWL}, and \citet{LGWSC}.
For a few \tiii\ lines in the background line list, experimental data and astrophysical determinations were used \citep{RHL,PTP,WLSC}.

Ti has five isotopes with non-negligible natural abundances
(see Table~\refTabIsotopes),
%(see Table~\tabIsotopes in \citealt{Heiter_etal_2020}),
%(see Table~\ref{tab:isotopes}),
including two with non-zero spin. Isotopic splittings are presented in \citet[their Sect.~6.2.2 and Table~2]{2015A&A...573A..26S} for two and six of the preselected \tii\ and \tiii\ lines, respectively, together with HFS constants. The largest wavelength shifts between two isotopes are 0.04~\AA\ for \tii, and $<$0.03~\AA\ for \tiii. None of the lines have HFS data for both levels. Therefore, neither isotopic nor HFS components were included in the Gaia-ESO line list for Ti.

\subsection{Vanadium (Z=23)}
\label{sect:V}

\begin{figure*}
   \begin{center}
      \resizebox{0.89\hsize}{!}{\includegraphics{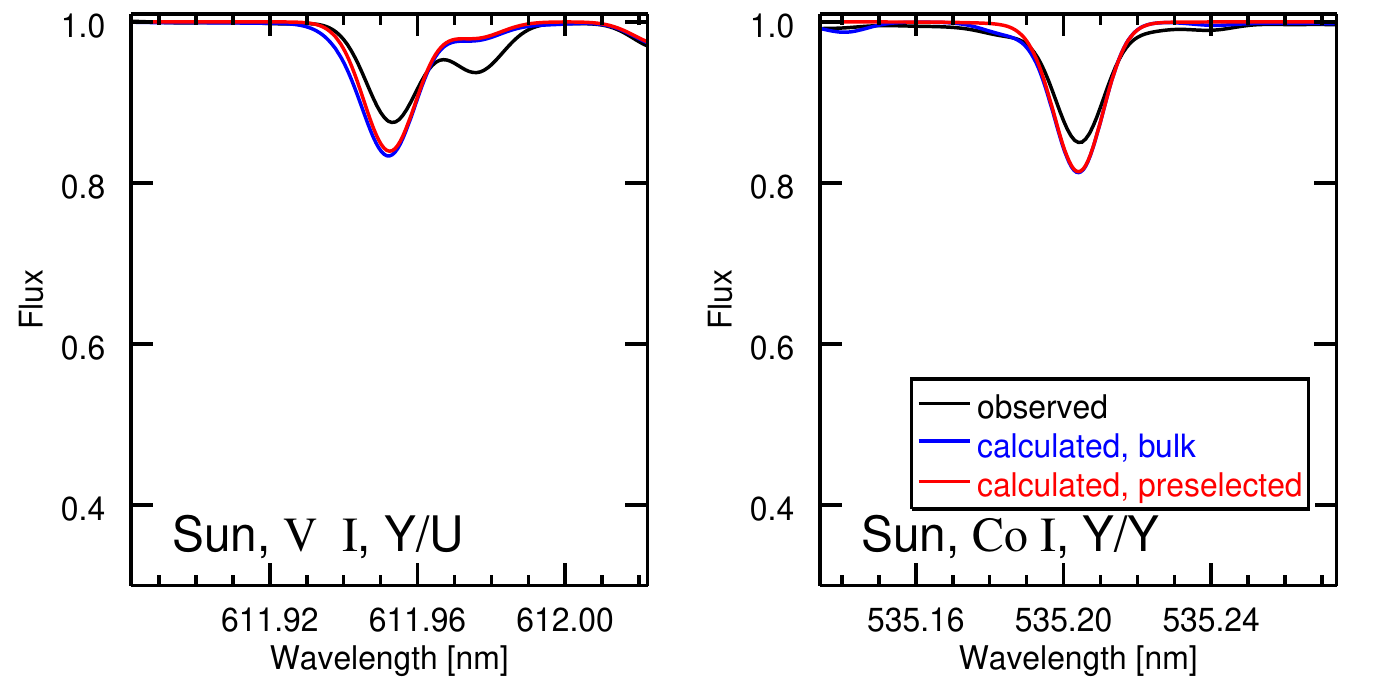}}
      \resizebox{0.89\hsize}{!}{\includegraphics{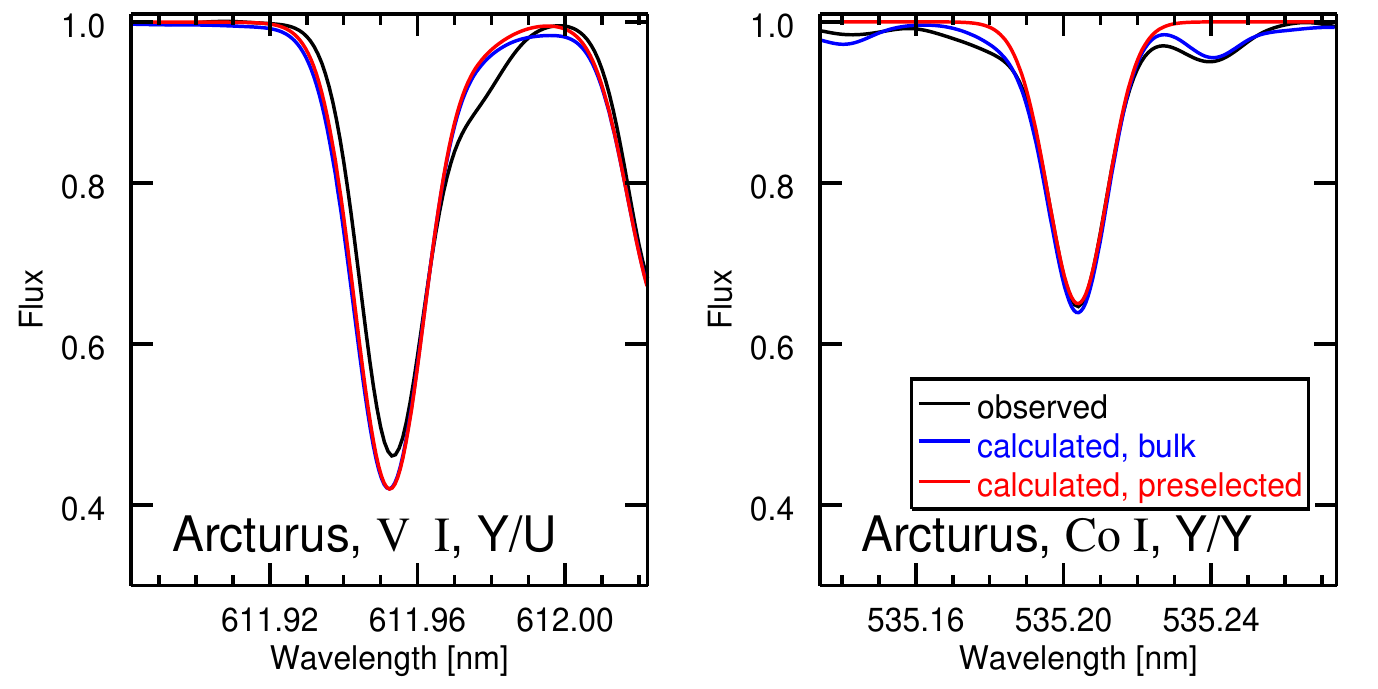}}
      \resizebox{0.89\hsize}{!}{\includegraphics{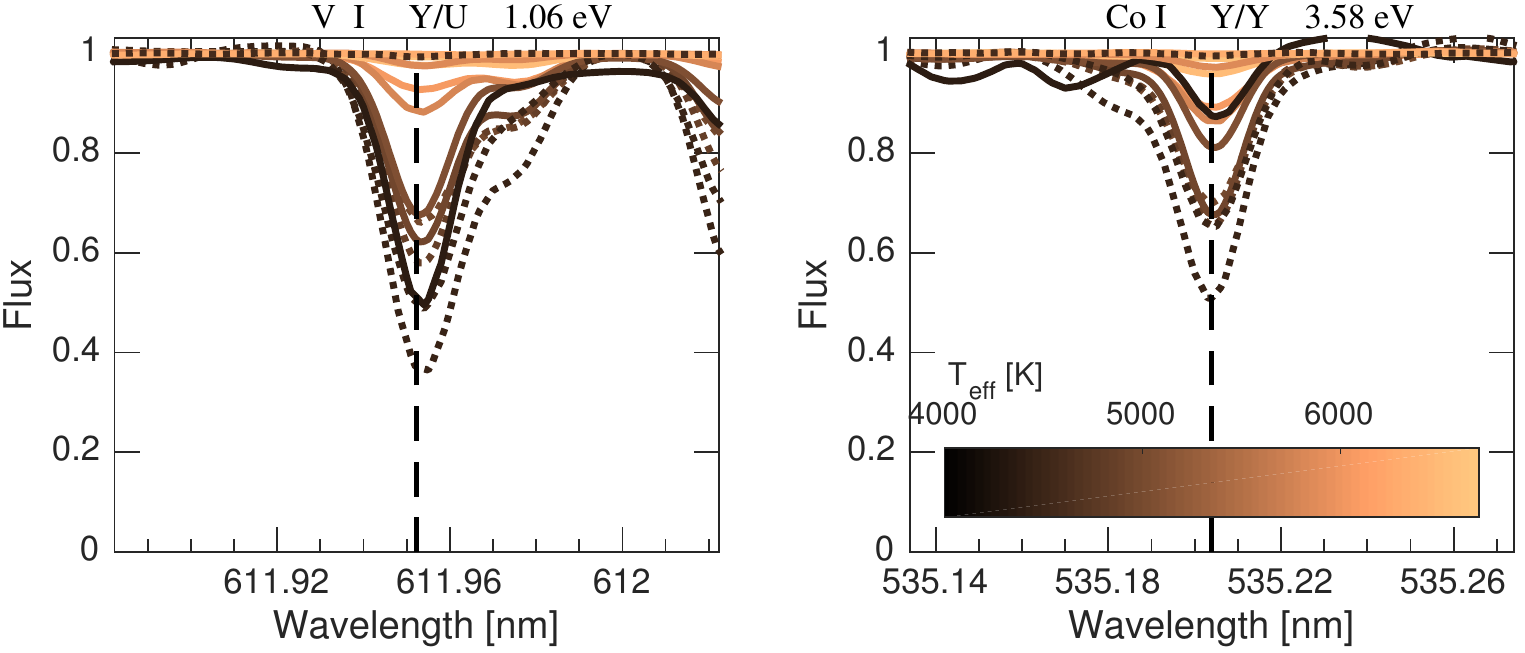}}
   \end{center}
   \caption{
   Observed and calculated line profiles around preselected lines of \vi\ (left) and \coi\ (right) for the Sun (top) and Arcturus (middle). Black lines: observations, red lines: calculations including preselected spectral lines only, blue lines: calculations including blends from background line list.
   Bottom: Line profiles generated from observed spectra of selected Gaia FGK benchmark stars.
   See Sect.~\refSectSpectraT.
   % See Sect.~\sectSpectra in \citet{Heiter_etal_2020}.
   % See Sect.~\ref{sect:spectra} for description.
   }
   \label{fig:V1Co1}
\end{figure*}

The work of \citet{1985AA...153..109W} %Whaling et al. (1985)
provides the most accurate measurements of \vi\ oscillator strengths to date, based on both FTS BFs and TRLIF life-times. Most of our preselected \vi\ lines are covered by this source (\gfflag=\Yes), and about 40\% of these lines are also unblended in the Sun and Arcturus (\synflag=\Yes).
Line profiles for the \vi\ line at 611.952~nm are shown in Fig.~\ref{fig:V1Co1} (left column) as a representative example. This line was used for the largest number of FGK dwarfs and giants in the determination of the V abundance for benchmark stars by \citet{2015A&A...582A..81J}
(see also Sect.~\refSectImpact).
%(see also Sect.~\sectImpact in \citealt{Heiter_etal_2020}).
%(see also Sect.~\ref{sect:impact}).

For 13 out of the 49 preselected \vi\ lines and most of the lines in the background line list, the $gf$-values come from the calculations of \citet[\gfflag=\No]{K09}, except when data are available in \citet{MFW}.
For \vii\ all $gf$-values come from the calculations of \citet[\gfflag=\No]{K10}, including the three preselected lines.

\paragraph{HFS data}
A large number of good experimental data on the HFS of \vi\ is available, with only few levels in common between different studies.
We followed the quality assessment of \citet{2015A&A...573A..26S}, who sorted the data into three groups of published works with no overlap between works within each group.
Data were then selected according to order of preference.
The group with the most preferred data comprises the studies of \citet[ABMR and LFS -- laser fluorescence spectroscopy -- data]{1979PhRvA..19..168C}, % Childs et al.
\citet[ABMR]{1992PhyB..179..103E} % El-Kashef & Ludwig
and \citet[ABMR]{1989ZPhyD..11..259U}. % Unkel et al.
The second-best group consists of measurements by \citet[FTS]{1995JPhB...28.3741P}, % Palmeri et al.
\citet[FTS]{2002PhyS...66..363L}, % Lefèbvre et al.
and \citet[crossed beam]{1998JPhB...31.2203C}. % Cochrane et al.
Data for levels not in the previous two groups were taken from \citet[LFS]{1989ZPhyD..11..259U}. % Unkel et al.
The complete set of HFS data is given in Table~\ref{HFSv}.

\subsection{Chromium (Z=24)}
\label{sect:Cr}

\citet{SLS} %Sobeck et al. (2007)
measured highly accurate \cri\ oscillator strengths
using FTS BFs normalised to TRLIF life-times of \citet{1997JQSRT..58...85C}. %Cooper et al. (1997)
Together with the data from \citet{1984MNRAS.207..533B} for one line, % Blackwell et al. (1984)
these were assigned the \gfflag=\Yes.
These data were complemented with experimental transition probabilities from \citet{MFW}, \citet{CSE}, and \citet{1968PhFl...11.1002W}, % Martin et al. (1988); Cocke et al. (1973); Wolnik et al. (1968)
and with semi-empirical $gf$-values from \citet{K10}, although with \gfflag=\No.

\begin{figure*}
   \begin{center}
      \resizebox{0.9\hsize}{!}{\includegraphics{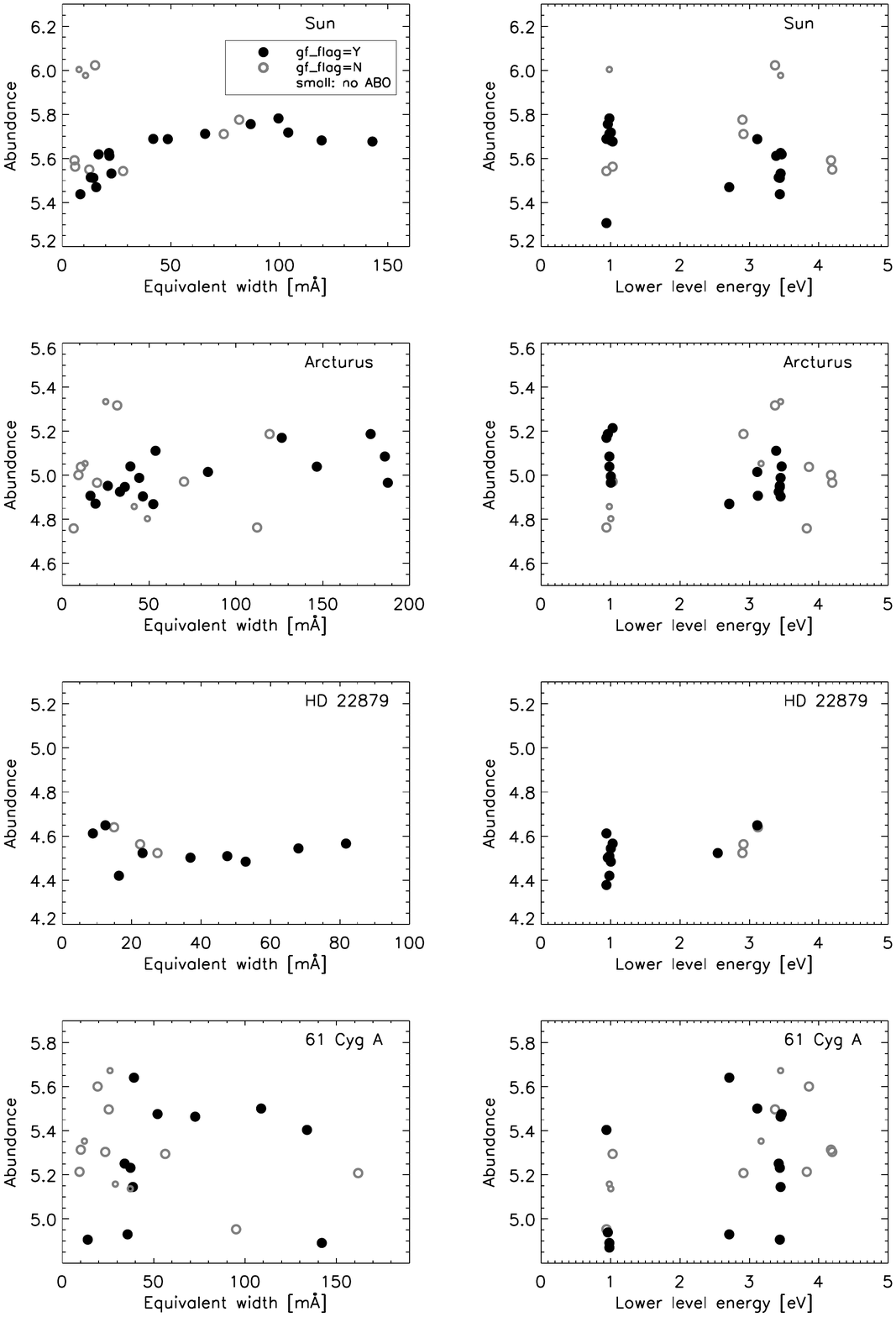}}
   \end{center}
\caption{Line-by-line \cri\ abundances determined for four benchmark stars for two groups of lines with different quality assessment of their transition probabilities, as a function of equivalent width and lower level energy. Abundances are given as $\log(\varepsilon_{\rm Cr})+12$, where $\varepsilon_{\rm Cr}=N_{\rm Cr}/N_{\rm H}$. Only lines with \synflag=\Yes\ or \Un\ and with equivalent widths $>5$~m\AA\ are included.
}
\label{fig:abuCr} 
\end{figure*}

Among the 35 lines with high-quality $gf$-values there are nine which also are blend-free in the Sun and Arcturus (\synflag=\Yes), and a further 15 lines with uncertain blend status (\synflag=\Un).
Line abundances derived for four benchmark stars
(see Sect.~\refSectSpectra)
%(see Sect.~\sectSpectra in \citealt{Heiter_etal_2020})
%(see Sect.~\ref{sect:spectra})
for these lines are compared to line abundances for \gfflag=\No\ lines in Fig.~\ref{fig:abuCr}, while mean abundances are given in Table~\ref{tab:abustat}.
For the Sun and Arcturus the scatter of the \gfflag=\No\ lines is twice as large as the scatter of the \gfflag=\Yes\ lines, while it is comparable for the cool dwarf star 61~Cyg~A for which all of the lines might be blended to some degree.

For a few of the preselected \crii\ lines experimental oscillator strengths were taken from \citet[\gfflag=\Yes]{PGBH}. % Pinnington et al. (1993)
For the remaining lines, we used the semi-empirical oscillator strengths from \citet[\gfflag=\No]{K10}. 
More than half of the \crii\ lines are heavily blended in the Sun and Arcturus (\synflag=\No), and among the remaining ones there is only one with \synflag=\Yes\ (530.585~nm), which unfortunately does not have an experimental $gf$-value
(see also Sect.~\refSectDataneeds).
%(see also Sect.~\sectDataneeds in \citealt{Heiter_etal_2020}).
%(see also Sect.~\ref{sect:dataneeds}).

In the background line list the majority of the \crii\ $gf$-values come from \citet{RU}, supplemented by data from \citet{SLd}, \citet{MFW}, and the two sources above.

Cr has four isotopes with non-negligible natural abundances
(see Table~\refTabIsotopes),
%(see Table~\tabIsotopes in \citealt{Heiter_etal_2020}),
%(see Table~\ref{tab:isotopes}),
including one with non-zero spin.
Isotopic splitting and HFS data for Cr are scarce, and are given for one \cri\ line in \citet[their Sect.~6.4.2 and Table~2]{2015A&A...573A..26S}, which is not among our preselected lines. The wavelength difference between the lightest and heaviest Cr isotope is just 0.003~\AA\ and can be ignored in the context of the GES.

\subsection{Manganese (Z=25)}
\label{sect:Mn}

For \mni\ lines we used accurate experimental $gf$-values from \citet{DLSSC} %Den Hartog et al. (2011)
and \citet{2007AA...472L..43B}. %Blackwell-Whitehead \& Bergemann (2007) 
For some lines data were taken from \citet{1984MNRAS.208..147B}. %Booth et al. (1984)
See \citet{2015A&A...573A..26S} for a more detailed discussion of these sources.
All of these data were assigned the \gfflag=\Yes\ (about three quarters of the preselected lines).
For the remaining preselected \mni\ lines we resorted to the calculations of \citet[\gfflag=\No]{K07}.
Note that for three \mni\ lines (601.349, 601.664, 602.179~nm) the upper level energy values in \citet{DLSSC} are incorrect in the fifth digit, leading to a shift in wavelength of 0.02 to 0.03~\AA\ \citep[see][]{2017A&A...601A..38J}. The wavelengths adopted in the line list are therefore taken from \citet{K07}.

\begin{figure*}
   \begin{center}
      \resizebox{0.89\hsize}{!}{\includegraphics{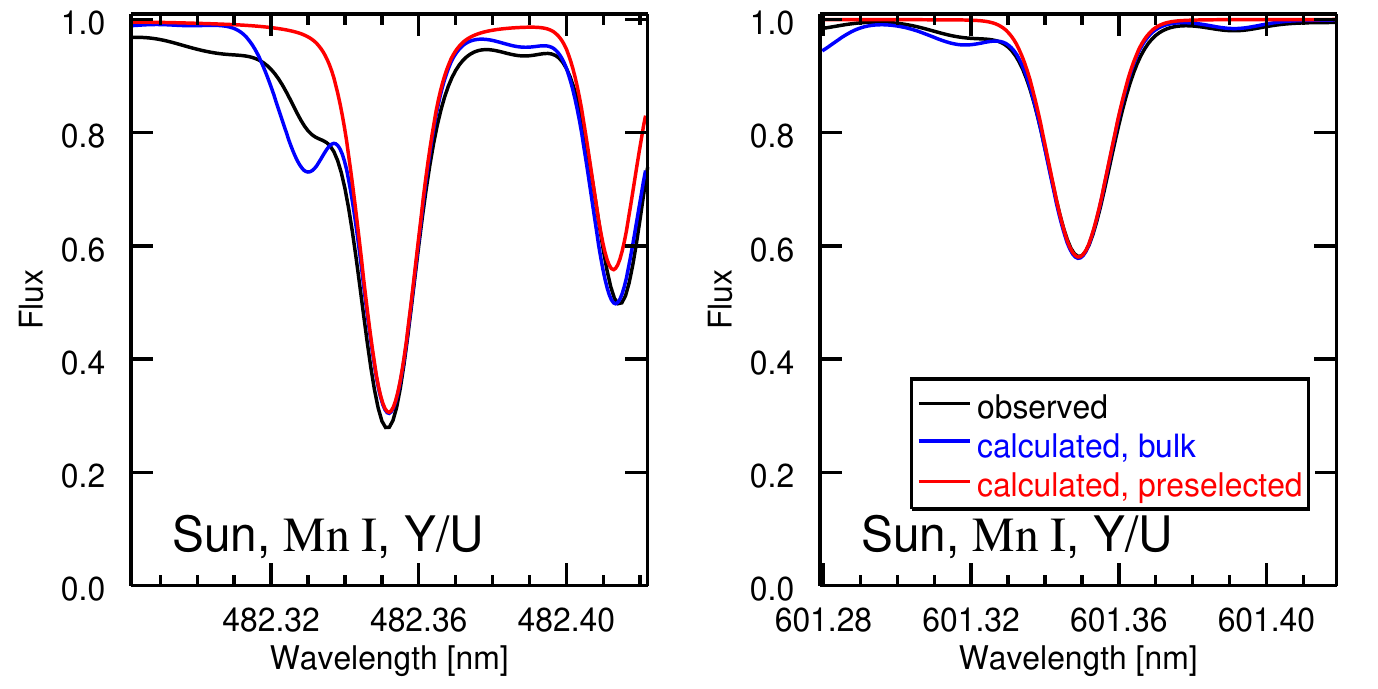}}
      \resizebox{0.89\hsize}{!}{\includegraphics{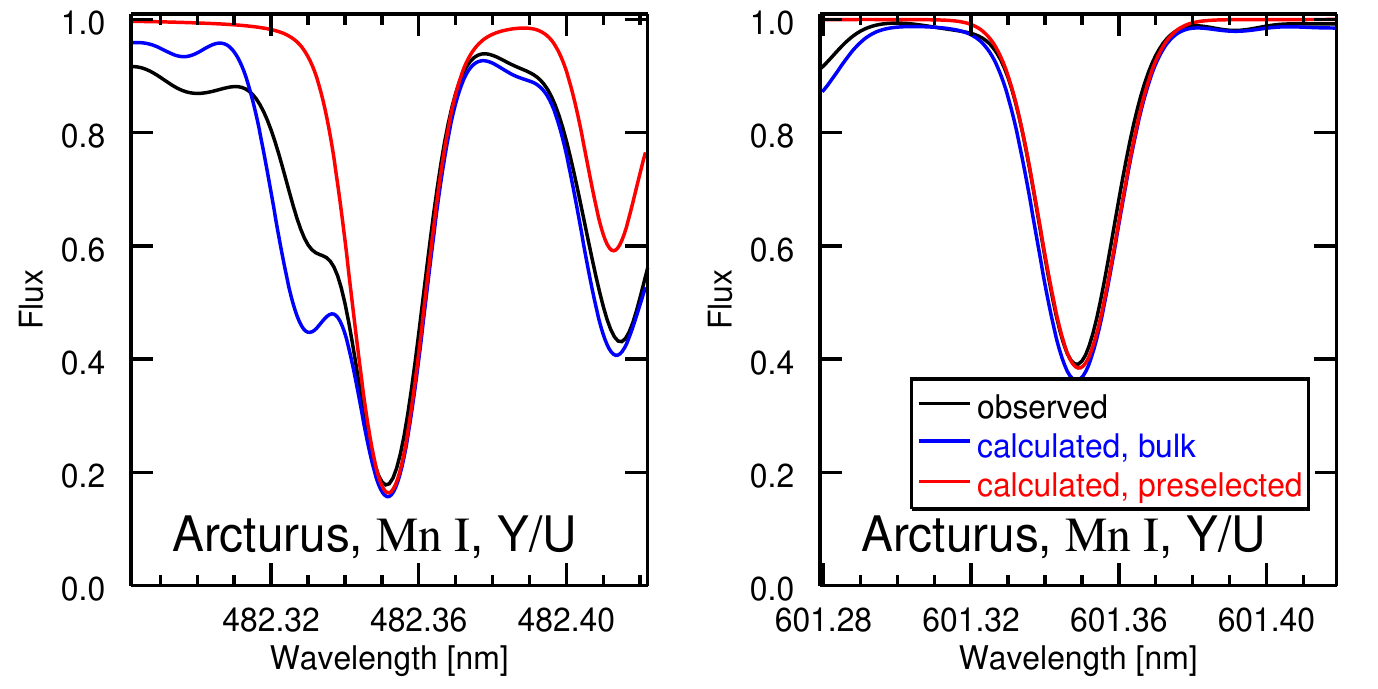}}
      \resizebox{0.89\hsize}{!}{\includegraphics{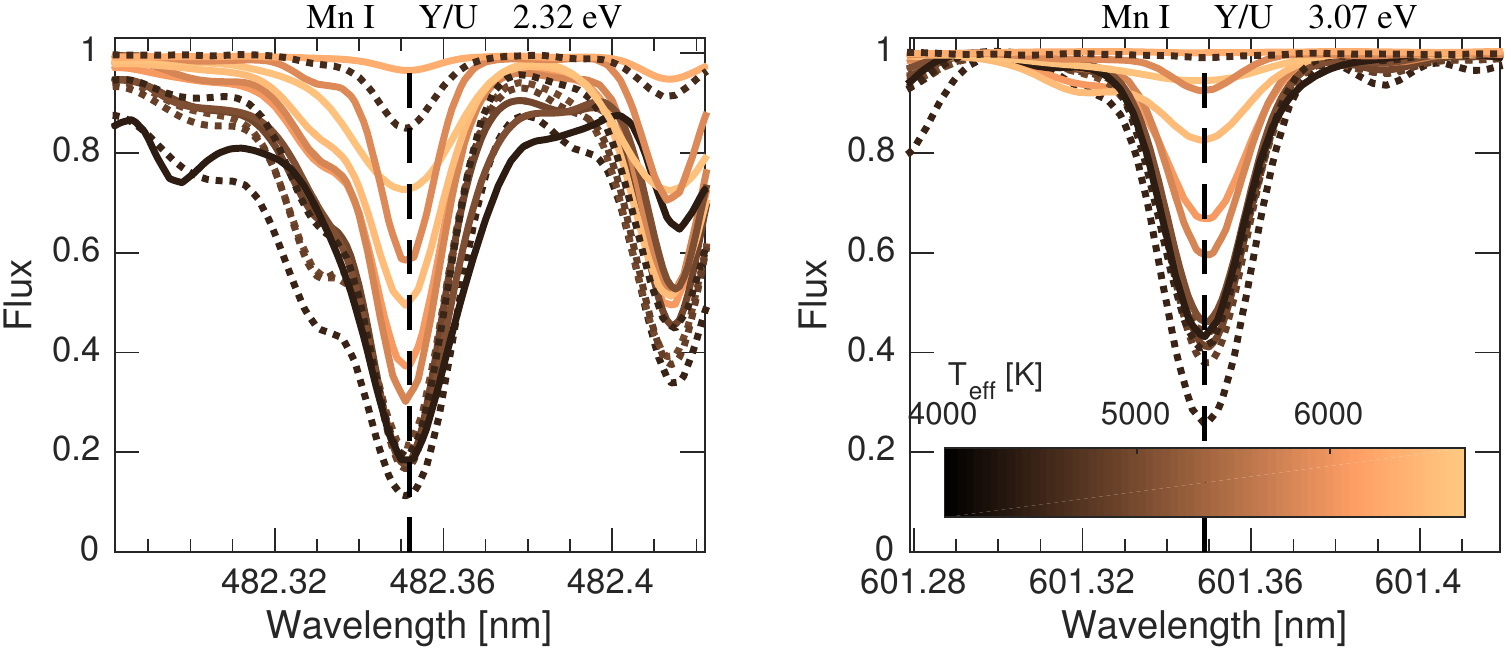}}
   \end{center}
   \caption{
   Observed and calculated line profiles around two preselected \mni\ lines for the Sun (top row) and Arcturus (middle row). Black lines: observations, red lines: calculations including preselected spectral lines only, blue lines: calculations including blends from background line list.
   Bottom row: Line profiles generated from observed spectra of selected Gaia FGK benchmark stars.
   See Sect.~\refSectSpectraT.
   % See Sect.~\sectSpectra in \citet{Heiter_etal_2020}.
   % See Sect.~\ref{sect:spectra} for description.
   }
   \label{fig:Mn1}
\end{figure*}

Only three of the \mni\ lines with high-quality $gf$-values have \synflag=\Yes\ (542.035, 539.467, 602.179~nm), but a further 15 may be useful for abundance analysis if a careful evaluation of blending lines is done (\synflag=\Un).
Two of these lines are shown in Fig.~\ref{fig:Mn1}. The line at 482.352~nm was used for most of the metal-poor stars in the abundance determination for benchmark stars by \citet{2015A&A...582A..81J}, and the line at 601.349~nm for most of the FGK dwarfs and giants in the same study
(see also Sect.~\refSectImpact).
%(see also Sect.~\sectImpact in \citealt{Heiter_etal_2020}).
%(see also Sect.~\ref{sect:impact}).

In addition, the background line list contains \mni\ data from \citet{MFW} and \citet{BXPNL}, and \mnii\ lines with calculated $gf$-values from \citet{K09}.

\paragraph{HFS data}

For the HFS data we again followed the quality assessment of \citet{2015A&A...573A..26S}, and selected accurate experimental data from several groups of publications, in the following order of preference:
1) \citet[spin-exchange]{1971PhRvA...3.1220D}, % Davis et al. 1971
\citet[ABMR]{1981ZPhyA.303....7J}, % Johann et al. 1981
\citet[ABMR]{1979ZPhyA.291..207D}, % Dembczynsky et al. 1979
and \citet[interference spectroscopy]{1987ZPhyD...7..161B}; % Brodzinski et al. 1987
2) \citet[optogalvanic spectroscopy]{2003PhyS...67..476B} for the $z^4D$ level, % Basar et al. 2003
\citet[FTS]{2005ApJS..157..402B} for all other levels; % Blackwell-Whitehead et al. 2005 
3) \citet[FTS]{2003AA...404.1153L}, % Lefebvre et al. 2003
\citet[FTS and Fabry-Perot spectrometry]{1972AA....18..209L}, % Luc \& Gerstenkorn (1972)
and \citet{1969PhLA...29..486H}. % Handrich et al. 1969
The complete set of HFS data is given in Table~\ref{HFSmn}.

% place all figures and tables here
%\clearpage

%
% Karin: Fe
%

\subsection{Iron (Z=26)}
\label{sect:Fe}

The first choice for the source of $gf$-values for \fei\ lines are the publications by three laboratory spectroscopy groups from the 1980s and 1990s.
The groups are those based in Oxford, UK \citep{BIPS,GESB79b,GESB82c,GESB82d,GESB86}, % Blackwell et al. 1979ab, 1982ab, 1986
Madison, Wisconsin \citep{BWL}, % O'Brian et al. (1991)
and
Hannover, Germany \citep{BKK,BK}. % Bard et al. (1991), Bard \& Kock (1994)
The Oxford group used absorption spectroscopy to determine very precise relative $gf$-values, which were put on an absolute scale using one line with an accurate ($\pm 0.02$\,dex) transition probability. The high-quality Madison and Hannover $gf$-values are based on emission spectroscopy, which provided life-times and BFs.

For the Madison dataset, the $gf$-values were calculated from the original published data (vacuum wavenumber $\bar\nu$ in cm$^{-1}$, transition probability $A_{ul}$ in s$^{-1}$, and upper level rotational quantum number $J_{\rm upp}$), using the following equation:
\begin{equation}
   gf = \frac{m_e c}{2\pi e^2 \cdot 4\pi10^{-7}c^2} \cdot (10^{-2}/\bar\nu)^2 \cdot (2 J_{\rm upp}+1) \cdot A_{ul}
\end{equation}
where $m_e$, $c$, and $e$ are the electron mass, the speed of light, and the electron charge, respectively, all in SI units (see e.g., \citealt{2008oasp.book.....G}).

Recently, \citet{2014MNRAS.441.3127R} and \citet{2014ApJS..215...23D} % Ruffoni et al., Den Hartog et al.
used a similar technique to determine transition probabilities for 142 and 203 \fei\ lines, respectively,
about 70 and 60 of which fall in the main Gaia-ESO wavelength regions (UVES-580 and GIRAFFE HR21 settings),
and 35 and 37 of which are among the preselected lines.
They focused on lines with high values of lower level energy, $3-6$~eV, with few measurements reported by the above mentioned groups.

For lines in common between all of the above sources, the $\log gf$-values were averaged, with weights based on the experimental uncertainties.
In total, we obtained data from these sources for 45\% of the 545 \fei\ lines % 248 lines
in the preselected line list, and assigned the \gfflag\ \Yes\ to them.

\begin{figure}
\centering
\includegraphics[width=7.5cm,viewport=7cm 0.5cm 27cm 20cm]{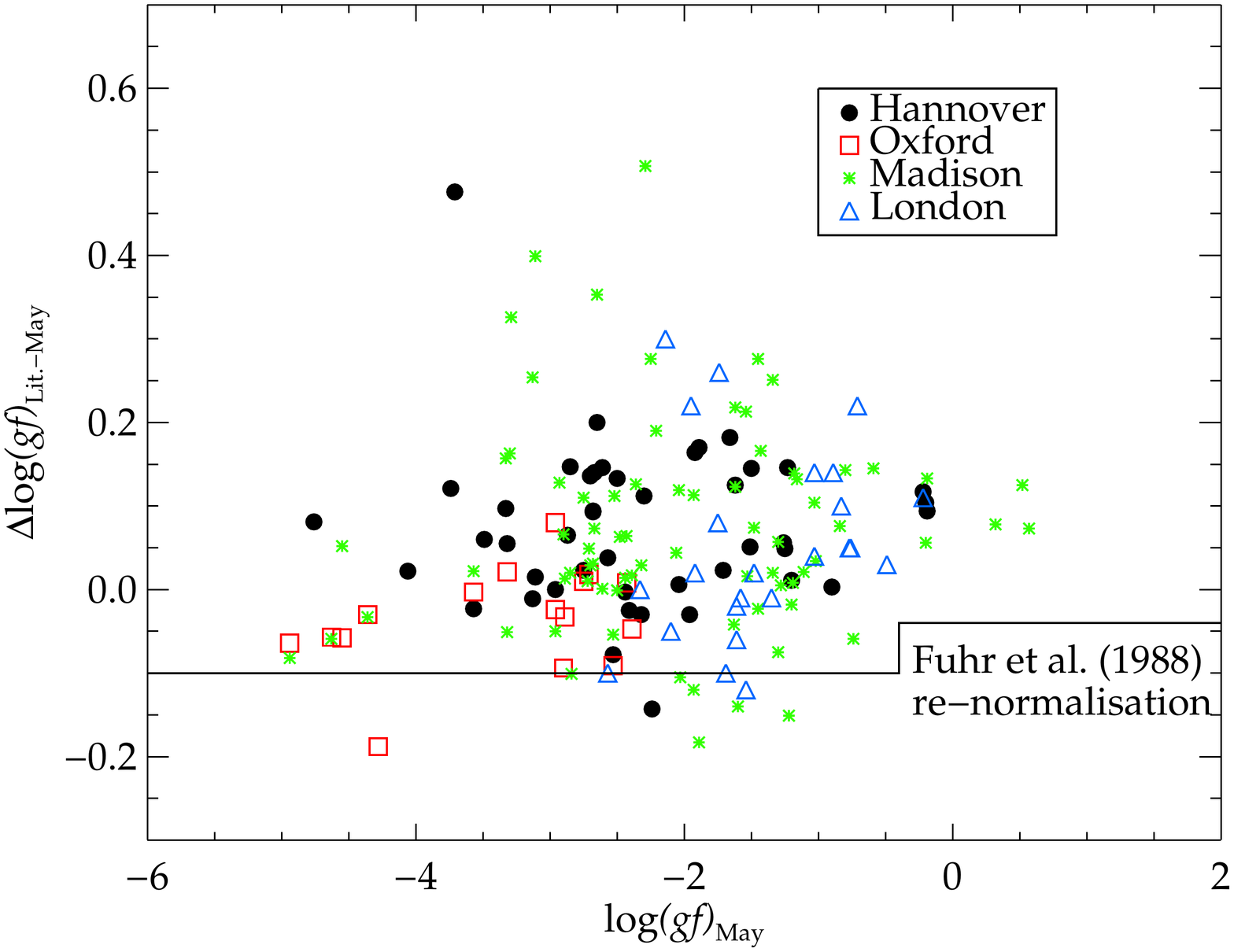}
\caption{The figure illustrates how \fei\ transition probabilities published by \citet{MRW} compare to newer, more accurate, data from the Oxford, Madison, Hannover, and London groups (see text for references). The solid line indicates the proposed normalisation of the \citet{MRW} values by \citet{FMW}, i.e., applying this normalisation would shift the zero-point to the solid line. All published lines in the wavelength region $476-895$~nm are shown.}
\label{fig:may}
\end{figure}

For lines not present in any of the above sources, $gf$-values from \citet{MRW} % May et al. (1974)
were used if available (28\% of the preselected lines). % 151 lines
These were assigned the \gfflag\ \Un, except for ten high-excitation lines with small uncertainties (0.05~dex or less), which are also unblended (\synflag\ \Yes\ or \Un, \gfflag=\Yes).
We stress here that we elected to adopt the original values of \citet{MRW} rather than those subsequently re-normalised by \citet{FMW} based on more recent life-time measurements. This choice was made after realising that, contrary to our expectations, the original $gf$-values agree significantly better with our four preferred sources for lines in common, as illustrated in Fig.\,\ref{fig:may}. 

For a few lines, % 17 lines
we used older laboratory data from \citet{1969AA.....2..274G}, \citet{1970AA.....9...37R} and \citet{1970ApJ...162.1037W,WBW}, choosing this time to trust the re-normalisation by \citet{FMW}. We have not rigorously investigated the performance of original and re-normalised data in this case, but assigned \gfflag\ \Un\ in all cases. The same ranking was made for four lines from \citet{BKor} and \citet{1984PhST....8...84K}, with absolute values re-normalised by \citet{2006JPCRD..35.1669F}.
For the remaining 25\% of the preselected lines, % 135 lines
we used the semi-empirical calculations from \citet[\gfflag=\No]{K07}. % Kurucz

\begin{figure*}
   \begin{center}
      \resizebox{0.9\hsize}{!}{\includegraphics{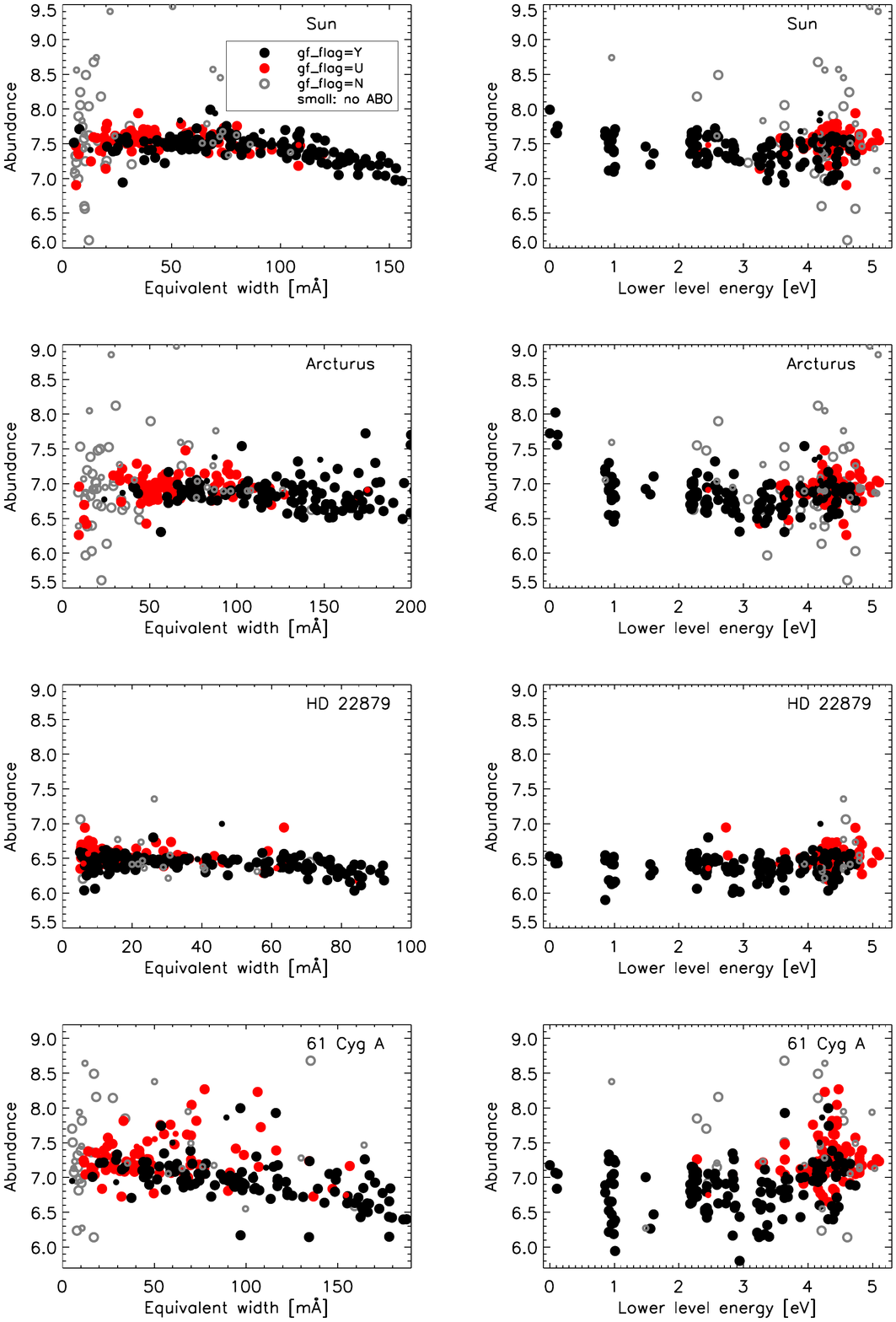}}
   \end{center}
\caption{Line-by-line \fei\ abundances determined for four benchmark stars for three groups of lines with different quality assessment of their transition probabilities, as a function of equivalent width and lower level energy. Abundances are given as $\log(\varepsilon_{\rm Fe})+12$, where $\varepsilon_{\rm Fe}=N_{\rm Fe}/N_{\rm H}$. Only lines with \synflag=\Yes\ or \Un\ and with equivalent widths $>5$~m\AA\ are included.
}
\label{fig:FeNUY} 
\end{figure*}

Fig.\,\ref{fig:FeNUY} illustrates how well lines with different \gfflag s perform with respect to each other in an abundance analysis.
Line abundances derived for four benchmark stars
(see Sect.~\refSectSpectra)
%(see Sect.~\sectSpectra in \citealt{Heiter_etal_2020})
%(see Sect.~\ref{sect:spectra})
for 240--290 \fei\ lines with \synflag\ \Yes\ or \Un\ and equivalent widths $>5$~m\AA\ are shown, while mean abundances for lines with equivalent widths $\le100$~m\AA\ are given in Table~\ref{tab:abustat}.
A significant increase in scatter can be seen for lines with the poorest ranking, which supports our recommendations.
Abundances for lines with \gfflag=\No\ deviate from the mean by up to 2~dex.
However, also among \gfflag=\Un\ and \Yes\ lines with small quoted laboratory uncertainties there are individual exceptions with poor astrophysical performance. Examples are the \gfflag=\Yes\ lines at 506.008~nm (0.0~eV, \synflag=\Un, equivalent widths of 68, 174, 16, 117~m\AA\ in the Sun, Arcturus, HD~22879, and 61~Cyg~A, respectively) and at 522.318~nm (3.635~eV, \synflag=\Yes, equivalent widths of 28, 56, 6, and 38~m\AA\ in the four stars), which result in $\sim$0.5~dex higher\footnote{Sun and Arcturus only} and lower abundances than other \gfflag=\Yes\ lines with similar equivalent widths, respectively.

The most accurate \feii\ $gf$-values come from the compilation of 
\citet{2009AA...497..611M}. % Melendez \& Barbuy (2009)
Their data are based on relative $gf$-values within each multiplet taken from theoretical works, and a calibration using experimental life-times \citep{1992A&A...259..301H,1999A&A...342..610S,2004A&A...414.1169S} % Hannaford et al. 1992, Schnabel et al. 1999, Schnabel et al. 2004
and BFs \citep[e.g.,][]{1987A&AS...67..225K,1990A&A...230..244H}. % Kroll \& Kock 1987, Heise \& Kock 1990
Solar calibrated values from \citet{2009AA...497..611M} were not considered here.
This source is the preferred choice and was used for 19 preselected lines, for which the \gfflag\ was set to \Yes.
For the remaining 23 preselected lines, the $gf$-values were taken from the theoretical work of \citet{RU}, % Raassen \& Uylings 1998
and the \gfflag\ was set to \Un.

In the background line list the majority of the \fei\ $gf$-values are based on the semi-empirical calculations by \citet{K07}. % Kurucz
They are supplemented by data from \citet{MRW}, \citet{FMW} and the Oxford, Madison, Hannover, and London groups.
We assigned a \gfflag=\Yes\ to 148 \fei\ lines in the background line list which lie in the UVES-580 and HR21 wavelength ranges (317 lines in the range from 420~nm to 920~nm used for the VALD extraction), which have data from the latter, ``first choice'' $gf$-value sources (while the \synflag\ is not defined for any of these lines).
We used theoretical data from \citet{K13} for about one third of the \feii\ $gf$-values, in addition to the data from \citet{RU}.

The overall situation of transition probabilities for Fe lines was investigated by \citet[their Sect.~2.2.2 and Fig.~4]{2017MNRAS.468.4311L}. % Lind et al.
They compared semi-empirical calculations made available by R.L. Kurucz in 2014 with available experimental data for about 2000 \fei\ lines (out of more than 500\,000 possible radiative bound-bound transitions) and found differences of up to $\pm$1~dex for transitions with log$gf\gtrsim-2$, which increased to $-3$~dex for transitions with log$gf\sim-4$ and high upper level energies. The compilation of the line list presented here was completed shortly before the new calculations by Kurucz for \fei\ became available, which is the reason why the previous calculations from 2007 were adopted. A comparison of the Kurucz data from 2007 with experimental data results in a figure which is very similar to Fig.~4 of \citet{2017MNRAS.468.4311L}, with only a few more deviating points appearing at log$gf\lesssim-5$. A direct comparison of the 2007 data with the 2014 data shows that the $gf$-values have changed by more than 0.5~dex for only about 50 out of 2000 lines.

Note that any line list is prone to being incomplete, as only about half of the lines visible in high-quality solar spectra have identified counterparts with good wavelengths according to \citet{2014dapb.book...39K}. % Kurucz
Many of these lines are expected to stem from iron, as indicated in the recent works by
\citet{2015ApJS..216....1P} % Peterson & Kurucz
and
\citet{2017ApJS..229...23P}, % Peterson et al.
who present \fei\ lines newly identified from an analysis of carefully selected stellar spectra. About 1200 of these lines fall in the wavelength range considered here, with astrophysically-determined $gf$-values included for about 20\% of them. While the nature of the GES speaks against the usage of astrophysical $gf$-values, as explained 
in Sect.~\refSectGfflag,
%in Sect.~\sectGfflag in \citet{Heiter_etal_2020},
%in Sect.~\ref{sect:gfflag},
these data should be included in future line lists of other projects, with the potential to provide more realistic synthetic spectra.

%
% Maria 2: Co
%

\subsection{Cobalt (Z=27)}
\label{sect:Co}

The most recent accurate \coi\ oscillator strengths were determined by \citet{1999ApJS..122..557N} %Nitz et al. (1999)
from FTS BFs put on an absolute scale using their own TRLIF life-time measurements \citep{1995JOSAB..12..377N}. %Nitz et al. (1995)
Reliable data are also available from \citet{1982ApJ...260..395C}. %Cardon et al. (1982)
These are the \gfflag=\Yes\ sources used for 70~\% of the preselected \coi\ lines, supplemented by calculations of \citet[\gfflag=\No]{K08}.
The overall quality status for preselected \coi\ lines is similar to that of \cri\ with eight \synflag=\Yes\ lines among those with high-quality $gf$-values, and a further 14 lines with \synflag=\Un\ 
(see Table~\refTabFlagstats).
%(see Table~\tabFlagstats in \citealt{Heiter_etal_2020}).
%(see Table~\ref{tab:flagstats}).
%
Line profiles for the \coi\ line at 535.204~nm are shown in Fig.~\ref{fig:V1Co1} (right column) as a representative example. This line was used for the largest number of FGK dwarfs and giants in the determination of the Co abundance for benchmark stars by \citet{2015A&A...582A..81J}
(see also Sect.~\refSectImpact).
%(see also Sect.~\sectImpact in \citealt{Heiter_etal_2020}).
%(see also Sect.~\ref{sect:impact}).

In addition, the background line list contains \coi\ data from \citet{MFW} and \citet{LWG}, and \coii\ lines with calculated $gf$-values from \citet{RPU} and \citet{K06}.

\paragraph{HFS data}

For \coi\ HFS we used the data measured by \citet{1996ApJS..107..811P}, who used $\gtrsim$1000 line profiles acquired with the high-resolution FTS at Imperial College. This yielded HFS $A$ and $B$ constants for 297 energy levels, almost all known \coi\ energy levels \citep{1996ApJS..107..811P}.
The complete set of HFS data is given in Table~\ref{HFSco}.

%
% Sarunas: Ni
%

\subsection{Nickel (Z=28)}
\label{sect:Ni}

Our preferred \nii\ oscillator strengths come from the recent work by
\citet{2014ApJS..211...20W}, % Wood et al. (2014)
who measured FTS and echelle spectrograph BFs and used the TRLIF life-times of 
\citet{1993JOSAB..10..794B}  % Bergeson \& Lawler (1993)
to provide the absolute scale.
These were available for 31 preselected \nii\ lines.
For the \nii\ line at 481.198~nm the high-quality $gf$-value from 
\citet{2003ApJ...584L.107J} % Johansson et al. (2003)
was used, which was derived from FTS BFs and a single TRLIF life-time.  
For the line at 542.465~nm the $gf$-value was taken from
\citet{1985JQSRT..33..307D}. % Doerr and Kock (1985)
Data from the above sources were assigned the \gfflag\ \Yes.
For four lines we used somewhat older experimental data from
\citet[\gfflag=\Un]{LWST}. % Lennard et al. (1975)
For the remaining 62 preselected lines we resorted to calculations by \citet[\gfflag=\No]{K08}. % Kurucz

About one third of the latter are blend-free in the Sun and Arcturus (\synflag=\Yes), and one half have uncertain blend status (\synflag=\Un). These lines should be given high priority for new laboratory measurements of $gf$-values
(see Sect.~\refSectDataneeds).
%(see Sect.~\sectDataneeds in \citealt{Heiter_etal_2020}).
%(see Sect.~\ref{sect:dataneeds}).
However, Ni abundances may be determined with confidence from the currently available data, as about half of the lines with high-quality $gf$-values also have \synflag=\Yes, and the other half has \synflag=\Un\ 
(see Table~\refTabFlagstats).
%(see Table~\tabFlagstats in \citealt{Heiter_etal_2020}).
%(see Table~\ref{tab:flagstats}).

\begin{figure*}
   \begin{center}
      \resizebox{0.9\hsize}{!}{\includegraphics{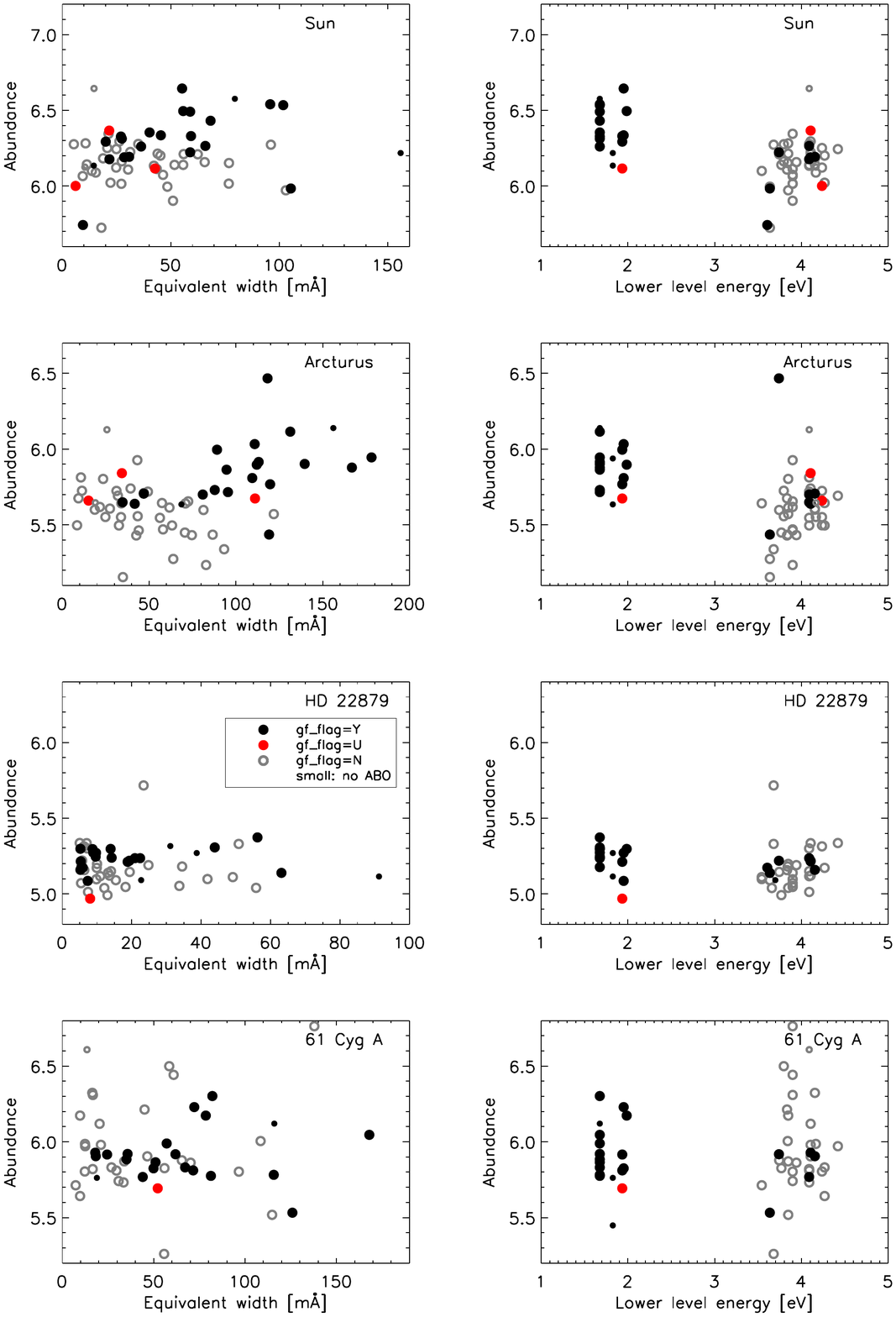}}
   \end{center}
\caption{Line-by-line \nii\ abundances determined for four benchmark stars for three groups of lines with different quality assessment of their transition probabilities, as a function of equivalent width and lower level energy. Abundances are given as $\log(\varepsilon_{\rm Ni})+12$, where $\varepsilon_{\rm Ni}=N_{\rm Ni}/N_{\rm H}$. Only lines with \synflag=\Yes\ or \Un\ and with equivalent widths $>5$~m\AA\ are included.
}
\label{fig:abuNi} 
\end{figure*}

Line abundances derived for four benchmark stars
(see Sect.~\refSectSpectra)
%(see Sect.~\sectSpectra in \citealt{Heiter_etal_2020})
%(see Sect.~\ref{sect:spectra})
for \nii\ lines with \synflag\ \Yes\ or \Un\ and equivalent widths $>5$~m\AA\ are shown in Fig.~\ref{fig:abuNi}, while mean abundances for lines with equivalent widths $\le100$~m\AA\ are given in Table~\ref{tab:abustat}.
For the Sun and Arcturus the scatter around the mean value is similar for the \gfflag=\Yes\ and \gfflag=\No\ lines, but the \No\ lines result in a systematically lower mean abundance by $\sim$0.1~dex than the \Yes\ lines. Note, however, that all the \No\ lines are high-excitation lines ($E_{\rm low} \sim$4~eV), while most of the \Yes\ lines are low-excitation lines ($E_{\rm low} \lesssim$2~eV). Therefore, the abundance difference may be related to modelling deficiencies rather than atomic data issues.
For HD~22879 and 61~Cyg~A the mean abundances of the different sets of lines are more similar, but the \No\ lines show a twice as large scatter than the \Yes\ lines.

The background line list contains \nii\ data from \citet{FMW} and \citet{WLa}, in addition to the sources above, as well as \niii\ data from \citet{K03}.

Ni consists of five stable isotopes, dominated by $^{58}$Ni and $^{60}$Ni in the Solar System
(see Table~\refTabIsotopes).
%(see Table~\tabIsotopes in \citealt{Heiter_etal_2020}).
%(see Table~\ref{tab:isotopes}).
The single isotope with non-zero spin contributes only 1\% to the natural Ni abundance, and thus HFS can be ignored.
Isotopic components were not included in the Gaia-ESO line list, but are listed for eight lines in Table~2 of \citet[see also their Sect.~6.8.2]{2015A&A...573A..26S}. Five of these lines are among our preselected lines, including three lines with both \gfflag\ and \synflag=\Yes\ (617.681, 622.398, and 637.825~nm).
The largest wavelength shift including all isotopes is 0.05~\AA, while the shifts for the dominating isotopes are $\lesssim$0.03~\AA.

%
% Martin 2: Cu, Zn
%

\subsection{Copper (Z=29)}
\label{sect:Cu}

For the six preselected \cui\ lines we adopted the experimental $gf$-values from \citet{KR}. % Kock \& Richter (1968) 
To be fully consistent with \citet{2015A&A...573A..27G} % Grevesse et al. (2015)
we re-normalised these to the improved life-times of \citet{1989ZPhyD..11..287C}, % Carlsson et al. (1989)
although in terms of $gf$-values the change is $<0.01$\,dex. 
The transition probabilities of these \cui\ lines are rated as accurate (\gfflag=\Yes).
For \cui\ lines only appearing in the background line list, we adopted transition probabilities from \citet{K12}, as well as from \citet{BIEMa} for a few lines.
In the Solar System, Cu consists to more than two thirds of $^{63}$Cu and almost one third $^{65}$Cu, both of which have non-zero nuclear spin and therefore causing hyperfine splitting
(see Table~\refTabIsotopes).
%(see Table~\tabIsotopes in \citealt{Heiter_etal_2020}).
%(see Table~\ref{tab:isotopes}).  
Table~\ref{HFScu} provides the adopted data for HFS and isotope shifts.

One of the six \cui\ lines for abundance purposes is considered largely clean (578.213\,nm, \synflag=\Yes) while the others are blended to varying degrees (\synflag=\Un).

\subsection{Zinc (Z=30)}
\label{sect:Zn}

For the two preselected \zni\ lines at 481.053 and 636.234\,nm the $gf$-values were taken from the theoretical work of \citet{1980AA....84..361B}. % Bi\'{e}mont \& Godefroid (1980)
In the case of 481.053\,nm, this was further slightly re-normalised to the more accurate life-times measured by \citet{1980ZPhyA.298..249K} % Kerkhoff et al. (1980)
as in \citet{2015A&A...573A..27G}. % Grevesse et al. (2015)
These transition probabilities are rated reliable (\gfflag=\Yes), but the lines are partly blended (\synflag=\Un), including 636.234\,nm which is located in a \cai\ autoionising line. 
The isotopes $^{64}$Zn, $^{66}$Zn, and $^{68}$Zn all contribute significantly to the Solar System abundance of Zn
(see Table~\refTabIsotopes),
%(see Table~\tabIsotopes in \citealt{Heiter_etal_2020}),
%(see Table~\ref{tab:isotopes}),
but data for isotopic shifts are not available.
Hyperfine splitting is of little importance owing to the low natural abundance of the single isotope of Zn with non-zero spin.

% place all figures and tables here
%\clearpage

%
% Maria 3: Sr, Y, Zr
%

\subsection{Strontium (Z=38)}
\label{sect:Sr}

Oscillator strengths for the preselected \sri\ lines were taken from \citet{GC} with \gfflag=\Yes\ (five lines), \citet{WGTG} and \citet{VGH} with \gfflag=\Un, and \citet{CB} with \gfflag=\No. % Garcia & Campos (1988); Werij et al. (1992); Vaeck et al. (1988); Corliss & Bozman (1962a)
\citet{GC} measured relative transition probabilities from emission-line spectrometry in a hollow cathode discharge lamp. These were placed on an absolute scale using previously published experimental life-times by other authors.
The data from \citet{WGTG} are derived from a combination of R-matrix, multichannel-quantum-defect-theory, and modified Coulomb-approximation calculations, plus branching-ratio measurements.

None of the preselected lines are blend-free in the Sun or Arcturus, and there are only two lines with \synflag=\Un\ (679.102~nm with \gfflag=\Yes\ and 654.678~nm with \gfflag=\Un). Both of these are very weak, and the remaining eight \sri\ lines have \synflag=\No.
The background line list also contains a few \srii\ lines, with data from \citet{Wc}.

Sr has three stable isotopes, one of them with non-zero spin
(see Table~\refTabIsotopes).
%(see Table~\tabIsotopes in \citealt{Heiter_etal_2020}).
%(see Table~\ref{tab:isotopes}).
Owing to the low Solar System abundance of the latter, we expect the effect of HFS to be negligible for Sr.
\citet{1957PhRv..105.1260H} % Hughes
presented measurements of isotope shifts for 13 \sri\ lines within the Gaia-ESO wavelength range. All of them have negligible shifts ($\lesssim$2~m\AA).

\subsection{Yttrium (Z=39)}
\label{sect:Y}

For the few \yi\ lines in the preselected line list only the semi-empirical calculations by \citet{K06} are available which were adopted for this work. Because the accuracy of these values is not known they were assigned a \gfflag=\No.
All of these lines are furthermore blended to some degree in the Sun and Arcturus
(see Table~\refTabFlagstats).
%(see Table~\tabFlagstats in \citealt{Heiter_etal_2020}).
%(see Table~\ref{tab:flagstats}).

The situation is much better for the more numerous \yii\ lines, all of which are contained in the recent experimental work by \citet{BBEHL}, %Biemont et al. (2011)
and were assigned a \gfflag=\Yes.
Four of these lines are also unblended in the Sun and Arcturus (\synflag=\Yes; 488.368, 508.742, 528.982, 572.889~nm), and seven additional lines have uncertain blend status (\synflag=\Un).
The background line list also contains data from \citet{K11} for \yii.

Y has only one stable isotope, with non-zero spin
(see Table~\refTabIsotopes).
%(see Table~\tabIsotopes in \citealt{Heiter_etal_2020}).
%(see Table~\ref{tab:isotopes}).
HFS data are listed in Table~2 of \citet[see also their Sect.~4.8]{2015A&A...573A..27G} for seven of our preselected \yii\ lines. However, these data result in HFS components with negligible wavelength difference ($\lesssim$5~m\AA).

\subsection{Zirconium (Z=40)}
\label{sect:Zr}

For the preselected \zri\ lines we adopted the $gf$-values of \citet{BGHL}, %Biemont et al. 1981
who measured life-times and branching ratios by laser-induced fluorescence and emission spectrometry.

For two of the three preselected \zrii\ lines we used the experimental transition probabilities of \citet[\gfflag=\Yes]{LNAJ}, %Ljung et al. (2006)
and for the third line the estimated oscillator strength of \citet[\gfflag=\Un]{CC} was used. %Cowley & Corliss (1983)

\begin{figure*}[ht]
   \begin{center}
      \resizebox{\hsize}{!}{\includegraphics{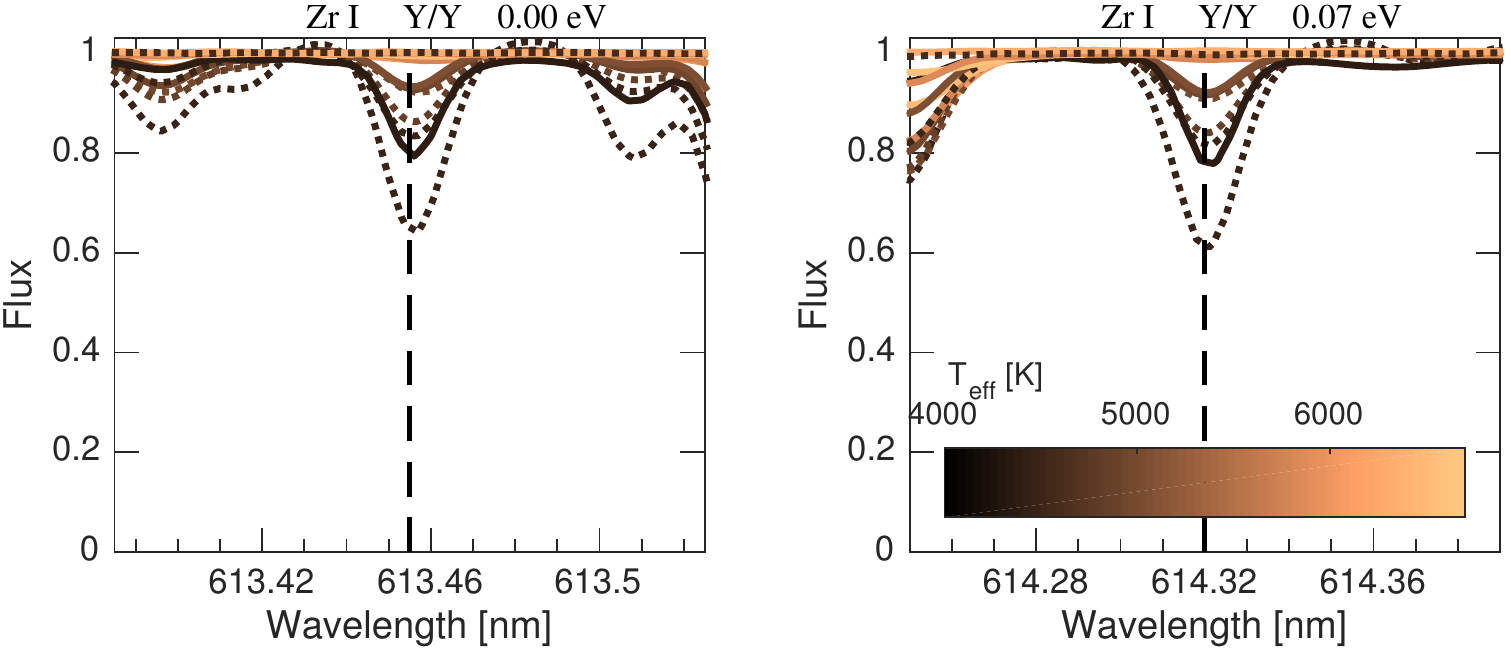}}
   \end{center}
   \caption{Line profiles for two high-quality preselected \zri\ lines, which are weak in solar-type stars but strong in cooler stars. Generated from observed spectra of selected Gaia FGK benchmark stars.
   See Sect.~\refSectSpectraT.
   % See Sect.~\sectSpectra in \citet{Heiter_etal_2020}.
   % See Sect.~\ref{sect:spectra} for description.
   }
   \label{fig:Zr1}
\end{figure*}

The Zr lines are all very weak in solar-type stars, but can be quite strong in cooler stars.
Two examples are shown in Fig.~\ref{fig:Zr1}.
Thus, a detailed blending analysis and line selection should be done whenever stars are analysed whose parameters differ from those of the Sun and Arcturus.
Five of the \zri\ lines seem to be good candidates for abundance analysis and were assigned \synflag=\Yes, with the remainder having \synflag=\Un.
The three \zrii\ lines all seem to be affected by blends
(see Table~\refTabFlagstats).
%(see Table~\tabFlagstats in \citealt{Heiter_etal_2020}).
%(see Table~\ref{tab:flagstats}).

In addition to the sources above, the background line list contains \zri\ data from \citet{CB}, and oscillator strengths estimated by \citet{MULT} for \zri\ and \zrii\ lines.

Zr has five stable isotopes which all contribute significantly to the Solar System abundance
(see Table~\refTabIsotopes).
%(see Table~\tabIsotopes in \citealt{Heiter_etal_2020}).
%(see Table~\ref{tab:isotopes}).
Three of them contribute at the 20\% level or more, and all of these have zero spin. Therefore we did not include any HFS components\footnote{HFS $A$ and $B$ constants can be found for 12 levels of \zri\ in \citealt{1998JaJAP..37.5049L}, which correspond to the lower levels of some of our preselected lines.}.
Isotopic shifts were measured by \citet{1994JOSAB..11..552L} for 330 \zri\ lines between 536 and 706~nm. The maximum shift between $^{90}$Zr and $^{94}$Zr in their list was 8~m\AA. Similar values have been reported by \citet{1987ZPhyD...7..129B} and \citet{1998JaJAP..37.5049L}. This level of isotopic splitting is negligible in the context of the GES.

%
% Martin 3: Nb, Mo, Ru
%

\subsection{Niobium (Z=41)}
\label{sect:Nb}

Accurate experimental oscillator strengths for the eight preselected \nbi\ lines were measured by \citet{1986JQSRT..35..281D}. % Duquette et al. (1986)
In addition, the background line list contains \nbi\ data from older measurements of \citet{CB,CBcor} and \citet{DLa}. % Corliss & Bozman (1962a,b) and Duquette & Lawler (1982)
Niobium has one isotope with a nuclear spin of 9/2 and can therefore be expected to show hyperfine structure. 
Because the Nb lines are all weak, HFS splitting has not been accounted for but relevant data can be found in \citet{NI} and \citet{NHEL}. % Nilsson & Ivarsson (2008) and Nilsson et al. (2010)
All of the \nbi\ lines are heavily blended (\synflag=\No).

\subsection{Molybdenum (Z=42)}
\label{sect:Mo}

\begin{figure*}[ht]
   \begin{center}
      \resizebox{\hsize}{!}{\includegraphics{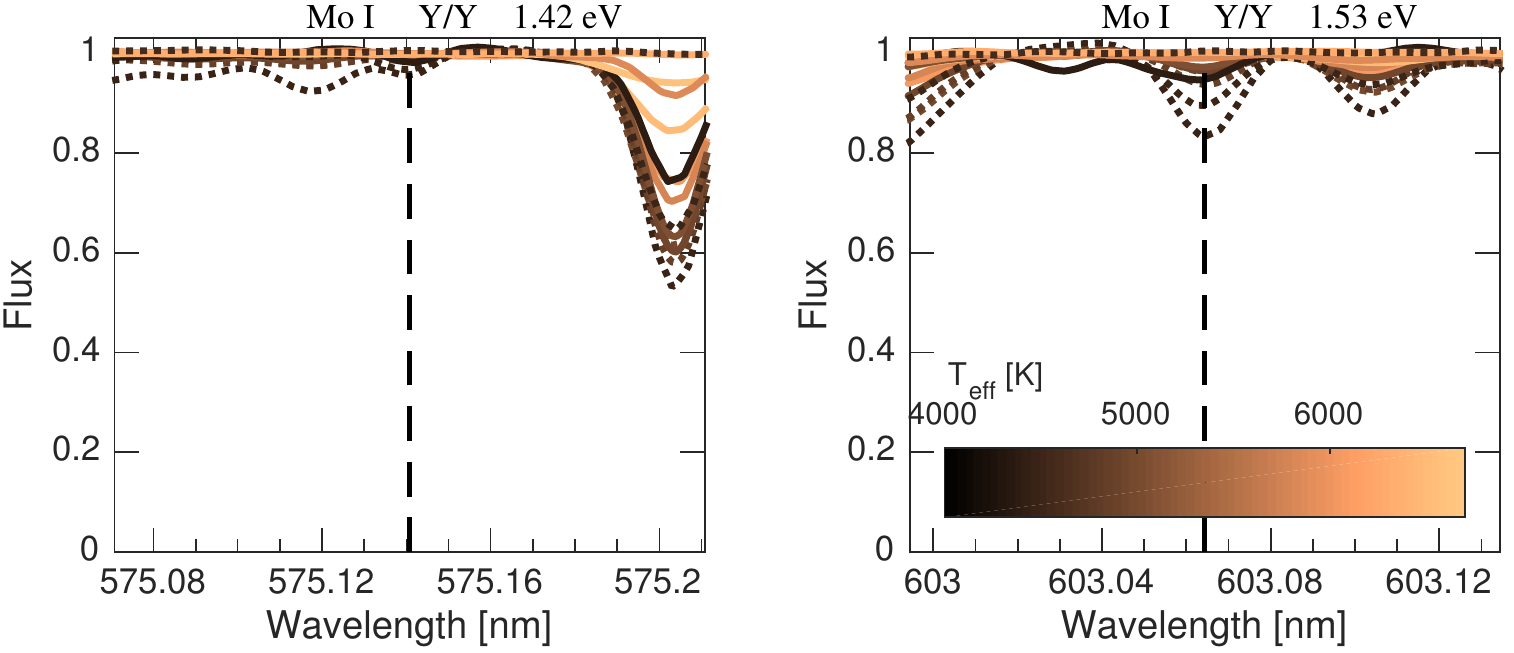}}
   \end{center}
   \caption{Line profiles for high-quality preselected \moi\ lines generated from observed spectra of selected Gaia FGK benchmark stars.
   See Sect.~\refSectSpectraT.
   % See Sect.~\sectSpectra in \citet{Heiter_etal_2020}.
   % See Sect.~\ref{sect:spectra} for description.
   }
   \label{fig:Mo1}
\end{figure*}

Experimental transition probabilities for the six preselected \moi\ lines are available from \citet{WBb}, % Whaling \& Brault (1988)
who used measured radiative life-times and branching ratios; these are rated accurate (\gfflag=\Yes).
The Mo lines are all very weak and therefore hyperfine and isotope splitting have not been accounted for.
The primary Mo abundance indicators are 575.141 and 603.064~nm (\synflag=\Yes), although they are too weak in dwarfs to be useful (see Fig.~\ref{fig:Mo1}).

\subsection{Ruthenium (Z=44)}
\label{sect:Ru}

We included one line of \rui\ in the preselected line list (486.9\,nm) with an experimental (life-time and branching ratio) $gf$-value from \citet{WSL}; % Wickliffe et al. (1994)
the line is not included in the more recent work of \citet{2009MNRAS.396.2124F}. % Fivet et al. (2009)
The adopted $gf$-value is considered accurate (\gfflag=\Yes) but the line is significantly blended (\synflag=\Un).
In addition, the background line list contains \rui\ data from \citet{CBcor}, \citet{BGKZ}, \citet{SLb}, and \citet{GUES}. % Corliss & Bozman (1962b), Biemont et al. (1984), Salih & Lawler (1985), Kurucz (1993)
The Ru lines are all very weak and therefore hyperfine and isotope splitting have not been accounted for.

%
% Ulrike: Ba, La, Ce, Pr, Nd, Sm, Eu, Gd, Dy
%

\subsection{Barium (Z=56)}
\label{sect:Ba}

\begin{figure*}[ht]
   \begin{center}
      \resizebox{\hsize}{!}{\includegraphics{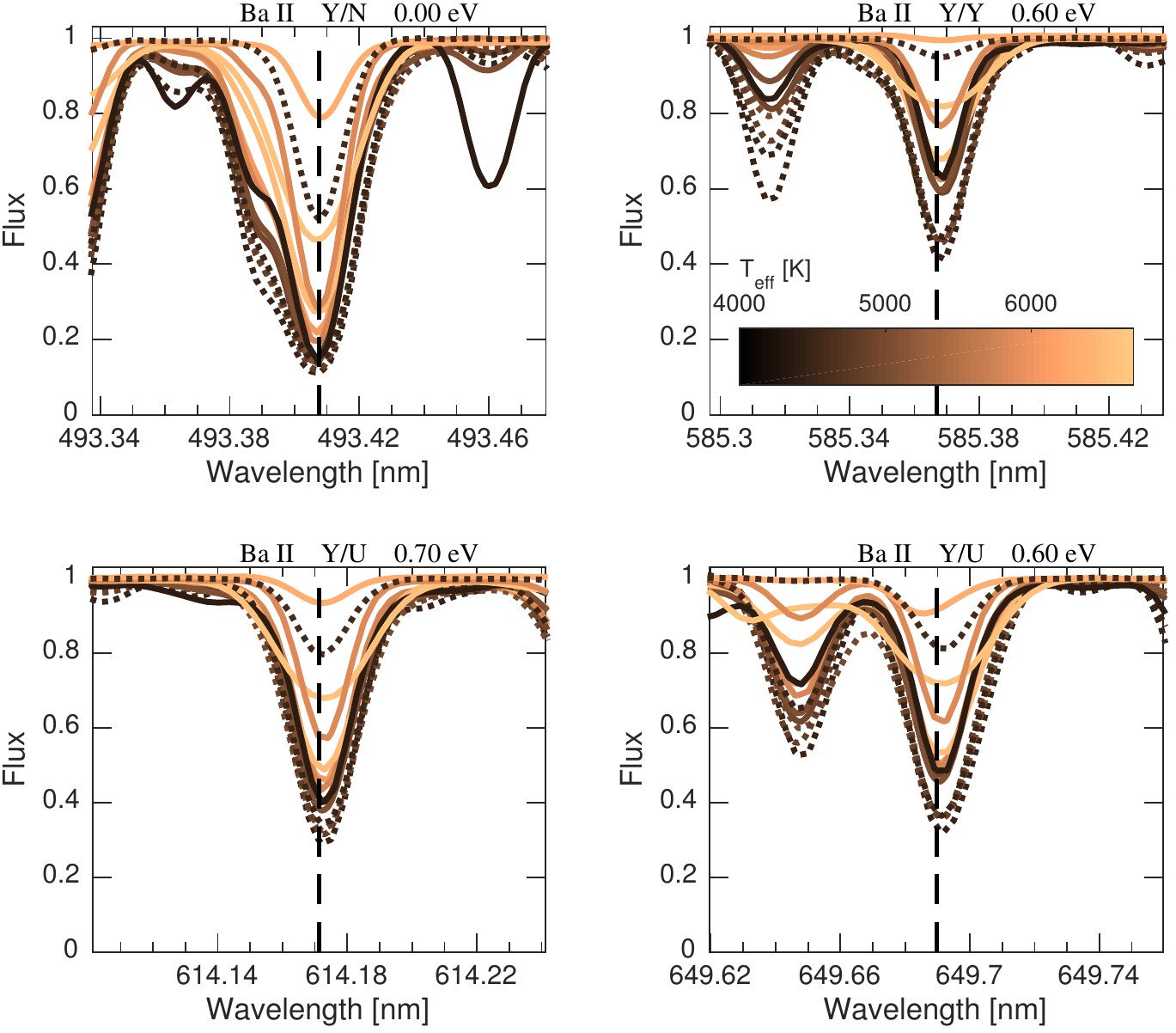}}
   \end{center}
   \caption{Line profiles for preselected \baii\ lines generated from observed spectra of selected Gaia FGK benchmark stars.
   See Sect.~\refSectSpectraT.
   % See Sect.~\sectSpectra in \citet{Heiter_etal_2020}.
   % See Sect.~\ref{sect:spectra} for description.
   }
   \label{fig:Ba2}
\end{figure*}

We used the experimental data from \citet{1992AA...255..457D} % Davidson et al. (1992)
for all four preselected \baii\ lines. Two of the lines are also included in the experimental work of \citet{2008PhRvA..77f0501K}. % Kurz et al. (2010)
The log$gf$ values from \citet{2008PhRvA..77f0501K} are lower than those from \citet{1992AA...255..457D} by $\sim$0.1~dex. For consistency, we adopted the source containing all four lines.
All four lines are strong in stellar spectra except for the most metal-poor stars. The line at 493.408~nm is strongly blended in all but the most metal-poor stars (see Fig.~\ref{fig:Ba2}).

In addition, the background line list contains \bai\ data from \citet{CB}, % Corliss & Bozman (1962a)
and data from \citet{MW} % Miles & Wiese (1969)
for \bai\ and \baii\ lines.

\paragraph{HFS and IS data}
Solar System material contains a mixture of several stable Ba isotopes, five of which contribute with more than 1\% to the Ba abundance
(Table~\refTabIsotopes).
% (Table~\tabIsotopes in \citealt{Heiter_etal_2020}).
% (Table~\ref{tab:isotopes}).
Isotopic splittings have been measured for the 493.408~nm line by \citet{1984ZPhyA.318..125W}, % Wendt et al. (1984)
for the 585.367 and 614.171~nm lines by \citet{1982JPhB...15.1805V}, % Van Hove et al. (1982)
and for the 649.690~nm line by \citet{1993JPhB...26.4289V}. % Villemoes et al. (1993)
The wavelength shifts are summarised in a table in the on-line database of R.L. Kurucz\footnote{\url{http://kurucz.harvard.edu/atoms/5601/isoshifts5601.dat}}, and the largest ones are 2~m\AA\ (between isotopes 134 and 138) for all lines.
Although this can be considered to have a negligible effect in the context of the GES and similar surveys all five isotopic components were included for the preselected \baii\ lines.

Two of the isotopes have non-zero nuclear spin leading to HFS in \baii\ lines with non-negligible effects on abundance analysis. Experimental determinations of the HFS $A$ and $B$ constants were found for all levels involved in the preselected \baii\ lines in \citet{1981pmfc.conf...99B}, % Becker et al. (1981)
\citet{1986PhRvA..33.2117S}, % Silverans et al. (1986)
and \citet{1993JPhB...26.4289V}. % Villemoes et al. (1993)
The complete set of HFS data is given in Table~\ref{HFSba}.

\subsection{Lanthanum (Z=57)}
\label{sect:La}

\begin{figure}[ht]
   \begin{center}
      \resizebox{\hsize}{!}{\includegraphics[trim=50 50 50 50,clip]{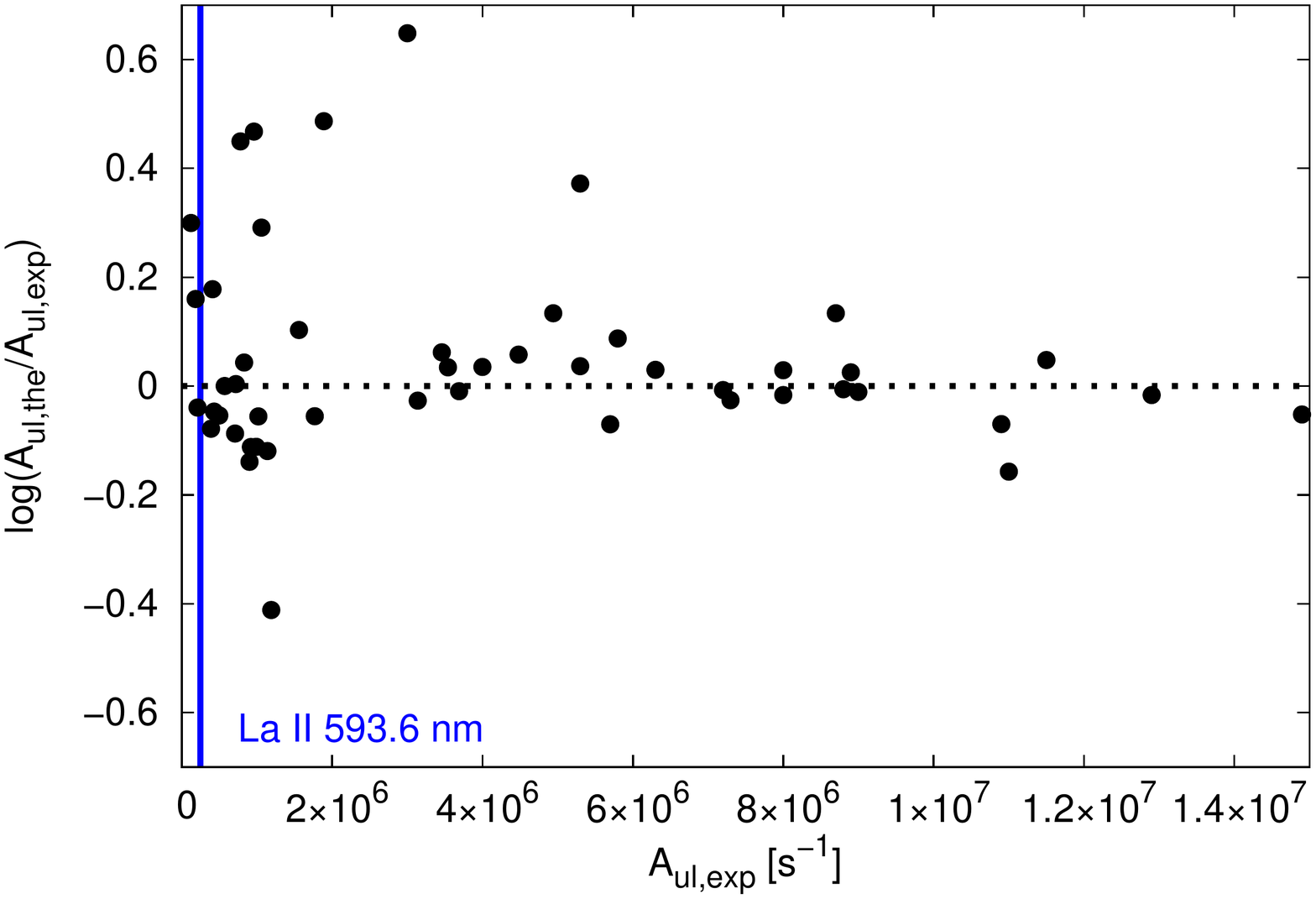}}
   \end{center}
   \caption{Comparison between \emph{the}oretical and \emph{exp}erimental transition probabilities $A_{ul}$ for \laii. Data are from \citet{2009JPhB...42r5002K} and \citet{LBS}, respectively. The blue vertical line indicates the location of the line at 593.6~nm.
   }
   \label{fig:La2}
\end{figure}

We used experimental data from \citet{LBS} % Lawler et al. 2001
for five of the preselected \laii\ lines (\gfflag=\Yes).
For the sixth line at 593.6~nm the only experimental data available are from \citet{CB}, which is known to have large, systematic uncertainties \citep[see, e.g.,][]{ABH,2010EAS....43...91W}.
We did not attempt to apply any corrections to the \citet{CB} data or adopt any previous such attempts. Instead, we resorted to the theoretical value from \citet{2009JPhB...42r5002K}. % Ku{\l}aga-Egger} and Migda{\l}ek 2009
To assess the uncertainty inherent in that work, we compared transition probabilities for lines in common with \citeauthor{LBS}, as shown in Fig.~\ref{fig:La2}.
The comparison shows a similar scatter as the comparison between the \citeauthor{LBS} data and data from \citet[see Figs.~1 to 3 in \citealt{LBS}]{1996MNRAS.278..997B}, which are based on a semi-empirical calibration of data from \citet{MC}. We assigned a \gfflag=\Un\ to the 593.6~nm line.
In addition, the background line list contains data for \lai\ and \laii\ by \citet{CB} and for \laii\ by \citet{ZZZ}.

\begin{figure*}[ht]
   \begin{center}
      \resizebox{0.9\hsize}{!}{\includegraphics[trim=0 0 40 0, clip]{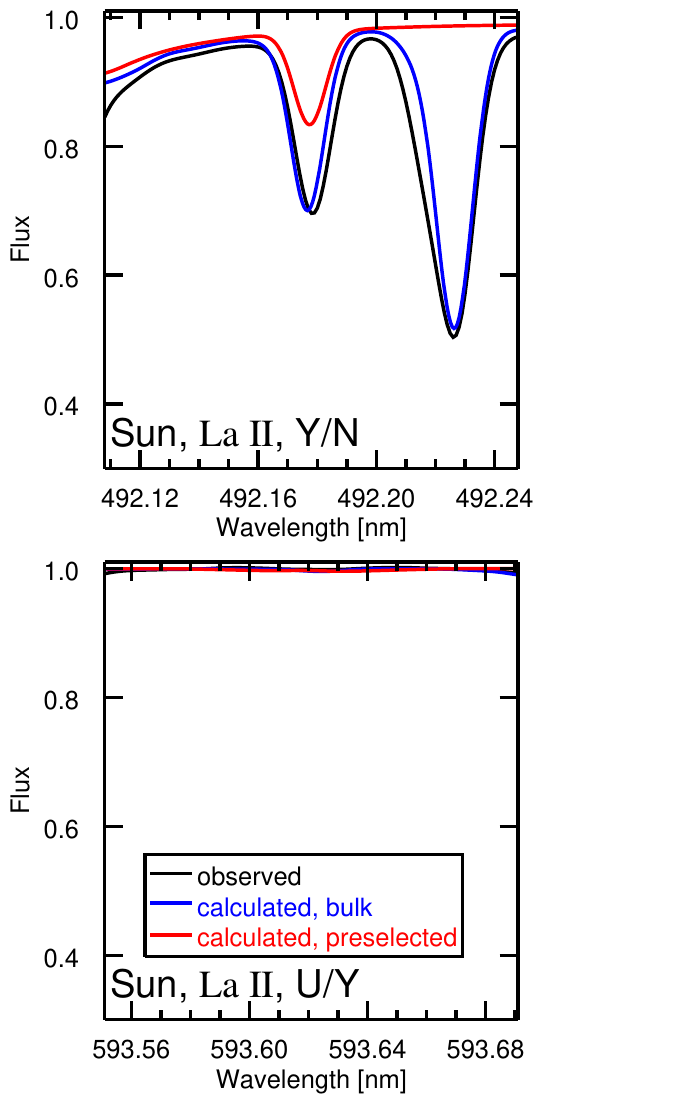}\includegraphics[trim=0 0 40 0, clip]{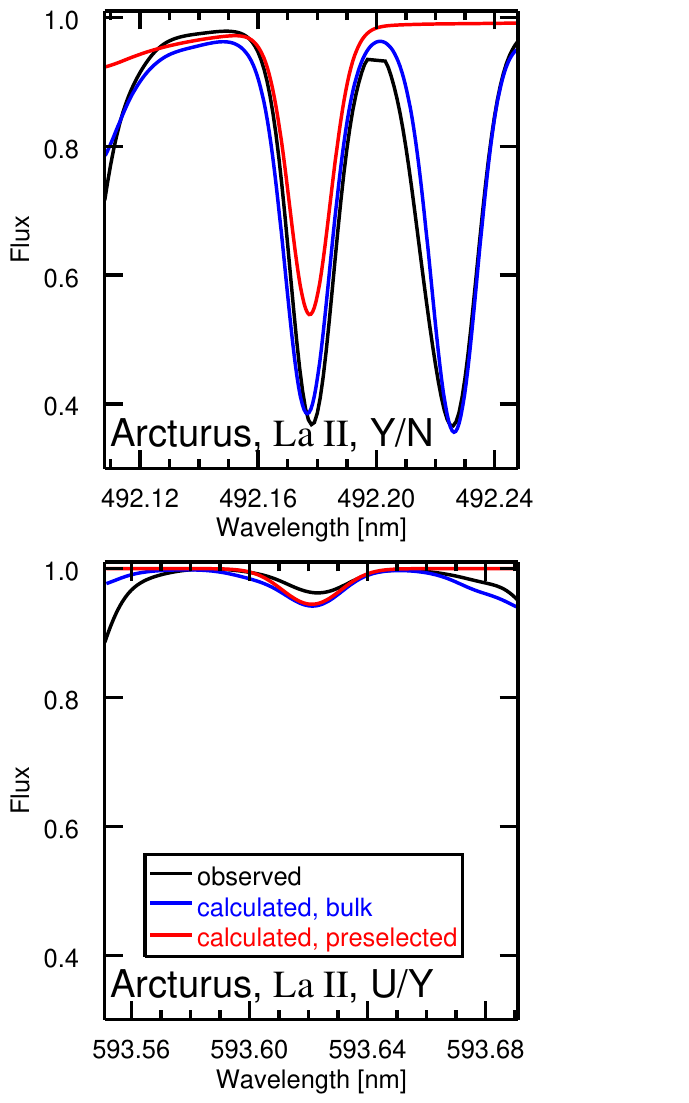}}
   \end{center}
   \caption{Comparison of observed and calculated line profiles around the two \laii\ lines at 492.2~nm (upper panels, \gfflag=\Yes, \synflag=\No), and at 593.6~nm (lower panels, \gfflag=\Un, \synflag=\Yes) for the Sun (left) and Arcturus (right). Black lines: observations, red lines: calculations including preselected spectral lines only, blue lines: calculations including blends from background line list.
   }
   \label{fig:La2_profiles_calc}
\end{figure*}

\begin{figure*}[ht]
   \begin{center}
      \resizebox{\hsize}{!}{\includegraphics{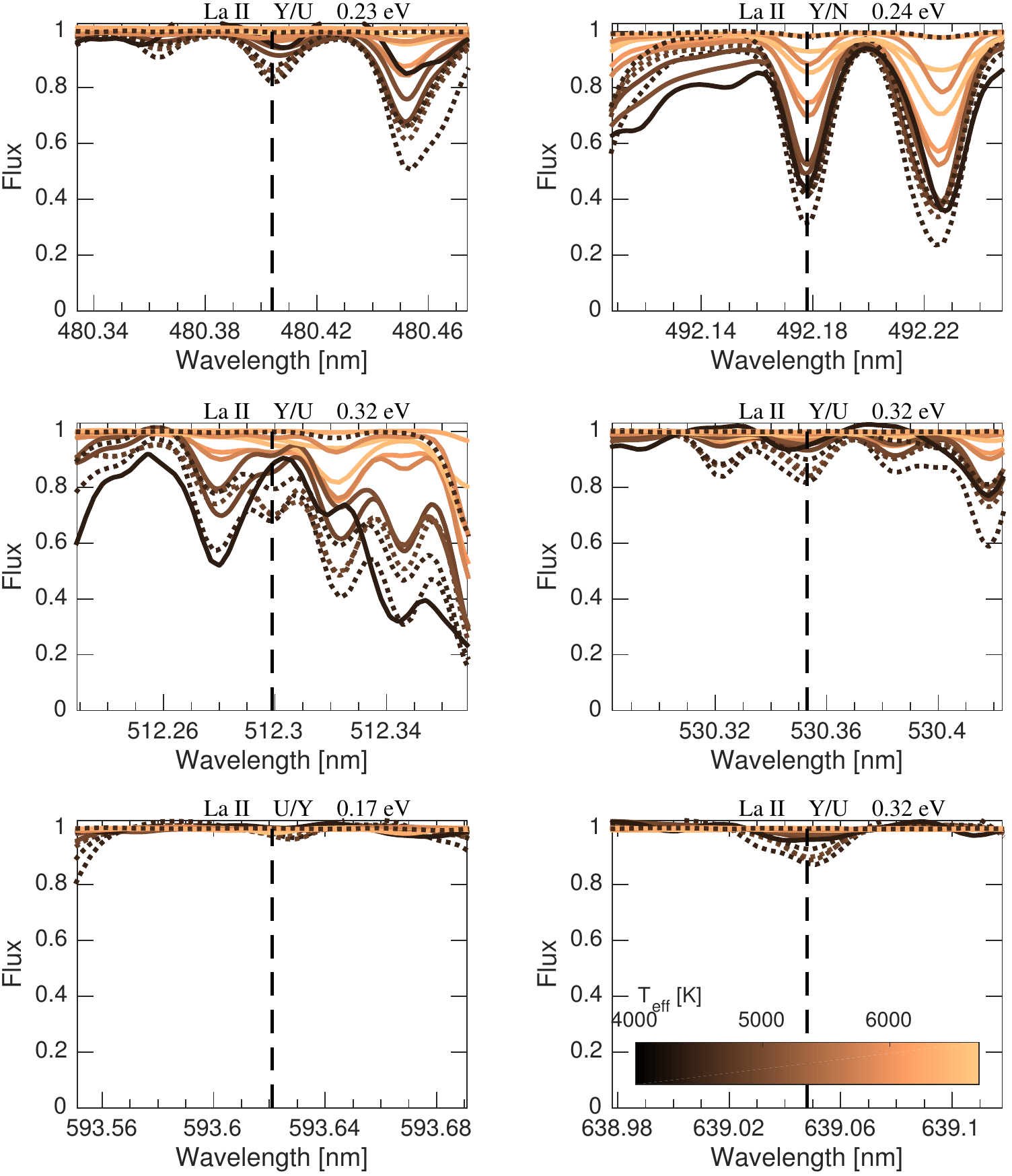}}
   \end{center}
   \caption{Line profiles for preselected \laii\ lines generated from observed spectra of selected Gaia FGK benchmark stars.
   See Sect.~\refSectSpectraT.
   % See Sect.~\sectSpectra in \citet{Heiter_etal_2020}.
   % See Sect.~\ref{sect:spectra} for description.
   }
   \label{fig:La2_profiles_GBS}
\end{figure*}

Most of the preselected lines are blended to different degrees in the spectra of the Sun, Arcturus, and other benchmark stars (see Figs.~\ref{fig:La2_profiles_calc} and \ref{fig:La2_profiles_GBS}).
The line at 492.2~nm appears rather clean in observed spectra, but a synthesis including surrounding lines from the background line list yields a significantly stronger line than a synthesis for the \laii\ line only. A look into the background line list reveals several lines of different species with similar strength at the same wavelength (\sii, \tii, \ceii, \yi). Hence, the line is not recommended for abundance analysis (\synflag=\No).
The only line with \synflag=\Yes\ at 593.6~nm is invisible in spectra of dwarf stars, and weak in spectra of giant stars, but might be stronger in spectra of r-process-rich, metal-poor stars.

\paragraph{HFS data}
The single dominant isotope $^{139}$La makes up 99.9\% of La in solar system material
(Table~\refTabIsotopes).
% (Table~\tabIsotopes in \citealt{Heiter_etal_2020}).
% (Table~\ref{tab:isotopes}).
Thus, La lines will not be affected by isotopic splitting.

On the other hand, HFS should be taken into account for the analysis of La lines.
$A$ and $B$ constants were extracted from the literature for each of the nine energy levels involved in the preselected \laii\ transitions, where possible. Most of the data were taken from the experimental work based on ion-beam-laser spectroscopy by \citet{1982ZPhyA.304..279H}. % Höhle et al. (1982), high-resolution spectroscopy on collinear laser-ion-beams
A similar type of measurement was found for one level in \citet{2001JaJAP..40.2508L}. % Li et al. (2001), Collinear Fast Ion-Beam-Laser Spectroscopy
HFS constants measured from FTS spectra for another two levels were found in \citet{LBS}. % Lawler et al. (2001), NSO FTS spectra, lanthanum-lined HCD lamps were used as a source of La II emission lines
Semi-empirical calculations of HFS data for the upper level of the line at 492.2~nm are provided in \citet{2008JPhB...41u5004F}, but were not included in the line list.
The complete set of HFS data is given in Table~\ref{HFSla}.

\subsection{Cerium (Z=58)}
\label{sect:Ce}

High-quality experimental data for \ceii\ lines from BFs and life-times are available from \citet[\gfflag=\Yes]{LSCI} for two thirds of the twelve preselected lines. % Lawler et al. (2009)
For the other lines we used theoretical data from \citet[\gfflag=\Un]{PQWB}.
The latter used two different approaches to calculate radiative life-times, designated ``Calculation A'' and ``Calculation B''. According to the authors, Calculation B is expected to be more accurate than Calculation A. \citet{PQWB} give $gf$-values using Calculation B for selected lines, including one of the lines considered here (511.7~nm).
The background line list contains numerous additional \ceii\ lines with data from these two sources, as well as \cei\ lines with data from \citet{CB, CBcor, CRC}.

The values for the three preselected NIR lines are taken from the on-line DREAM database\footnote{\url{http://hosting.umons.ac.be/html/agif/databases/dream.html}}, which appears to provide the results of Calculation A of \citet{PQWB}.
In addition to having uncertain $gf$-values, none of these three lines is recommended for abundance analysis due to their unfavourable blending properties (\synflag=\Un\ or \No).

\begin{figure*}[ht]
   \begin{center}
      \resizebox{\hsize}{!}{\includegraphics{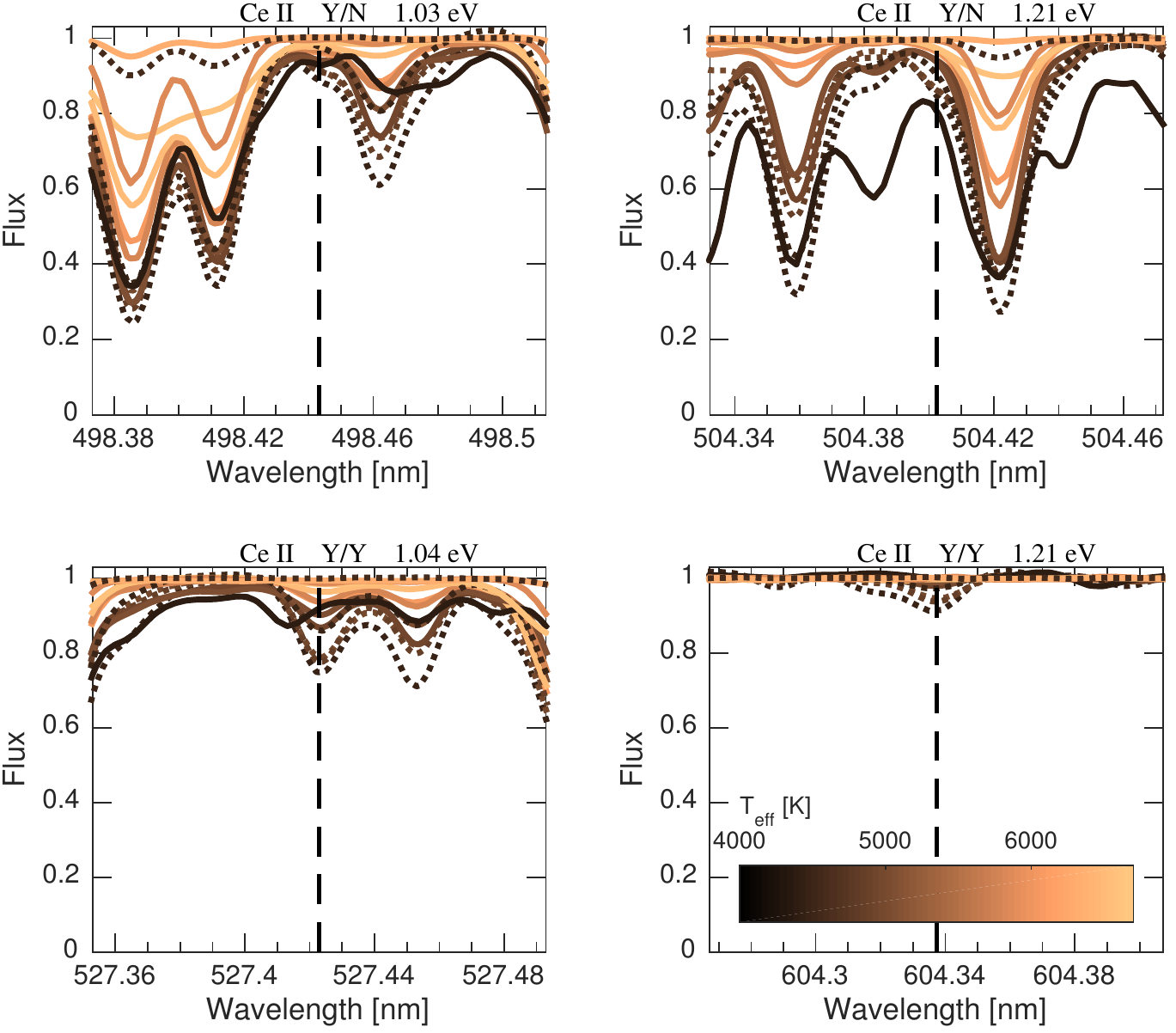}}
   \end{center}
   \caption{Line profiles for preselected \ceii\ lines generated from observed spectra of selected Gaia FGK benchmark stars.
   See Sect.~\refSectSpectraT.
   % See Sect.~\sectSpectra in \citet{Heiter_etal_2020}.
   % See Sect.~\ref{sect:spectra} for description.
   }
   \label{fig:Ce2_profiles_GBS}
\end{figure*}

Among the optical lines two appear unblended in the spectra of the Sun, Arcturus, and other benchmark stars (Fig.~\ref{fig:Ce2_profiles_GBS}, lower panels), while two others are strongly blended and cannot be recommended for abundance analysis (Fig.~\ref{fig:Ce2_profiles_GBS}, upper panels). The remaining lines were assigned a \synflag=\Un\ and their usage needs to be decided on a case-by-case basis.

% IS and HFS
Ce has four stable isotopes, all with zero nuclear spin. Two of them contribute to the Solar System abundance with more than 1\%
(Table~\refTabIsotopes).
% (Table~\tabIsotopes in \citealt{Heiter_etal_2020}).
% (Table~\ref{tab:isotopes}).
However, the isotope splitting is negligible for \ceii\ lines \citep{LSCI}. % Lawler et al. (2009)

\subsection{Praseodymium (Z=59)}
\label{sect:Pr}

\begin{figure*}[ht]
   \begin{center}
      \resizebox{0.89\hsize}{!}{\includegraphics{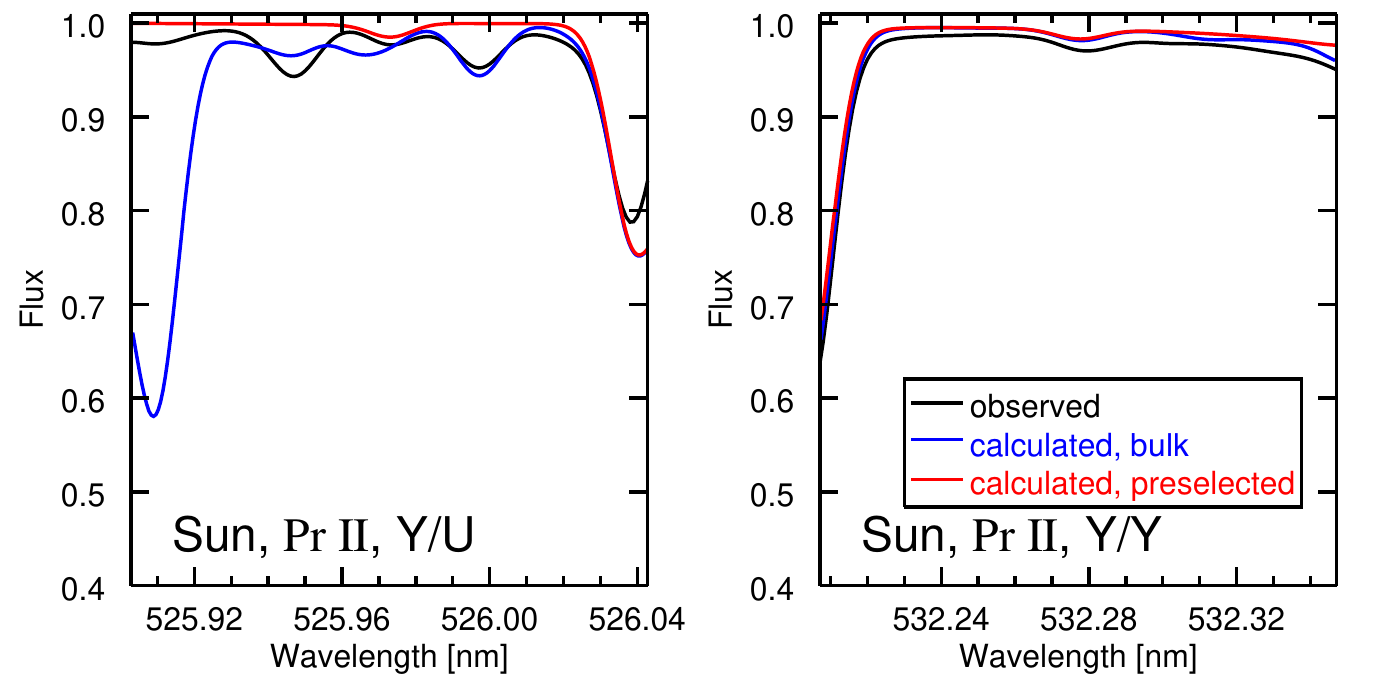}}
      \resizebox{0.89\hsize}{!}{\includegraphics{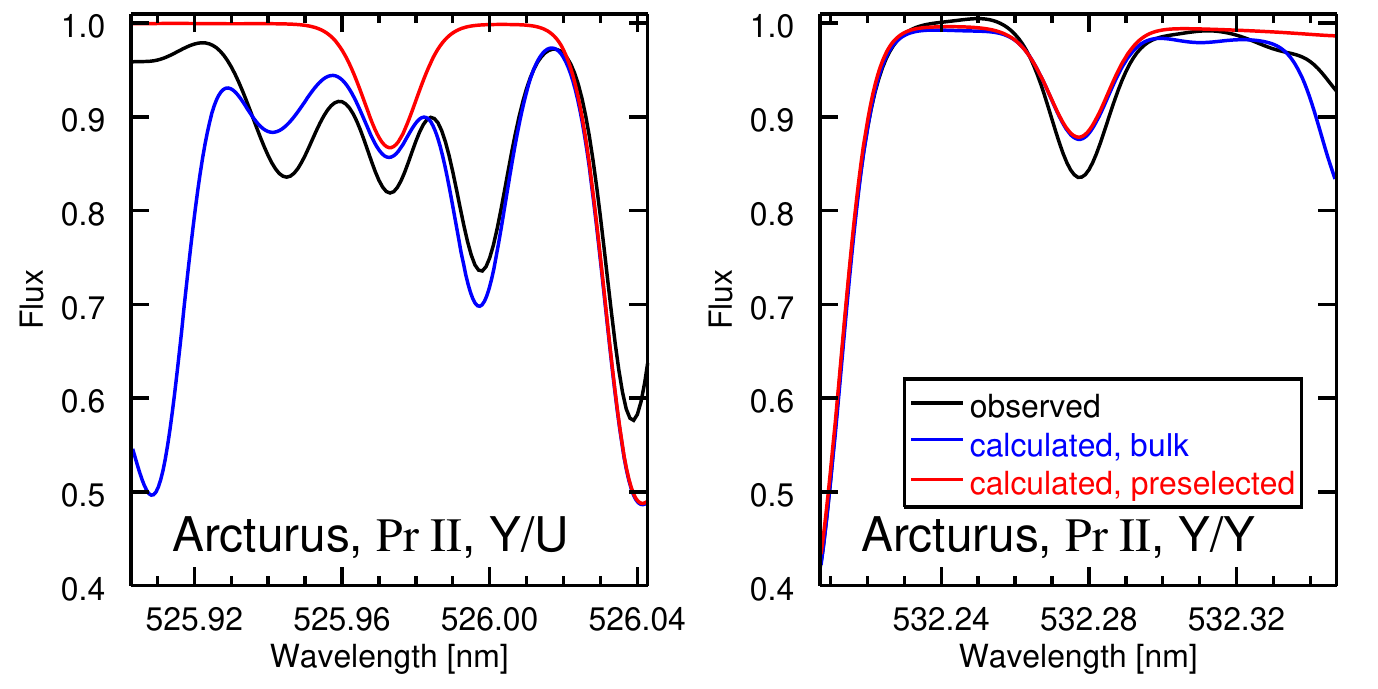}}
      \resizebox{0.89\hsize}{!}{\includegraphics{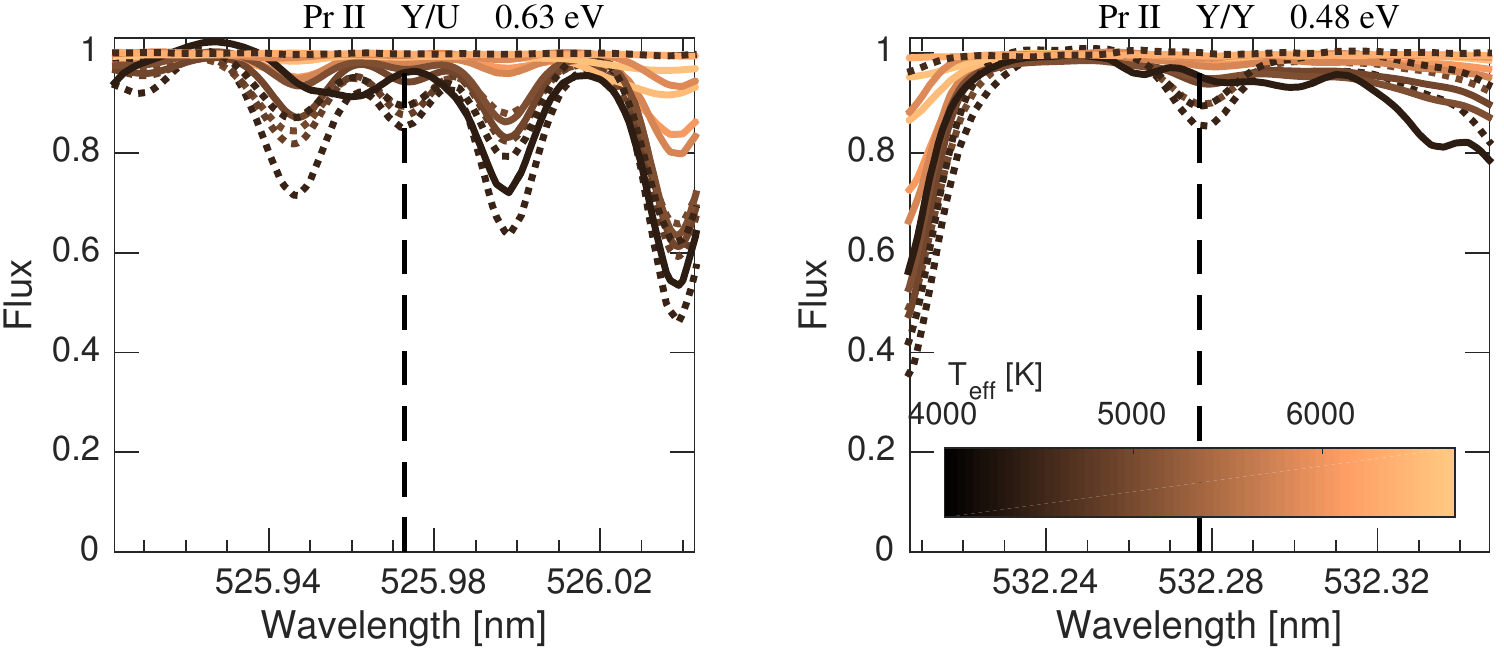}}
   \end{center}
   \caption{
   Observed and calculated line profiles around two preselected \prii\ lines for the Sun (top row) and Arcturus (middle row). Black lines: observations, red lines: calculations including preselected spectral lines only, blue lines: calculations including blends from background line list.
   Bottom row: Line profiles generated from observed spectra of selected Gaia FGK benchmark stars.
   See Sect.~\refSectSpectraT.
   % See Sect.~\sectSpectra in \citet{Heiter_etal_2020}.
   % See Sect.~\ref{sect:spectra} for description.
   }
   \label{fig:Pr2_profiles}
\end{figure*}

Experimental $gf$-values are available for all seven preselected \prii\ lines from \citet[six lines]{ILW} and from \citet[four lines]{2007PhyS...76..577L}, with three lines in common. In general, the data from the two publications agree well \citep[see Fig. 2 of][]{2009ApJS..182...80S}. An exception is the line at 532.3~nm (0.2~dex difference in log$gf$). \citet{2009ApJS..182...80S} found that the higher value from \citet{2007PhyS...76..577L} results in a more consistent solar Pr abundance.
Synthetic spectra calculated for Arcturus with scaled solar abundances and the \citeauthor{2007PhyS...76..577L} $gf$-value also support the validity of the higher value (see Fig.~\ref{fig:Pr2_profiles}).
In summary, we used the data from \citet{ILW} for five lines, and data from \citet{2007PhyS...76..577L} for two lines (521.7~nm and 532.3~nm), all with \gfflag=\Yes. 
Note that the $gf$-values given in Table~2 of \citet{2007PhyS...76..577L} do not agree with the values obtained by using their Eq.~(5), with differences of more than 0.005~dex in some cases. We used the calculated values instead of the tabulated ones (the differences calculated--tabulated are +0.017~dex and $-$0.018~dex for the two lines, respectively).

In addition, the background line list contains \prii\ data from \citet{BLQS}, data from \citet{MC} for \pri\ and \prii\ lines, and data from \citet{ISAN} for a few \priii\ lines.

Most of the preselected \prii\ lines are invisible and/or severely blended in the spectra of the Sun, Arcturus, and other benchmark stars. However, two lines, at 526.0~nm and 532.3~nm have non-negligible strengths and seem to be more or less unblended (\synflag=\Un\ and \Yes, respectively, see Fig.~\ref{fig:Pr2_profiles}).
Five of the lines (the previous two and 517.4, 522.0, 529.3~nm) have been used by \citet{2009ApJS..182...80S} to derive the Pr abundance in r-process-rich, metal-poor ([Fe/H]$\approx-3$~dex) stars, in which these lines are stronger than in the Sun.

\paragraph{HFS data}
Pr has a single stable isotope, $^{141}$Pr, with non-zero nuclear spin, giving rise to HFS in \prii\ lines.
HFS $A$ constants are available in the literature for each of the twelve energy levels involved in the preselected \prii\ transitions.
Electric quadrupole interaction is considered negligible for the HFS of Pr \citep[e.g.,][]{1989PhyS...39..694G,2009ApJS..182...80S}, and thus the HFS $B$ constants were set to zero for all levels.
Five of the levels are included in the experimental work based on fast-ion-beam laser spectroscopy by \citet{2002CaJPh..80..557R}. % Rivest et al. (2002)
Data for the remaining levels were taken from the work of \citet{1989PhyS...39..694G}, % Ginibre (1989)
who determined HFS $A$ constants from an analysis of FTS spectra.
The complete set of HFS data is given in Table~\ref{HFSpr}.

\subsection{Neodymium (Z=60)}
\label{sect:Nd}

\begin{figure}[ht]
   \begin{center}
      \resizebox{\hsize}{!}{\includegraphics[trim=50 40 50 50,clip]{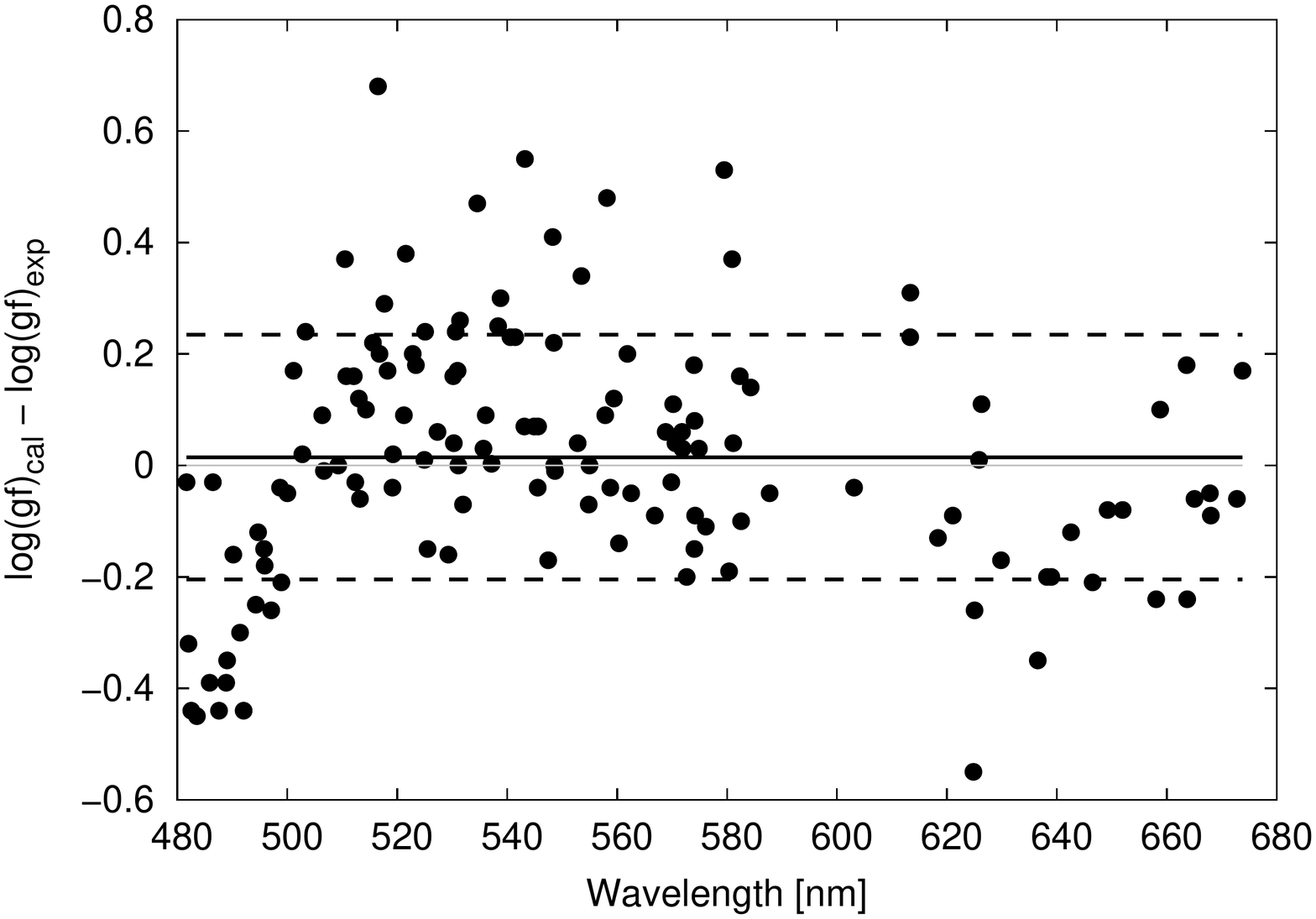}}
   \end{center}
   \caption{Comparison between \emph{cal}ibrated and \emph{exp}erimental transition probabilities for \ndii\ lines in the UVES-580 wavelength range. The data are from \citet{1995KurCD..23.....K} based on \citet{MC}, and from \citet{HLSC}, respectively. The solid black line indicates the mean difference in log$gf$, and the dashed lines indicate the standard deviation.}
   \label{fig:Nd2}
\end{figure}

With over 50 lines, \ndii\ lines are the most numerous among all rare earth species in the preselected line list. For the majority of these lines, including one of the four lines in the NIR, we used the experimental $gf$-values determined by \citet[\gfflag=\Yes]{HLSC} % Den Hartog et al. (2003)
from a combination of LIF measurements of radiative life-times and FTS measurements of BFs.
The $gf$-values for the remaining lines (reference code MC in the electronic table) are from \citet{1995KurCD..23.....K}. % Kurucz & Bell (1995)
They are derived from a calibration of measured line intensities from \citet{MC} % Meggers et al. (1975), NBS Monograph 145
following the procedure outlined by \citet[T.A. Ryabchikova and R.L. Kurucz, priv. comm.]{CC}. % Cowley & Corliss (1983)
A comparison of the MC-derived data to the \citet{HLSC} data for 133 lines in common within the UVES-580 wavelength range is shown in Fig.~\ref{fig:Nd2}. The comparison is characterised by a standard deviation of 0.22~dex, and additional systematic differences varying with wavelength. We assigned a \gfflag=\Un\ to the $gf$-values from this source.

For five lines with MC-derived data there are also experimental $gf$-values by \citet{2007CaJPh..85.1343L} available (which are between 0.04~dex larger and 0.16~dex smaller), but these were not included in the line list.
In addition, the background line list contains data from \citet{MC} for \ndi\ lines, from \citet{XSCL} for \ndii\ lines, and from \citet{RRKB} for \ndiii\ lines.

\begin{figure*}[ht]
   \begin{center}
      \resizebox{0.80\hsize}{!}{\includegraphics{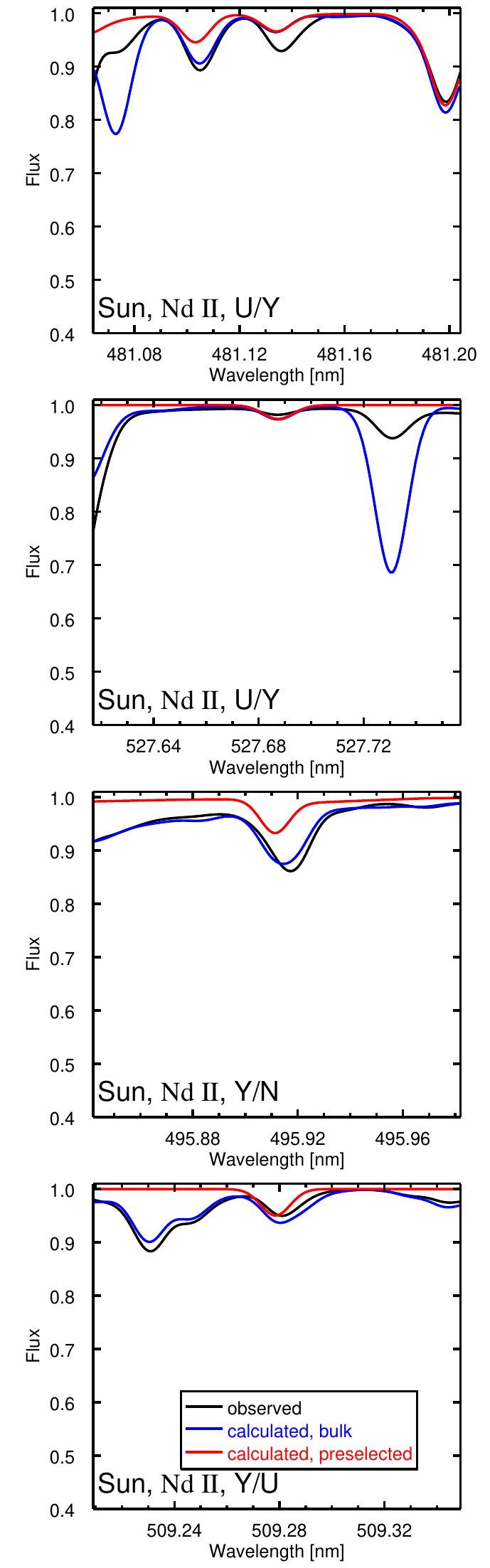}\includegraphics{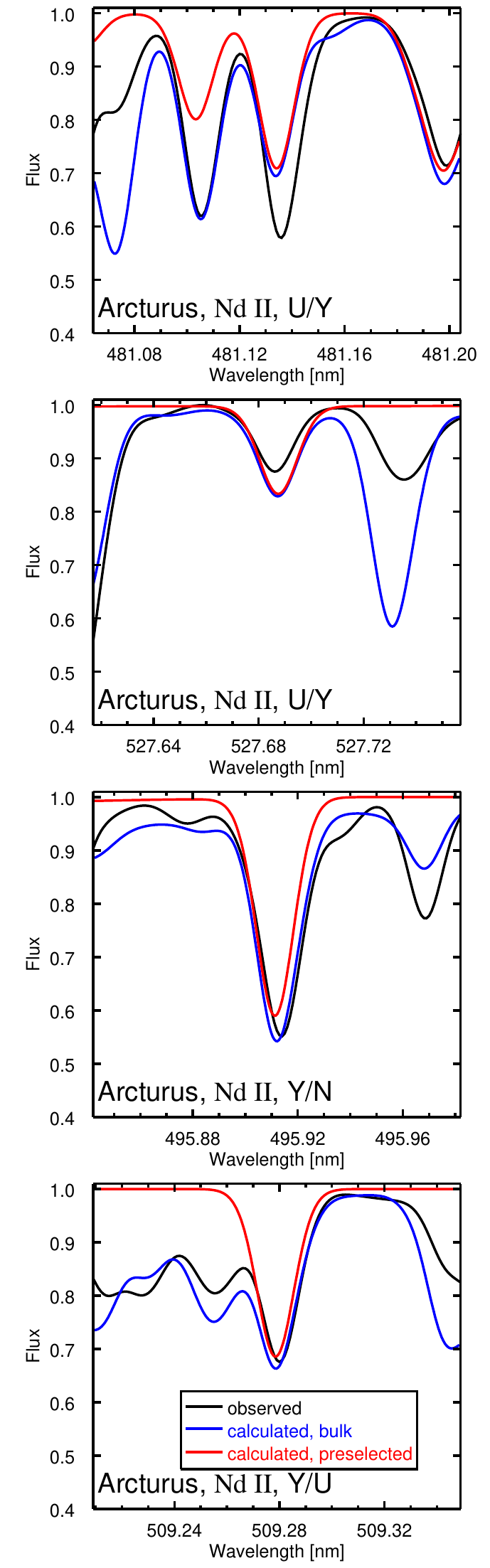}}
   \end{center}
   \caption{Comparison of observed and calculated line profiles around four of the preselected \ndii\ lines for the Sun (left) and Arcturus (right). Black lines: observations, red lines: calculations including preselected spectral lines only, blue lines: calculations including blends from background line list.
   }
   \label{fig:Nd2_profiles_calc}
\end{figure*}

\begin{figure*}[ht]
   \begin{center}
      \resizebox{\hsize}{!}{\includegraphics{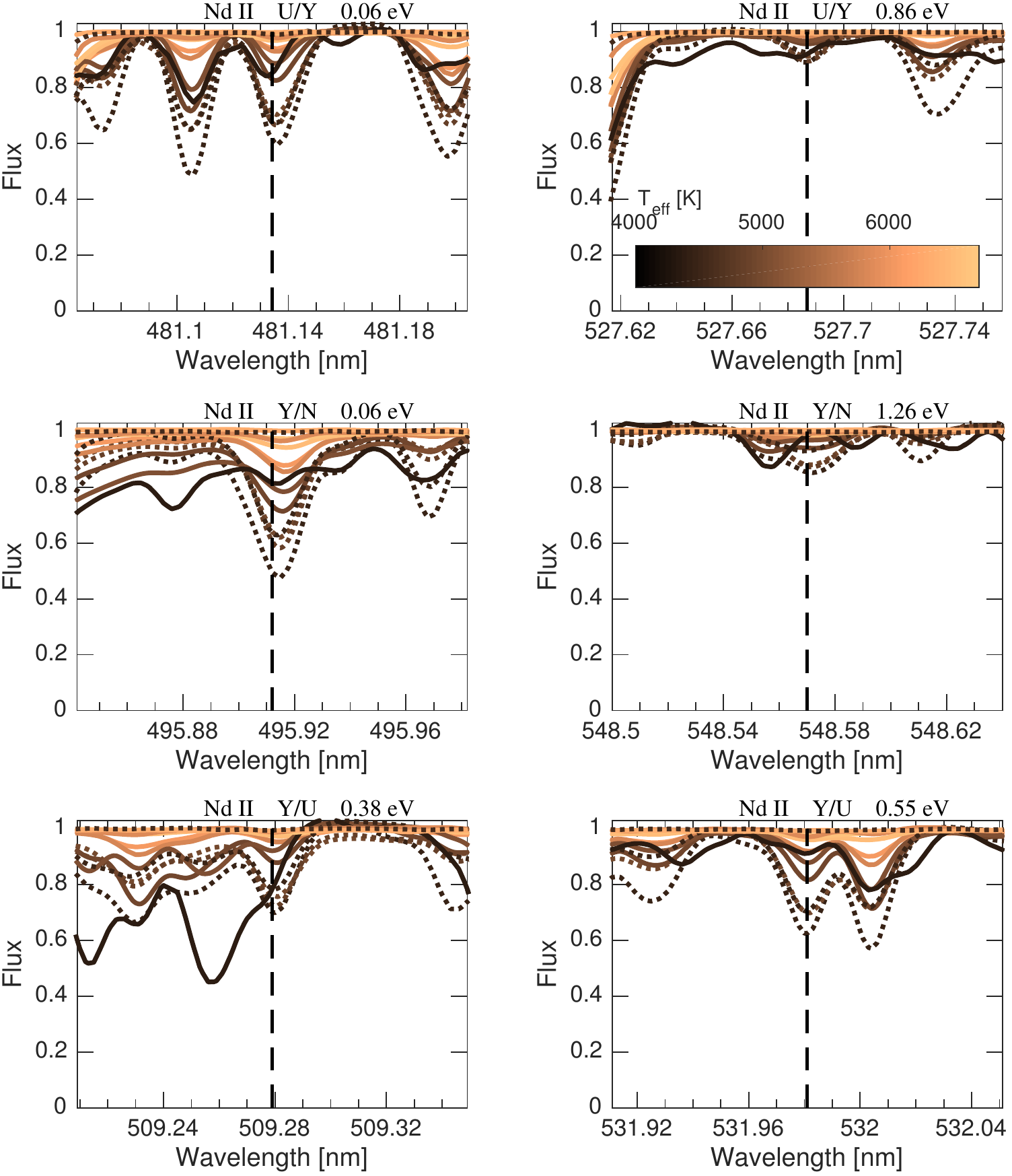}}
   \end{center}
   \caption{Line profiles for preselected \ndii\ lines generated from observed spectra of selected Gaia FGK benchmark stars.
   See Sect.~\refSectSpectraT.
   % See Sect.~\sectSpectra in \citet{Heiter_etal_2020}.
   % See Sect.~\ref{sect:spectra} for description.
   }
   \label{fig:Nd2_profiles_GBS}
\end{figure*}

The four preselected \ndii\ lines in the NIR are invisible in solar-like and giant stars, with the possible exception of one line in metal-rich giants such as $\mu$~Leo.
Only two of the lines in the UVES-580 range appear unblended in solar- or Arcturus-like stars (\synflag=\Yes). However, their $gf$-values are uncertain, and at least one of them could be affected by unidentified blends.
About half of the remaining lines are clearly blended (\synflag=\No), and the other half are border-line cases or lines for which the blending properties strongly depend on stellar parameters (\synflag=\Un).
A few selected cases are illustrated in Figs.~\ref{fig:Nd2_profiles_calc} and \ref{fig:Nd2_profiles_GBS}.
Note that this assessment is not directly applicable to r-process-rich, metal-poor stars, where the relative strengths of \ndii\ lines and blending lines might be significantly different.

\paragraph{HFS and IS data}
Nd has seven stable isotopes with rather evenly distributed Solar System abundance ratios
(Table~\refTabIsotopes).
% (Table~\tabIsotopes in \citealt{Heiter_etal_2020}).
% (Table~\ref{tab:isotopes}).
IS data are available for all of the preselected \ndii\ lines for the isotope pair 144--150 and for 41 of the lines for the isotope pair 142--144. For the remaining isotopes, the wavelengths were set to those of isotope 150.
The shifts in level energy for the 142--144 pair were measured by \citet{1981AcSpe..36..943A} % Ahmad and Saksena (1981)
on hollow cathode lamp spectra recorded by a Fabry-Perot spectrometer.
For the 144--150 pair most of the required levels were measured by \citet{1997ZPhyD..42...71N} % Nakhate et al. (1997)
using the same technique.
Complementary data for the remaining levels were taken from the extensive FTS work by \citet{1984PhyS...29..119B}. % Blaise et al. (1984) 
The complete set of IS data used in the Gaia-ESO line list is given in Table~\ref{ISnd}.

\citet{2005AcSpe..60..447K} % Koczorowski et al. (2005), Table 4
measured isotopic shifts for 27 levels of \ndii\ for all pairs of Nd isotopes with even baryon number, using the LIF method (including 13 of the lowest levels and eight of the upper levels of the preselected transitions). Their results are consistent with the sources cited above, for the corresponding isotope pairs. The derived shift values are similar for all pairs, although the values for the 150--148 pair tend to be larger than for the others.

Two of the Nd isotopes have odd baryon numbers with non-zero spin, and thus hyperfine structure.
HFS $A$ and $B$ constants were measured by \citet{2005CaJPh..83..841R} % Rosner et al. (2005)
for eleven even-parity low-energy levels, and 64 odd-parity high-energy levels of $^{143}$Nd using the collinear fast-ion-beam laser spectroscopy technique.
$A$ and $B$ constants for $^{145}$Nd were then calculated by dividing the constants obtained for $^{143}$Nd by the ratios between the nuclear magnetic moments and between the nuclear electric quadrupole moments of $^{143}$Nd and $^{145}$Nd, respectively. These ratios were determined by \citeauthor{2005CaJPh..83..841R} as 1.60884~(2) and 1.919~(33), respectively.
These measurements cover 32 out of the 69 \ndii\ levels involved in the preselected transitions.
For two further lower levels we used data from \citet{2004ADNDT..86....3M} % Ma and Yang (2004)
measured by the same technique.
\citet{2004ADNDT..86....3M} determined $A$ and $B$ constants for both $^{143}$Nd and $^{145}$Nd for ten levels in total, and their ratios of $A_{143}/A_{145}$ vary between 1.60 and 1.63, in reasonable agreement with the theoretical expectation. Their ratios of $B_{143}/B_{145}$ vary between 1.7 and 2.3, for levels with significant $B$ values, comparable to the expected value of 1.9. However, for two of the levels the $B$ constant changes sign between isotope 143 and 145, including the level at 0.745~eV used here.
Together, these sources provide complete HFS data for 16 of the preselected \ndii\ lines, while about half of the lines have data for only one level, and for a few lines there are no data at all (see Table~\ref{ISnd}).
The complete set of available HFS data is given in Table~\ref{HFSnd}.

\subsection{Samarium (Z=62)}
\label{sect:Sm}

For the five preselected \smii\ lines we used the TRLIF/FTS experimental $gf$-values from \citet[\gfflag=\Yes]{LD-HS}. % Lawler et al. (2006)
Four of these lines are also contained in the work of \citet{2006CaJPh..84..723R}, % Rehse et al. (2006)
who used a substantially different experimental method (fast-ion-beam LIF). Their $gf$-values are within 0.05~dex of those by \citeauthor{LD-HS} for three lines, while their $gf$-value for the 492.956~nm line is 0.13~dex larger. 
A detailed comparison between the two studies for over 300 lines in common is given in \citet{2008CaJPh..86.1033L}. % Lawler et al. (2008)
They found good agreement in general, in particular for the radiative life-time measurements, for which the differences are confined within half of the combined uncertainties quoted by the authors. However, the transition probabilities showed a mean difference of 0.4 times the combined uncertainties, and a standard deviation larger than expected from the uncertainties. This led them to conclude that the uncertainties on the BFs from one or both experiments are larger than estimated by the authors.
In addition to \citeauthor{LD-HS} the background line list contains data from \citet{XSQG} for \smii\, and from \citet{MC} for \smi\ and \smii\ lines.

\begin{figure*}[ht]
   \begin{center}
      \resizebox{0.89\hsize}{!}{\includegraphics{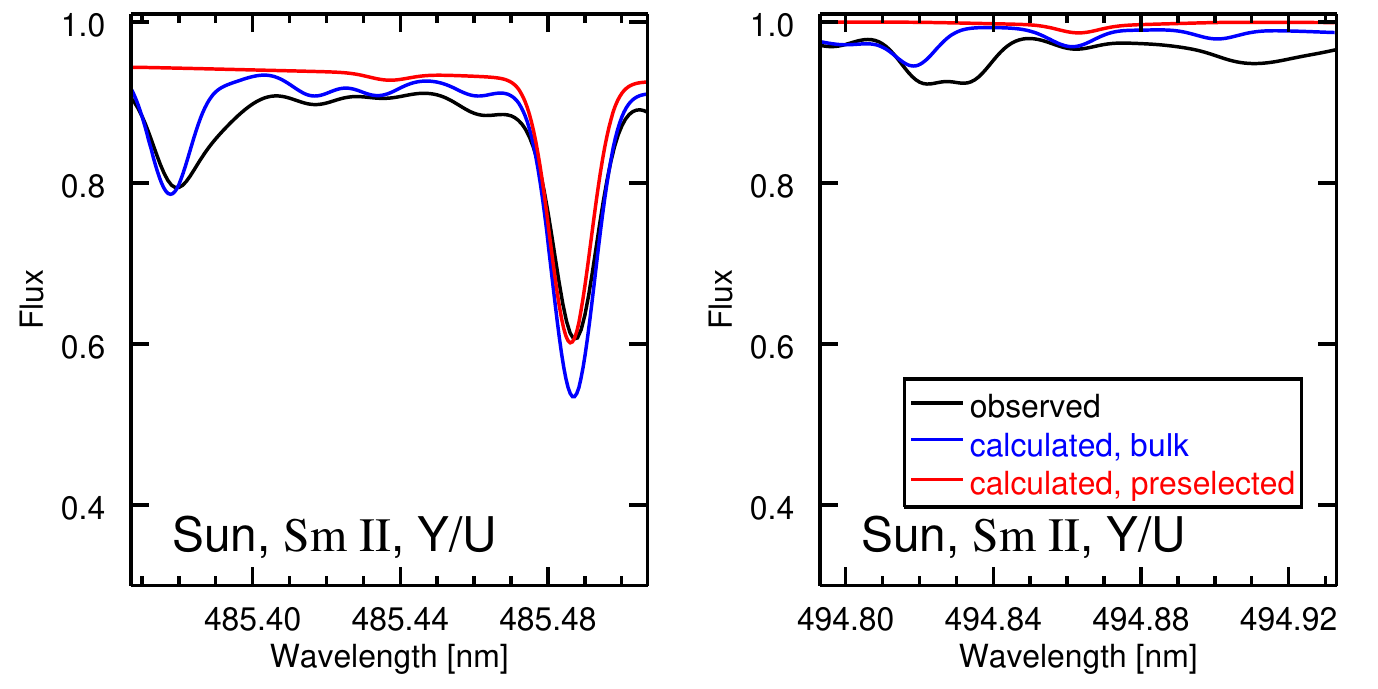}}
      \resizebox{0.89\hsize}{!}{\includegraphics{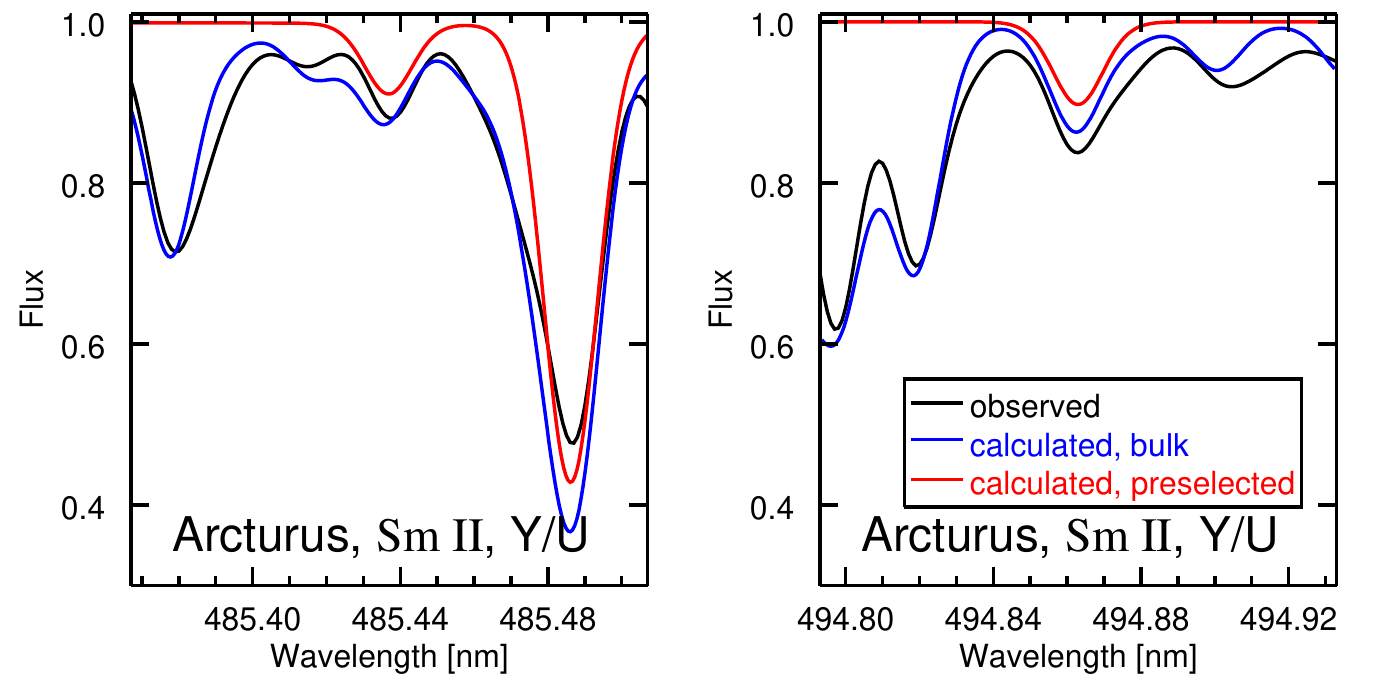}}
      \resizebox{0.89\hsize}{!}{\includegraphics{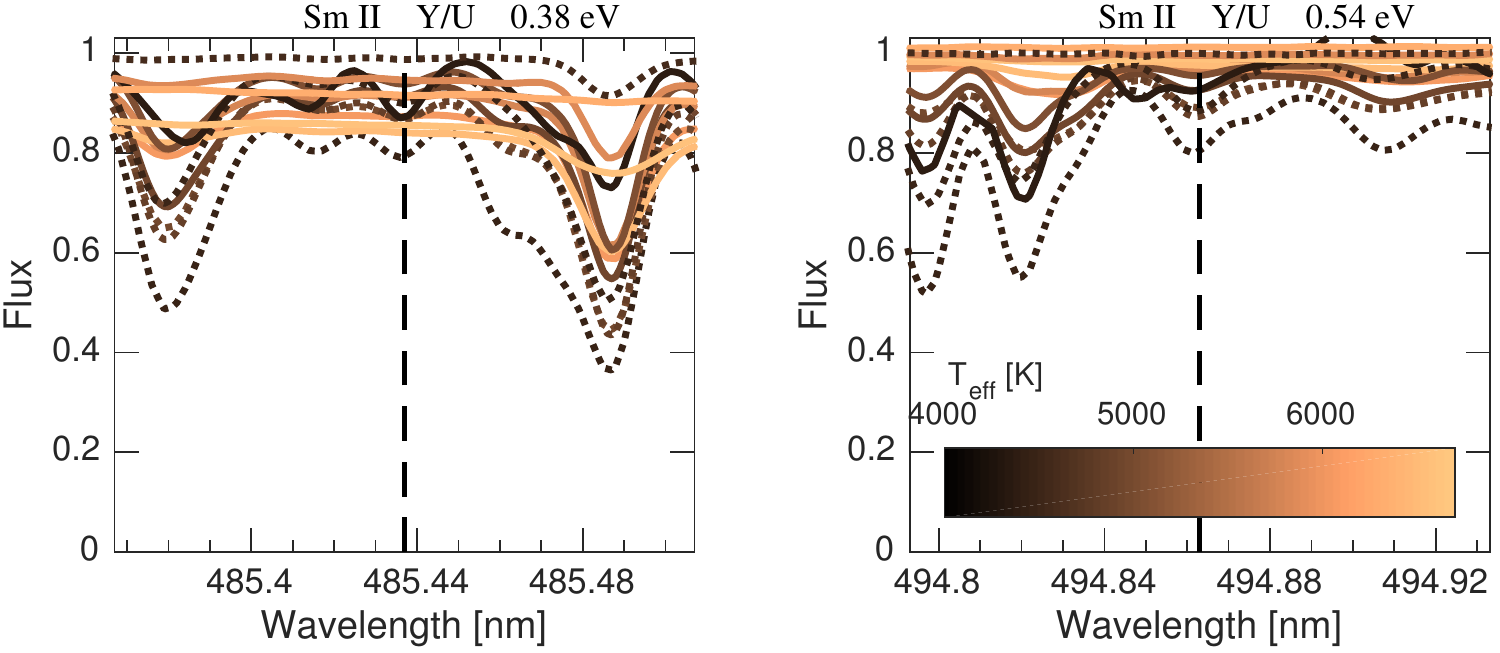}}
   \end{center}
   \caption{
   Observed and calculated line profiles around two preselected \smii\ lines for the Sun (top row) and Arcturus (middle row). Black lines: observations, red lines: calculations including preselected spectral lines only, blue lines: calculations including blends from background line list.
   Bottom row: Line profiles generated from observed spectra of selected Gaia FGK benchmark stars.
   See Sect.~\refSectSpectraT.
   % See Sect.~\sectSpectra in \citet{Heiter_etal_2020}.
   % See Sect.~\ref{sect:spectra} for description.
   }
   \label{fig:Sm2_profiles}
\end{figure*}

All of the preselected \smii\ lines are very weak and affected by blends. The strongest line (481.6~nm) is severely blended (\synflag=\No), and the line at 493.0~nm is invisible in all normal stars. Two examples for lines with \synflag=\Un\ are shown in Fig.~\ref{fig:Sm2_profiles}.
The one at 485.4~nm is located in the blue wing of the H$\beta$ line.
\citet{LD-HS} list further 130 \smii\ lines in the UVES-580 wavelength range (and 17 in the HR21 wavelength range), for which a blending analysis could be done, preferably including the parameters of an r-process-rich, metal-poor star. % e.g., from \citet{LD-HS}.

\paragraph{HFS and IS data}
Sm has seven stable isotopes, each of them contributing significantly to the Solar System abundance of Sm
(Table~\refTabIsotopes).
% (Table~\tabIsotopes in \citealt{Heiter_etal_2020}).
% (Table~\ref{tab:isotopes}).
IS measurements are not available for any of the preselected \smii\ lines. However, isotopic shifts for \smii\ lines may usually be neglected compared to other line broadening mechanisms. For example, the largest shift between $^{147}$Sm and $^{154}$Sm given by \citet{2003CaJPh..81.1389M} % Masterman et al. (2003)
for 87 lines with wavelengths between 418 and 465~nm is 0.05~\AA\ (somewhat larger shifts were measured for the rare isotope $^{144}$Sm).
Note that individual lines for each of the seven isotopes are included in the line list (with equal wavelengths and $gf$-values for those without HFS), to allow proper scaling according to isotopic abundances together with the HFS-split isotopic lines.

For the two Sm isotopes with non-zero spin HFS $A$ and $B$ constants are available
from the experimental work by \citet{2003CaJPh..81.1389M} % Masterman et al. (2003)
based on collinear fast-ion-beam laser spectroscopy.
Their data cover all of the ten levels involved in the preselected \smii\ transitions except one (the upper level of the line at 483.7~nm), and are summarised in Table~\ref{HFSsm}.
Note that \citeauthor{2003CaJPh..81.1389M} write that they fixed the ratios of $A_{147}/A_{149}$ and $B_{147}/B_{149}$ for each level to the values corresponding to the ratios of the nuclear magnetic and electric quadrupole moments (1.213 and $-3.5$), respectively. This does not appear to be the case for the $B$ constants, however, for which the ratios for the levels of interest vary between $-3.4$ and $-3.7$.

\subsection{Europium (Z=63)}
\label{sect:Eu}

For the preselected \euii\ lines, we rely on the $gf$-values measured by \citet{LWHS} % Lawler et al. (2001)
based on experimental life times and BFs.
In addition, the background line list contains data by \citet{DHWL} for \eui, by \citet{ZLLZ} for \euii, and by \citet{MC} for both ions.

\begin{figure*}[ht]
   \begin{center}
      \resizebox{0.89\hsize}{!}{\includegraphics{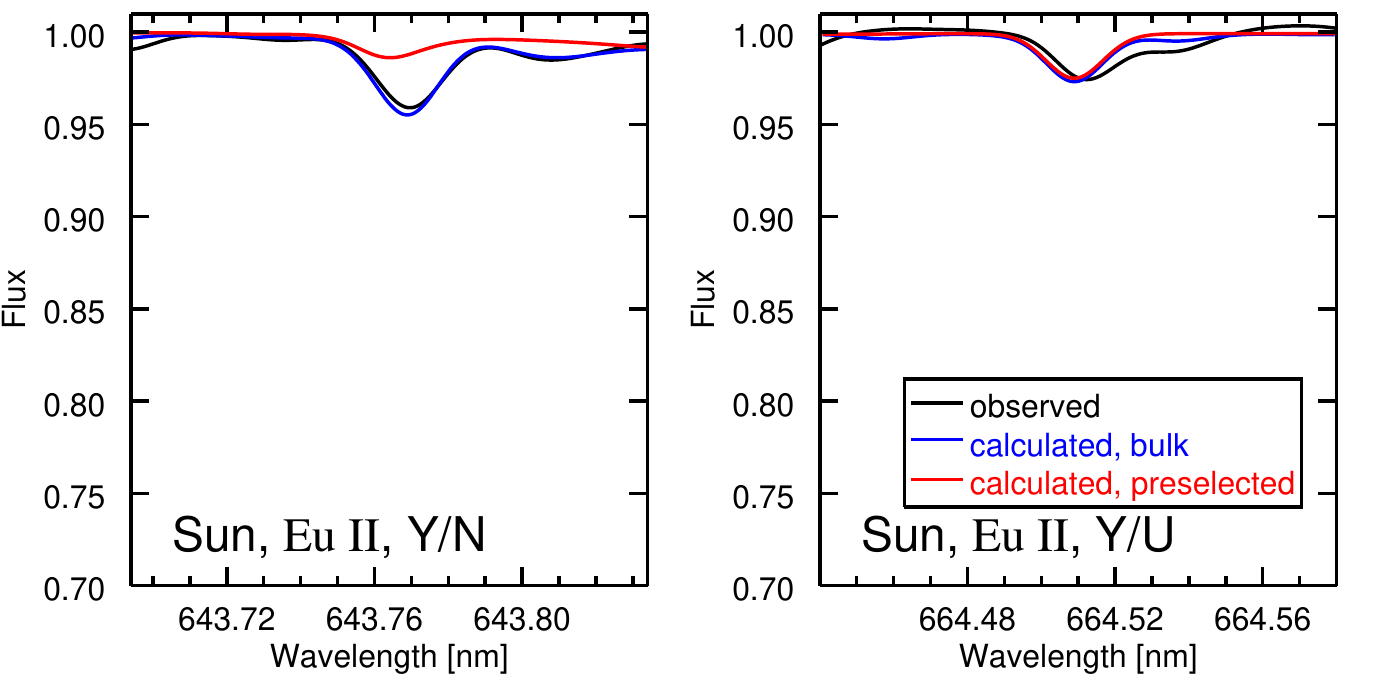}}
      \resizebox{0.89\hsize}{!}{\includegraphics{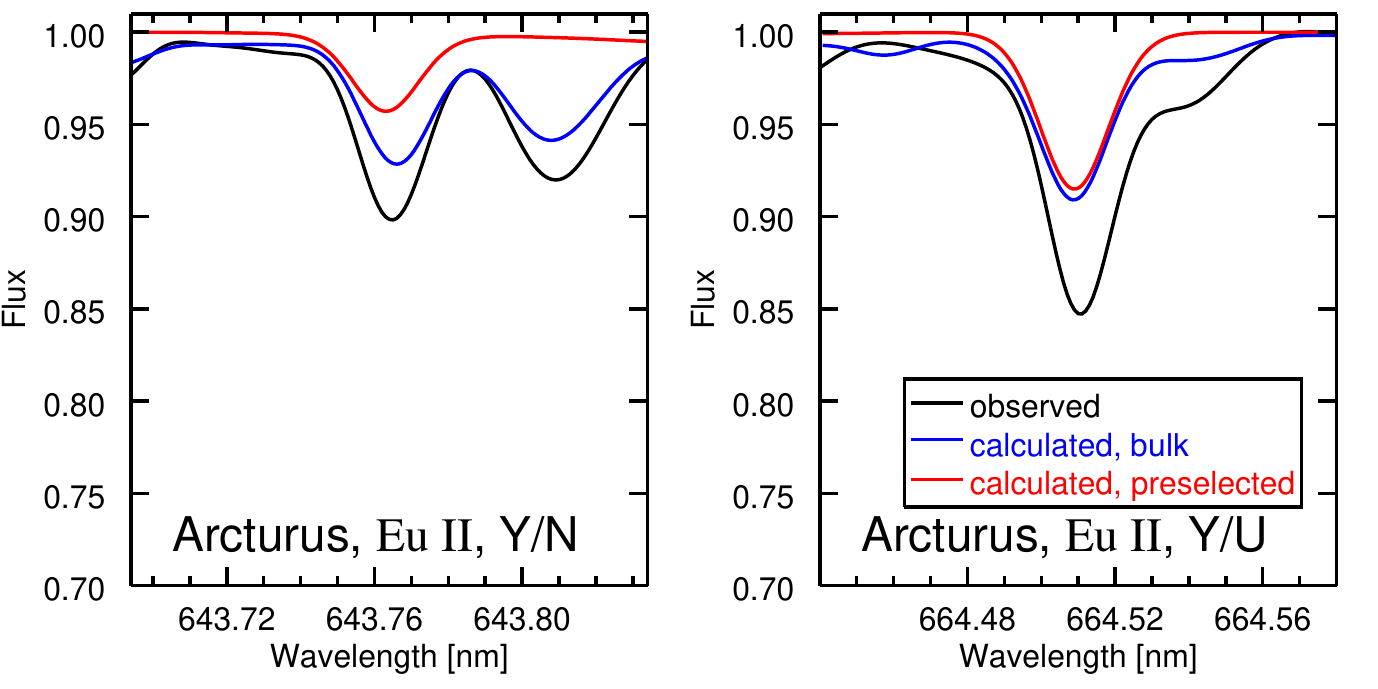}}
      \resizebox{0.89\hsize}{!}{\includegraphics{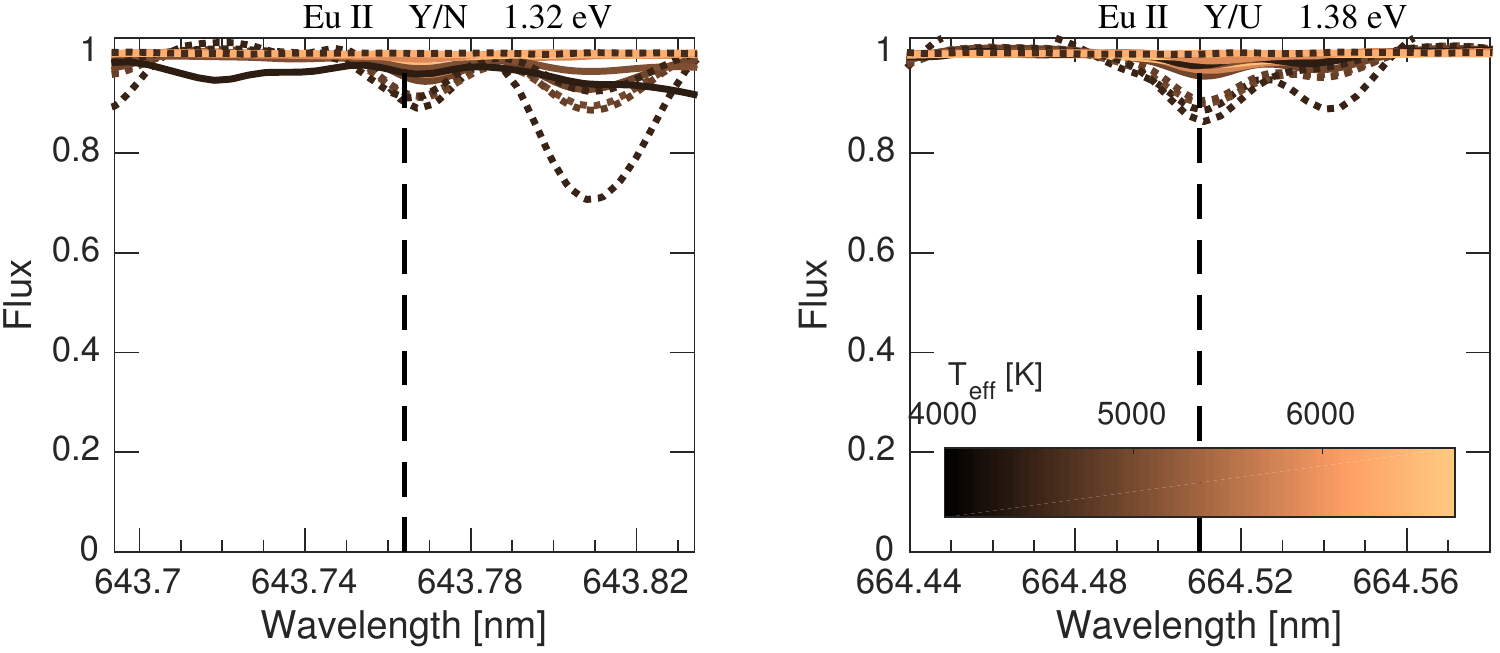}}
   \end{center}
   \caption{
   Observed and calculated line profiles around two preselected \euii\ lines for the Sun (top row) and Arcturus (middle row). Black lines: observations, red lines: calculations including preselected spectral lines only, blue lines: calculations including blends from background line list.
   Bottom row: Line profiles generated from observed spectra of selected Gaia FGK benchmark stars.
   See Sect.~\refSectSpectraT.
   % See Sect.~\sectSpectra in \citet{Heiter_etal_2020}.
   % See Sect.~\ref{sect:spectra} for description.
   }
   \label{fig:Eu2_profiles}
\end{figure*}

The five preselected \euii\ lines are rather weak and partly blended in the spectra of the Sun, Arcturus, and other benchmark stars.
Fig.~\ref{fig:Eu2_profiles} shows the line profiles of the two strongest lines, of which the one at 643.8~nm is clearly blended (\synflag=\No). The line at 664.5~nm might be useful for giant stars at high signal-to-noise ratios (to be decided on a case-by-case basis, \synflag=\Un).

\paragraph{HFS and IS data}

Eu occurs in the form of two isotopes ($^{151}$Eu and $^{153}$Eu, with non-zero nuclear spin), each contributing about equally to the Solar System Eu abundance
(Table~\refTabIsotopes).
% (Table~\tabIsotopes in \citealt{Heiter_etal_2020}).
% (Table~\ref{tab:isotopes}).
Isotopic shifts for 24 optical \euii\ transitions were compiled by \citet[their Table~5]{LWHS}. % Lawler et al. (2001)
These are in general smaller than for \smii\ lines, and particularly small for the preselected \euii\ lines ($\lesssim$0.02~\AA, not resolved in FTS spectra).
Therefore IS data were not included for \euii\ lines in the Gaia-ESO line list.

For four energy levels involved in the preselected \euii\ transitions, we used the experimental HFS $A$ and $B$ constants by \citet{Huhn:1992} % Hühnermann et al. 1992
and \citet{1993PhRvL..70..541M}, % B: Möller et al. 1993
respectively. These are based on the LIF method, and
are in excellent agreement with the measurements by \citet{1987PhRvA..36.1983S} % Sen & Childs 1987
obtained from laser radio frequency double resonance spectroscopy
(see compilation by \citealt[their Table~4]{LWHS}). % Lawler et al. (2001)
For the remaining three levels, we used the LIF data from \citet{1992PhLA..162..178V}, % Villemoes et 1992
who give essentially the same values as \citet{Huhn:1992} and \citet{1993PhRvL..70..541M} for these levels.
The complete set of HFS data is given in Table~\ref{HFSeu}.

\subsection{Gadolinium (Z=64) and Dysprosium (Z=66)}
\label{sect:Gd}\label{sect:Dy}

\gdii\ and \dyii\ are the heaviest species included in the preselected line list, with one line each (at 486.504~nm and 516.969~nm, respectively).
Laboratory transition probabilities based on the TRLIF/FTS technique have been published for both lines, by \citet{DLSC} % Den Hartog et al. (2006)
and \citet{WLN}, % Wickliffe et al. (2000)
respectively, where the latter provide data for \dyi\ lines as well.
In addition to these sources, the background line list contains data for \gdi, \gdii, \dyi, and \dyii\ from \citet{MC}.

\begin{figure*}[ht]
   \begin{center}
      \resizebox{0.88\hsize}{!}{\includegraphics{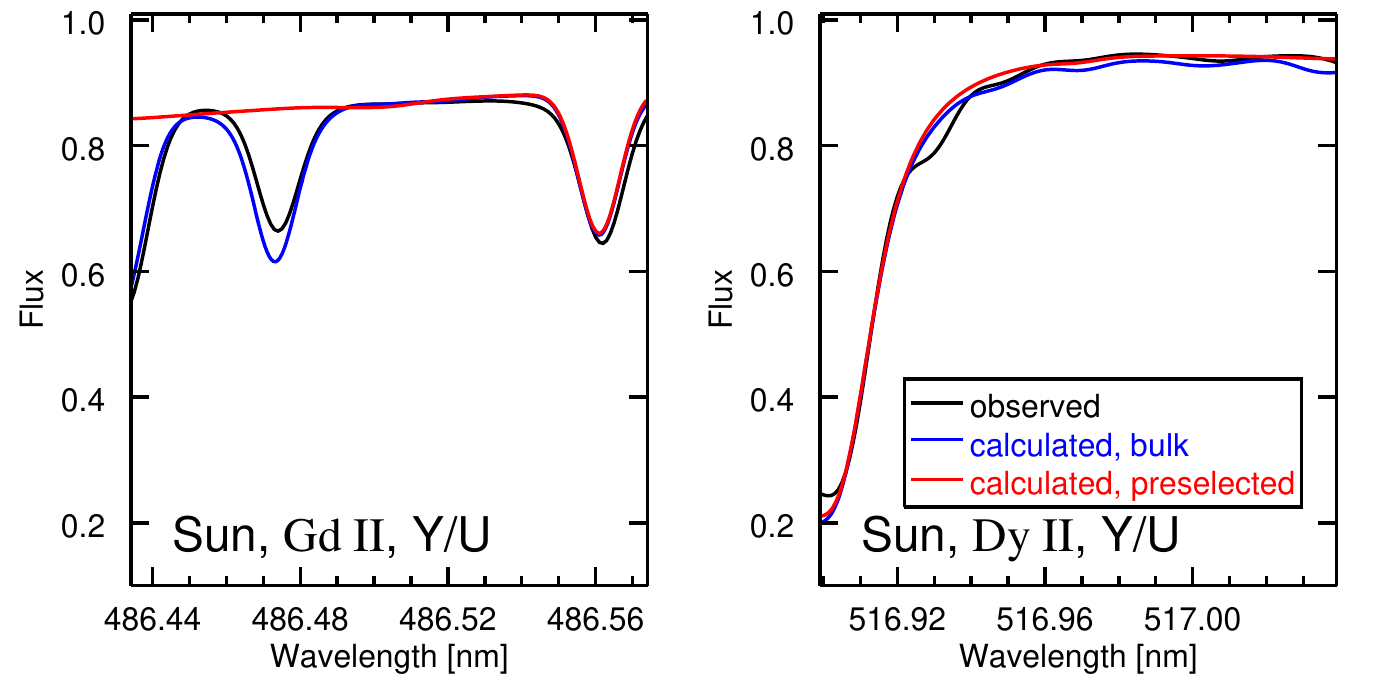}}
      \resizebox{0.88\hsize}{!}{\includegraphics{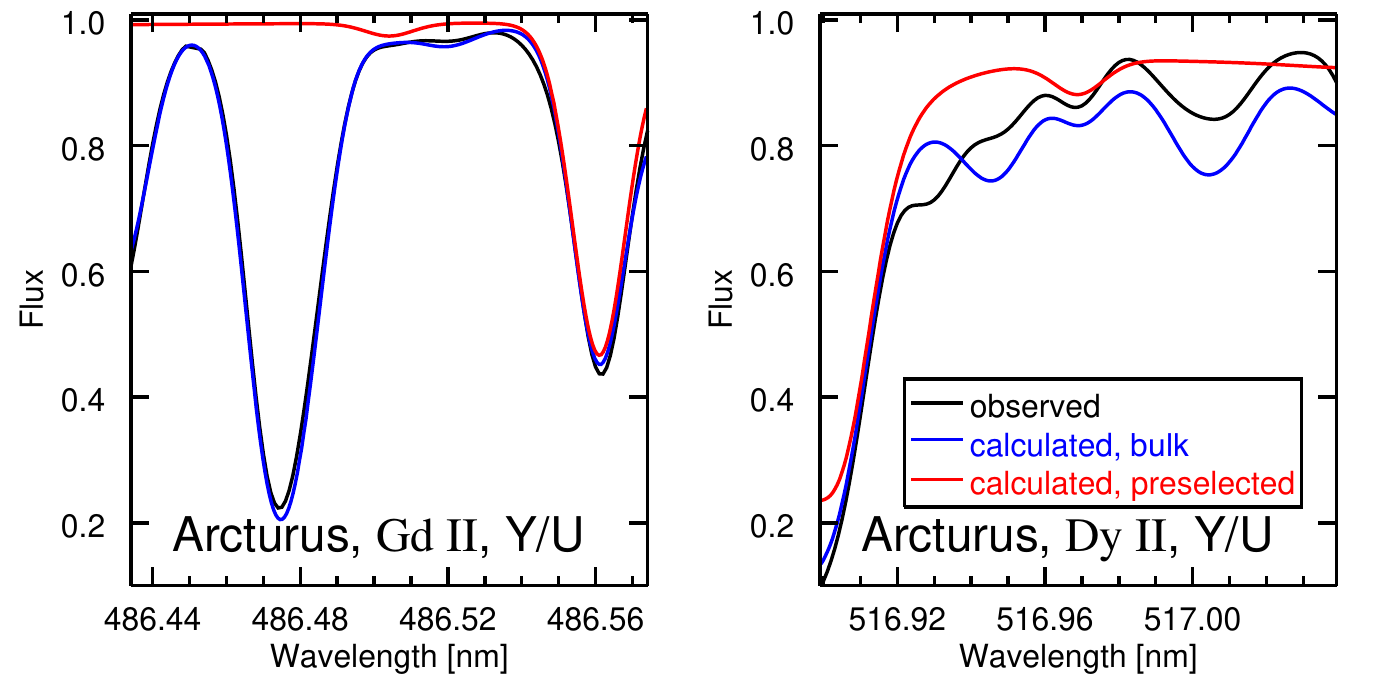}}
      \resizebox{0.88\hsize}{!}{\includegraphics{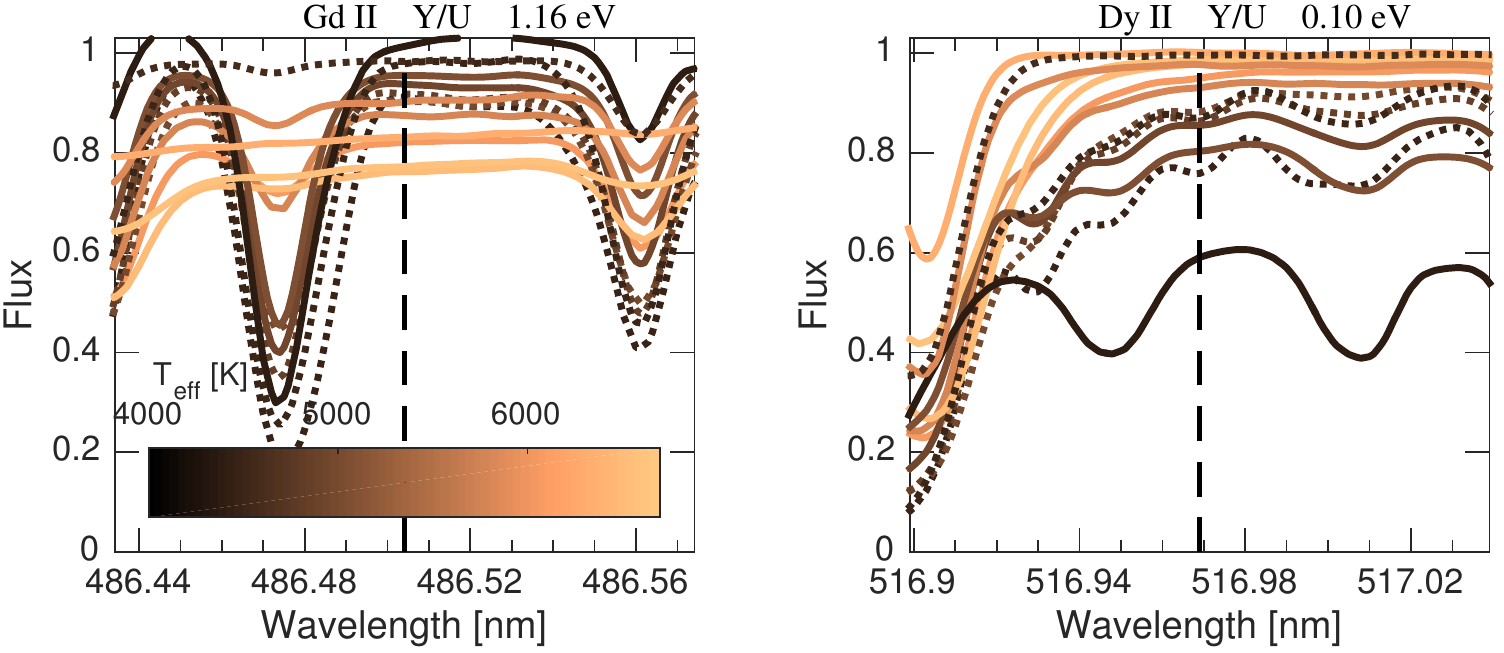}}
   \end{center}
   \caption{
   Observed and calculated line profiles around the preselected \gdii\ and \dyii\ lines for the Sun (top row) and Arcturus (middle row). Black lines: observations, red lines: calculations including preselected spectral lines only, blue lines: calculations including blends from background line list.
   Bottom row: Line profiles generated from observed spectra of selected Gaia FGK benchmark stars.
   See Sect.~\refSectSpectraT.
   % See Sect.~\sectSpectra in \citet{Heiter_etal_2020}.
   % See Sect.~\ref{sect:spectra} for description.
   }
   \label{fig:Gd2Dy2_profiles}
\end{figure*}

Both of the preselected lines are very weak and blended in the spectra of the Sun, Arcturus, and other benchmark stars, and thus of questionable use for a meaningful abundance analysis (\synflag=\Un).
In dwarf stars, the \gdii\ line is located in the red wing of the H$\beta$ line.
The line profiles for both transitions are illustrated in Fig.~\ref{fig:Gd2Dy2_profiles}.
However, \citet{DLSC} and \citet{WLN} list about 150 and 30 \gdii\ and \dyii\ lines, respectively, in the UVES-580 wavelength range (and about 10 lines each in the HR21 range), for which a blending analysis could be done. This analysis should include the parameters of an r-process-rich, metal-poor star.

% IS and HFS
Gd and Dy have several stable isotopes, of which five and four contribute rather evenly to the Solar System Gd and Dy abundances, respectively
(Table~\refTabIsotopes).
% (Table~\tabIsotopes in \citealt{Heiter_etal_2020}).
% (Table~\ref{tab:isotopes}).
Both elements have two isotopes with non-zero nuclear spin, giving rise to HFS.
\citet{DLSC} % Den Hartog et al. (2006)
observed partially resolved isotopic and hyperfine structure in the profiles for a few of the \gdii\ lines in their highest resolution FTS data, and
\citet{WLN} % Wickliffe et al. (2000)
noticed partially resolved structure for many of the \dyii\ lines in their FTS spectra.
However, most of the lines of both species were considered sufficiently narrow to be treated as single component lines in stellar abundance work \citep{LSCI}. % Lawler et al. (2009)

% place all figures and tables from this section here
\clearpage

%----------------------------------------------------------------------------
% Appendix: HFS and IS tables
%----------------------------------------------------------------------------

\section{Hyperfine structure and isotopic splitting data}

This appendix provides the HFS data and isotopic shifts presented in Sect.~\ref{sect:gf} for several elements in the form of tables. If not stated otherwise, each table lists level energies $E$ in eV, HFS constants $A$ and $B$ in units of cm$^{-1}$, representing magnetic dipole and electric quadrupole interactions for each level, as well as level designations and references. If applicable, isotopic shifts (IS) are given, in units of cm$^{-1}$. Subscripts indicate baryon numbers for different isotopes.

\begin{table*}
\renewcommand{\tabcolsep}{1.0mm}
\caption[HFS constants \sci\ levels]{HFS constants for 12 \sci\ levels.}
\label{HFSsc}
\centering
\begin{tabular}{ccccl}
\hline\hline\noalign{\smallskip}
 $E$ (eV) & $A$ & $B$ & Level &  Reference \\
\noalign{\smallskip}\hline\noalign{\smallskip}
0.000 &  0.0090 & -0.0009  &  4s$^2$ $^2$D & \citet{1971PhRvA...4.1767C} \\ % Childs et al. 1971
1.440 &  0.0083 & -0.0003  &  4s $^4$F  & \citet{1976PhLA...55..405Z} \\ % Zeiske et al 1976
1.448 &  0.0095 & -0.0005  &  4s $^4$F  & \citet{1976ZPhyA.276....9E} \\ % Ertmer \& Hofer 1976
1.851 &  0.0100 &  0.0000  &  4s $^2$F  & \citet{2004PhyS...69..189B} \\ % Basar et al. 2004
1.865 & -0.0008 &  0.0000  &  4s $^2$F  & \citet{2004PhyS...69..189B} \\ % Basar et al. 2004
1.996 & -0.0115 &  0.0000  &  4p $^2$D  & \citet{2004PhyS...69..189B} \\ % Basar et al. 2004
3.619 &  0.0029 &  0.0000  &  4p $^4$G  & \citet{2004PhyS...69..189B} \\ % Basar et al. 2004
3.633 &  0.0015 &  0.0000  &  4p $^4$G  & \citet{2004PhyS...69..189B} \\ % Basar et al. 2004
3.878 &  0.0019 &  0.0000  &  4p $^4$F  & \citet{1402-4896-75-5-006}  \\ % Öztürk et al.
4.110 &  0.0034 &  0.0000  &  4p $^2$G  & \citet{1402-4896-75-5-006}  \\ % Öztürk et al.
4.111 &  0.0071 &  0.0000  &  4p $^2$F  & \citet{1402-4896-75-5-006}  \\ % Öztürk et al.
4.179 &  0.0015 &  0.0000  &  4p $^2$D  & \citet{2004PhyS...69..189B} \\ % Basar et al. 2004
\noalign{\smallskip}\hline\hline
\end{tabular}
\end{table*}

\begin{table*}
\renewcommand{\tabcolsep}{1.0mm}
\caption[HFS constants for \vi\ levels]{HFS constants for 53 \vi\ levels.}
\label{HFSv}
\centering
\begin{tabular}{lcccl}
\hline\hline\noalign{\smallskip}
 $E$ (eV) & $A$ & $B$ & Level &  Reference \\
\noalign{\smallskip}\hline\noalign{\smallskip}
 0.040  &   0.0083 &  0.0002  &  a $^4$F &  \citet{1979PhRvA..19..168C} \\ % Childs et al. 1979     
 0.069  &   0.0076 &  0.0003  &  a $^4$F &  \citet{1979PhRvA..19..168C} \\ % Childs et al. 1979     
 0.262  &   0.0251 &  0.0000  &  a $^6$D &  \citet{1998JPhB...31.2203C} \\ % Cochrane et al. 1998   
 0.267  &   0.0135 & -0.0003  &  a $^6$D &  \citet{1998JPhB...31.2203C} \\ % Cochrane et al. 1998   
 0.275  &   0.0125 & -0.0002  &  a $^6$D &  \citet{1998JPhB...31.2203C} \\ % Cochrane et al. 1998   
 0.287  &   0.0128 &  0.0001  &  a $^6$D &  \citet{1998JPhB...31.2203C} \\ % Cochrane et al. 1998   
 0.301  &   0.0136 &  0.0005  &  a $^6$D &  \citet{1998JPhB...31.2203C} \\ % Cochrane et al. 1998   
 1.043  &   0.0426 &  0.0000  &  a $^4$D &  \citet{1979PhRvA..19..168C} \\ % Childs et al. 1979     
 1.051  &   0.0002 &  0.0000  &  a $^4$D &  \citet{1979PhRvA..19..168C} \\ % Childs et al. 1979     
 1.064  &  -0.0048 &  0.0002  &  a $^4$D &  \citet{1979PhRvA..19..168C} \\ % Childs et al. 1979     
 1.081  &  -0.0053 &  0.0005  &  a $^4$D &  \citet{1979PhRvA..19..168C} \\ % Childs et al. 1979     
 1.183  &  -0.0118 &  0.0000  &  a $^4$P &  \citet{1989ZPhyD..11..259U} \\ % Unkel 89               
 1.195  &   0.0061 &  0.0003  &  a $^4$P &  \citet{1989ZPhyD..11..259U} \\ % Unkel 89               
 1.218  &   0.0038 & -0.0004  &  a $^4$P &  \citet{1989ZPhyD..11..259U} \\ % Unkel 89               
 1.712  &   0.0002 &  0.0000  &  a $^2$P &  \citet{2002PhyS...66..363L} \\ % Lefebvre et al. 2002   
 1.849  &   0.0011 &  0.0000  &  a $^4$H &  \citet{1989ZPhyD..11..259U} \\ % Unkel 89               
 1.853  &   0.0087 & -0.0004  &  a $^4$H &  \citet{1989ZPhyD..11..259U} \\ % Unkel 89               
 1.868  &   0.0148 & -0.0003  &  a $^4$H &  \citet{1989ZPhyD..11..259U} \\ % Unkel 89               
 2.242  &   0.0314 &  0.0000  &  z $^6$D &  \citet{1998JPhB...31.2203C} \\ % Cochrane et al. 1998   
 2.247  &   0.0198 & -0.0001  &  z $^6$D &  \citet{1998JPhB...31.2203C} \\ % Cochrane et al. 1998   
 2.247  &  -0.0117 &  0.0000  &  z $^6$F &  \citet{1998JPhB...31.2203C} \\ % Cochrane et al. 1998   
 2.253  &   0.0096 & -0.0001  &  z $^6$F &  \citet{1998JPhB...31.2203C} \\ % Cochrane et al. 1998   
 2.256  &   0.0179 & -0.0001  &  z $^6$F &  \citet{1998JPhB...31.2203C} \\ % Cochrane et al. 1998   
 2.264  &   0.0130 &  0.0000  &  z $^6$F &  \citet{1998JPhB...31.2203C} \\ % Cochrane et al. 1998   
 2.269  &   0.0172 &  0.0000  &  z $^6$D &  \citet{1998JPhB...31.2203C} \\ % Cochrane et al. 1998   
 2.278  &   0.0145 &  0.0001  &  z $^6$F &  \citet{1998JPhB...31.2203C} \\ % Cochrane et al. 1998   
 2.286  &   0.0168 &  0.0001  &  z $^6$D &  \citet{1998JPhB...31.2203C} \\ % Cochrane et al. 1998   
 2.295  &   0.0156 &  0.0004  &  z $^6$F &  \citet{1998JPhB...31.2203C} \\ % Cochrane et al. 1998   
 2.316  &   0.0165 &  0.0006  &  z $^6$F &  \citet{1998JPhB...31.2203C} \\ % Cochrane et al. 1998   
 2.359  &   0.0133 & -0.0001  &  b $^2$H &  \citet{1992PhyB..179..103E} \\ % 
 2.374  &   0.0054 & -0.0004  &  b $^2$H &  \citet{1992PhyB..179..103E} \\ % 
 2.582  &   0.0182 &  0.0000  &  z $^4$D &  \citet{1995JPhB...28.3741P} \\ % Palmeri et al. 1995    
 2.608  &   0.0204 &  0.0000  &  z $^4$D &  \citet{1995JPhB...28.3741P} \\ % Palmeri et al. 1995    
 3.071  &   0.0000 &  0.0000  &  z $^4$P &  none                        \\ %                        
 3.089  &   0.0000 &  0.0000  &  z $^4$P &  none                        \\ %                        
 3.116  &   0.0000 &  0.0000  &  z $^4$P &  none                        \\ %                        
 3.131  &   0.0000 &  0.0000  &  y $^6$F &  none                        \\ %                        
 3.215  &   0.0212 &  0.0000  &  y $^4$F &  \citet{1995JPhB...28.3741P} \\ % Palmeri et al. 1995    
 3.224  &   0.0072 &  0.0000  &  y $^4$F &  \citet{2002PhyS...66..363L} \\ % Lefebvre et al. 2002   
 3.239  &   0.0058 &  0.0000  &  y $^4$F &  \citet{1995JPhB...28.3741P} \\ % Palmeri et al. 1995    
 3.245  &   0.0028 &  0.0011  &  y $^4$F &  \citet{1989ZPhyD..11..259U} \\ % Unkel et al. 1989      
 3.246  &   0.0012 &  0.0000  &  y $^4$D &  \citet{1995JPhB...28.3741P} \\ % Palmeri et al. 1995    
 3.255  &   0.0002 &  0.0000  &  y $^4$D &  \citet{1995JPhB...28.3741P} \\ % Palmeri et al. 1995    
 3.266  &   0.0001 &  0.0000  &  z $^2$G &  \citet{1995JPhB...28.3741P} \\ % Palmeri et al. 1995    
 3.267  &   0.0000 &  0.0000  &  y $^4$D &  \citet{1995JPhB...28.3741P} \\ % Palmeri et al. 1995    
 3.283  &   0.0000 &  0.0000  &  y $^4$D &  \citet{1995JPhB...28.3741P} \\ % Palmeri et al. 1995    
 3.798  &   0.0002 & -0.0002  &  y $^4$G &  \citet{1989ZPhyD..11..259U} \\ % Unkel et al. 1989      
 3.806  &   0.0106 & -0.0011  &  y $^4$G &  \citet{1989ZPhyD..11..259U} \\ % Unkel et al. 1989      
 3.827  &   0.0186 & -0.0002  &  y $^4$G &  \citet{1989ZPhyD..11..259U} \\ % Unkel et al. 1989      
 3.963  &   0.0000 &  0.0000  &  y $^2$S &  \citet{2002PhyS...66..363L} \\ % Lefebvre et al. 2002   
 4.609  &   0.0000 &  0.0000  &  v $^2$G &  none                        \\ %                        
 4.739  &   0.0000 &  0.0000  &  x $^2$H &  none                        \\ %                        
 5.251  &   0.0000 &  0.0000  &  e $^6$G &  none                        \\ %                        
\noalign{\smallskip}\hline\hline
\end{tabular}
\end{table*}

\begin{table*}
\renewcommand{\tabcolsep}{1.0mm}
\caption[HFS constants for \mni\ levels]{HFS constants for 35 \mni\ levels.}
\label{HFSmn}
\centering
\begin{tabular}{rcccl}
\hline\hline\noalign{\smallskip}
 $E$ (eV) & $A$ & $B$ & Level &  Reference \\
\noalign{\smallskip}\hline\noalign{\smallskip}
   0.000 & -0.0024 &  0.0     &   a $^6$S & \citet{1971PhRvA...3.1220D} \\ % Davis et al. 1971
   2.143 &  0.0153 &  0.0007  &   a $^6$D & \citet{1979ZPhyA.291..207D} \\ % Dembczynsky et al. 1979
   2.164 &  0.0146 & -0.0015  &   a $^6$D & \citet{1979ZPhyA.291..207D} \\ % Dembczynsky et al. 1979
   2.178 &  0.0157 & -0.0022  &   a $^6$D & \citet{1979ZPhyA.291..207D} \\ % Dembczynsky et al. 1979
   2.187 &  0.0294 &  0.0     &   a $^6$D & \citet{1979ZPhyA.291..207D} \\ % Dembczynsky et al. 1979
   2.282 &  0.0191 &  0.0009  &   z $^8$P & \citet{1987ZPhyD...7..161B} \\ % Brodzinski et al. 1987
   2.298 &  0.0182 & -0.0034  &   z $^8$P & \citet{1987ZPhyD...7..161B} \\ % Brodzinski et al. 1987
   2.319 &  0.0152 &  0.0016  &   z $^8$P & \citet{1987ZPhyD...7..161B} \\ % Brodzinski et al. 1987
   2.888 & -0.0054 &  0.0     &   a $^4$D & \citet{2005ApJS..157..402B} \\ % Blackwell-Whitehead et al. 2005
   2.920 & -0.0046 &  0.0     &   a $^4$D & \citet{2005ApJS..157..402B} \\ % Blackwell-Whitehead et al. 2005
   2.941 &  0.0017 &  0.0     &   a $^4$D & \citet{2005ApJS..157..402B} \\ % Blackwell-Whitehead et al. 2005
   2.953 &  0.0506 &  0.0     &   a $^4$D & \citet{2005ApJS..157..402B} \\ % Blackwell-Whitehead et al. 2005
   3.072 &  0.0191 &  0.0004  &   z $^6$P & \citet{1969PhLA...29..486H} \\ % Handrich et al. 1969
   3.073 &  0.0156 & -0.0025  &   z $^6$P & \citet{1969PhLA...29..486H} \\ % Handrich et al. 1969
   3.075 &  0.0143 &  0.0021  &   z $^6$P & \citet{1969PhLA...29..486H} \\ % Handrich et al. 1969
   3.133 &  0.0135 &  0.0     &   a $^4$G & \citet{1981ZPhyA.303....7J} \\ % Johann et al. 1981
   3.134 &  0.0199 &  0.0     &   a $^4$G & \citet{1981ZPhyA.303....7J} \\ % Johann et al. 1981
   3.772 &  0.0096 &  0.0     &   b $^4$D & \citet{2005ApJS..157..402B} \\ % Blackwell-Whitehead et al. 2005
   3.844 & -0.0203 &  0.0025  &   z $^4$P & \citet{1987ZPhyD...7..161B} \\ % Brodzinski et al. 1987
   3.853 & -0.0271 & -0.0013  &   z $^4$P & \citet{1987ZPhyD...7..161B} \\ % Brodzinski et al. 1987
   4.425 & -0.0324 &  0.0006  &   y $^6$P & \citet{1972AA....18..209L} \\ % Luc \& Gerstenkorn 1972
   4.429 & -0.018  & -0.0023  &   y $^6$P & \citet{1972AA....18..209L} \\ % Luc \& Gerstenkorn 1972
   4.435 & -0.013  & -0.0062  &   y $^6$P & \citet{1972AA....18..209L} \\ % Luc \& Gerstenkorn 1972
   4.889 &  0.0246 &  0.0016  &   e $^8$S & \citet{1987ZPhyD...7..161B} \\ % Brodzinski et al. 1987
   5.133 &  0.027  &  0.0     &   e $^6$S & \citet{1987ZPhyD...7..161B} \\ % Brodzinski et al. 1987
   5.396 &  0.0046 &  0.0     &   z $^6$F & \citet{2003AA...404.1153L} \\ % Lefebvre et al. 2003
   5.491 &  0.0044 &  0.0     &   z $^4$F & \citet{2005ApJS..157..402B} \\ % Blackwell-Whitehead et al. 2005
   5.520 &  0.0057 &  0.0     &   z $^4$F & \citet{2005ApJS..157..402B} \\ % Blackwell-Whitehead et al. 2005
   5.542 &  0.0095 &  0.0     &   z $^4$F & \citet{2005ApJS..157..402B} \\ % Blackwell-Whitehead et al. 2005
   5.556 &  0.0223 &  0.0     &   z $^4$F & \citet{2005ApJS..157..402B} \\ % Blackwell-Whitehead et al. 2005
   5.696 &  0.0027 &  0.0     &   z $^4$D & \citet{2003PhyS...67..476B} \\ % Basar et al. 2003
   5.853 &  0.0155 &  0.0     &   e $^6$D & \citet{1972AA....18..209L} \\ % Luc \& Gerstenkorn 1972
   5.854 &  0.0158 &  0.0     &   e $^6$D & \citet{1972AA....18..209L} \\ % Luc \& Gerstenkorn 1972
   5.854 &  0.0176 &  0.0     &   e $^6$D & \citet{1972AA....18..209L} \\ % Luc \& Gerstenkorn 1972
   6.149 & -0.0505 &  0.0     &   e $^4$S & \citet{2003AA...404.1153L} \\ % Lefebvre et al. 2003
\noalign{\smallskip}\hline\hline
\end{tabular}
\end{table*}

\begin{table*}
\renewcommand{\tabcolsep}{1.0mm}
\caption[HFS constants for \coi\ levels]{HFS constants for 47 \coi\ levels measured by \citet{1996ApJS..107..811P}.}
\label{HFSco}
\centering
\begin{tabular}{lccc}
\hline\hline\noalign{\smallskip}
 $E$ (eV) & $A$ & $B$ & Level \\
\noalign{\smallskip}\hline\noalign{\smallskip}
   1.710 &  0.0059 & -0.008   &  a $^4P$   \\
   1.740 &  0.0106 &  0.004   &  a $^4P$   \\
   1.785 & -0.0236 &  0.0     &  a $^4P$   \\
   1.883 &  0.0374 &  0.005   &  b $^4P$   \\
   1.956 &  0.0154 & -0.002   &  b $^4P$   \\ 
   2.042 &  0.0205 &  0.002   &  a $^2G$   \\
   2.080 &  0.0463 &  0.004   &  a $^2D$   \\
   2.137 &  0.028  & -0.0032  &  a $^2G$   \\
   2.280 &  0.0112 &  0.004   &  a $^2P$   \\
   2.328 &  0.0201 &  0.0     &  a $^2P$   \\
   3.117 &  0.0253 &  0.008   &  z $^6G$   \\
   3.216 &  0.0206 &  0.005   &  z $^6G$   \\
   3.252 &  0.0183 &  0.006   &  z $^6G$   \\
   3.298 &  0.0049 &  0.004   &  z $^6G$   \\
   3.514 &  0.0270 & -0.002   &  z $^4F$   \\ 
   3.576 &  0.0258 &  0.007   &  z $^4G$   \\
   3.629 &  0.0173 &  0.006   &  z $^4G$   \\
   3.632 &  0.0251 &  0.00    &  z $^4D$   \\
   3.713 &  0.0232 &  0.001   &  z $^4D$   \\
   3.775 &  0.0235 &  0.001   &  z $^4D$   \\
   3.812 &  0.0275 &  0.0     &  z $^4D$   \\ 
   3.952 &  0.015  &  0.006   &  z $^2F$   \\
   3.971 &  0.016  &  0.005   &  y $^4D$   \\
   4.021 &  0.0100 &  0.0     &  y $^4G$   \\ 
   4.025 &  0.0111 &  0.0     &  y $^4G$   \\ 
   4.049 &  0.0155 &  0.00    &  y $^4D$   \\
   4.064 &  0.0349 &  0.0     &  z $^2F$   \\
   4.110 &  0.0197 &  0.00    &  y $^4D$   \\
   4.149 &  0.0154 &  0.00    &  z $^2D$   \\
   4.149 &  0.0129 &  0.0     &  y $^4G$   \\
   4.232 &  0.0248 & -0.004   &  y $^2G$   \\
   4.240 &  0.0397 & -0.002   &  y $^4F$   \\
   4.259 &  0.0461 &  0.002   &  z $^2D$   \\
   4.475 &  0.0164 &  0.00    &  y $^2D$   \\
   4.572 &  0.042  &  0.00    &  y $^2D$   \\
   5.552 &  0.0134 &  0.00    &  e $^4F$   \\
   5.663 &  0.0335 &  0.005   &  e $^6F$   \\
   5.791 &  0.0287 & -0.001   &  e $^6F$   \\
   5.839 &  0.0257 & -0.001   &  e $^6F$   \\
   5.873 &  0.0194 & -0.001   &  e $^6F$   \\
   5.892 &  0.0360 &  0.005   &  f $^4F$   \\
   5.976 &  0.0283 &  0.003   &  f $^4F$   \\
   6.341 &  0.0000 &  0.0     &  e $^4H$   \\
   6.345 &  0.0000 &  0.00    &  e $^4H$   \\
   6.348 &  0.0000 &  0.00    &  e $^4G$   \\
   6.462 &  0.0000 &  0.00    &  e $^4H$   \\
   6.543 &  0.0000 &  0.00    &  e $^4G$   \\
\noalign{\smallskip}\hline\hline
\end{tabular}
\end{table*}

\begin{table*}
\renewcommand{\tabcolsep}{1.0mm}
\caption[HFS constants for \cui\ levels]{HFS constants and isotopic shifts relative to the ground state ($4s \enspace ^2S$) for five \cui\ levels.}
\label{HFScu}
\centering
\begin{tabular}{l ccc ccc cll}
\hline\hline\noalign{\smallskip}
 $E$ (eV) & $A_{63}$ & $B_{63}$ & IS$_{63}$ & $A_{65}$ & $B_{65}$ & IS$_{65}$ & Level & Ref. (HFS) &  Ref. (IS)\tablefootmark{$a$} \\
\noalign{\smallskip}\hline\noalign{\smallskip}
1.389 & 0.0250 &   0.0062 &   0.0168 & 0.0268 &  0.0058 &  -0.0377  &  4s$^2$ $^2$D & \citet{1967ZPhy..200..158F} & Kurucz (2012), \citet{1955ZPhy..141..122W} \\ % Fischer et al 1967, Wagner 1955
1.642 & 0.0619 &   0.0046 &   0.0170 & 0.0661 &  0.0043 &  -0.0381  &  4s$^2$ $^2$D & \citet{1989ZPhyD..13..203B} & Kurucz (2012), \citet{1955ZPhy..141..122W} \\ % Bergström et al. 1989, Wagner 1955
3.786 & 0.0169 &   0.0    &  -0.0059 & 0.0181 &  0.0    &   0.0132  &  4p $^2$P     & \citet{1993AcSpe..48.1259H} & Kurucz (2012), \citet{1955ZPhy..141..122W} \\ % Hermann et al. 1993, Wagner 1955
3.817 & 0.0065 &  -0.0010 &  -0.0062 & 0.0069 & -0.0009 &   0.0139  &  4p $^2$P     & \citet{1993AcSpe..48.1259H} & Kurucz (2012), \citet{1955ZPhy..141..122W} \\ % Hermann et al. 1993, Wagner 1955
6.191 & 0.0    &   0.0    &  -0.006  & 0.0    &  0.0    &   0.013   &  4d $^2$D     & none                        & Kurucz (2012) \\
\noalign{\smallskip}\hline\hline
\end{tabular}
\tablefoot{
\tablefoottext{$a$}{\url{http://kurucz.harvard.edu/atoms/2900/ab290063.dat}, \url{http://kurucz.harvard.edu/atoms/2900/ab290065.dat}}
}
\end{table*}

\begin{table*}
\renewcommand{\tabcolsep}{1.0mm}
\caption[HFS constants for \baii\ levels]{HFS constants for five \baii\ levels, for isotopes with non-zero nuclear spin.}
\label{HFSba}
\centering
\begin{tabular}{l cc cc cl}
\hline\hline\noalign{\smallskip}
 $E$ (eV) & $A_{135}$ & $B_{135}$ &  $A_{137}$ & $B_{137}$ & Level & Reference \\
\noalign{\smallskip}\hline\noalign{\smallskip}
0.000  &  0.1198 &  0.0    &   0.1341 &  0.0     &  6s $^2$S & \citet{1981pmfc.conf...99B} \\ % Becker et al. 1981, Table 1 (or 1981ecap.conf..208B)
0.604  &  0.0057 &  0.0010 &   0.0063 &  0.0015  &  5d $^2$D & \citet{1986PhRvA..33.2117S} \\ % Silverans et al. 1986  
0.704  & -0.0004 &  0.0013 &  -0.0004 &  0.0020  &  5d $^2$D & \citet{1986PhRvA..33.2117S} \\ % Silverans et al. 1986  
2.512  &  0.0222 &  0.0    &   0.0248 &  0.0     &  6p $^2$P & \citet{1993JPhB...26.4289V} \\ % Villemoes et al. 1993  
2.722  &  0.0038 &  0.002  &   0.0042 &  0.0031  &  6p $^2$P & \citet{1993JPhB...26.4289V} \\ % Villemoes et al. 1993  
\noalign{\smallskip}\hline\hline
\end{tabular}
\end{table*}

\begin{table*}
\renewcommand{\tabcolsep}{0.3mm}
\caption[HFS constants for \laii\ levels]{HFS constants for nine \laii\ levels.}
\label{HFSla}
\centering
\begin{tabular}{rcccl}
\hline\hline\noalign{\smallskip}
 $E$ (eV) & $A$ & $B$ & Level &  Reference \\
\noalign{\smallskip}\hline\noalign{\smallskip}
0.173 & $ 3.165 \times 10^{-2}$  &  $1.66\times 10^{-3}$ & 6s     a$^1$D     & \citet{1982ZPhyA.304..279H} \\ % Höhle et al. (1982)
0.235 & $-3.760 \times 10^{-2}$  &  $1.66\times 10^{-3}$ & 6s     a$^3$D     & \citet{1982ZPhyA.304..279H} \\ % Höhle et al. (1982)
0.244 & $-6.19  \times 10^{-4}$  &  $1.25\times 10^{-3}$ & 5d$^2$ a$^3$F     & \citet{1982ZPhyA.304..279H} \\ % Höhle et al. (1982)
0.321 & $-3     \times 10^{-4}$  &  $1.89\times 10^{-3}$ & 6s     a$^3$D     & \citet{1982ZPhyA.304..279H} \\ % Höhle et al. (1982)
2.261 & $ 4.96  \times 10^{-3}$  &  $1.5 \times 10^{-4}$ & 5d     y$^3$F$^o$ & \citet{1982ZPhyA.304..279H} \\ % Höhle et al. (1982)
2.658 & $ 1.44  \times 10^{-2}$  &  --                   & 5d     z$^3$D$^o$ & \citet{LBS}                 \\ % Lawler et al. (2001)
2.741 & $ 4.26  \times 10^{-3}$  &  $9   \times 10^{-5}$ & 5d     z$^3$D$^o$ & \citet{2001JaJAP..40.2508L} \\ % Li et al. (2001)
2.763 & $ 4.5   \times 10^{-3}$  &  $2   \times 10^{-3}$ & 5d     z$^3$G$^o$ & \citet{2008JPhB...41u5004F}\tablefootmark{$a$} \\ % Furmann et al. (2008)
2.815 & $ 2.6   \times 10^{-3}$  &  --                   & 5d     z$^3$P$^o$ & \citet{LBS}                 \\ % Lawler et al. (2001)
\noalign{\smallskip}\hline\hline
\end{tabular}
\tablefoot{
\tablefoottext{$a$}{not included in the line list}
}
\end{table*}

\begin{table*}
\renewcommand{\tabcolsep}{1.0mm}
\caption[HFS constant for \prii\ levels]{HFS constant $A$ for twelve \prii\ levels.}
\label{HFSpr}
\centering
\begin{tabular}{rccl}
\hline\hline\noalign{\smallskip}
 $E$ (eV) & $A$ & Level &  Reference \\
\noalign{\smallskip}\hline\noalign{\smallskip}
0.483 & $ 3.01  \times 10^{-2}$ & 5d $^5$L$^o$ & \citet{1989PhyS...39..694G} \\ % Ginibre (1989)
0.508 & $ 3.25  \times 10^{-2}$ & 5d $^5$K$^o$ & \citet{1989PhyS...39..694G} \\ % Ginibre (1989)
0.633 & $ 2.22  \times 10^{-2}$ & 5d $^5$L$^o$ & \citet{1989PhyS...39..694G} \\ % Ginibre (1989)
0.648 & $ 2.40  \times 10^{-2}$ & 5d $^5$K$^o$ & \citet{1989PhyS...39..694G} \\ % Ginibre (1989)
0.795 & $ 1.88  \times 10^{-2}$ & 5d $^5$K$^o$ & \citet{1989PhyS...39..694G} \\ % Ginibre (1989)
0.796 & $ 1.82  \times 10^{-2}$ & 5d $^5$L$^o$ & \citet{1989PhyS...39..694G} \\ % Ginibre (1989)
0.968 & $ 1.50  \times 10^{-2}$ & 5d $^5$L$^o$ & \citet{1989PhyS...39..694G} \\ % Ginibre (1989)
2.811 & $ 2.554 \times 10^{-2}$ & 6p $^5$K     & \citet{2002CaJPh..80..557R} \\ % Rivest et al. (2002)
2.884 & $ 1.949 \times 10^{-2}$ & 5d$^2$       & \citet{2002CaJPh..80..557R} \\ % Rivest et al. (2002)
2.990 & $ 2.740 \times 10^{-2}$ & 6p $^5$K     & \citet{2002CaJPh..80..557R} \\ % Rivest et al. (2002)
3.170 & $ 2.301 \times 10^{-2}$ & 6p $^5$K     & \citet{2002CaJPh..80..557R} \\ % Rivest et al. (2002)
3.363 & $ 1.973 \times 10^{-2}$ & 6p $^5$K     & \citet{2002CaJPh..80..557R} \\ % Rivest et al. (2002)
\noalign{\smallskip}\hline\hline
\end{tabular}
\end{table*}

\begin{table*}
\renewcommand{\tabcolsep}{1.0mm}
\caption[Isotopic splitting for \ndii\ lines]{Wavelength shifts for isotopes with baryon numbers 150 and 142 relative to isotope 144 for the 53 preselected \ndii\ lines.}
\label{ISnd}
\centering
\begin{tabular}{l rr l >{\raggedleft}p{4em}c>{\raggedleft\arraybackslash}p{4em}}
\hline\hline\noalign{\smallskip}
\multicolumn{1}{c}{$\lambda$\tablefootmark{$a$}} & \multicolumn{1}{c}{$E_{\rm low}$} & \multicolumn{1}{c}{$E_{\rm upp}$} & HFS\tablefootmark{$b$} & \multicolumn{1}{c}{IS(144$-$150)\tablefootmark{$c$}} & Ref.(IS)\tablefootmark{$c$} & \multicolumn{1}{c}{IS(142$-$144)\tablefootmark{$d$}} \\
% units
\multicolumn{1}{c}{[\AA]} & \multicolumn{1}{c}{[eV]} & \multicolumn{1}{c}{[eV]} &  & \multicolumn{1}{c}{[\AA]} &  & \multicolumn{1}{c}{[\AA]} \\
\noalign{\smallskip}\hline\noalign{\smallskip}
4811.342 & 0.064 & 2.640 &      $E_{\rm low}$ &      $ 0.0088$ &   1 1  &       $ 0.0025$ \\
4849.062 & 0.471 & 3.028 &               both &      $ 0.0052$ &   1 1  &       $ 0.0012$ \\
4859.03  & 0.321 & 2.871 &               both &      $-0.0040$ &   1 1  &       $-0.0012$ \\
4902.04  & 0.064 & 2.592 &      $E_{\rm low}$ &      $ 0.0202$ &   1 1  &       $ 0.0063$ \\
4914.38  & 0.380 & 2.902 &               both &      $ 0.0138$ &   1 1  &       $ 0.0041$ \\
4943.899 & 0.205 & 2.712 &      $E_{\rm low}$ &      $-0.0007$ &   1 1  &       $-0.0005$ \\
4947.02  & 0.559 & 3.065 &               both &      $ 0.0198$ &   1 1  &       $ 0.0054$ \\
4959.12  & 0.064 & 2.563 &      $E_{\rm low}$ &      $ 0.0224$ &   1 1  &       $ 0.0066$ \\
4961.387 & 0.631 & 3.129 &      $E_{\rm low}$ &      $-0.0005$ &   1 2  &       $       $ \\
4970.91  & 0.321 & 2.814 &               both &      $ 0.0237$ &   1 1  &       $ 0.0067$ \\
4987.16  & 0.742 & 3.227 &               both &      $ 0.0134$ &   1 1  &       $ 0.0040$ \\
4989.95  & 0.631 & 3.115 &      $E_{\rm low}$ &      $-0.0105$ &   1 1  &       $-0.0032$ \\
4998.541 & 0.471 & 2.951 &      $E_{\rm low}$ &      $-0.0065$ &   1 1  &       $ 0.0025$ \\
5089.832 & 0.205 & 2.640 &      $E_{\rm low}$ &      $ 0.0086$ &   1 1  &       $ 0.0023$ \\
5092.79  & 0.380 & 2.814 &               both &      $ 0.0249$ &   1 1  &       $ 0.0065$ \\
5130.59  & 1.304 & 3.720 &               none &      $-0.0061$ &   1 1  &       $-0.0024$ \\
5132.33  & 0.559 & 2.975 &               both &      $ 0.0137$ &   1 1  &       $ 0.0037$ \\
5143.34  & 0.182 & 2.592 &      $E_{\rm low}$ &      $ 0.0225$ &   1 1  &       $ 0.0069$ \\
5165.13  & 0.680 & 3.080 &      $E_{\rm upp}$ &      $-0.0667$ &   1 1  &       $-0.0198$ \\
5200.121 & 0.559 & 2.943 &      $E_{\rm low}$ &      $ 0.0162$ &   1 2  &       $       $ \\
5212.36  & 0.205 & 2.583 &      $E_{\rm low}$ &      $ 0.0139$ &   1 1  &       $ 0.0041$ \\
5234.19  & 0.550 & 2.918 &               both &      $-0.0411$ &   1 1  &       $-0.0118$ \\
5249.58  & 0.976 & 3.337 &      $E_{\rm upp}$ &      $-0.0174$ &   2 1  &       $       $ \\
5250.81  & 0.745 & 3.105 &               none &      $-0.0532$ &   1 1  &       $-0.0154$ \\
5255.51  & 0.205 & 2.563 &      $E_{\rm low}$ &      $ 0.0238$ &   1 1  &       $ 0.0069$ \\
5276.869 & 0.859 & 3.208 &               both &      $-0.0518$ &   1 1  &       $-0.0153$ \\
5293.16  & 0.823 & 3.165 &      $E_{\rm upp}$ &      $-0.0193$ &   1 1  &       $-0.0056$ \\
5306.46  & 0.859 & 3.195 &      $E_{\rm low}$ &      $-0.0462$ &   1 1  &       $-0.0135$ \\
5310.04  & 1.136 & 3.471 &      $E_{\rm upp}$ &      $-0.0671$ &   1 1  &       $-0.0195$ \\
5311.45  & 0.986 & 3.319 &      $E_{\rm upp}$ &      $-0.0395$ &   1 1  &       $-0.0116$ \\
5319.81  & 0.550 & 2.880 &               both &      $-0.0394$ &   1 1  &       $-0.0110$ \\
5356.97  & 1.264 & 3.578 &      $E_{\rm upp}$ &      $-0.0413$ &   1 1  &       $-0.0123$ \\
5361.17  & 0.559 & 2.871 &               both &      $-0.0049$ &   1 1  &       $-0.0020$ \\
5361.467 & 0.680 & 2.992 &      $E_{\rm upp}$ &      $-0.0411$ &   1 1  &       $-0.0115$ \\
5416.374 & 0.859 & 3.148 &               both &      $-0.0681$ &   1 1  &       $-0.0203$ \\
5431.52  & 1.121 & 3.403 &      $E_{\rm upp}$ &      $-0.0469$ &   1 1  &       $-0.0121$ \\
5442.264 & 0.680 & 2.958 &      $E_{\rm upp}$ &      $-0.0587$ &   1 1  &       $-0.0175$ \\
5485.70  & 1.264 & 3.524 &      $E_{\rm upp}$ &      $-0.0238$ &   1 1  &       $-0.0072$ \\
5533.82  & 0.559 & 2.799 &      $E_{\rm low}$ &      $ 0.0211$ &   1 1  &       $ 0.0061$ \\
5548.45  & 0.550 & 2.784 &               both &      $-0.0585$ &   1 1  &       $-0.0182$ \\
5581.59  & 0.859 & 3.080 &               both &      $-0.0779$ &   1 1  &       $-0.0231$ \\
5618.99  & 1.773 & 3.979 &               none &      $-0.0531$ &   2 1  &       $       $ \\
5740.86  & 1.160 & 3.319 &      $E_{\rm upp}$ &      $-0.0455$ &   1 1  &       $-0.0135$ \\
5811.57  & 0.859 & 2.992 &               both &      $-0.0483$ &   1 1  &       $-0.0135$ \\
5842.366 & 1.282 & 3.403 &      $E_{\rm upp}$ &      $-0.0540$ &   2 1  &       $       $ \\
5882.786 & 0.559 & 2.666 &      $E_{\rm low}$ &      $ 0.0284$ &   1 2  &       $       $ \\
6365.54  & 0.933 & 2.880 &      $E_{\rm upp}$ &      $-0.0564$ &   1 1  &       $-0.0158$ \\
6385.154 & 1.160 & 3.101 &               none &      $-0.0779$ &   1 1  &       $       $ \\
6637.19  & 1.452 & 3.319 &      $E_{\rm upp}$ &      $-0.0621$ &   2 1  &       $       $ \\
8530.545 & 0.064 & 1.517 &      $E_{\rm low}$ &      $ 0.0539$ &   1 2  &       $       $ \\
8594.883 & 1.140 & 2.583 &               none &      $-0.1293$ &   2 1  &       $       $ \\
8643.48  & 1.200 & 2.634 &               none &      $-0.2542$ &   2 2  &       $       $ \\
8691.303 & 1.350 & 2.776 &               none &      $-0.1738$ &   2 1  &       $       $ \\
\noalign{\smallskip}\hline\hline
\end{tabular}
\tablefoot{
\tablefoottext{$a$}{Average wavelengths.}
\tablefoottext{$b$}{Indicates if HFS data are available for both levels, for $E_{\rm low}$ or $E_{\rm upp}$ only (see Table~\ref{HFSnd}), or for none.}
\tablefoottext{$c$}{References for level energy shifts between isotopes 144 and 150 used to calculated the wavelength shifts are given in column ``Ref.(IS)'': 1) \citet{1997ZPhyD..42...71N}, 2) \citet{1984PhyS...29..119B}. The two numbers in each row refer to $E_{\rm low}$ and $E_{\rm upp}$, respectively.}
\tablefoottext{$d$}{Level energy shifts between isotopes 142 and 144 were taken from \citet{1981AcSpe..36..943A}.}
}
\end{table*}

\begin{table*}
\renewcommand{\tabcolsep}{1.0mm}
\caption[HFS constants for \ndii\ levels]{HFS constants for 34 out of the 69 \ndii\ levels involved in the preselected transitions.}
\label{HFSnd}
\centering
\begin{tabular}{l rr rr c}
\hline\hline\noalign{\smallskip}
 $E$ (eV) & \multicolumn{1}{c}{$A_{143}$} & \multicolumn{1}{c}{$B_{143}$} & \multicolumn{1}{c}{$A_{145}$} & \multicolumn{1}{c}{$B_{145}$} & Level \\
\noalign{\smallskip}\hline\noalign{\smallskip}
0.064 & $-1.123\times 10^{-2}$ &  $ 4.39\times 10^{-3}$ &  $-6.979\times 10^{-3}$ &  $ 2.29\times 10^{-3}$ & 6s  $^6$I \\
0.182 & $ 1.028\times 10^{-2}$ &  $ 3.57\times 10^{-3}$ &  $ 6.390\times 10^{-3}$ &  $ 1.86\times 10^{-3}$ & 6s  $^6$I \\
0.205 & $-1.981\times 10^{-3}$ &  $ 4.24\times 10^{-3}$ &  $-1.231\times 10^{-3}$ &  $ 2.21\times 10^{-3}$ & 6s  $^4$I \\
0.321 & $-9.405\times 10^{-3}$ &  $ 1.00\times 10^{-5}$ &  $-5.846\times 10^{-3}$ &  $ 5.21\times 10^{-6}$ & 6s  $^6$I \\
0.380 & $-2.578\times 10^{-4}$ &  $ 3.08\times 10^{-3}$ &  $-1.603\times 10^{-4}$ &  $ 1.61\times 10^{-3}$ & 6s  $^4$I \\
0.471 & $-8.608\times 10^{-3}$ &  $ 7.27\times 10^{-3}$ &  $-5.350\times 10^{-3}$ &  $ 3.79\times 10^{-3}$ & 6s  $^6$I \\
0.550 & $-4.736\times 10^{-3}$ &  $ 9.34\times 10^{-4}$ &  $-2.943\times 10^{-3}$ &  $ 4.87\times 10^{-4}$ & 5d  $^6$L \\
0.559 & $ 3.936\times 10^{-4}$ &  $ 1.35\times 10^{-3}$ &  $ 2.447\times 10^{-4}$ &  $ 7.04\times 10^{-4}$ & 6s  $^4$I \\
0.631 & $-8.096\times 10^{-3}$ &  $ 6.50\times 10^{-3}$ &  $-5.032\times 10^{-3}$ &  $ 3.39\times 10^{-3}$ & 6s  $^6$I \\
0.742 & $ 5.551\times 10^{-4}$ &  $ 3.47\times 10^{-3}$ &  $ 3.450\times 10^{-4}$ &  $ 1.81\times 10^{-3}$ & 6s  $^4$I \\
0.745\tablefootmark{$a$} & $-6.671\times 10^{-3}$ &  $-3.88\times 10^{-3}$ &  $-4.140\times 10^{-3}$ &  $ 2.23\times 10^{-3}$ & 5d  $^6$K \\
0.859\tablefootmark{$a$} & $-5.030\times 10^{-3}$ &  $ 1.82\times 10^{-3}$ &  $-3.119\times 10^{-3}$ &  $ 9.37\times 10^{-4}$ & 5d  $^6$K \\
2.784 & $-6.301\times 10^{-3}$ &  $-2.45\times 10^{-3}$ &  $-3.917\times 10^{-3}$ &  $-1.28\times 10^{-3}$ &     $^o$\\
2.814 & $-5.092\times 10^{-3}$ &  $ 5.57\times 10^{-3}$ &  $-3.165\times 10^{-3}$ &  $ 2.90\times 10^{-3}$ & 5d$^2$ $^4$I$^o$ \\
2.871 & $-6.492\times 10^{-3}$ &  $-3.80\times 10^{-3}$ &  $-4.036\times 10^{-3}$ &  $-1.98\times 10^{-3}$ &     $^o$\\
2.880 & $-6.639\times 10^{-3}$ &  $-3.38\times 10^{-3}$ &  $-4.127\times 10^{-3}$ &  $-1.76\times 10^{-3}$ & 6p  $^6$K$^o$ \\
2.902 & $-4.765\times 10^{-3}$ &  $ 4.80\times 10^{-4}$ &  $-2.962\times 10^{-3}$ &  $ 2.50\times 10^{-4}$ &     $^o$\\
2.918 & $-6.016\times 10^{-3}$ &  $-1.60\times 10^{-3}$ &  $-3.740\times 10^{-3}$ &  $-8.34\times 10^{-4}$ &     $^o$\\
2.958 & $-3.959\times 10^{-3}$ &  $-8.21\times 10^{-4}$ &  $-2.461\times 10^{-3}$ &  $-4.28\times 10^{-4}$ &     $^o$\\
2.975 & $-4.892\times 10^{-3}$ &  $ 6.00\times 10^{-3}$ &  $-3.041\times 10^{-3}$ &  $ 3.13\times 10^{-3}$ &     $^o$\\
2.992 & $-4.930\times 10^{-3}$ &  $ 3.94\times 10^{-4}$ &  $-3.064\times 10^{-3}$ &  $ 2.05\times 10^{-4}$ &     $^o$\\
3.028 & $-4.817\times 10^{-3}$ &  $-1.83\times 10^{-3}$ &  $-2.994\times 10^{-3}$ &  $-9.56\times 10^{-4}$ &     $^o$\\
3.065 & $-4.463\times 10^{-3}$ &  $-2.57\times 10^{-3}$ &  $-2.774\times 10^{-3}$ &  $-1.34\times 10^{-3}$ &     $^4$I$^o$ \\
3.080 & $-3.629\times 10^{-3}$ &  $-1.00\times 10^{-4}$ &  $-2.255\times 10^{-3}$ &  $-5.21\times 10^{-5}$ &     $^o$\\
3.148 & $-6.468\times 10^{-3}$ &  $ 3.70\times 10^{-3}$ &  $-4.020\times 10^{-3}$ &  $ 1.93\times 10^{-3}$ &     $^o$\\
3.165 & $-4.835\times 10^{-3}$ &  $ 1.47\times 10^{-3}$ &  $-3.005\times 10^{-3}$ &  $ 7.65\times 10^{-4}$ & 6p  $^6$K$^o$ \\
3.208 & $-6.537\times 10^{-3}$ &  $ 3.40\times 10^{-3}$ &  $-4.063\times 10^{-3}$ &  $ 1.77\times 10^{-3}$ & 6p  $^6$I$^o$ \\
3.227 & $-4.928\times 10^{-3}$ &  $-6.67\times 10^{-5}$ &  $-3.063\times 10^{-3}$ &  $-3.48\times 10^{-5}$ &     $^o$\\
3.319 & $-5.091\times 10^{-3}$ &  $ 2.60\times 10^{-3}$ &  $-3.164\times 10^{-3}$ &  $ 1.36\times 10^{-3}$ & 6p  $^6$I$^o$ \\
3.337 & $-4.404\times 10^{-3}$ &  $-1.33\times 10^{-3}$ &  $-2.737\times 10^{-3}$ &  $-6.95\times 10^{-4}$ & 6p  $^6$K$^o$ \\
3.403 & $-4.174\times 10^{-3}$ &  $ 4.67\times 10^{-4}$ &  $-2.594\times 10^{-3}$ &  $ 2.43\times 10^{-4}$ & 6p  $^o$\\
3.471 & $-5.273\times 10^{-3}$ &  $-3.67\times 10^{-3}$ &  $-3.278\times 10^{-3}$ &  $-1.91\times 10^{-3}$ &     $^o$\\
3.524 & $-4.053\times 10^{-3}$ &  $-3.14\times 10^{-3}$ &  $-2.519\times 10^{-3}$ &  $-1.63\times 10^{-3}$ & 6p  $^6$K$^o$ \\
3.578 & $-4.180\times 10^{-3}$ &  $-1.30\times 10^{-3}$ &  $-2.598\times 10^{-3}$ &  $-6.78\times 10^{-4}$ & 6p  $^6$I$^o$ \\
\noalign{\smallskip}\hline\hline
\end{tabular}
\tablefoot{
The data are from \citet{2005CaJPh..83..841R}, except as noted for two levels. The subscripts indicate baryon numbers for isotopes $^{143}$Nd and $^{145}$Nd with nuclear spin $I = 3.5$.
\tablefoottext{$a$}{data from \citet{2004ADNDT..86....3M}}
}
\end{table*}

\begin{table*}
\renewcommand{\tabcolsep}{1.0mm}
\caption[HFS constants for \smii\ levels]{HFS constants for 9 out of the 10 \smii\ levels involved in the preselected transitions, for isotopes with non-zero nuclear spin. The data are from \citet{2003CaJPh..81.1389M}.}
\label{HFSsm}
\centering
\begin{tabular}{l rr rr c}
\hline\hline\noalign{\smallskip}
 $E$ (eV) & \multicolumn{1}{c}{$A_{147}$} & \multicolumn{1}{c}{$B_{147}$} & \multicolumn{1}{c}{$A_{149}$} & \multicolumn{1}{c}{$B_{149}$} & Level \\
\noalign{\smallskip}\hline\noalign{\smallskip}
0.104 &  $-1.058 \times 10^{-2}$ & $-2.37 \times 10^{-3}$ & $-8.719 \times 10^{-3}$ & $ 7.00 \times 10^{-4}$ &  6s  $^8$F \\
0.185 &  $-9.026 \times 10^{-3}$ & $-1.97 \times 10^{-3}$ & $-7.438 \times 10^{-3}$ & $ 5.67 \times 10^{-4}$ &  6s  $^8$F \\
0.378 &  $-7.759 \times 10^{-3}$ & $ 2.74 \times 10^{-3}$ & $-6.394 \times 10^{-3}$ & $-8.01 \times 10^{-4}$ &  6s  $^8$F \\
0.544 &  $ 2.845 \times 10^{-3}$ & $ 3.54 \times 10^{-3}$ & $ 2.345 \times 10^{-3}$ & $-1.03 \times 10^{-3}$ &  6s  $^6$F \\
0.659 &  $ 2.151 \times 10^{-3}$ & $ 7.34 \times 10^{-3}$ & $ 1.775 \times 10^{-3}$ & $-2.13 \times 10^{-3}$ &  6s  $^6$F \\
2.758 &  $-6.401 \times 10^{-3}$ & $ 1.20 \times 10^{-3}$ & $-5.277 \times 10^{-3}$ & $ -3.3 \times 10^{-4}$ &  6p  $^o$  \\
2.932 &  $-5.811 \times 10^{-3}$ & $ 3.67 \times 10^{-4}$ & $-4.787 \times 10^{-3}$ & $ -1.0 \times 10^{-4}$ &  6p  $^o$  \\
3.049 &  $-5.814 \times 10^{-3}$ & $ 1.13 \times 10^{-3}$ & $-4.793 \times 10^{-3}$ & $ -3.3 \times 10^{-4}$ &  6p  $^o$  \\
3.174 &  $-2.342 \times 10^{-3}$ & $ 7.34 \times 10^{-4}$ & $-1.931 \times 10^{-3}$ & $ -2.0 \times 10^{-4}$ &  6p  $^o$  \\
\noalign{\smallskip}\hline\hline
\end{tabular}
\end{table*}

\begin{table*}
\renewcommand{\tabcolsep}{1.0mm}
\caption[HFS constants for \euii\ levels]{HFS constants for seven \euii\ levels.}
\label{HFSeu}
\centering
\begin{tabular}{l rr rr cl}
\hline\hline\noalign{\smallskip}
 $E$ (eV) & \multicolumn{1}{c}{$A_{151}$} & \multicolumn{1}{c}{$B_{151}$} & \multicolumn{1}{c}{$A_{153}$} & \multicolumn{1}{c}{$B_{153}$} & Level & Reference \\
\noalign{\smallskip}\hline\noalign{\smallskip}
1.230 & $-0.0178$  & $ 0.0036$  &  $-0.0079$  & $ 0.0092$   &  a $^9$D  & $A$: \citet{Huhn:1992}, $B$: \citet{1993PhRvL..70..541M} \\ % A: Hühnermann et al. 1992, B: Möller et al. 1993
1.279 & $-0.0038$  & $-0.0072$  &  $-0.0017$  & $-0.0183$   &  a $^9$D  & $A$: \citet{Huhn:1992}, $B$: \citet{1993PhRvL..70..541M} \\ % A: Hühnermann et al. 1992, B: Möller et al. 1993
1.320 & $-0.0024$  & $ 0.0007$  &  $-0.0011$  & $ 0.0020$   &  a $^9$D  & \citet{1992PhLA..162..178V} \\ % Villemoes et 1992
1.380 & $-0.0024$  & $ 0.0120$  &  $-0.0011$  & $ 0.0302$   &  a $^9$D  & \citet{1992PhLA..162..178V} \\ % Villemoes et 1992
3.245 & $ 0.0048$  & $ 0.0155$  &  $ 0.0022$  & $ 0.0399$   &  z $^9$P  & \citet{1992PhLA..162..178V} \\ % Villemoes et 1992
3.328 & $ 0.0019$  & $ 0.0052$  &  $ 0.0008$  & $ 0.0131$   &  z $^7$P  & $A$: \citet{Huhn:1992}, $B$: \citet{1993PhRvL..70..541M} \\ % A: Hühnermann et al. 1992, B: Möller et al. 1993
3.361 & $-0.0043$  & $-0.0166$  &  $-0.0019$  & $-0.0425$   &  z $^7$P  & $A$: \citet{Huhn:1992}, $B$: \citet{1993PhRvL..70..541M} \\ % A: Hühnermann et al. 1992, B: Möller et al. 1993
\noalign{\smallskip}\hline\hline
\end{tabular}
\end{table*}

% place all figures and tables from this section here
\clearpage

%----------------------------------------------------------------------------
% Appendix: Background line list references
%----------------------------------------------------------------------------

\section{Species occurring in the background line list only}
\label{sect:background}

The background line list includes a number of lines for species which do not occur in the preselected line list. The references for the $gf$-values of those lines in the UVES-580 and GIRAFFE HR21 wavelength ranges with an estimated line depth in Arcturus greater than 1\% are given in Table~\ref{tab:master} (omitting species with less than five lines).

\begin{table}[h]
\caption{References for $gf$-values for species which occur only in the background line list.}
\label{tab:master}
\centering
\begin{tabular}{lrlrl}
\hline\hline\noalign{\smallskip}
Name       & $Z$ & Species   & $N$ & References \\
\noalign{\smallskip}\hline\noalign{\smallskip}
 Phosphor  & 15 & \phosphori &  35 & KP, GUES, LAW, MRB\\
 Potassium & 19 & \ki\       &  40 & K12, WSM\\
 Arsenic   & 33 & \asi       &  10 & GUES\\
 Rubidium  & 37 & \rbi       &   5 & GHR\\
 Rhodium   & 45 & \rhi       &  70 & CB, DLb, KZBa, SDL\\
 Palladium & 46 & \pdi       &  20 & CB\\
 Terbium   & 65 & \tbii      &  20 & MC, LWCS\\
 Holmium   & 67 & \hoi       &  30 & MC\\
 Erbium    & 68 & \eri       & 110 & MC\\
           &    & \erii      &  80 & MC, LSCW, XJZD\\
 Thulium   & 69 & \tmi       &  10 & MC, PK, WL\\
           &    & \tmii      &  30 & QPB, WL\\
 Ytterbium & 70 & \ybi       &  15 & PGK, PK\\
           &    & \ybii      &  40 & BDMQ\\
 Lutetium  & 71 & \lui       &  10 & FDLP, WV\\
           &    & \luii      &  15 & DCWL, QPBM\\
 Hafnium   & 72 & \hfi       &  75 & CB, CBcor, DSLb\\
           &    & \hfii      &  50 & CB, LDLS, LNWLX\\
 Tantalum  & 73 & \tai       & 140 & CB, CBcor, SDL\\
 Rhenium   & 75 & \rei       &  40 & CB, DSLc\\
 Osmium    & 76 & \osi       &  90 & CBcor, IAN, KZB\\
 Iridium   & 77 & \iri       &  25 & CBcor, GHcor\\
 Platinum  & 78 & \pti       &  20 & CB, DHL, GHLa, LGb\\
 Thorium   & 90 & \thi       & 110 & MC\\
           &    & \thii      & 205 & MC, NZL\\
\noalign{\smallskip}\hline\hline
\end{tabular}
\tablefoot{$N$ is the approximate number of lines with an estimated line depth in Arcturus greater than 0.01.
References:
BDMQ  \ldots  \citet{BDMQ}
CB    \ldots  \citet{CB},
CBcor \ldots  \citet{CBcor},
DCWL  \ldots  \citet{DCWL},
DLb   \ldots  \citet{DLb},
DHL   \ldots  \citet{DHL},
DSLb  \ldots  \citet{DSLb},
DSLc  \ldots  \citet{DSLc},
FDLP  \ldots  \citet{FDLP},
GHcor \ldots  \citet{GHcor},
GHLa  \ldots  \citet{GHLa},
GHR   \ldots  \citet{GHR},
GUES  \ldots  \citet{GUES},
IAN   \ldots  \citet{IAN},
K12   \ldots  \citet{K12},
KP    \ldots  \citet{KP},
KZB   \ldots  \citet{KZB},
LAW   \ldots  \citet{LAW},
LDLS  \ldots  \citet{LDLS},
LGb   \ldots  \citet{LGb},
LNWLX \ldots  \citet{LNWLX},
LSCW  \ldots  \citet{LSCW}
LWCS  \ldots  \citet{LWCS},
MC    \ldots  \citet{MC},
MRB   \ldots  \citet{MRB},
NZL   \ldots  \citet{NZL},
PGK   \ldots  \citet{PGK}
PK    \ldots  \citet{PK}
QPB   \ldots  \citet{QPB}
QPBM  \ldots  \citet{QPBM},
SDL   \ldots  \citet{SDL},
WL    \ldots  \citet{WL}
WSM   \ldots  \citet{WSM},
WV    \ldots  \citet{WV},
XJZD  \ldots  \citet{XJZD}.
}
\end{table}

%----------------------------------------------------------------------------
% Appendix: Collisional broadening due to neutral hydrogen - 4 sections
%----------------------------------------------------------------------------

\newpage
\section{Theoretical background for collisional broadening data}
\label{sect:theory}

 We discuss three commonly used and presently available options for the calculation of spectral line broadening by collisions with ground state neutral hydrogen atoms. These are:
\begin{enumerate}
   \item The \emph{Unsöld recipe}. Lindholm-Foley theory employing the van der Waals long-range interaction potential $C_6/R^6$ as a function of internuclear distance $R$, where $C_6$ is calculated approximately as formulated by \citet{1955QB461.U55......}. % Unsöld
   \item \emph{Kurucz calculations}. Lindholm-Foley theory as in 1, where $C_6$ is calculated in more detail, as described by \citet[p. 76]{1981SAOSR.390.....K}. % Kurucz
   \item The \emph{ABO theory}. Detailed scattering calculations based on potentials from perturbation theory as described by \citet{1991MNRAS.253..549A}, % Anstee & O’Mara
         and successive expansions by P.S.\ Barklem and collaborators. 
\end{enumerate}
All three methods employ the impact theory \citep[see for example Sect.~4.1 of][and references therein]{2016A&ARv..24....9B}, and thus the collisionally broadened line profile is described by a Lorentz profile. The long-range dispersion interaction between atoms is given by the van der Waals potential, which may be written in terms of the frequency shift $\Delta \nu = C_6/R^6$. In Lindholm-Foley theory \citep{lindholm_pressure_1946,foley_pressure_1946}, assuming such a form for the state-dependent dispersive part of the interaction that leads to line broadening, the line width (full width at half-maximum) is given in angular frequency units by 
\begin{equation}
   \gamma = 17.0 \cdot (C_{6,\rm upp}-C_{6,\rm low})^{2/5} \cdot \bar \varv^{3/5} \cdot N_H,
   \label{eq:gamma6U}
\end{equation}
where $\bar \varv$ is the mean relative speed, $N_H$ the number density of hydrogen atom perturbers, and $C_{6,\rm low}$ and $C_{6,\rm upp}$ are the relevant constants for the interactions between the ground state hydrogen atoms and the atom of interest in the lower and upper states of the transition, respectively. The dispersion coefficient $C_6$ for the interaction between atoms A and B is given by \citep[e.g.,][Sect.~C.I.a]{1939RvMP...11....1M} % Margenau
\begin{equation}
C_6 = \frac{3}{2} \frac{e^2 \hbar^2}{m_e} \sum_{k^\prime\neq k}\sum_{l^\prime\neq l}
\frac{f^A_{kk^\prime} f^B_{ll^\prime}}
{(E_k^A + E_l^B - E^A_{k^\prime} - E^B_{l^\prime})
(E_k^A - E^A_{k^\prime} ) (E_l^B - E^B_{l^\prime})},
\label{eqn:C6def}
\end{equation}
where $E^X_i$ is the energy of level $i$ in atom X, and $f^X_{ij}$ is the oscillator strength for the transition between levels $i$ and $j$, and the remaining symbols have their usual meaning.

If one of the atoms is a ground state hydrogen atom, then the energy level separations are at least 10.2~eV, typically much larger than those in the other atom (especially for neutrals), and thus the sums can be separated and the hydrogen atom part written in terms of the static dipole polarisability of hydrogen in its ground state $\alpha_H$ (see \citealt{1981SAOSR.390.....K}, p. 76):
\begin{equation}
   C_{6,k} = \frac{e^2 \alpha_H}{h} \cdot \frac{3 \hbar^2}{2 m_e}\sum_{k'} \frac{f_{kk'}}{E_k-E_{k'}}.
   \label{eq:C6K}
\end{equation}
This is the expression used by Kurucz and the sums are computed over all possible transitions (of the electrons in outer shells), using calculated $gf$-values calibrated by fits to observed energy levels. These sums are in general not complete, but they ``should be complete or nearly so for most strong lines''\footnote{see \url{http://kurucz.harvard.edu/atoms/PROGRAMS/expand3007.for}}. Furthermore, for the calculation of the line width, Eq.~(\ref{eq:gamma6U}) is slightly modified -- the maximum value of $C_{6,\rm upp}$ and $C_{6,\rm low}$ is used instead of the difference.

The Unsöld recipe further approximates the summation over all states of the atom of interest by use of the completeness relation (noting that the oscillator strength is related by definition to the dipole operator), and the equation reduces to \citep{1955QB461.U55......}
\begin{equation}
C_{6,k} = - \frac{e^2 \alpha_H}{h} \langle p^2 \rangle,
\end{equation}
where $\langle p^2 \rangle$ is the mean square radial coordinate of the optical electron on the atom of interest. The value of $\langle p^2 \rangle$ can be estimated from the result in a hydrogenic atom, which gives $\langle p^2 \rangle$ as a simple analytic expression as a function of the effective principle quantum number $n^*$ and the orbital angular momentum quantum number $l$ \citep[see][p.~111]{Bates1949}. Adopting this expression and neglecting the dependence on $l$ for simplicity, one obtains:
\begin{equation}
   C_{6,k} = - \frac{e^2 \alpha_H}{h} \cdot 2.5 a_0^2 \frac{n^{*4}_k}{(Z-N_e+1)^2},
   \label{eq:C6U}
\end{equation}
where $a_0$ is the Bohr radius, $Z$ is the atomic number, and $N_e$ is the total number of electrons for a given species (i.e., $Z-N_e+1$ = 1, 2, \ldots\ for neutral, singly ionised, \ldots\ species).

The effective principal quantum number is defined by the hydrogenic formula for the binding energy of the electron $nl$, $E_{nl} = -(Z-N_e+1)^2/(n^*)^2$ Rydbergs. For a series of (core)-$nl$ levels, the binding energy of the electron is given by $E_{nl} = E_{\rm limit}-E_{i}$, where $E_{\rm limit}$ is the series limit energy for the appropriate core.
For the case of a single excited electron outside a core in the ground state of the next higher ionisation stage, $E_{\rm limit}$ is equal to the ionisation potential $E_{\rm ion}$ of the absorbing species. However, the parent configuration of the state of interest may correspond to an excited state of the core. In that case, the energy of the excited core state $EP$ (relative to the core ground state) must be added to the ionisation potential when calculating $n^*$, i.e., $E_{\rm limit} = E_{\rm ion} + EP$, in order to obtain the correct binding energy for the optical electron. Thus, for level $i$ with energy $E_{i}$, the effective principal quantum number should be calculated according to $n^*_i= (Z-N_e+1) \left( \frac{E_H}{E_{\rm limit}-E_{i}} \right)^{1/2}$, where $E_H$ is the ionisation potential of hydrogen. Correct calculation of the binding energy can be very important in complex atoms such as Ca and Fe \citep[see, e.g.,][]{2000A&A...363.1091B}, but requires detailed information about the electron configurations. As far as we know, this has never been taken into account in the context of the Unsöld recipe, probably in part due to the fact that then the simplicity of the calculation is largely lost, as one needs electron configuration information. In this work we adopt the same principle, and always adopt $E_{\rm limit} = E_{\rm ion}$ for the Unsöld recipe calculations; however, the correct values are adopted when employing the ABO theory, which is now described.

The ABO theory is a modification of the line-broadening theory by \citet{1971ApJ...169..621B}, and is described in \citet{1991MNRAS.253..549A}. The main improvement with respect to theories based on the van der Waals potential, is that interactions at intermediate internuclear distances are more accurately described, i.e., the calculation of the interaction energy does not employ the multipole expansion and thus is more appropriate for distances where the electronic wavefunctions overlap moderately. Intermediate-range interactions make an important contribution to the line-broadening cross sections.
The second-most important improvement is that each $m$ state ($|m| = 0, 1, \ldots l$) is treated, thus removing the averaging over phases found in the Lindholm-Foley theory. ABO theory applies a model of the collision dynamics that includes the $m$-substates explicitly, and also includes the relative rotation of the atoms during the collision. \citeauthor{1991MNRAS.253..549A} showed that neglect of these effects leads to an overestimate of the broadening by of order 30\%.
See \citet{1991MNRAS.253..549A} for more details, and \citet{2016A&ARv..24....9B} for more references and a more detailed review of the differences between Lindholm-Foley and ABO theory.

For \emph{neutral species} ABO theory can be applied to calculate cross-sections in general for transitions between different $l$ states. 
Cross-sections for transitions between $l=0, 1, 2, 3$ $(s, p, d, f)$ states are presented in the publications listed in Table~\ref{tab:ABO}.
These tabulate the line-broadening cross sections $\sigma$ for a specific value of the relative speed, $\varv_0=10^4$~ms$^{-1}$, as a function of the effective principal quantum numbers $n^*$ of the relevant levels. The dependence of $\sigma$ on the collision speed $\varv$ is parametrized by a power-law exponent $\alpha$, which is also tabulated for the different $n^*$ combinations. We note that the ABO theory only applies to transitions that obey the one electron selection rule $\Delta l = \pm 1$. However, in complex atoms one often finds situations where the optical electron (the outermost electron that determines the interaction potential) in the lower and upper states does not obey this rule, and one has transitions where the optical electron is in the same $l$ state throughout the transition, e.g., $p-p$. In such cases, since the upper state dominates the broadening, data can be estimated by taking a corresponding transition with the same upper state, e.g., $s-p$ instead of $p-p$. 

\citet{BPM} % Barklem et al. 2000
determined effective principal quantum numbers, accounting correctly for core excitations, for 4872 lines of 23 neutral species from Li to Ni, and obtained a list of broadening data by interpolation in the tables published earlier (see Table~\ref{tab:ABO}). The list was constructed for all neutral species in the wavelength region between 230 and 1300~nm, for which all of the required information was available in the NIST Atomic Spectra Database at that time, and for which the transition type and the $n^*$ values were within the table limits. 
An additional line selection criterion was $\log gf \ge -5$, as lines with lower $gf$-values were assumed to be unaffected by this type of broadening.

\begin{table}
\renewcommand{\tabcolsep}{1.0mm}
\caption{Publications of line-broadening data from ABO theory and applicable ranges of effective principal quantum numbers for neutral species.}
\label{tab:ABO}
\centering
\begin{tabular}{lll}
\hline\hline
% table header
Transition type & $n^*$ limits & Reference \\
                & lower \ldots upper & \\
\hline
% table data
$s-p$ and $p-s$ & $s$: 1.0 \ldots 3.0 & \citet{1995MNRAS.276..859A} \\
                & $p$: 1.3 \ldots 3.0 &  \\
$p-d$ and $d-p$ & $d$: 2.3 \ldots 4.0 & \citet{1997MNRAS.290..102B} \\
$d-f$ and $f-d$ & $f$: 3.3 \ldots 5.0 & \citet{1998MNRAS.296.1057B} \\
\hline\hline
\\
\end{tabular}
\end{table}

For \emph{ionised species} it is not possible to tabulate the broadening data in a general way, while retaining the same accuracy as for neutrals.
Broadening data were calculated for a few strong lines of \mgii\ and \caii\ by \citet{1998MNRAS.300..863B}, % Barklem & O’Mara (1998), 2+5
of \beii, \srii, and \baii\ by \citet{2000MNRAS.311..535B}, % Barklem & O’Mara (2000), 2+5+5
and these data were included in \citet{BPM}. % Barklem et al. 2000
Since then, over 24\,000 \feii\ lines were calculated by \citet{BA-J}, % Barklem & Aspelund-Johansson (2005), 24188
and over 13\,000 \crii\ lines by P.S.\ Barklem\footnote{\url{http://www.astro.uu.se/~barklem/data.html\#ad}}. % 13165

All of the published broadening data (cross sections $\sigma$ and velocity parameters $\alpha$) for specific lines of neutral and ionised species from the ABO data are made available through the VALD database. Using the speed dependence of $\sigma$ parametrized by $\alpha$ together with a Maxwell distribution, the integration over velocity leads to the following expression for line width:
\begin{equation}
   \gamma = 2\cdot (4/\pi)^{\alpha/2} \ \Gamma(2-\alpha/2) \ \varv_0  \ \sigma(\varv_0)  \ (\bar \varv/\varv_0)^{1-\alpha} \cdot N_H,
   \label{eq:gammaABO}
\end{equation}
where $\Gamma(x)$ is the Gamma function and $\varv_0=10^4$~ms$^{-1}$.

Finally, we note that in all cases, collisions with ground state neutral helium can additionally be taken into account in an approximate way by replacing $N_H$ with $N_H$+$cN_{He}$, where $N_{He}$ is the number density of helium atoms, and $c\approx0.41$, which accounts for differences in the polarisability and for the mass (and thus mean velocity) of He.

%----------------------------------------------------------------------------

\section{Broadening data comparisons}
\label{sect:comparison}

The ABO theory has been extensively tested through comparisons with the solar spectrum \citep[e.g., all papers in Table~\ref{tab:ABO}, as well as][]{Barklem2001} and with more detailed calculations where available \citep[e.g.,][]{Kerkeni2004}. Based on these comparisons the ABO theory is estimated to be accurate to better than 20 to 30\%, though in some cases it may be significantly better \citep[see][]{1995MNRAS.276..859A}. On the other hand, the Unsöld recipe, based on the Lindholm-Foley theory and the van der Waals interaction, is well-known to be typically only reliable to of order a factor of 2. See \citet{2016A&ARv..24....9B} for a more detailed discussion and comparison of the various theories and the evidence for their accuracies. However, the conclusion is that the evidence indicates that the ABO theory is the most reliable of the three methods considered here.

%
% Figure for Fe I
%
\begin{figure}
   \begin{center}
      \resizebox{\hsize}{!}{\includegraphics[trim=50 50 50 50]{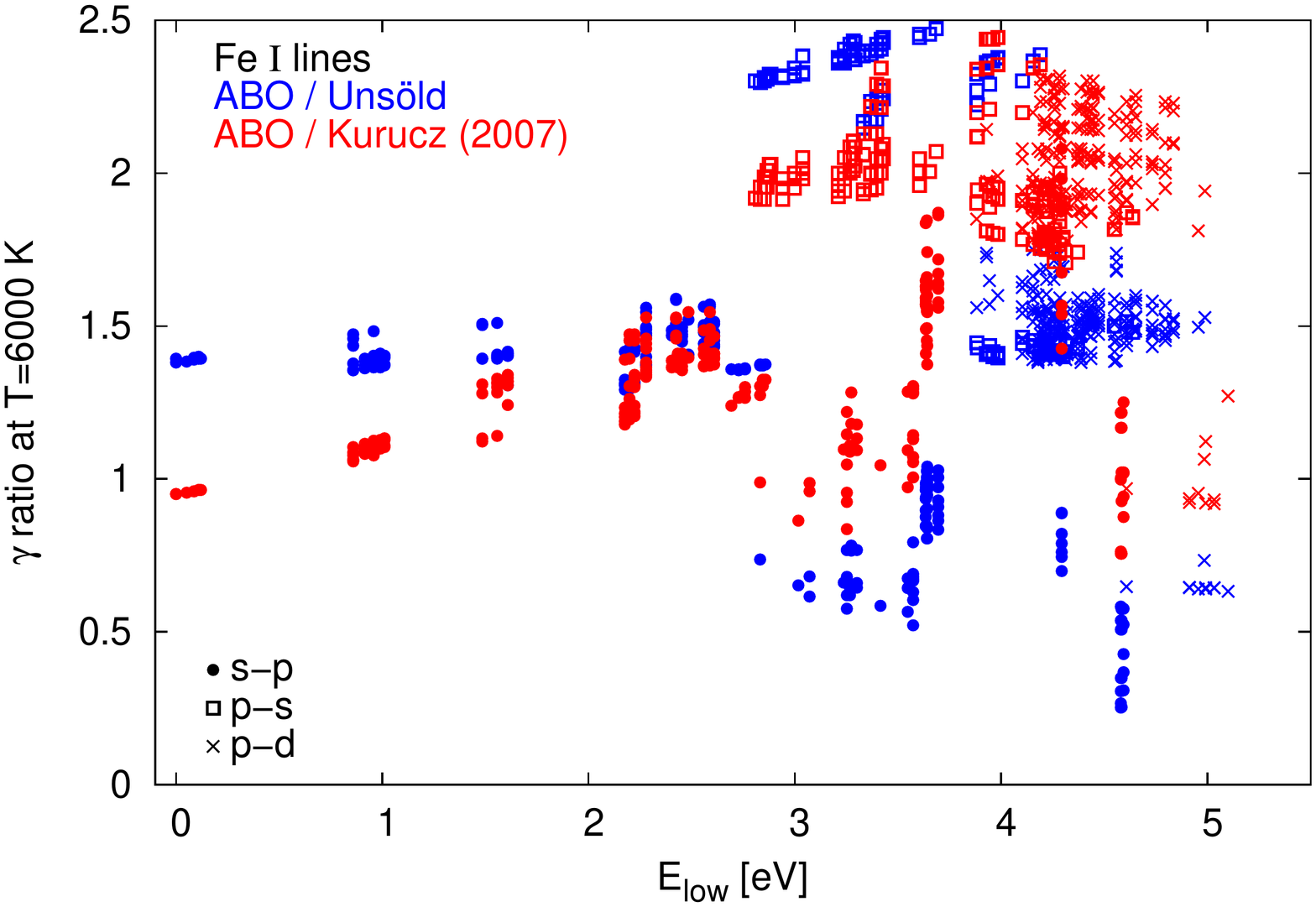}}
   \end{center}
   \caption{Ratio of collisional line width $\gamma$ calculated from ABO theory to line width from van der Waals theory using Kurucz or Unsöld $C_6$ values, for \fei\ lines of transition types $s-p$, $p-s$, and $p-d$, at a temperature of 6000~K, as a function of lower level energy $E_{\rm low}$.}
   \label{fig:gammaFe1}
\end{figure}

In Fig.~\ref{fig:gammaFe1}, we compare line widths computed with the three different options described in Sect.~\ref{sect:theory}, for lines of a representative \emph{neutral species}--\fei, at a representative temperature of 6000~K. Lines for which both Kurucz calculations and ABO data are available were extracted from the VALD database in the wavelength range 480 to 680~nm (on 2 Jul 2012). The figure shows the ratio of the ABO width to the width from the Kurucz line list, and to the Unsöld width (the latter calculated from Eqs.~\ref{eq:C6U} and \ref{eq:gamma6U}).
On the basis that the ABO value represents the most realistic line width, the ratio can be seen as an ``enhancement factor'' to be applied to the Kurucz or Unsöld $\gamma$-values.
The ratio is shown as a function of lower level energy, and different symbols represent different transition types (combinations of $l$ quantum numbers).

A large spread in the ratios is evident for high values of the excitation energy ($E_{\rm low} \gtrsim 2.8$~eV), for both Kurucz and Unsöld values.
For $p-s$ and $p-d$ transitions, the Kurucz ratios cluster around 2, while the Unsöld ratios cluster at two different values (close to 2.5 for $p-s$ and 1.5 for $p-d$), and a few $p-s$ and $p-d$ lines are seen at 1.5 and 0.5, respectively.
For the $s-p$ transitions with $E_{\rm low} \gtrsim 2.8$~eV, all Unsöld ratios are less than 1, implying that a ``reduction factor'' would be required for Unsöld values with respect to ABO theory.
Low ABO values are obtained when the transition takes place involving one or two levels with excited cores. In such a case, the value of $E_{\rm limit}$ in the calculation of the effective principal quantum number $n^*$ is increased with respect to if $E_{\rm ion}$ is used (as typically done in the Unsöld recipe), which implies a smaller $n^*$ value and hence a decreased $\sigma$.
On the other hand, the Kurucz ratios cluster around 1 and 1.5.
Finally, for lines with $E_{\rm low} < 2.8$~eV ($s-p$ transitions), the Kurucz ratios gradually increase from 1 to 1.5, while the Unsöld ratios are constant at around 1.5.

%
% Figures for Fe II
%
\begin{figure}
   \begin{center}
      \resizebox{\hsize}{!}{\includegraphics[trim=50 50 50 50]{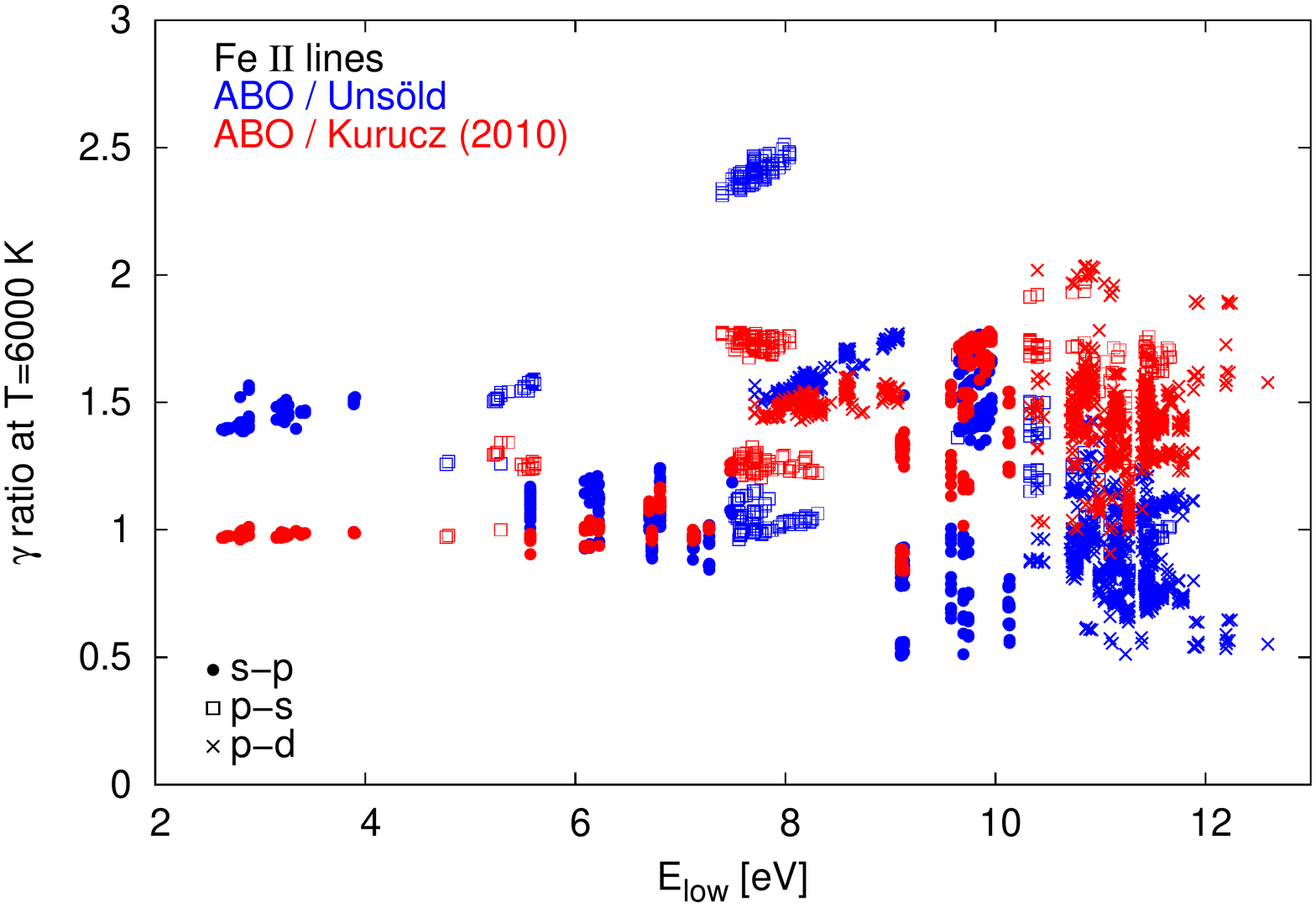}}
      \resizebox{\hsize}{!}{\includegraphics[trim=50 50 50 50]{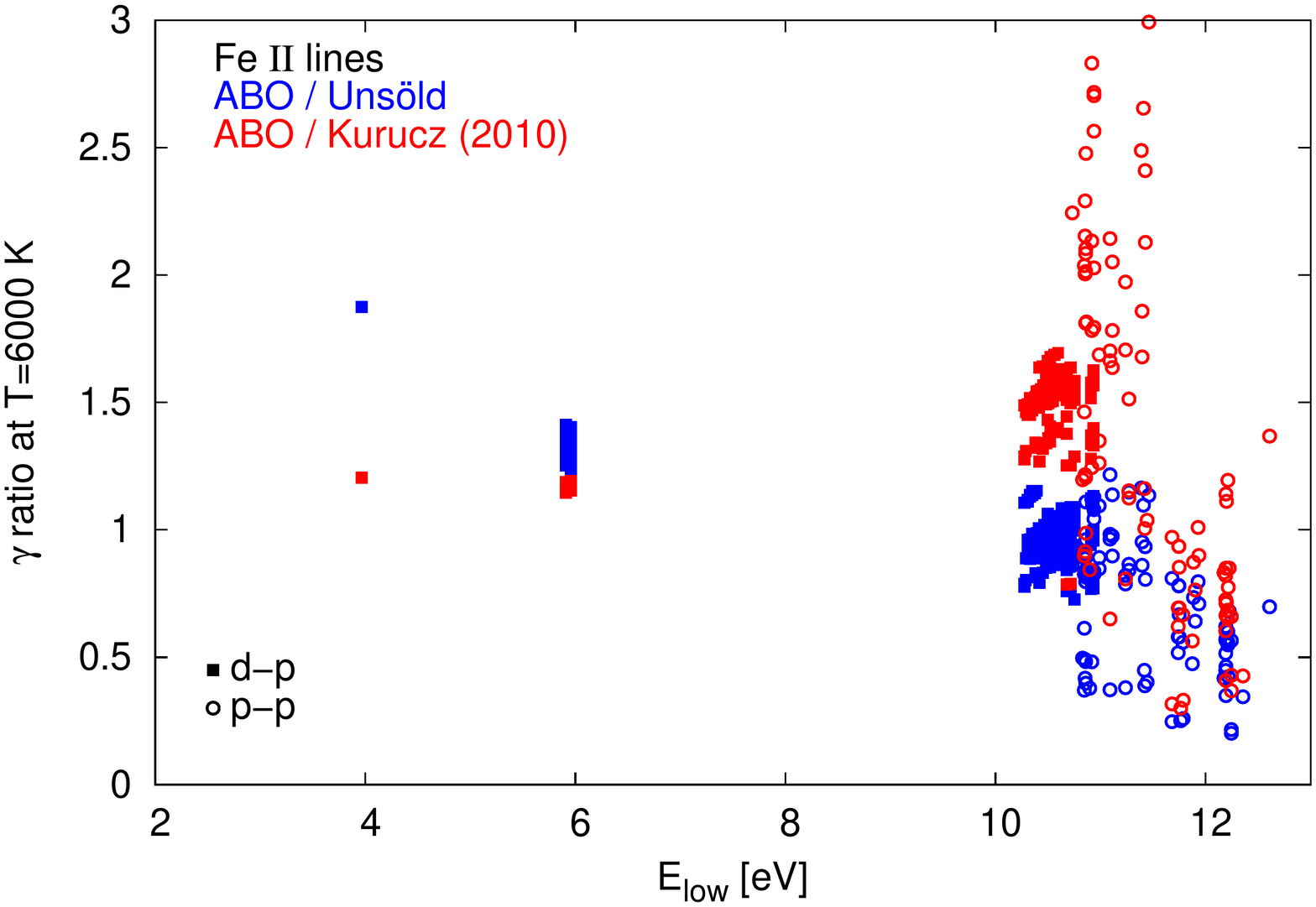}}
   \end{center}
   \caption{Ratio of collisional line width $\gamma$ calculated from ABO theory to line width from van der Waals theory using Kurucz or Unsöld $C_6$ values, for \feii\ lines of transition types $s-p$, $p-s$, $p-d$ (upper panel), $d-p$ and $p-p$ (lower panel), at a temperature of 6000~K, as a function of lower level energy $E_{\rm low}$.}
   \label{fig:gammaFe2}
\end{figure}

Fig.~\ref{fig:gammaFe2} shows a similar comparison for lines of a representative \emph{ionised species}--\feii. We see an equally complex pattern of enhancements of line widths, in particular for Unsöld values at high excitation energies ($E_{\rm low} \gtrsim 7.3$~eV). Kurucz ratios are close to 1 for $E_{\rm low} < 7.3$~eV (all of these are $s-p$, $p-s$, or $d-p$ transitions).

%----------------------------------------------------------------------------

\section{Broadening data for preselected lines}
\label{sect:broad}

\subsection{Neutral species}
\label{sect:data1}

In view of the comparisons presented in Sect.~\ref{sect:comparison}, it seems impossible to define a universal enhancement factor for broadening data calculated with the Unsöld recipe. Thus, it is desirable to use ABO theory data whenever possible.
ABO data were available from \citet{BPM} % Barklem et al. 2000
for about 70\% of the preselected lines in the Gaia-ESO line list
(see
\refSectPresel\ 
%\citealt{Heiter_etal_2020}
and
\refSectGf\ 
%\citealt{Heiter_etal_2020_S1}
for a description of the preselected line list).
We aimed at increasing the percentage at least for the lines with otherwise high-quality data. New ABO data were determined for most of the \Yes/\Yes\ lines and a few \Yes/\Un, \Un/\Yes, and \Un/\Un\ lines lacking such data.
These were mostly lines of \fei, and a few lines of \ci, \sii, \cai, \sci, \cri, \zni, \zri, and \moi.

For these lines we extracted the atomic data from the NIST Atomic Spectra Database required to calculate the effective principal quantum number $n^*$. In particular, term designations and electron configurations of the levels involved in the transitions were obtained.
We also determined the electron configuration of the parent terms and checked if they deviated from the ground states of the corresponding singly ionised species listed in Table~\ref{tab:IonEnergy}. In that case, the excitation energy of the parent term was determined and added to the ionisation energy in the calculation of the effective principal quantum number. All of these input data are given in Table~\ref{tab:GES-clean-neutrals-1}.

Where possible, we then interpolated in the cross-section tables of the ABO publications listed in Table~\ref{tab:ABO}.
The interpolation was done using the code \texttt{widthcomp} (version 2.0) from the \emph{abo-cross} package \citep{barklem_abocross_2015}.
For a few lines, the $n^*$ values (given in Table~\ref{tab:GES-clean-neutrals-2}) fall outside the range of the relevant cross-section interpolation table. In these cases, the cross sections were calculated individually for each line. % by PSB
For example, this occurs for eight of the \fei\ lines, which are $d-p$ transitions.
In all cases both states have a parent configuration corresponding to an excited state of \feii\ (see Table~\ref{tab:GES-clean-neutrals-1}), resulting in $n^*_d\approx1.8$, and $n^*_p \lesssim 1.9$ (see Table~\ref{tab:GES-clean-neutrals-2}). The cross sections calculated directly for these lines are about 300--400 atomic units.
The results for all lines are given in Table~\ref{tab:GES-clean-neutrals-2}. In total, we determined new ABO values for 41 lines: 28 Fe lines ($s-p$, $p-s$, $p-d$, $d-p$), 2 C lines ($p-p$, $p-d$), 3/2/1 $s-p$ lines of Si/Sc/Cr, and 5 lines for elements with $Z>28$ ($s-p$ and $p-s$ lines of Zn, Zr, and Mo).
For the two Ca lines it was not possible to determine reliable ABO values.

\begin{table}
\caption{Ionization energies $E_{\rm ion}$ of neutral species listed in Tables~\ref{tab:GES-clean-neutrals-1} and \ref{tab:GES-clean-neutrals-2} and ground states of corresponding singly ionised species \citep{IonEnergy,NISTASD}.}
\label{tab:IonEnergy}
\centering
\begin{tabular}{rlrrr}
\hline\hline
% table header
$Z$ & Element & $E_{\rm ion}$ [eV] & Configuration & Term \\
\hline
% table data
 6  & C   & 11.2603  & $2s^2 2p$     & $^2P^o$ \\
14  & Si  &  8.1517  & $3s^2 3p$     & $^2P^o$ \\
20  & Ca  &  6.1132  & $3p^6 4s$     & $^2S$ \\
21  & Sc  &  6.5615  & $3p^6 3d4s$   & $^3D$ \\
24  & Cr  &  6.7665  & $3d^5$        & $a^6S$ \\
26  & Fe  &  7.9024  & $3d^6(^5D)4s$ & $a^6D$ \\
30  & Zn  &  9.3942  & $3d^{10}4s$   & $^2S$ \\
40  & Zr  &  6.6339  & $4d^2(^3F)5s$ & $a^4F$ \\
42  & Mo  &  7.0924  & $4d^5$        & $a^6S$ \\
\hline\hline
\end{tabular}
\end{table}

\begin{table*}
   \caption{Atomic data for high-quality lines of neutral species in the preselected Gaia-ESO line list which are \emph{not} included in \citet{BPM}.}
\label{tab:GES-clean-neutrals-1}
\centering
\begin{tabular}{rlrrrrrrrr}
\hline\hline\noalign{\smallskip}
% header
$Z$ & El & Wavelength [\AA] & $T_{\rm low}$ & $T_{\rm upp}$ & Type & $PT_{\rm low}$ & $EP_{\rm low}$ [eV] & $PT_{\rm upp}$ & $EP_{\rm upp}$ [eV] \\
\noalign{\smallskip}\hline\noalign{\smallskip}
% data
 6 & C  & 6587.6100 & $ ^1P  $ & $ ^1P^o$ & $p-d$ & $ ^2P^o$ & 0        & $ ^2P^o$ & 0 \\
 6 & C  & 8727.1260 & $ ^1D  $ & $ ^1S  $ & $p-p$ & $ ^2P^o$ & 0        & $ ^2P^o$ & 0 \\
14 & Si & 5690.4250 & $ ^3P^o$ & $ ^3P  $ & $s-p$ & $ ^2P^o$ & 0        & $ ^2P^o$ & 0 \\
14 & Si & 5701.1040 & $ ^3P^o$ & $ ^3P  $ & $s-p$ & $ ^2P^o$ & 0        & $ ^2P^o$ & 0 \\
14 & Si & 5793.0730 & $ ^3P^o$ & $ ^3D  $ & $s-p$ & $ ^2P^o$ & 0        & $ ^2P^o$ & 0 \\
20 & Ca & 5512.9800 & $ ^1P  $ & $ ^1S  $ & $p-p$ & $ ^2S  $ & 0        & $ ^2S  $ & 0 \\
20 & Ca & 5867.5620 & $ ^1P  $ & $ ^1S  $ & $p-s$ & $ ^2S  $ & 0        & $ ^2S  $ & 0 \\
21 & Sc & 5356.0910 & $ ^2F  $ & $ ^2D^o$ & $s-p$ & $ ^3F  $ & 0.595480 & $ ^3F  $ & 0.595480 \\
21 & Sc & 5484.6260 & $ ^2F  $ & $ ^2F^o$ & $s-p$ & $ ^3F  $ & 0.595480 & $ ^3F  $ & 0.595480 \\
24 & Cr & 5788.3820 & $b^5D  $ & $y^5D^o$ & $s-p$ & $b^4D  $ & 3.103779 & $a^6D  $ or $a^4D$ & 1.95213\tablefootmark{$a$}\\
26 & Fe & 4802.8797 & $b^3D  $ & $w^3P^o$ & $s-p$ & $a^6D  $ & 0        & $a^2P  $ & 2.2764 \\
26 & Fe & 4962.5719 & $y^5F^o$ & $e^3H  $ & $p-d$ & $a^4F$   & 0.2322   & $a^4F$   & 0.2322 \\
26 & Fe & 4985.2529 & $z^3D^o$ & $e^3D  $ & $p-s$ & $a^6D  $ & 0        & $a^4D  $ & 0.9863 \\
26 & Fe & 5060.0780 & $a^5D  $ & $z^7D^o$ & $s-p$ & $a^6D$   & 0        & $a^6D$   & 0      \\
26 & Fe & 5293.9588 & $c^3F  $ & $u^3D^o$ & $d-p$ & $a^4F$   & 0.2322   & $a^2P$   & 2.2764 \\
26 & Fe & 5365.3990 & $a^1H  $ & $z^1G^o$ & $s-p$ & $a^2H  $ & 2.5219   & $a^2G  $ & 1.9645 \\
26 & Fe & 5464.2796 & $c^3F  $ & $y^1D^o$ & $d-p$ & $a^4F  $ & 0.2322   & $a^2P  $ & 2.2764 \\
26 & Fe & 5494.4626 & $c^3F  $ & $x^3H^o$ & $d-p$ & $a^4F$   & 0.2322   & $a^4G$ or $b^2G$ &  3.4602\tablefootmark{$a$} \\
26 & Fe & 5655.1760 & $x^5F^o$ & $g^5G  $ & $p-d$ & $a^6D$   & 0        & $a^6D$   & 0      \\
26 & Fe & 5662.5160 & $y^5F^o$ & $g^5D  $ & $p-s$ & $a^4F  $ & 0.2322   & $a^4D  $ & 0.9863 \\
26 & Fe & 5775.0806 & $y^5F^o$ & $g^5D  $ & $p-s$ & $a^4F  $ & 0.2322   & $a^4D  $ & 0.9863 \\
26 & Fe & 5853.1483 & $a^3F  $ & $z^5P^o$ & $s-p$ & $a^4F$   & 0.2322   & $a^6D$   & 0      \\
26 & Fe & 6027.0509 & $c^3F  $ & $v^3G^o$ & $d-p$ & $a^4F  $ & 0.2322   & $a^2G  $ & 1.9645 \\
26 & Fe & 6120.2460 & $a^5F  $ & $z^7P^o$ & $s-p$ & $a^4F  $ & 0.2322   & $a^6D  $ & 0 \\
26 & Fe & 6127.9060 & $c^3F  $ & $y^3H^o$ & $d-p$ & $a^4F$   & 0.2322   & $a^2G$   & 1.9645 \\
26 & Fe & 6157.7279 & $c^3F  $ & $w^3F^o$ & $d-p$ & $a^4F$   & 0.2322   & $a^2G$   & 1.9645 \\
26 & Fe & 6165.3600 & $c^3F  $ & $v^3G^o$ & $d-p$ & $a^4F  $ & 0.2322   & $a^2G  $ & 1.9645 \\
26 & Fe & 6301.5000 & $z^5P^o$ & $e^5D  $ & $p-s$ & $a^6D$   & 0        & $a^6D$   & 0      \\
26 & Fe & 6315.8110 & $c^3F  $ & $y^1G^o$ & $d-p$ & $a^4F  $ & 0.2322   & $b^2H$   & 3.2447\\
26 & Fe & 6380.7430 & $c^3F  $ & $w^3F^o$ & $d-p$ & $a^4F  $ & 0.2322   & $a^2G  $ & 1.9645 \\
26 & Fe & 6400.3170 & $a^5F  $ & $z^7F^o$ & $s-p$ & $a^4F  $ & 0.2322   & $a^6D  $ & 0 \\
26 & Fe & 6574.2270 & $a^5F  $ & $z^7F^o$ & $s-p$ & $a^4F$   & 0.2322   & $a^6D$   & 0      \\
26 & Fe & 6648.0800 & $a^5F  $ & $z^7F^o$ & $s-p$ & $a^4F  $ & 0.2322   & $a^6D  $ & 0 \\
26 & Fe & 8515.1084 & $b^3G  $ & $z^3G^o$ & $s-p$ & $a^6D  $ & 0        & $a^4F  $ & 0.2322 \\
26 & Fe & 8582.2574 & $b^3G  $ & $z^3G^o$ & $s-p$ & $a^6D  $ & 0        & $a^4F  $ & 0.2322 \\
26 & Fe & 8621.6007 & $b^3G  $ & $z^3G^o$ & $s-p$ & $a^6D  $ & 0        & $a^4F  $ & 0.2322 \\
26 & Fe & 8699.4540 & $x^5D^o$ & $f^5F  $ & $p-d$ & $a^6D  $ & 0        & $a^6D  $ & 0 \\
26 & Fe & 8922.6500 & $x^5F^o$ & $f^5F  $ & $p-d$ & $a^6D  $ & 0        & $a^6D  $ & 0 \\
30 & Zn & 4810.5280 & $ ^3P^o$ & $ ^3S  $ & $p-s$ & $ ^2S  $ & 0        & $ ^2S  $ & 0 \\
40 & Zr & 6127.4750 & $a^3F  $ & $z^3F^o$ & $s-p$ & $a^4F  $ & 0        & $a^4F  $ & 0 \\
40 & Zr & 6134.5850 & $a^3F  $ & $z^3F^o$ & $s-p$ & $a^4F  $ & 0        & $a^4F  $ & 0 \\
40 & Zr & 6140.5350 & $a^3P  $ & $z^3P^o$ & $s-p$ & $a^4F  $ & 0        & $a^4F  $ & 0 \\
42 & Mo & 6030.6440 & $a^5D  $ & $z^5P^o$ & $s-p$ & $a^6D  $ & 1.460950 & $a^6S  $ & 0 \\
\noalign{\smallskip}\hline\hline
\end{tabular}
\tablefoot{
$T_{\rm low}$ and $T_{\rm upp}$ are the terms of the lower and upper levels. Type is the transition type, i.e. the orbital angular momentum quantum numbers of the optical electrons. $PT_{\rm low}$ and $PT_{\rm upp}$ are the parent terms of the lower and upper levels (to be compared with Table~\ref{tab:IonEnergy}), with excitation energies $EP_{\rm low}$ and $EP_{\rm upp}$. Term designations are from \citet{NISTASD}.
Level energies and rotational quantum numbers can be found in the general line-list table.
\tablefoottext{$a$}{average energy of the two possible terms}
}
\end{table*}

\begin{table}
   \caption{Effective principal quantum numbers $n^*$, and ABO theory broadening data, for high-quality lines of neutral species, which are \emph{not} included in \citet{BPM}.}
\label{tab:GES-clean-neutrals-2}
\centering
\begin{tabular}{rlrllrrl}
\hline\hline\noalign{\smallskip}
% header
$Z$ & El & Wavelength & $n^*_{\rm low}$ & $n^*_{\rm upp}$ & $\sigma$ & $\alpha$ & \\
    &    & [\AA]      &                 &                 & [$a_0^2$] &      &       \\
\noalign{\smallskip}\hline\noalign{\smallskip}
% data
 6 & C  & 6587.6100 & 2.23  & 4.02  & 1953 & 0.320 & \\
 6 & C  & 8727.1260 & 1.17  & 1.26  &  154 & 0.266 & $a$ \\
14 & Si & 5690.4250 & 2.05  & 3.61  & 1770 & 0.220 & $b$ \\
14 & Si & 5701.1040 & 2.05  & 3.60  & 1770 & 0.220 & $b$ \\
14 & Si & 5793.0730 & 2.05  & 3.54  & 1700 & 0.230 & $b$ \\
20 & Ca & 5512.9800 & 2.07  & 3.83  &      &       & $c$ \\
20 & Ca & 5867.5620 & 2.07  & 3.57  &      &       & $d$ \\
21 & Sc & 5356.0910 & 1.603 & 2.137 &  412 & 0.271 & \\
21 & Sc & 5484.6260 & 1.601 & 2.113 &  401 & 0.281 & \\
24 & Cr & 5788.3820 & 1.408 & 1.953 &  316 & 0.264 & \\
26 & Fe & 4962.5719 & 1.854 & 3.052 &  905 & 0.278 & \\
26 & Fe & 4802.8797 & 1.794 & 1.862 &  356 & 0.244 & \\
26 & Fe & 4985.2529 & 1.857 & 2.360 &  742 & 0.240 & \\
26 & Fe & 5060.0780 & 1.312 & 1.579 &  205 & 0.252 & \\
26 & Fe & 5293.9588 & 1.846 & 1.918 &  290 & 0.250 & $b$ \\
26 & Fe & 5365.3990 & 1.412 & 1.855 &  283 & 0.261 & \\
26 & Fe & 5464.2796 & 1.846 & 1.900 &  380 & 0.250 & $b$ \\
26 & Fe & 5494.4626 & 1.830 & 1.644 &  330 & 0.290 & $b$ \\
26 & Fe & 5655.1760 & 2.189 & 4.583 & 2010 & 0.250 & $b$ \\
26 & Fe & 5662.5160 & 1.862 & 2.337 &  724 & 0.235 & \\
26 & Fe & 5775.0806 & 1.872 & 2.337 &  720 & 0.231 & \\
26 & Fe & 5853.1483 & 1.430 & 1.778 &  264 & 0.242 & \\
26 & Fe & 6027.0509 & 1.830 & 1.908 &  380 & 0.250 & $b$ \\
26 & Fe & 6120.2460 & 1.376 & 1.661 &  229 & 0.252 & $e$ \\
26 & Fe & 6127.9060 & 1.846 & 1.917 &  290 & 0.250 & $b$ \\
26 & Fe & 6157.7279 & 1.830 & 1.897 &  375 & 0.255 & $b$ \\
26 & Fe & 6165.3600 & 1.846 & 1.914 &  380 & 0.250 & $b$ \\
26 & Fe & 6301.5000 & 1.789 & 2.441 &  832 & 0.243 & \\
26 & Fe & 6315.8110 & 1.830 & 1.682 &  410 & 0.250 & $b$ \\
26 & Fe & 6380.7430 & 1.856 & 1.907 &  380 & 0.250 & $b$ \\
26 & Fe & 6400.3170 & 1.376 & 1.646 &  226 & 0.253 & \\
26 & Fe & 6574.2270 & 1.380 & 1.645 &  227 & 0.254 & \\
26 & Fe & 6648.0800 & 1.385 & 1.650 &  229 & 0.254 & $e$ \\
26 & Fe & 8515.1084 & 1.674 & 1.936 &  356 & 0.242 & \\
26 & Fe & 8582.2574 & 1.669 & 1.925 &  352 & 0.244 & \\
26 & Fe & 8621.6007 & 1.662 & 1.913 &  347 & 0.246 & \\
26 & Fe & 8699.4540 & 2.148 & 2.989 &  817 & 0.272 & \\
26 & Fe & 8922.6500 & 2.173 & 3.022 &  839 & 0.278 & \\
30 & Zn & 4810.5280 & 1.599 & 2.228 &  676 & 0.238 & \\
40 & Zr & 6127.4750 & 1.449 & 1.747 &  260 & 0.244 & \\
40 & Zr & 6134.5850 & 1.432 & 1.717 &  251 & 0.248 & \\
40 & Zr & 6140.5350 & 1.491 & 1.822 &  287 & 0.242 & \\
42 & Mo & 6030.6440 & 1.392 & 1.969 &  322 & 0.262 & \\
\noalign{\smallskip}\hline\hline
\end{tabular}
\tablefoot{
ABO theory broadening data are given as cross section $\sigma$ in atomic units and velocity exponent $\alpha$.
Broadening data were either interpolated in the tables of \citet{1995MNRAS.276..859A} and \citet{1997MNRAS.290..102B} or calculated by PSB.
($a$) $p-p$ transition; $s-p$ table was used; $\log gf=-8$.
($b$) new calculation or estimate by PSB.
($c$) $p-p$ transition; outside range of $s-p$ table.
($d$) $\sigma=2170$~a.u.\ at $\varv=10^4$~ms$^{-1}$, $\alpha$ could not be determined because of poor dependence of $\sigma$ on $\varv$.
($e$) $\log gf<-5$.
}
\end{table}

\subsection{Ionised species}
All lines of \crii\ and \feii\ in the preselected line list have ABO data \citep{BA-J}. % Barklem & Aspelund-Johansson (2005)
Also, all four \baii\ lines
(\refSectBa)
%(Sect.~\sectBa in \citealt{Heiter_etal_2020_S1})
%(Sect.~\ref{sect:Ba})
have ABO data \citep{1998MNRAS.300..863B,2000MNRAS.311..535B}.
For \caii\ 
(\refSectCa),
%(Sect.~\sectCa in \citealt{Heiter_etal_2020_S1}),
%(Sect.~\ref{sect:Ca}),
only the three IR triplet lines have ABO data \citep{1998MNRAS.300..863B}. For the other preselected \caii\ lines, calculations by  \citet{K10} are available. These are $p-d$, $d-f$, and $f-g$ transitions with rather high excitation potentials ($E_{\rm low} > 7$~eV).

No ABO theory calculations are available for the remaining ionised species.
For the two preselected \siii\ lines ($s-p$ transitions with $E_{\rm low}=8.1$~eV), calculations by \citet{K12} are available.
For the lines of \scii\ ($s-p$, $d-p$), \tiii\ ($s-p$, $d-p$), and \yii\ ($d-p$), calculations by \citet{K09,K10,K11} are available, respectively.
The remaining species and transitions have no data.
These are \zrii\ ($s-p$, $d-p$), and the rare-earth elements \laii\ ($s-d$, $d-d$), \ceii\ ($s-p$, $d-p$, $d-d$), \prii\ ($d-p$, $d-d$), \ndii\ ($s-s$, $s-d$, $s-?$, $d-p$, $d-d$, $d-?$), \smii\ ($s-s$, $s-p$), \euii\ ($d-p$), \gdii\ ($s-s$), and \dyii\ ($s-p$). Here ``?'' means that there are some transitions for which the upper level $l$ state is unknown.
All lines of the ionised species without ABO data (except for \siii\ and \caii) have low excitation potentials ($E_{\rm low} \lesssim 3$~eV).

%----------------------------------------------------------------------------

\section{Recommendations and conclusions for collisional broadening data}
\label{sect:broadrec}

For \emph{neutral species} in the preselected line list, we recommend to use the ABO theory cross-sections provided with the Gaia-ESO line list. These include the newly derived values given in Table~\ref{tab:GES-clean-neutrals-2}, and should be used with Eq.~(\ref{eq:gammaABO}) to compute line widths. The number of lines with ABO data per species and for various quality classifications is given 
in Table~\refTabFlagstatsT.
%in Table~\tabFlagstats in \citet{Heiter_etal_2020}.
%in Table~\ref{tab:flagstats}.
%
For future revisions of the line list, the ABO theory interpolation work should be extended to the remaining preselected neutral lines.
These are 103 \fei\ lines
and about 200 lines of other species % 203
(e.g., 42 \sii\ and 21 \tii\ lines,
and about 10 lines each of neutral Na, Mg, S, Cr, Mn, Co, Ni, Sr, Zr, Nb). % 7-15 lines each = 101 lines
% + 39 lines for 11 other species (1-6 each)

For \emph{ionised species}, Fig.~\ref{fig:gammaFe2} indicates that broadening calculations by Kurucz are close to the ABO theory data for low excitation energies. Although we do not know the limit for $E_{\rm low}$ for species other than \feii, we recommend to use Kurucz data for the preselected \scii, \tiii, and \yii\ lines, based on their rather low $E_{\rm low}$ values\footnote{Line broadening widths per hydrogen atom calculated by Kurucz are included in the line list in logarithmic form ($\log\gamma_K$) for a temperature $T_0=10\,000~K$. These need to be scaled to the local temperature $T$: $\gamma/N_H = 10^{\log\gamma_K} (T/T_0)^{0.3}$.}.
For the rare-earth species, we recommend to use the Unsöld recipe (Eqs.~\ref{eq:C6U} and \ref{eq:gamma6U}) with an enhancement factor for the line width of 1.5, again based on Fig.~\ref{fig:gammaFe2}.

For all lines without broadening data the parameter value was set to zero in the line list.
Finally, note that collisional broadening does not significantly affect the equivalent widths $W$ of spectral lines with low ratios of $\log W/\lambda$. As a rough guideline, we can derive a limit of $\log W/\lambda \lesssim -4.8$ from Fig.~2 in \citet{1967MNRAS.136..381W}, % Warner (1967)
i.e., $W\lesssim$75--110~m\AA\ for the UVES-580 wavelength range and $W\lesssim$130--145~m\AA\ for the wavelength range of the HR21 setting.
For such lines, the value of the line broadening constant or the theory used should be less relevant.

% place all figures and tables from this section here
\clearpage

%----------------------------------------------------------------------------
%\end{comment}

\end{appendix}

\end{document}